\documentclass[aps,pra,amssymb,twocolumn,groupedaddress,showpacs,eqsecnum,floatfix]{revtex4}
\usepackage{epsfig}

\def\sech{{\text{sech}}}

\def\psib{\bar\psi}
\def\phib{\bar\phi}
\def\bphib{{\bf\bar\phi}}
\def\bphi{{\bf\phi}}
\def\Bb{\bar B}
\def\bBb{{\bf\bar B}}
\def\bB{{\bf B}}
\def\bu{{\bf u}}
\def\bv{{\bf v}}
\def\w{\hat{\omega}}
\def\dl{\hat{\delta}}
\def\a{\hat{a}_1}

\def\r{{\bf r}}

\def\bk{{\bf k}}

\def\bp{{\bf p}}
\def\ah{\hat{a}}

\def\phdag{{\phantom{\dagger}}}
\def\q{{\bf q}}
\def\p{{\bf p}}
\def\k{{\bf k}}
\def\u{{\bf u}}
\def\v{{\bf v}}
\def\kf{k_F}
\def\as{a}
\def\bh{\hat{b}}
\def\eh{{\hat{\epsilon}}}
\def\kh{{\hat{k}}}
\def\mh{\hat{\mu}}
\def\El{{\hat{E}_\Lambda}}
\def\uh{{\hat{\bf u}}}
\def\vh{{\hat{\bf v}}}
\def\uhh{{\hat{ u}}}
\def\vhh{{\hat{ v}}}
\def\lambdah{\hat{\lambda}}
\def\8{\infty}
\def\oh{\frac{1}{2}}

\def\d{\partial}
\def\ve{\varepsilon}

\def\undertext#1{\vtop{\hbox{#1}\kern 1pt \hrule}}

\def\VEV#1{\langle\,#1\,\rangle}
\def\tr{\hbox{Tr}\,}

\def\pp#1{\frac{\partial}{\partial#1}}
\def\pbyp#1#2{\frac{\partial#1}{\partial#2}}

\def\be{\begin{equation}}
\def\ee{\end{equation}}
\def\bea{\begin{eqnarray}}
\def\eea{\end{eqnarray}}

\def\rf#1{(\ref{#1})}

\def\rf#1{(\ref{#1})}
\def\t{\tilde}

\def\rfs#1{Eq.~\rf{#1}}
\def\sign{{\rm sign}}
\def\grsq{\nabla^2}
\def\curO{{\mathcal O}}
\def\p{{\bf p}}

\newcommand{\muh}{\hat{\mu}}
\newcommand{\Deltah}{\hat{\Delta}}
\newcommand{\deltah}{\hat{\omega}_0}
\newcommand{\ef}{\epsilon_{\rm F}}
\newcommand{\width}{\Gamma_0}

\newcommand{\Deltahbcs}{\Deltah_{\rm BCS}}

\begin{document}


\title{Resonantly-paired fermionic superfluids}


\author{V. Gurarie}
\author{L. Radzihovsky}
\affiliation{Department of Physics, University of Colorado,
Boulder CO 80309}


\date{\today}

\begin{abstract}
  We present a theory of a degenerate atomic Fermi gas, interacting
  through a narrow Feshbach resonance, whose position and therefore
  strength can be tuned experimentally, as demonstrated recently in
  ultracold trapped atomic gases.  The distinguishing feature of the
  theory is that its accuracy is controlled by a dimensionless
  parameter proportional to the ratio of the width of the resonance to
  Fermi energy. The theory is therefore quantitatively accurate for a narrow
  Feshbach resonance. In the case of a narrow $s$-wave resonance, our
  analysis leads to a {\em quantitative} description of the crossover
  between a weakly-paired BCS superconductor of overlapping Cooper
  pairs and a strongly-paired molecular Bose-Einstein condensate of
  diatomic molecules. In the case of pairing via a $p$-wave resonance,
  that we show is always narrow for a sufficiently low density, we
  predict a detuning-temperature phase diagram, that in the course of
  a BCS-BEC crossover can exhibit a host of thermodynamically-distinct
  phases separated by quantum and classical phase transitions. For an
  intermediate strength of the dipolar anisotropy, the system exhibits
  a $p_x+ip_y$ paired superfluidity that undergoes a topological phase
  transition between a weakly-coupled gapless ground state at large
  positive detuning and a strongly-paired fully-gapped molecular
  superfluid for a negative detuning. In two dimensions the former
  state is characterized by a Pfaffian ground state exhibiting
  topological order and non-Abelian vortex excitations familiar from
  fractional quantum Hall systems.

\end{abstract}
\pacs{03.75.Ss, 67.90.+z, 74.20.Rp}

\maketitle \tableofcontents

\section{Introduction}
\label{Introduction}
\subsection{Weakly- and strongly-paired fermionic superfluids}

Paired superfluidity in Fermi systems is a rich subject with a long
history dating back to the discovery of superconductivity (charged
superfluidity) in mercury by Kamerlingh Onnes in 1911. Despite
considerable progress on the phenomenological level and many
experimental realizations in other metals that followed, a detailed
microscopic explanation of superconductivity had to await
seminal breakthrough by Bardeen, Cooper and Schrieffer (BCS) (for
the history of the subject see, for example, Ref.~\cite{Schrieffer}
and references therein). They discovered that in a degenerate,
finite density system, an arbitrarily weak fermion attraction
destabilizes the Fermi sea (primarily in a narrow shell around the
Fermi energy) to a coherent state of strongly overlaping ``Cooper
pairs'' composed of weakly correlated time-reversed pairs of
fermions.

In contrast, superfluidity in systems (e.g., liquid $^4$He), where
constituent fermions (neutrons, protons, electrons) are strongly
bound into a nearly point-like bosonic atom, was readily
qualitatively identified with the strongly interacting liquid limit
of the Bose-Einstein condensation of composite bosonic $^4$He atoms
(for a review, see for example Ref.~\cite{Khalatnikov}).

While such weakly- and strongly-paired fermionic $s$-wave
superfluids were well understood by early 1960's, the relation
between them and a quantitative treatment of the latter remained
unclear until Eagles's~\cite{Eagles1969} and later
Leggett's~\cite{Leggett1980}, and Nozi\`eres and
Schmitt-Rink's~\cite{Nozieres1985} seminal works. Working with the
mean-field BCS model, that is quantitatively valid only for a weak
attraction and high density (a superconducting gap much smaller than
the Fermi energy), they boldly applied the model outside its
quantitative range of validity \cite{Levinsen2006} to fermions with
an arbitrarily strong attraction. Effectively treating the BCS state
as a variational ground state, such approach connected in a concrete
mean-field model the two types of $s$-wave paired superfluids,
explicitly demonstrating that they are two extreme regimes of the
same phenomenon, connected by a smooth (analytic) crossover as the
strength of attractive interaction is varied from weak to strong.
This lack of qualitative distinction between a ``metallic'' (BCS)
and ``molecular'' (BEC) $s$-wave superfluids, both of which are
characterized by a complex scalar (bosonic) order parameter $\Psi$,
was also  anticipated much earlier based on symmetry grounds by the
Ginzburg-Landau theory \cite{Schrieffer}.

Nevertheless, the two types of superfluids regimes exhibit drastically
(quantitatively~\cite{commentBCSBEC}) distinct phenomenologies
\cite{Nozieres1985,deMelo1993}.  While in a weakly-paired BCS
superconductor the transition temperature $T_c$ nearly coincides with
the Cooper-pair binding (dissociation) energy, that is exponentially
small in the pairing potential, in the strongly-paired BEC superfluid
$T_c$ is determined by the density, set by the Fermi temperature, and
is nearly independent of the attractive interaction between fermions.
In such strongly coupled systems the binding energy, setting the
temperature scale $T_*$ above which the composite boson dissociates
into its constituent fermions (e.g., of order eV in $^4$He) can
therefore be orders of magnitude larger than the actually condensation
temperature $T_c\ll T_*$.  This large separation between $T_c$ and
$T_*$ is reminiscent of the phenomenology observed in the
high-temperature superconductors (with the range $T_c < T < T_*$
referred to as the ``pseudo-gap'' regime), rekindling interest in the
BCS-BEC crossover in the mid-90's~\cite{deMelo1993} and more
recently~\cite{Levin2005}.

With a discovery of novel superconducting materials (e.g.,
high-T$_c$'s, heavy fermion compounds), and superfluids ($^3$He),
that are believed to exhibit finite angular momentum pairing, the
nature of strongly- and weakly-paired superfluids has received even
more attention. It was soon appreciated~\cite{VolovikBook,Read2000,VolovikBook1}
that, in contrast to the $s$-wave case, strongly and weakly paired
states at a finite angular momentum are {\em qualitatively}
distinct. This is most vividly illustrated in three dimensions,
where for weak attraction a two-particle bound state is absent, the
pairing is stabilized by a Fermi surface and therefore necessarily
exhibits nodes and gapless excitations in the finite angular
momentum paired state.  In contrast, for strong attraction a
two-particle bound state appears, thereby exhibiting a fully-gapped
superfluidity with concomitant drastically distinct low temperature
thermodynamics. Other, more subtle topological distinctions, akin to
quantum Hall states, between the two types of paired grounds states
also exist and have been investigated~\cite{Read2000,VolovikBook}.
Consequently, these qualitative distinctions require a genuine
quantum phase transition (rather than an analytic crossover, as in
the case of $s$-wave superfluid) to separate the weakly and
strongly-paired states. This transition should be accessible if the
pairing strength were experimentally tunable.

\subsection{Paired superfluidity via a Feshbach resonance}
\label{validity}

The interest in paired superfluidity was recently revived by the
experimental success in producing degenerate (temperature well below
Fermi energy) trapped atomic Fermi gases of $^6$Li and $^{40}$K
~\cite{DeMarco1999,Hulet2003,Jin2004,Ketterle2004}. A remarkable new
experimental ingredient is that the atomic two-body interactions in
these systems can be tuned by an external magnetic field to be
dominated by the so-called Feshbach resonant
(FR)~\cite{Feshbach1959,StoofFeshbach} scattering through an intermediate
molecular (virtual or real bound) state.

\begin{figure}[bt]
\includegraphics[height=3in]{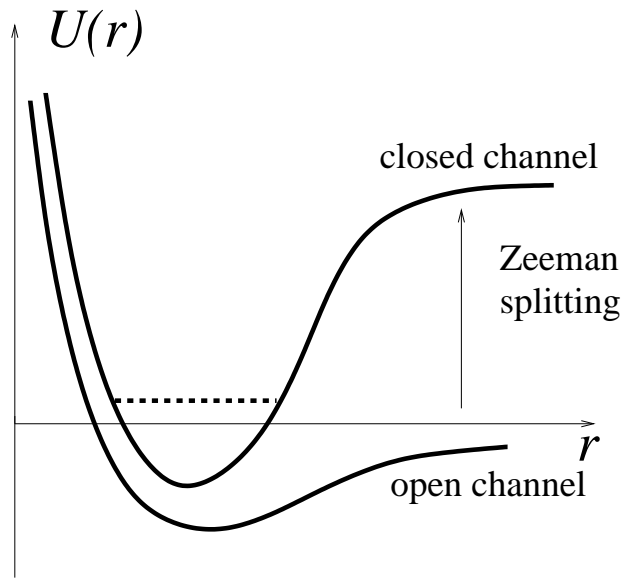}
\caption{Interactions between the atoms
  generically depends on their mutual spin state. This figure depicts
  two potentials corresponding to two spin states of the pairs of
  atoms.  One of them (usually referred to as an ``open channel") is
  too weak to support a bound state, while the other (a ``closed
  channel") supports a bound state, but is energetically unfavorable
  at large distances. The closed channel potential can be moved
  vertically with respect to the open channel potential via the Zeeman
  effect, by changing an external magnetic field.}
\label{Fig-Feshbach}
\end{figure}

As depicted in Fig.~\ref{Fig-Feshbach}, such tunable Feshbach
resonance~\cite{Timmermans1999} arises in a system where
the interaction potential between two atoms depends on their total
electron spin state, admitting a bound state in one spin channel
(usually referred to as the ``closed channel'', typically an
approximate electron spin-singlet). The interaction in the
second~\cite{simpleFR} ``open'' channel (usually electron spin
triplet, that is too shallow to admit a bound state) is then
dominated by a scattering through this closed channel
resonance~\cite{commentHF}. Since the two channels, coupled by the
hyperfine interaction, generically have distinct magnetic moments,
their relative energies (the position of the Feshbach resonance) and
therefore the open-channel atomic interaction can be tuned via an
external magnetic field through the Zeeman splitting, as depicted in
Fig.~\ref{Fig-Feshbach}.

\begin{figure}[bt]
\includegraphics[height=3in]{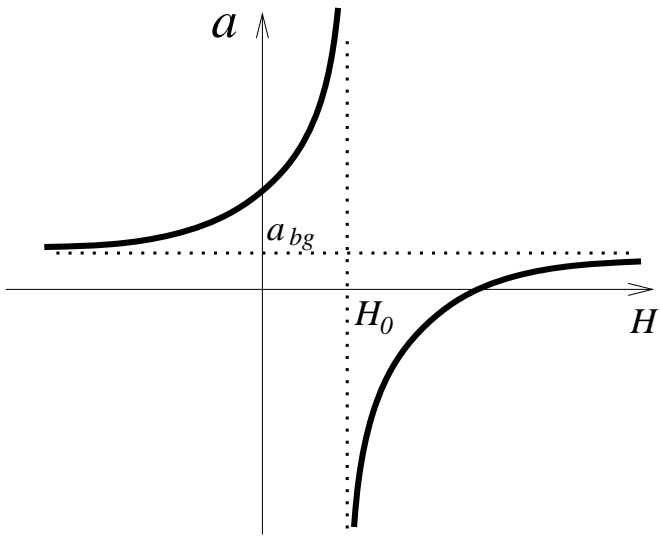}
\caption{A schematic of a typical, experimentally observed behavior
  of an $s$-wave scattering length $a(H)$ as a function of magnetic
  field $H$ in a vicinity of a Feshbach resonance.}
\label{fig:as}
\end{figure}

In the dilute, two-body limit the low-energy $s$-wave Feshbach
resonant scattering is characterized by an $s$-wave scattering
length, that, as illustrated in Fig.\ref{fig:as}, is observed to
behave according to~\cite{Ketterle1998,Moerdijk1995}
\begin{equation}
a(H) = a_{bg} \left( 1- \frac{H_w}{H-H_0}\right), \label{as}
\end{equation}
diverging as the magnetic field passes through a (system-dependent)
field $H_0$, corresponding to a tuning of the resonance through zero
energy. (Analogously, a $p$-wave resonance is characterized by a
scattering volume $v(H)$, as discussed in detail in
Sec.~\ref{sec:pwavescat}). In above, the experimentally measurable
parameters $a_{bg}$ and $H_w$ are, respectively, the background (far
off-resonance) scattering length and the so-called (somewhat
colloquially; see ~\cite{commentExpWidth}) magnetic ``resonance
width''.

An $s$-wave Feshbach resonance is also characterized by an
additional length scale, the so-called effective range $r_0$, and a
corresponding energy scale
\begin{equation}
\Gamma_0=\frac{4 \hbar^2}{m r_0^2},
\end{equation} that only
weakly depend on $H$.  This important scale measures the intrinsic
energy width of the two-body resonance and is related to the
measured magnetic-field width $H_w$ via
\begin{equation}
\Gamma_0\approx 4m\mu_B^2 a_{bg}^2 H_w^2/\hbar^2, \label{Gamma0Bw}
\end{equation}
with $\mu_B$ the Bohr magneton. $\Gamma_0$ sets an energy crossover
scale between two regimes of (low- and intermediate-energy) behavior
of two-atom $s$-wave scattering amplitude.

A key observation is that, independent of the nature of the
complicated atomic interaction leading to a Feshbach resonance, its
resonant physics outside of the short microscopic (molecular size of
the closed-channel) scale can be correctly captured by a
pseudo-potential with an identical low-energy two-body scattering
amplitude, that, for example, can be modeled by a far simpler
potential exhibiting a minimum separated by a large barrier, as
illustrated in Fig.~\ref{Fig-pot}. The large barrier suppresses the
decay rate of the molecular quasi-bound state inside the well,
guaranteeing its long lifetime even when its energy is tuned above
the bottom of the continuum of states.

Although such potential scattering, Fig.~\ref{Fig-pot} is
microscopically quite distinct from the Feshbach resonance,
Fig.~\ref{Fig-Feshbach}, this distinction only appears at high
energies. As we will see, the low energy physics of a shallow
resonance is controlled by a nearly universal scattering amplitude,
that depends only weakly on the microscopic origin of the resonance.
Loosely speaking, for a large barrier of a potential scattering
depicted on Fig.~\ref{Fig-pot} one can associate (quasi-) bound
state inside the well with the closed molecular channel, the outside
scattering states with the open channel, and the barrier height with
the hyperfine interactions-driven hybridization of the open and closed channels of
the Feshbach resonant system.  The appropriate theoretical model was
first suggested in Ref.~\cite{Timmermans1999}, and in turn exhibits
two-body physics identical to that of the famous Fano-Anderson model
\cite{Fano1961} of a single level inside a continuum (see Appendix~\ref{appendixFano}).

A proximity to a Feshbach resonance allows a high tunability
(possible even in ``real'' time) of attractive atomic interactions
in these Feshbach-resonant systems, through a resonant control of
the $s$-wave scattering length $a(H)$, Eq.~(\ref{as}) via a magnetic
field. As we will discuss in Sec.~\ref{PWaveChapter}, a $p$-wave Feshbach
resonance similarly permits studies of $p$-wave interacting systems
with the interaction tunable via a resonant behavior of the
scattering volume $v(H)$.  This thus enables studies of paired
superfluids across the full crossover between the BCS regime of
weakly-paired, strongly overlapping Cooper pairs, and the BEC regime
of tightly bound (closed-channel), weakly interacting diatomic
molecules. More broadly, it allows access to interacting atomic
many-body systems in previously unavailable highly coherent and even
nonequilibrium regimes~\cite{Wieman2001,Levitov2004,Andreev2004},
unimaginable in more traditional solid state systems.

\subsection{Narrow vs wide resonances and model's validity}

An atomic gas at a finite density $n$ (of interest to us here)
provides an additional length, $n^{-1/3}\sim k_F^{-1}$ and
corresponding energy, $\epsilon_F=\hbar^2 k_F^2/2m$ scales. For the
$s$-wave resonance, these scales, when combined with the length $r_0$
or the resonance width $\Gamma_0$, respectively, allow us to define an
$s$-wave dimensionless parameter (with numerical factor chosen for later convenience)
\begin{equation}
\gamma_s=\frac{\sqrt{8}}{\pi} \sqrt{\Gamma_0\over \epsilon_F}=\frac{8}{\pi}{\hbar \over k_F |r_0|},
\end{equation}
that measures the width of the resonance or equivalently the strength
of the Feshbach resonance coupling (hybridization of an atom-pair with
a molecule) relative to Fermi energy. For a $p$-wave (and higher angular momentum) resonance a
similar dimensionless parameter can be defined (see below). The key
resonance-width parameter $\gamma$~\cite{commentExpWidth} naturally
allows a distinction between two types of finite density
Feshbach-resonant behaviors, a narrow ($\gamma\ll 1$) and broad ($\gamma\gg
1$).  Physically, these are distinguished by how the width $\Gamma_0$
compares with a typical atomic kinetic energy $\epsilon_F$.
Equivalently, they are contrasted by whether upon growth near the
resonance, the scattering length $a(H)$ first reaches the effective
range $|r_0|$ (broad resonance) or the atom spacing $\ell$ (narrow
resonance).

Systems exhibiting a {\em narrow resonant} pairing are extremely
attractive from the theoretical point of view.  As was first
emphasized in Ref.~\onlinecite{Andreev2004} and detailed in this
paper, such systems can be accurately modeled by a simple two-channel
Hamiltonian characterized by the small dimensionless parameter
$\gamma$, that remains small (corresponding to long-lived molecules)
throughout the BCS-BEC crossover. Hence, while nontrivial and strongly
interacting, narrow Feshbach resonant systems allow a {\em
  quantitative} analytical description, detailed below, that can be
made arbitrarily accurate (exact in the zero resonance width limit),
with corrections controlled by powers of the small dimensionless
parameter $\gamma$, computable through a systematic perturbation
theory in $\gamma$. The ability to treat narrowly resonant systems
perturbatively physically stems from the fact that such an
interaction, although arbitrarily strong at a particular energy, is
confined only to a narrow energy window around a resonant energy.

\begin{figure}[bt]
\includegraphics[height=4in]{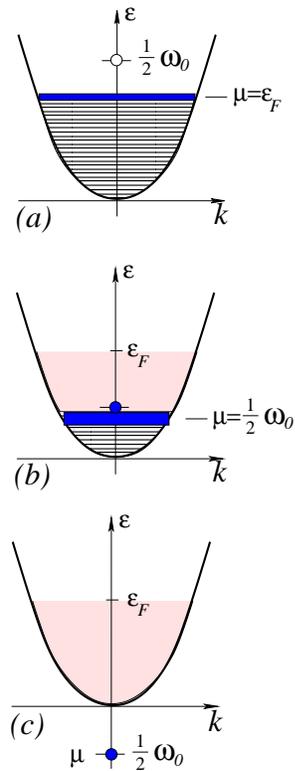}
\caption{\label{gBCSBEC} An illustration of the BCS-BEC crossover in
  the limit of a narrow Feshbach resonance width, $\gamma_s\ll 1$. The
  evolution with detuning $\omega_0$ is illustrated, with (a) the BCS
  regime of $\omega_0 > 2\epsilon_F$, where particles are
  predominantly open-channel atoms forming a Cooper-paired Fermi sea,
  (b) the crossover regime of $0 < \omega_0 < 2\epsilon_F$, where a
  fraction of atoms between $\omega_0$ and $\epsilon_F$ have converted
  into a BEC of bosonic (closed-channel) molecules, with the rest
  forming a Cooper-paired Fermi sea at a chemical potential $\mu$, and
  (c) the BEC regime of $\omega_0 < 0$, where (to order $\gamma_s\ll
  1$) only Bose-condensed molecules are present.}
\end{figure}

\begin{figure}[bt]
\includegraphics[height=3in]{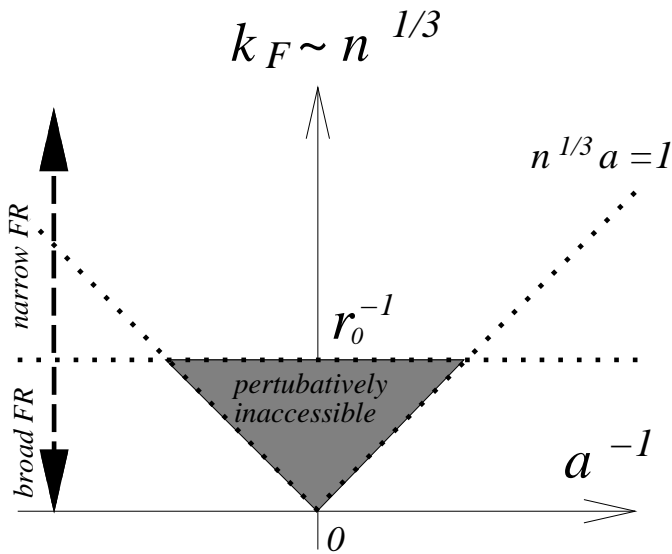}
\caption{An illustration of perturbatively accessible and inaccessible
  (grey) regions in the inverse particle spacing vs inverse scattering
  length, $n^{1/3}$--$a^{-1}$ plane around a Feshbach resonace, where
  $a$ diverges. Note that outside the grey region, even for a broad
  Feshbach resonance there is a small parameter that is either the gas
  parameter or Feshbach resonance coupling, or both, and hence the
  system can be analyzed perturbative.} \label{fig:perturbative}.
\end{figure}

\begin{figure}[bt]
\includegraphics[height=1.5in]{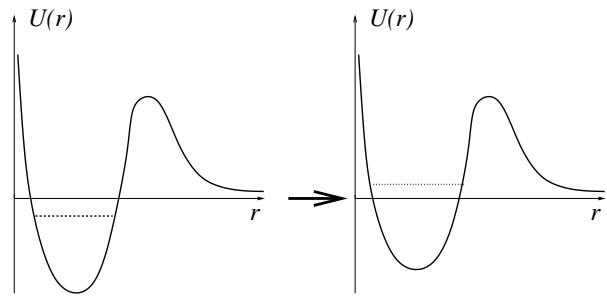}
\caption{\label{Fig-pot} A potential with a low-energy bound state
  whose energy is shown by a dashed line. If the potential is modified
  to make it more shallow, the bound state disappears altogether,
  replaced by a virtual bound state. If the potential is made even
  more shallow, a resonance - a state with positive energy and finite
  lifetime - appears.}
\end{figure}

As we will show in this paper~\cite{Andreev2004}, such narrow resonant
systems exhibit a following simple picture of a pairing superfluid
across the BCS-BEC crossover, illustrated in Fig.~\ref{gBCSBEC}.  For
a Feshbach resonance tuned to a positive (detuning) energy the
closed-channel state is a resonance~\cite{Resonance}, that generically
leads to a negative scattering length and an effective attraction
between two atoms in the open-channel.  For detuning larger than twice
the Fermi energy, most of the atoms are in the open-channel, forming a
weakly BCS-paired Fermi sea, with exponentially small molecular
density, induced by a weak Feshbach resonant (2-)atom-molecule
coupling (hybridization). The BCS-BEC crossover initiates as the
detuning is lowered below $2\epsilon_F$, where a finite density of
atoms binds into Bose-condensed (at $T=0$) closed-channel
quasi-molecules, stabilized by the Pauli principle. The formed
molecular (closed-channel) superfluid coexists with the
strongly-coupled BCS superfluid of (open-channel) Cooper pairs, that,
while symmetry-identical and hybridized with it by the Feshbach
resonant coupling is physically distinct from it. This is made
particularly vivid in highly anisotropic, one dimensional traps, where
the two distinct (molecular and Cooper-pair) superfluids can actually
decouple due to quantum fluctuations suppressing the Feshbach coupling
at low energies~\cite{SheehyDecouple}. The crossover to BEC superfluid
terminates around zero detuning, where conversion of open-channel
atoms (forming Cooper pairs) into closed-channel molecules is nearly
complete.  In the asymptotic regime of a negative detuning a true
bound state appears in the closed-channel, leading to a positive
scattering length and a two-body repulsion in the open-channel. In
between, as the position of the Feshbach resonance is tuned through
zero energy, the system is taken through (what would at zero density
be) a strong unitary scattering limit, corresponding to a divergent
scattering length, that is nevertheless quantitatively accessible in a
narrow resonance limit, where $\gamma_s \sim 1/(k_F r_0)$ plays the
role of a small parameter.
This contrasts strongly with systems interacting via a {\em
featureless} attractive (e.g., short-range two-body) potential, where
due to a lack of a large potential barrier (see Fig.~\ref{Fig-pot-NS})
no well-defined (long-lived) resonant state exists at positive energy
and a parameter $\gamma_s$ (proportional to the inverse of effective
range $r_0$) is effectively infinite. For such broad-resonance
systems, a gas parameter $n^{1/3} \left| a(H) \right|$ is the only
dimensionless parameter. Although for a dilute gas ($n^{1/3}a_{bg}\ll
1$) a controlled, perturbative analysis (in a gas parameter) of such
systems is possible away from the resonance, where $n^{1/3} \left|
a(H) \right| \ll 1$, a description of the gas (no matter how dilute),
sufficiently close to the resonance, such that $n^{1/3}|a(H)| > 1$ is
quantitatively intractable in broad-resonance
systems~\cite{EpsLargeN}. This important distinction between the
narrow and broad Feshbach resonances and corresponding
perturbatively-(in)accessible regions in the $k_F$~ -~ $a^{-1}$ plane
around a Feshbach resonance are illustrated in
Fig.~\ref{fig:perturbative}.

Nevertheless, because of their deceiving simplicity and experimental
motivation (most current experimental systems are broad), these
broad-resonance systems (exhibiting no long-lived positive energy
resonance~\cite{Resonance}) were a focus of the aforementioned
earlier studies~\cite{Eagles1969,Leggett1980,Nozieres1985,deMelo1993} that
provided a valuable {\em qualitative} elucidation of the BCS-BEC
crossover into the strongly-paired BEC superfluids. However, (recent
refinements, employing enlightening but uncontrolled approximations
notwithstanding~\cite{Timmermans1999,Holland2001,Griffin2002,Levin2005}) these
embellished mean-field descriptions are {\em quantitatively}
untrustworthy outside of the BCS regime, where weak interaction
(relative to the Fermi energy) provides a small parameter justifying a
mean-field treatment \--- and outside of the BEC regime where, although mean-field
techniques break down, a treatment perturbative in $n^{1/3}|a| \ll 1$ is still
possible~\cite{Levinsen2006}. The inability to quantitatively treat the
crossover regime for generic (non-resonant) interactions is not an
uncommon situation in physics, where quantitative analysis of the
intermediate coupling regime requires an exact or numerical
solution~\cite{Bulgac2006}.  By integrating out the virtual molecular
state, systems interacting through a {\em broad} (large $\gamma$)
resonance can be reduced to a nonresonant two-body interaction of
effectively infinite $\gamma$, and are therefore, not surprisingly,
also do not allow a quantitatively accurate perturbative analysis
outside of the BCS weak-coupling regime~\cite{EpsLargeN}.

The study of a fermionic gas interacting via a {\em broad} resonance reveals
the following results.  If $a<0$ (the interactions are attractive but
too weak to support a bound state) and $n^{1/3} |a| \ll 1$, such a
superfluid is the standard BCS superconductor described accurately by
the mean-field BCS theory. If $a>0$ (the interactions are attractive
and strong enough to support a bound state) and $n^{1/3} a \ll 1$, the
fermions pair up to form molecular bosons which then Bose condense.
The resulting molecular Bose condensate can be studied using $n^{1/3}
a$ as a small parameter. In particular, in a very interesting regime
where $a \gg |r_0|$ (even though $a \ll n^{-1/3} $) the scattering
length of the bosons becomes approximately $a_b\approx 0.6a$
\cite{Petrov2005}, and the Bose condensate behaves as a weakly
interacting Bose gas with that scattering length~\cite{LL9}, as shown
in Ref.~\cite{Levinsen2006}. Finally, when $|a| n^{1/3} \gg 1$, the
mean-field theory breaks down, the superfluid is said to be in the
BCS-BEC crossover regime, and its properties so far could for the most
part be only studied numerically, although with some encouraging
recent analytical progress in this direction~\cite{EpsLargeN}. Much
effort is especially concentrated on understanding the $|a| n^{1/3}
\rightarrow \infty$ unitary regime~\cite{Bulgac2006} (so called because the fermion
scattering proceeds in the unitary limit and the behavior of the
superfluid becomes universal, independent of anything but its
density).

In this paper we concentrate solely on resonantly-paired superfluids
with {\em narrow} resonances, amenable to an accurate treatment by
mean-field theory regardless of the scattering length $a$.  The
identification of a small parameter~\cite{Andreev2004,EpsLargeN}, allowing a quantitative
treatment of the BCS-BEC crossover in resonantly-paired superfluids in
itself constitutes a considerable theoretical progress.  In practice
most $s$-wave Feshbach resonances studied up to now correspond to
$\gamma_s\simeq 10$, which is consistent with the general consensus in
the literature that they are wide.  Yet one notable exception is the
very narrow resonance discussed in Ref.~\cite{Hulet2003} where we
estimate $\gamma_s \simeq 0.1$; for a more detailed discussion of
this, see Sec.~\ref{sec:CWE}.

Even more importantly is the observation that the perturbative
parameter $\gamma$ is density- (Fermi energy, $\epsilon_F$) dependent,
scaling as $\gamma_s\sim 1/\sqrt{\epsilon_F}$, $\gamma_p \sim \sqrt{\epsilon_F}$ for an $s$-wave
and $p$-wave Feshbach
resonant pairing respectively. Hence, even resonances that are classified as broad
for currently achievable densities can in principle be made narrow by
working at higher atomic densities.

\subsection{Finite angular momentum resonant pairing: $p$-wave superfluidity}

We also study a $p$-wave paired superfluidity driven by a $p$-wave
Feshbach resonance, where the molecular (closed-channel) level is in
the angular momentum $\ell=1$ state. While in degenerate atomic
gases a $p$-wave superfluidity has not yet been experimentally
demonstrated, the existence of a $p$-wave Feshbach resonance at a
two-body level has been studied in exquisite experiments in $^{40}$K and $^6$Li ~\cite{Ticknor2004,Schunck2005}. Recently,
these have duly attracted considerable theoretical
attention~\cite{Ho2005,Ohashi2005,Botelho2005,Gurarie2005,Yip2005}.

One might worry that at low energies, because of the centrifugal
barrier, $s$-wave scattering will always dominate over a finite
angular momentum pairing. However, this is easily avoided by working
with a single fermion species, in which the Pauli exclusion
principle prevents identical fermionic atoms from scattering via an
$s$-wave channel, with a $p$-wave scattering therefore dominating~\cite{commentSvanish}.
Being the lowest angular momentum channel in a single-species
fermionic gas not forbidden by the Pauli exclusion principle, a
$p$-wave interaction is furthermore special in that, at low energies
it strongly dominates over the higher (than $\ell=1$) angular
momentum channels.

There is a large number of features special to $p$-wave resonant
superfluids that make them extremely interesting, far more so than
their $s$-wave cousins. Firstly, as we will show in
Sec.~\ref{sec:pwavescat} and \ref{sec:pwave2Cfinitedensity}, $p$-wave
(and higher angular-momentum) resonances are naturally narrow, since
at finite density a dimensionless measure of their width scales as
$\gamma_p\equiv\gamma_1\sim\epsilon_F^{1/2}$
($\gamma_\ell\sim\epsilon_F^{\ell -1/2}$ in the $\ell$ angular
momentum channel), that in contrast to the $s$-wave case can be made
arbitrarily narrow by simply working at sufficiently low densities
(small $\epsilon_F$). Consequently, a {\em narrow} $p$-wave
Feshbach-resonant superfluid, that can be described arbitrarily
accurately~\cite{c2largeComment} at sufficiently low densities for any
value of detuning, is, in principle, experimentally realizable.

Secondly, superfluids paired at a {\em finite} angular-momentum are
characterized by richer order parameters (as exemplified by a $p$-wave
paired $^3$He, heavy-fermion compounds, and $d$-wave high-T$_c$
superconductors) corresponding to different projections of a finite
angular momentum and distinct symmetries, and therefore admit sharp
quantum (and classical) phase transitions between qualitatively
distinct $\ell$-wave paired superfluid ground states.  In fact, as we
will show, even purely topological (non-symmetry changing) quantum
phase transitions at a critical value of detuning are
possible~\cite{Volovik2004,Read2000,Gurarie2005,Botelho2005}. This
contrasts qualitatively with a smooth (analytic) BCS-BEC crossover
(barring an ``accidental'' first-order transition), guaranteed by the
aforementioned absence of a {\em qualitative} difference between BCS
and BEC paired superfluidity.

Thirdly, some of the $p$-wave (and higher angular momentum) paired
states are isomorphic to the highly nontrivial fractional quantum
Hall effect ground states (e.g., the Pfaffian Moore-Read state) that
have been demonstrated to display a topological order and excitations
(vortices) that exhibit non-Abelian statistics~\cite{Read2000}.
Since these features are necessary ingredients for topological
quantum computing~\cite{Kitaev2003}, a resonant $p$-wave paired
atomic superfluid is an exciting new candidate~\cite{Gurarie2005} for this approach to
fault-tolerant quantum computation.

Finally, a strong connection to unconventional finite angular
momentum superconductors in solid-state context, most notably the
high-temperature superconductors provides an additional motivation
for our studies.

\subsection{Outline}

This paper, while quite didactic, presents considerable elaboration
and details on our results reported in two recent
Letters~\cite{Andreev2004,Gurarie2005}. The rest of it is organized as
follows. We conclude this Introduction section with a summary of our
main experimentally relevant results. In Sec.~\ref{sec:RRST} we
present general, model-independent features of a low and intermediate
energy $s$-wave and $p$-wave scattering, with and without low energy
resonances present. In Sec.~\ref{sec:MC} we discuss general features
of the microscopic models of scattering, tying various forms of
scattering amplitudes discussed in Sec.~\ref{sec:RRST} to concrete
scattering potentials.  We introduce one- and two-channel models of
$s$-wave and $p$-wave Feshbach
resonances~\cite{Timmermans1999,Holland2001} in Sec.~\ref{OneCM} and
Sec.~\ref{TwoCM}, compute exactly the corresponding two-body
scattering amplitudes measured in experiments, and use them to fix the
parameters of the two corresponding model Hamiltonians. These models
then by construction reproduce exactly the experimentally-measured
two-body physics. In the Sec.~\ref{SWaveChapter} we use the resulting
$s$-wave Hamiltonian to study the $T=0$ narrow resonance BCS-BEC
crossover in an $s$-wave resonantly-paired superfluid, and compute as
a function of detuning the molecular condensate fraction, the atomic
(single-particle) spectrum, the 0th-sound velocity, and the condensate
depletion. In the Sec.~\ref{Sec:CritTemp} contained within the
Sec.~\ref{SWaveChapter}, we extend these results to a finite
temperature.  In Sec.~\ref{PWaveChapter} we use the $p$-wave
two-channel model Hamiltonian to analytically determine the $p$-wave
paired ground state, the spectrum and other properties of the
corresponding atomic gas interacting through an idealized {\em
  isotropic} $p$-wave resonance. We extend this analysis to a
physically realistic {\em anisotropic} $p$-wave resonance, split into
a doublet by dipolar interactions. We demonstrate that such a system
undergoes quantum phase transitions between different types of
$p$-wave superfluids, details of which depend on the magnitude of the
FR dipolar splitting. We work out the ground-state energy and the
resulting phase diagram as a function of detuning and dipolar
splitting. In Sec.~\ref{Sec:TopPhaseTrans} we discuss the topological
phases and phase transitions occurring in the $p$-wave condensate and
review recent suggestions to use them as a tool to observe non-Abelian
statistics of the quasiparticles and build a decoherence-free quantum
computer. In Sec.~\ref{sec:CWE} we discuss the connection between
experimentally measured resonance width $H_w$ and a dimenionless
parameter $\gamma_s$ and compute the value of $\gamma_s$ for a couple
of prominent experimentally realized Feshbach resonances.  Finally, we
conclude in Sec.~\ref{conclusion} with a summary of our results.

Our primarily interest is in a many-body physics of degenerate atomic
gases, rather than in (a possibly interesting) phenomena associated
with the trap. Consequently, throughout the manuscript we will focus
on a homogeneous system, confined to a ``box'', rather than an
inhomogeneous (e.g., harmonic) trapping potential common to realistic
atomic physics experience. An extension of our analysis to a trap are
highly desirable for a more direct, quantitative comparison with
experiments, but is left for a future research.

We recognize that this paper covers quite a lot of material. We spend
considerable amount of time studying various models, not all of which
are subsequently used to understand the actual behavior of
resonantly paired superfluids. This analysis is important, as it
allows us to choose and justify the correct model to properly describe
resonantly interacting Fermi gas under the conditions of interest to
us. Yet, these extended models development and the scattering theory
analysis can be safely omitted at a first reading, with the main
outcome of the analysis being that the ``pure" two-channel model
(without any additional contact interactions) is sufficient for our
purposes.  Thus, we would like to suggest that for basic understanding
of the $s$-wave BCS-BEC crossover one should read Sections~\ref{swave},
\ref{Sec:SWaveTCM}, \ref{sec:inrl}, and
\ref{ztbcsbeccrossover}.

\subsection{Summary of results}
\label{results}

Our results naturally fall into two classes of the $s$-wave and
$p$-wave Feshbach resonant pairing for two and one species of
fermionic atoms, respectively. For the first case of an $s$-wave
resonance many results (see \cite{Levin2005} and references therein)
have appeared in the literature, particularly while this lengthy
manuscript was under preparation. However, as described in the
Introduction, most of these have relied on a mean-field approximation
that is not justified by any small parameter and is therefore not
quantitatively trustworthy in the strong-coupling regime outside of
the weakly-coupled BCS regime. One of our conceptual contribution is
the demonstration that the two-channel model of a narrow resonance is
characterized by a small dimensionless parameter $\gamma$, that
controls the validity of a convergent expansion about an
exactly-solvable mean-field $\gamma=0$ limit. For a small $\gamma$,
the perturbative expansion in $\gamma$ gives results that are
quantitatively trustworthy throughout the BCS-BEC crossover.  For
$s$-wave and $p$-wave resonances these key dimensionless parameters
are respectively given by:
\begin{eqnarray} \label{eq:gamma_s1}
\gamma_s &=& {m^2 g_s^2\over n^{1/3}} \frac{1}{(3 \pi^8)^{1/3}},\\
\label{gamma_p} \gamma_p &=& m^2 g_p^2 n^{1/3} \frac{2^{1/3}}{(3\pi^2)^{2/3}},
\end{eqnarray}
where $n$ is the atomic density, $g_s$ and $g_p$ are the closed-open channels
coupling in $s$-wave and $p$-wave resonances, controlling the width of the resonance and $m$ an atom's
mass. The numerical factors in Eqs.~\rf{eq:gamma_s1}, \rf{gamma_p} are chosen for purely for later convenience.

The many-body study of the corresponding finite density systems is
expressible in terms of physical parameters that are experimentally
determined by the two-body scattering measurements. Hence to define
the model we work out the {\em exact} two-body scattering amplitude
for the $s$-wave~\cite{Timmermans1999,Holland2002} and $p$-wave
two-channel models, demonstrating that they correctly capture the
low-energy resonant phenomenology of the corresponding Feshbach
resonances. We find that the scattering amplitude in the $s$-wave
case is
\begin{eqnarray} \label{eq:swaveintro}
f_s(k)&=& -\frac{1}{- a^{-1} + \oh r_0 k^2 - i k}  \cr
&=& -\frac{1}{\sqrt{m}} \frac{\sqrt{\Gamma_0}}{E - \omega_0 + i
\sqrt{\Gamma_0 E}},
\end{eqnarray}
where $\omega_0 \approx 2 \mu_B \left(H-H_0 \right)$ is the magnetic
field-controlled detuning (in energy units), $E= k^2/m$, and
$\Gamma_0$, introduced in \rfs{Gamma0Bw}, is the width of the
resonance. $a$ and $r_0$, which can be expressed in terms of
$\Gamma_0$ and $\omega_0$, represent standard notations in the
scattering theory~\cite{LL} and are the scattering length and the
effective range. We note that $r_0<0$ which reflects that the
scattering represented by \rfs{eq:swaveintro} is resonant. Our
analysis gives $a$ and $r_0$ in terms of the channel coupling $g_s$
and detuning $\omega_0$
\begin{equation} \label{eq:control}
a=-\frac{m g_s^2}{4\pi \omega_0}, \  r_0=-\frac{8 \pi}{m^2 g_s^2}.
\end{equation}

In the $p$-wave case, the scattering amplitude is found to be
\begin{equation}
f_p(k) =-\frac{k^2}{-v^{-1}+\oh k_0 k^2 -ik^3},
\end{equation}
where $v$ is the magnetic field controlled scattering volume, and
$k_0$ is a parameter with dimensions of inverse length which controls
the width of the resonance appearing at negative scattering volume.
$v$ and $k_0$ can in turn be further expressed in terms of
interchannel coupling $g_p$ and detuning $\omega_0$
\begin{eqnarray}
v&=&-\frac{m g_p^2} {6\pi \left(1+\frac{m^2 g_p^2 \Lambda}{3\pi^2}
\right)
\omega_0 },\\
k_0&=&-\frac{12 \pi}{m^2 g_p^2} \left( 1+ \frac{m^2 g_p^2 \Lambda}{3
\pi^2} \right).
\end{eqnarray}
In contrast to our many-body predictions (that are only quantitatively
accurate in a narrow resonance limit), above two-body results are {\em
  exact} in the low-energy limit, with corrections vanishing as ${\cal
  O}(p/\Lambda)$, where $\Lambda \sim 1/d$ is the ultra-violet cutoff
set by the inverse size $d$ of the closed-channel molecular bound
state. We establish that at the two-body level this model is identical
to the extensively studied Fano-Anderson~\cite{Fano1961} of a
continuum of states interacting through (scattering on) a localized
level~(see Appendix~\ref{appendixFano}). For completeness and to put the two-channel model in
perspective, we also calculate the two-body scattering amplitude for
two other models that are often studied in the literature, one
corresponding to a purely local, $\delta$-function two-body
interaction and another in which both a local and resonant
interactions are included. By computing the exact scattering
amplitudes of these two models we show that the low-energy scattering
of the former corresponds to $r_0\rightarrow 0$ limit of the
two-channel model. More importantly, we demonstrate that including a
local interaction in addition to a resonant one, as so often done in
the literature~\cite{Holland2001,Griffin2002,Levin2005} is
superfluous, as it can be cast into a purely resonant
model~\cite{Timmermans1999} with redefined parameters, that, after all
are experimentally determined.

\begin{figure}[bt]
\includegraphics[height=2.4in]{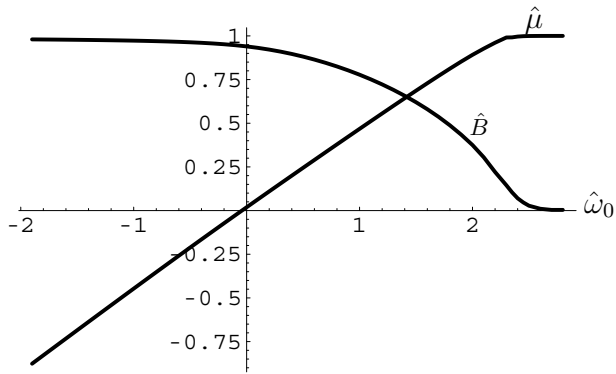}
\caption{\label{Fig:cond-det} Normalized condensate
  order parameter $\hat B=\sqrt{2 n_b/n}$ and normalized chemical
  potential $\hat \mu = \mu/\epsilon_F$ as a function of normalized
  detuning $\hat \omega_0 = \omega_0/\epsilon_F$. }
\end{figure}

For the $s$-wave resonance we predict the zero-temperature molecular
condensate density, $n_b=|B(\omega_0)|^2$. In the BCS regime of
$\omega_0\gg 2 \epsilon_F + \gamma_s \epsilon_F$ we find
\begin{equation}
n_b(\omega_0) \approx \frac{48 n}{e^4 \gamma_s}
\exp \left( - 2  \frac{ \omega_0 - 2
\epsilon_F }{\gamma_s \epsilon_F} \right),
\end{equation}
and in the BEC regime of $\omega_0 \ll - \epsilon_F$
\begin{equation}
n_b(\omega_0) = \frac{n}{2}\left( 1 -\frac{\pi \gamma_s }{4 \sqrt{2}}  \sqrt{\frac{\epsilon_F}{\left|
\omega_0 \right|}} \right),
\end{equation}
where $n$ is the total density of the original fermions. The full
form of $n_b$ is plotted in Fig.~\ref{Fig:cond-det}.

\begin{figure}[bt]
\includegraphics[height=1.7in]{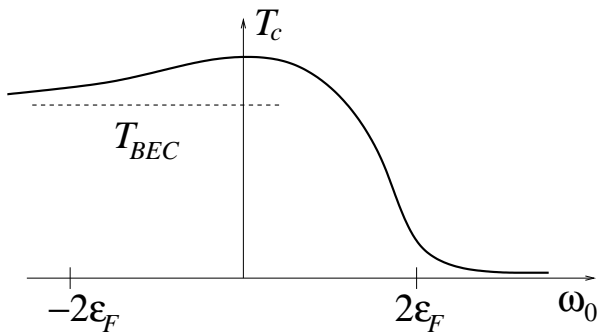}
\caption{\label{Fig-crit} A sketch of the critical temperature
  $T_c(\omega_0)$ as a function of detuning $\omega_0$, displaying a
  maximum at intermediate $\omega_0$. $T_{BEC}$ denotes the
  asymptotics of $T_c$ at large negative $\omega_0$. }
\end{figure}

Following Ref.~\cite{LevitovUnpublished} we also compute the zeroth
sound velocity
and find that it interpolates between the deep BCS value of
\begin{equation}
v_s^{BCS}=\frac{v_F}{\sqrt{3}},
\end{equation}
where $v_F=\sqrt{ 2 \epsilon_F /m }$ is the Fermi velocity, and the
BEC value of
\begin{equation}
v_s^{BEC}=\frac{\gamma_s \epsilon_F^{5/4} \sqrt{\pi}}{2^{5/4}\sqrt{6m}}
\frac{1}{|\omega_0|^{3/2}}.
\end{equation}
The BEC speed of sound quoted here should not be confused with the BEC
speed of sound of the $s$-wave condensate undergoing {\sl wide}
resonance crossover, which was computed in Ref.~\cite{Levinsen2006}.
The crossover in the speed of sound as function of detuning $\omega_0$ should in principle be
observable through Bragg spectroscopy.  Extending our analysis to
finite $T$, we predict the detuning-dependent transition temperature
$T_c(\omega_0)$ to the $s$-wave resonant superfluid. In the BCS regime
\begin{equation}
\label{eq:BCSttintro}
T_c = \frac{8 e^{C-2}}{ \pi}\epsilon_F~
\exp\left( -   \frac{ \omega_0 - 2 \epsilon_F }{\gamma_s
\epsilon_F} \right),\ \ \omega_0\gg 2\epsilon_F,
\end{equation}
where $C$ is the Euler constant, $\ln C \approx 0.577$. In the BEC
regime $T_c(\omega_0)$ quickly approaches the standard BEC transition
temperature for a Bose gas of density $n/2$ and of particle mass $2m$
\begin{equation}
\label{eq:BECttinto} T_c = {\pi \over m} \left( {n \over 2
\zeta\left( {3 \over 2} \right)} \right)^{2/3},
\ \ \omega_0\ll -\epsilon_F.
\end{equation}
Taking into account bosonic fluctuations reviewed for a Bose gas in
Ref.~\cite{Andersen2004}, we also observe that $T_c$ is approached
from above, as $\omega_0$ is decreased. The full curve is plotted in
Fig.~\ref{Fig-crit}. In the broad-resonance limit of
$\gamma_s\rightarrow\infty$ this coincides with earlier predictions of
Refs.~\cite{Leggett1980,Nozieres1985,Holland2002,Griffin2002,Levin2005}.

For a single-species $p$-wave resonance we determine the nature of the
$p$-wave superfluid ground state. Since the $p$-wave resonance is
observed in a system of effectively spinless fermions (all atoms are
in the same hyperfine state), two distinct phases of a condensate are
available: $p_x+ ip_y$ phase which is characterized by the molecular
angular momentum $m=\pm 1$ and a $p_x$ whose molecular angular
momentum is equal to $m=0$.

We show that in the idealized case of {\em isotropic } resonance, the ground
state is always a $p_x+ip_y$ superfluid regardless of whether the
condensate is in BCS or BEC regime. In the BCS limit of large positive
detuning this reproduces the seminal result of Anderson and
Morel~\cite{Anderson1961} for pairing in a spin-polarized (by strong
magnetic field) triplet pairing in $^3$He, the so-called $A_1$ phase.
Deep in the BCS regime we predict that the ratio of the condensation
energy $E_{p_x+ip_y}$ of this $m=1$ state to the $E_{p_x}$ of the competitive $m=0$
$p_z$ state is given by $R=E_{p_x+ip_y}/E_{p_x}=e/2$, exactly.

A much more interesting, new and experimentally relevant are our
predictions for a Feshbach resonance split into a doublet of $m=\pm 1$
and $m=0$ resonances by dipolar anisotropy $\delta$ ~\cite{Ticknor2004}. Our
predictions in this case strongly depend on the strength of the
dipolar splitting, $\delta$ and the resonance detuning, $\omega_0$.
The three regimes of small, intermediate and large value of splitting
(to be defined more precisely below) are summarized respectively by
phase diagrams in Figs.~\ref{Fig-phasediaglow},
\ref{Fig-phasediaginter}, and \ref{Fig-phasediaghigh}.

\begin{figure}[bt]
\includegraphics[height=1.5in]{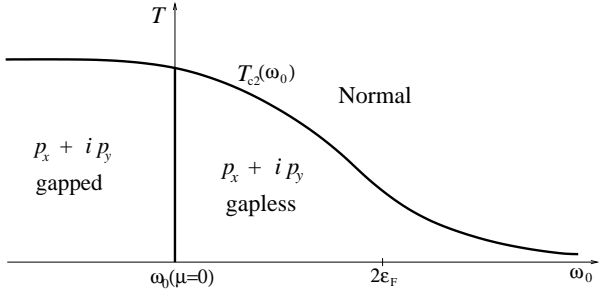}
\caption{\label{Fig-phasediagno} Temperature vs detuning phase diagram
  of a $p$-wave resonant Fermi gas, for the case of no resonance
  splitting, $\delta=0$, i.e., isotropic system.  This phase
  diagram is also expected to describe a resonance with a splitting
  much larger than the Fermi energy for $\omega_0$ tuned to the $m=\pm
  1$ resonance doublet.}
\end{figure}

\begin{figure}[bt]
\includegraphics[height=1.5in]{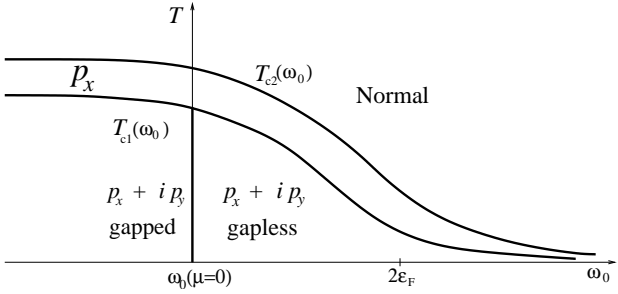}
\caption{\label{Fig-phasediaglow} Temperature vs detuning phase diagram
  of a $p$-wave resonant Fermi gas, for the case of a small resonance
  splitting, $0<\delta < \delta_c^{BEC}$.}
\end{figure}

\begin{figure}[bt]
\includegraphics[height=1.5in]{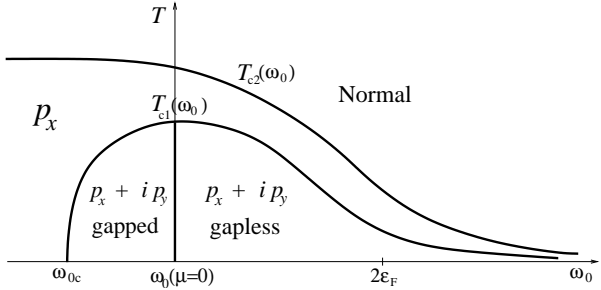}
\caption{\label{Fig-phasediaginter} Temperature vs detuning phase diagram
  of a $p$-wave resonant Fermi gas, for the case of an intermediate
  resonance splitting, $\delta_c^{BEC} < \delta < \delta_c^{BCS}$. The
  critical temperature $T_{c1}(\omega_0)$ vanishes in a universal way
  at the quantum critical point $\omega_{0c}$, according to
  \rfs{Tc1vanish}.}
\end{figure}

\begin{figure}[bt]
\includegraphics[height=1.5in]{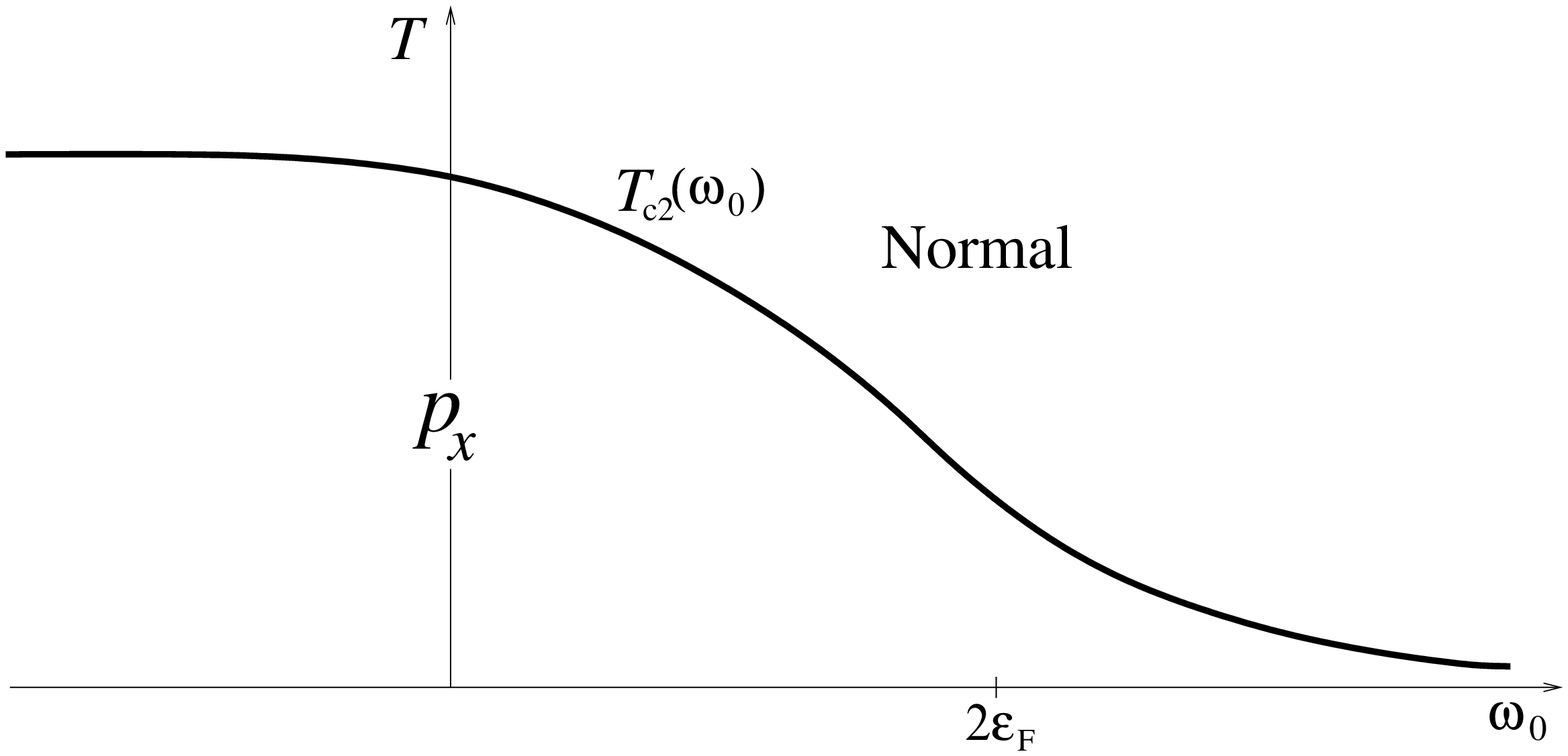}
\caption{\label{Fig-phasediaghigh} Temperature vs detuning phase diagram
  of the $p$-wave resonant Fermi gas for the case of a high resonance
  splitting, $\delta > \delta_c^{BCS}$. This phase diagram is also
  expected to describe a resonance with a splitting much larger than
  the Fermi energy for $\omega_0$ tuned to the $m=0$ resonance.}
\end{figure}

Consistent with above result of vanishing splitting, for weak dipolar
splitting, $0 < \delta <\delta_{c}^{BEC}$ we find that the $p$-wave
$m=1$ superfluid ground state is stable, but slightly deformed to $p_x
+ i\alpha p_y$, with function $\alpha(\delta,\omega_0)$ that we
compute. For an intermediate dipolar splitting,
$\delta_c^{BEC}<\delta<\delta_c^{BCS}$ the ground is a $p_x+i
p_y$-superfluid ($m=1$) in the BCS regime and is a $p_x$-superfluid
($m=0$) in the BEC regime. We therefore predict a quantum phase
transition at $\omega_{0c}$ between these two $p$-wave superfluids for
intermediate range of dipolar splitting~\cite{Gurarie2005,commentYip}.
For a large Feshbach-resonance splitting, $\delta > \delta_c^{BSC}$ the
ground state is a stable $p_x$-superfluid for all detuning. We show
that in all these anisotropic cases the $p_x$-axis of the $p$-wave
condensate order parameter is aligned along the external magnetic
field.  Finally, we expect that for an extremely large dipolar
splitting, much bigger than $\epsilon_F \gamma_p$ (which could quite
well be the current experimental situation), the system can be
independently tuned into $m=0$ and $m=\pm 1$ resonances, and may
therefore display the $p_x+ip_y$ and $p_x$ states separately,
depending on to which of the $m=0$ or $m=\pm1$ resonances the system
is tuned. Thus even in the case of an extremely large dipolar
splitting, phase diagrams in Fig.~\ref{Fig-phasediagno} and in
Fig.~\ref{Fig-phasediaghigh} will be separately observed for tuning
near the $m=1$ and $m=0$ resonances, respectively.

As illustrated in the phase diagrams above, we have also extended
these results to a finite temperature, using a combination of detailed
microscopic calculation of the free energy with more general
Landau-like symmetry arguments. We show quite generally that for a
dipolar-split (anisotropic) resonant gas, the normal to a $p$-wave
superfluid transition at $T_{c2}(\omega_0,\delta)$ is always into a
$p_x$-superfluid, that, for an intermediate dipolar splitting is
followed by a $p_x$-superfluid to $p_x+i p_y$ superfluid transition at
$T_{c1}(\omega_0,\delta)$. The ratio of these critical temperatures is
set by
\begin{equation}
{T_{c2}\over T_{c1}}\sim e^{\delta/a_1},
\end{equation}
where $a_1$, given in \rfs{eq:a1}, is an energy scale that we derive. As seen from the
corresponding phase diagram, Fig.~\ref{Fig-phasediaginter},
we predict that $T_{c1}(\omega_0)$ vanishes in a universal way according to
\begin{equation}
\label{Tc1vanish}
T_{c1}(\omega_0)\sim |\omega_0-\omega_{0c}|^{1/2},
\end{equation}
at a quantum critical point $\omega_{0c}$, that denotes a $T=0$
quantum phase transition between $p_x$ and $p_x+i p_y$ superfluids.

In addition to these conventional quantum and classical phase
transitions, we predict that a $p$-wave resonant superfluid can
exhibit as a function of detuning, $\omega_0$ quite unconventional
(non-Landau type) phase transitions between a weakly-paired (BCS
regime of $\mu > 0$) and a strongly-paired (BEC regime of $\mu < 0$)
versions of the $p_x$ and $p_x+i p_y$
superfluids~\cite{VolovikBook,VolovikBook1,Volovik2004,VolovikReview}.
In three dimensions these are clearly distinguished by a gapless (for
$\mu >0$) and a gapped (for $\mu < 0$) quasiparticle spectra, and
also, in the case of a $p_x +i p_y$ superfluid via a topological
invariant that we explicitly calculate.

While the existence of such transitions at $\mu=0$ have been
previously noted in the literature
~\cite{VolovikBook,VolovikBook1,Volovik2004,VolovikReview} our
analysis demonstrates that these (previously purely theoretical
models) can be straightforwardly realized by a $p$-wave resonant Fermi
gas by varying the Feshbach resonance detuning, $\omega_0$.

Moreover, if the condensate is confined to two dimensions, at a
positive chemical potential this state is a Pfaffian, isomorphic to
the Moore-Read ground state of a fraction quantum Hall ground state believed to
describe the ground state of the plateau at the filling fraction
$\nu=5/2$. This state has been shown to exhibit topological
order\cite{Read2000,VolovikBook1}, guaranteeing a $4$-fold ground
state degeneracy on the torus and vortex excitations  that
exhibit non-Abelian statistics.

As was shown by Read and Green ~\cite{Read2000}, despite the fact that
both weakly- and strongly-paired $p$-wave superfluid states are gapped
in the case of a $p_x + i p_y$- (but not $p_x$-) superfluid the
topological order classification and the associated phase transition
at $\mu=0$ remains. Consistent with the existence of such order, we
also show~\cite{Gurarie2006} (via an explicit construction) that for $\mu >0$, an odd
vorticity vortex in a $p_x + i p_y$-superfluid will generically
exhibit a single zero mode localized on it. In an even vorticity
vortex such zero-energy solutions are absent.

In the presence of far separated vortices, these zero-modes will
persist (up to exponential accuracy), leading to a degenerate
many-particle ground state, and are responsible for the non-Abelian
statistics of associated
vortices~\cite{Read2000,VolovikBook1,VolovikReview,Gurarie2006}. This
new concrete realization of a topological ground state with
non-Abelian excitations, may be important (beyond the basic physics
interest) in light of a recent observation that non-Abelian
excitations can form the building blocks of a ``topological quantum
computer'', free of decoherence~\cite{Kitaev2003}.  We thus propose a
Feshbach resonant Fermi gas, tuned to a $p_x+ip_y$-superfluid ground
state as a potential system to realize a topological quantum
computer~\cite{Gurarie2005,Nayak2006a}.

\section{Resonant Scattering Theory: Phenomenology}
\label{sec:RRST}

A discussion of a two-body scattering physics, that defines our system
in a dilute limit, is a prerequisite to a formulation of a proper
model and a study of its many-body phenomenology. We therefore first
focus on a two-particle quantum mechanics, that, for short-range
interaction is fully characterized by a scattering amplitude
$f(\k,\k')$, where $\pm\k$ and $\pm\k'$ are scattering momenta before
and after the collision, respectively, measured in the center of mass
frame.  In the case of a centrally symmetric interaction potential
$U(r)$, the scattering amplitude $f(k,\theta)$ only depends on the
magnitude of the relative momentum, namely energy
$$E=\frac{k^2}{2m_r}$$ (with $m_r=m_1 m_2/(m_1+m_2)$ the reduced mass)
and the scattering angle $\theta$ (through
$\k\cdot\k'=k^2\cos\theta$), and therefore can be expanded in Legendre
polynomials, $P_\ell(\cos\theta)$
\begin{equation} \label{eq:partialwaves}
f(k,\theta)=\sum_{\ell=0}^\infty
(2\ell+1)f_\ell(k)P_\ell(\cos\theta).
\end{equation}
The scattering amplitude is related to the differential scattering cross
section, the probability density of scattering into a solid angle
$\Omega$, by a standard relation $d\sigma/d\Omega=|f|^2$.  The
$\ell$-th partial-wave scattering amplitude $f_\ell(k)$ measures the
scattering in the angular momentum channel $\ell$, conserved by the
spherically symmetric potential $U(r)$. For later convenience, when we
focus on $s$- and $p$-wave channels, we denote $\ell=0$ and $\ell=1$
quantities with subscripts $s$ and $p$, respectively, as in
\begin{eqnarray}
f_s&\equiv& f_{\ell=0}\\
f_p&\equiv& f_{\ell=1}.
\label{fsp_notation}
\end{eqnarray}
In terms of the scattering matrix $S_\ell=e^{i2\delta_\ell}$ in
channel $\ell$, defined by a phase shift $\delta_\ell$, the scattering
amplitude is given by $f_\ell=(e^{i2\delta_\ell}-1)/(2i k)$.

Analyticity and unitarity of the scattering matrix, $|S_\ell|=1$, then
restrict the scattering amplitude to a generic form
\begin{equation}
\label{eq:scatamp} f_{\ell} (k)={1 \over k^{-2 \ell} F_\ell(k^2)-i
k},
\end{equation}
where $F_{\ell}(k^2)$ is a real function Taylor expandable in powers
of its argument \cite{LL}. It is directly related to the scattering
phase shifts $\delta_\ell(k)$ through the scattering matrix
$S_\ell=e^{i2\delta_\ell}$ via $k^{2 \ell+1} \cot
\delta_\ell(k)=F_\ell(k^2)$. Notice that at small $k$,
\begin{equation} \label{eq:scatamp1} f_{\ell}(k) \sim k^{2
\ell}.
\end{equation}

Important information is contained in the poles $E_{\rm pole}$ of
scattering amplitude (defined by $f_\ell^{-1}(E_{\rm pole})=0$), when
it is studied as a function of complex energy $E$.  Poles in
$f_\ell(E)$ correspond to discrete eigenstates
with different boundary conditions that can be obtained
without explicitly solving the corresponding Schrodinger equation.
However, because $k=\sqrt{2m_r E}$, the scattering amplitude, while a
single-valued function of the momentum is a multi-valued function of
the energy, and one must be careful to specify the branch on which a
pole is located in identifying it with a particular eigenstate of a
Schrodinger equation.  Starting with a branch where $E>0$ and $k>0$,
negative energy $E<0$ can be approached from the positive real axis
either via the upper or lower half complex plane.  A pole which lies
on the negative real axis, approached via the upper half plane is
equivalent to $k=+i\sqrt{2m_r |E|}$, i.e., ${\rm Im}(k) > 0$, and
therefore corresponds to a true bound state of the potential $U(r)$,
with a wavefunction $\psi(r)\sim e^{-|k|r}$ that properly decays at
long distances. On the other hand, a pole on the negative real axis,
approached via the lower half plane is not associated with a bound
state, since it corresponds to $k=-i\sqrt{2m_r |E|}$, i.e., ${\rm
Im}(k) < 0$ and therefore to an unphysical wavefunction that grows at
large distances as $\psi(r)\sim e^{|k|r}$. Although it reflects a real
low-energy feature of a scattering amplitude $f_\ell(E)$, the
so-called virtual bound state \cite{LL} does not correspond to any
physical bound state solution of a Schrodinger equation as it does not
satisfy decaying boundary conditions demanded of a physical bound
state.

On the other hand a pole
\begin{equation}
E_{\rm pole}=E_r - i\Gamma/2,
\label{EpoleDefine}
\end{equation}
of $f_\ell(E)$, with ${\rm Re}~ E_{\rm pole}\equiv E_r>0$, ${\rm
Im}~E_{\rm pole}\equiv -\Gamma/2 < 0$ is a resonance, that corresponds
to a long-lived state with a positive energy $E_r={\rm Re}~E_{\rm
pole}$ and width $\Gamma=-2~ {\rm Im}~E_{\rm pole}$, latter
characterizing the lifetime $\tau=1/\Gamma$ for this state to decay
into a continuum. A complex conjugate pole that always appears along
with this resonance pole, corresponds to an eigenstate that is time
reversal of the resonance solution \cite{LL,Resonance}.

Coming back to the scattering amplitude \rfs{eq:scatamp}, a
low-energy scattering (small $k$) is characterized by a first few
low-order Taylor expansion coefficients of $F_\ell(k^2)$, and
therefore only weakly depends on details of the interaction
potential $U(r)$. This observation is at the heart of our ability to
capture with a simple model Hamiltonian (see
Sec~\ref{TwoCM}, below) the experimentally determined
two-body phenomenology, governed by a complicated atomic interaction
potential $U(r)$ or even multi-channel model as in the case of a
Feshbach resonant systems. To do this we next specialize our
discussion to a particular angular momentum channel.

\subsection{Low energy $s$-wave scattering}
\label{swave}

We first concentrate on $s$-wave ($\ell=0$) scattering that, by virtue
of Eq.~\rf{eq:scatamp1}, is the channel, that, for two fermion species
dominates at low energies.

\subsubsection{Scattering in the asymptotically low energy limit}

Scattering at low energies can be analyzed by expanding the amplitude
$F_s(k^2)$ in powers of its argument, that to lowest order leads to a
simple form
\begin{equation}
\label{eq:scatlength} f_s(k) = - {1 \over a^{-1}+i k},
\end{equation}
with $a=-1/F_s(0)$, where $a$ is called the $s$-wave scattering
length. The latter can be identified with particle effective
interaction (in Born approximation proportional to a Fourier transform
of the potential), with $a>0$ ($a < 0$) generally (but not always)
corresponding to a repulsive (attractive) potential. We observe that
at zero momentum the scattering amplitude is simply equal to the
scattering length, $f(0)=-a$, leading to $\sigma=4\pi a^2$ scattering
cross section.

We can now give a physical interpretation to the only pole  of
\rfs{eq:scatlength} located at
\begin{eqnarray}
\label{eq:pole_p} k_{\rm pole}&=&i a^{-1},\\
\label{eq:pole_E} E_{\rm pole} &=& -{1\over 2 m_r a^2}.
\end{eqnarray}
The key observation at this stage is that by virtue of \rfs{eq:pole_p}
and the fact that to be a physical bound state $\psi\sim e^{i k_{\rm
    pole} r}$ must {\em decay} at large $r$, the pole \rfs{eq:pole_p}
corresponds to the true bound state with energy, $E_{\rm pole}$ {\em
  only if} $a>0$. In contrast, for $a<0$ the scattering amplitude pole
corresponds to a wavefunction that {\em grows} exponential with $r$
and therefore, despite having a negative $E_{\rm pole}$, is {\em not}
a physical bound state or a resonance solution of a Schrodinger
equation, but is what is called a virtual bound state.\cite{LL} Hence,
a physical bound state characterized by a binding energy $1/(2 m_r
a^2)$, that vanishes with $a^{-1}\rightarrow 0$, only exists for a
positive scattering length $a$ and disappears for a negative $a$.

\begin{figure}[bt]
\includegraphics[height=1.5in]{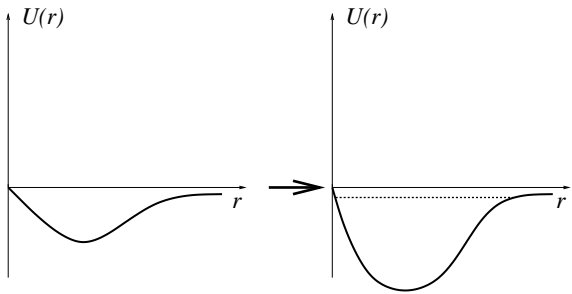}
\caption{\label{Fig-pot-NS} A weakly attractive potential is
progressively made more and more attractive, until a bound state
appears, indicated in the figure by a dashed line. Although such
potential leads to strong resonant scattering, when a bound state is
close to or has just appeared, in contrast to a potential in
Fig.\ref{Fig-pot} it does {\em not} exhibit a resonance in a sense of
a long-lived state with a positive energy and finite
width~\cite{Resonance}.}
\end{figure}

Thus, lacking any other poles at this lowest order of approximation,
the scattering amplitude \rf{eq:scatlength}, while capturing the
asymptotic low-energy bound states of the potential $U(r)$, does not
exhibit any resonances~\cite{Resonance}, i.e., states with a positive energy and a
finite lifetime.  $f_s(k)$ in \rfs{eq:scatlength} corresponds to a
scattering from a relatively featureless potential of the form
illustrated in Fig.~\ref{Fig-pot-NS}, where for a sufficiently deep
well, there is a bound state and $a>0$, but only a continuum of states
with $a<0$ and no resonance for well more shallow than a critical
depth. This is despite the existence of an (unphysical) virtual bound
state for $a < 0$, with a negative energy $E=-1/(2m_r a^2)$ identical
to that of a true bound state (only present for $a > 0$). This point
is, unfortunately often missed in the discussions of
\rfs{eq:scatlength} that have appeared in the literature. As we
discuss in detail below, this scattering phenomenology is captured by
a featureless short-ranged attractive $2$-body interaction
(pseudo-potential) such as the commonly used delta-function four-Fermi
many-body interaction.

We notice that in order to be able to trust Eqs.~\rf{eq:pole_p} and
\rf{eq:pole_E}, all higher order terms in the expansion of $F_s(k^2)$
calculated at this value of energy have to be negligible when compared
with $|k|_{\rm pole}=a^{-1}$. In other words, $a$ has to be
sufficiently large, and $|E|_{\rm pole}$ sufficiently small, with
precise criteria determined by the details of the scattering potential
and the corresponding coefficients of higher order terms in the Taylor
expansion of $F_s(k^2)$.

\subsubsection{Intermediate energy resonant scattering}
\label{swaveMidE}

In order to capture the resonant states (absent in the approximation
\rfs{eq:scatlength}), which could be present in the potential $U(r)$,
$F_s(k^2)$ in $f_s(k)$ must be expanded to the next order in $k^2$,
\begin{equation}
\label{eq:swaveexp} F_s(k^2)=-a^{-1}+\oh r_0 k^2,
\end{equation}
with parameter $r_0$ usually called the effective range of the
interaction potential. For a generic, everywhere attractive potential,
$U(r)<0$, $r_0$ can be shown to be positive \cite{LL}, and moreover,
to roughly coincide with the spatial extent of $U(r)$, hence the name
``effective range". However, as is clear from physical considerations
and an analysis of pole structure of $f_s(k^2)$, a potential which is
attractive everywhere cannot support a resonance. In order to be able
to capture a positive energy quantum particle for a significant amount
of time, the potential must be attractive at short scales and exhibit
a positive energy barrier at intermediate scales, of a generic form
illustrated in Fig.~\ref{Fig-pot}. It can be shown that for such a
potential, $r_0$ is in fact negative, with its magnitude having
nothing to do with the range of $U(r)$. Instead for such resonant
$U(r)$ as shown on Fig.~\ref{Fig-pot}, $|r_0|^{-1}$ reflects the
barrier transmission coefficient, with the higher barrier
corresponding to a longer resonance lifetime and larger $|r_0|$.   Therefore,
focusing on resonant potentials, we will take $r_0<0$, keeping in mind
that $|r_0|$ can be much longer than the actual microscopic range of the
scattering potential, $d\equiv2\pi/\Lambda$. In short, to leave open
the possibility for the scattering to go in the presence of low-energy
resonances, in addition to bound states and virtual bound states,
$r_0$ must be negative and ``anomalously" large, a condition that will
be assumed throughout the rest of this paper.

At this higher level of approximation, the scattering amplitude is
given by
\begin{equation} \label{eq:amplitudeswaver0}
f_s(k) = -{1 \over - \oh r_0 k^2 + a^{-1} + i k},
\label{fsk}
\end{equation}
Equivalently, in terms of energy $E=k^2/2m_r$ (in a slight abuse of
notation) $f_s$ takes the form
\begin{equation}
f_s(E) =
-{1\over\sqrt{2m_r}}{\Gamma_0^{1/2}\over E-\omega_0+i\Gamma_0^{1/2}E^{1/2}},
\label{fsE}
\end{equation}
in which
\begin{equation}
\omega_0\equiv{1\over m_r r_0 a}=\oh\Gamma_0\frac{r_0}{a},
\label{Er-r0}
\end{equation}
and, as discussed in the Introduction, Sec.~\ref{validity}, a
characteristic energy scale
\begin{equation}
\Gamma_0\equiv{2\over m_r r_0^2}
\label{Gamma0-r0}
\end{equation}
is made explicit, with $r_0=-\sqrt{2/m_r\Gamma_0}$. It marks a
crossover energy scale between a low- and intermediate-energy
behaviors of $f_s(E)$. Also, as we will see below, $\Gamma_0$ defines
an energy scale for the low-energy pole above (below) which, $1/m_r
a^2 \gtrsim \Gamma_0$ ($1/m_r a^2 \lesssim \Gamma_0$) a resonant state
appears (disappears).

The poles of the scattering amplitude are given by
\begin{eqnarray}
\label{eq:poleR_p} k_{\rm pole}^{\pm} &=& {i \over r_0} \pm
{\sqrt{2 a r_0-a^2} \over a r_0},\\
\label{eq:poleR_E} E_{\rm pole} &=& {1 \over m_r r_0^2} \left( {r_0
\over a} - 1 + \sqrt{1-2 {r_0 \over a}} \right),\nonumber\\
&=&\omega_0-\frac{1}{2}\Gamma_0\left(1-\sqrt{1-4\omega_0/\Gamma_0}\right),
\end{eqnarray}
where in $E_{\rm pole}$, Eq.\rf{eq:poleR_E} we only kept the ``minus''
pole (with the minus sign in front of the square-root of) $k_{\rm
pole}^-$, as the other pole $k_{\rm pole}^+$ (with a plus sign)
corresponds to an unphysical virtual bound state (regardless of the
sign of the scattering length $a$), and therefore will be ignored in
all further discussions.

\begin{figure}[bt]
\includegraphics[height=2.5in]{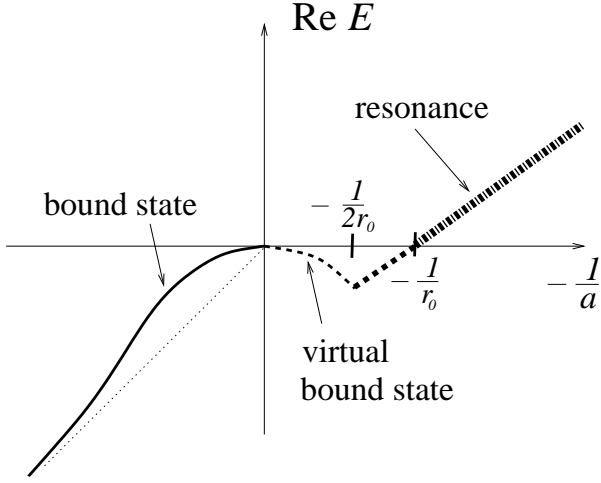}
\caption{\label{Fig-polesEr} The pole of the scattering amplitude
  $f_s(E)$, \rfs{eq:poleR_E} as a function of $-1/a$ for $r_0<0$. As discussed
  in the text, only a bound state and a resonance correspond to
  physical solutions of the Schrodinger's equation with proper
  boundary conditions. The thin dotted  line indicates asymptotic
  linear behavior of the bound state for small positive $a$.}
\end{figure}

\begin{figure}[bt]
\includegraphics[height=3in]{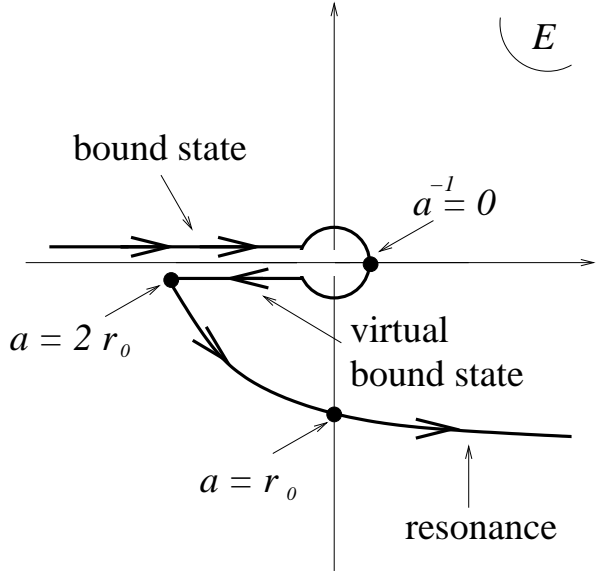}
\caption{\label{Fig-polesE} The pole of the scattering amplitude
$f_s(k)$, \rfs{eq:poleR_E}, shown in a complex plane of $E$. The
arrows indicate pole's motion as $-1/a$ is increased.}
\end{figure}

The real part of the energy $E_{\rm pole}$, \rfs{eq:poleR_E}, as a
function of $-a^{-1}$ (with $r_0<0$) is illustrated in
Fig.~\ref{Fig-polesEr}.  As $-a^{-1}$ is changed from $-\infty$ to
$+\infty$, the pole first represents a bound state, then a virtual
bound state (plotted as dotted curve), and finally a resonance. This
is further illustrated in Fig.~\ref{Fig-polesE}, where the position of
the pole $E_{\rm pole}$ is shown in a complex plane of energy $E$,
with arrows on the figure indicating its motion with increasing
$-a^{-1}$. The bound state and virtual bound state correspond to Im
$E_{\rm pole}=0^\pm$, respectively, with the former (latter)
approaching negative real axis from above (below) the branch cut.  The
resonance, on the other hand, corresponds to Im $E_{\rm pole}<0$ and a
positive real part of the energy, Re $E_{\rm pole} > 0$.

We note that for $1/|a| \ll 1/|r_0|$, the \rfs{eq:poleR_E}, lying
close to zero, approximately coincides with \rfs{eq:pole_E}, as
expected, since the higher order term $\oh r_0 k_{\rm pole}^2$ in
$f_s(k)$ is subdominant to $i k_{\rm pole}$. In other words, at a
sufficiently large scattering length, the scattering is
well-approximated by the asymptotic low-energy (scattering-length)
approximation of the previous section. For such large positive $a\gg
|r_0|$ it gives bound state energy, \rfs{eq:pole_E}, that grows
quadratically with $1/a$. Further away from the resonance, on the
positive $a$ side, where the scattering length drops significantly
below the effective range, $a\ll |r_0|$, the bound state energy
crosses over to a linear dependence on $1/a$, as illustrated in
Fig.\ref{Fig-polesEr} and summarized by
\bea
\label{Ebound_cross}
\hspace{-1.5cm}E_{\rm bound}(a)
&=&
\cases{-\frac{1}{2m_r a^2},
& \text{for $|r_0|\ll a > 0$}\cr
-\frac{1}{m_r |r_0| a},
& \text{for $|r_0|\gg a > 0$}.\cr}
\eea

More importantly, however, on the other side of the resonance, where
the scattering length is negative, unlike \rfs{eq:pole_E},
\rfs{eq:poleR_E} also describes a resonant state, that appears for $a
< 0$ and shorter than the effective range $r_0$, i.e., for $-\frac 1 a
> - \frac 1 {r_0}$, when the real part of the energy $E_{\rm pole}$
becomes positive. The resonant state is characterized by a peak at
energy $E_r=E_{\rm resonance}$ and a width $\Gamma_s$ given by
\begin{eqnarray}
\label{eq:Eres} E_{\rm resonance} &=&{1 \over m_r r_0^2} \left({
{r_0\over a} -1} \right),\\
&=& \omega_0-\oh\Gamma_0,\nonumber\\
\label{eq:Gamma}\Gamma_s&=&{2 \over m_r r_0^2} \sqrt{{2r_0 \over a} -
1}\\
&=&\Gamma_0\sqrt{{4\omega_0\over\Gamma_0}-1}\nonumber.
\end{eqnarray}
Hence, we find that in the $s$-wave resonant case, generically, even
potentials that exhibit a resonance for small $|a|$ (high energy),
lose that resonance and therefore reduce to a nonresonant case for
sufficiently large and negative $a$ (low energy).

The transition from a bound state to a resonance as a function of $a$
is exhibited by scattering via a generic resonant potential
illustrated in Fig.\ref{Fig-pot}. A sufficiently deep well will
exhibit a true bound state, whose energy will vanish with decreasing
depth and correspondingly increasing $a$, according to $E_{\rm
bound}=-1/(2m_r a^2)$, Eq.\rf{eq:pole_E}. We note, however, that as
the potential is made even more shallow, $1/a$ crosses $0$ and the
true bound state disappears (turning into an unphysical virtual bound
state), the resonant state (positive energy and finite lifetime) does
not appear until a {\em later} point at which scattering length
becomes shorter than the effective range, i.e., until $|a| < |r_0|$.

This somewhat counterintuitive observation can be understood by noting
that the lifetime of an $s$-wave resonant state is finite, given by
the inverse probability of tunneling through a finite barrier, which
only weakly depends on the energy of the state as long as the
potential in Fig.~\ref{Fig-pot} is not too long-ranged. Thus, even
when the energy of the resonance goes to zero, its width remains
finite. Hence, since the bound state's width is exactly zero, small
deepening of the potential cannot immediately change a resonance into
a bound state, simply by reasons of continuity. There has to be some
further deepening of the potential $U(r)$ (range of $a$), over which
the resonance has already disappeared, but the bound state has not yet
appeared. During this intermediate range of potential depth
corresponding to $0 < 1/|a| < 1/|r_0|$, when the potential is not deep
enough to support a true bound state but not yet shallow enough to
exhibit a resonance, the scattering is dominated by a virtual bound
state pole, as illustrated in Figs.~\ref{Fig-polesEr},~\ref{Fig-polesE}.

\subsection{$P$-wave scattering}
\label{sec:pwavescat}

As remarked earlier, in a low-energy scattering of a particle off a
potential $U(r)$, the $s$-wave ($\ell=0$) channel dominates over
higher angular momentum $\ell\neq 0$ contributions, that by virtue of
the generic form of the scattering amplitude, \rfs{eq:scatamp} vanish
as $k^{2 \ell}$, \rfs{eq:scatamp1}.  This suppression for $\ell\neq 0$
arises due to a long-ranged centrifugal barrier, that at low energies
prevents a particle from approaching the origin where the short range
scattering potential $U(r)$ resides.

Hence, in the case of a Feshbach resonance of two hyperfine species
Fermi gas, where the scattered particles are distinguishable (by their
hyperfine state), at low energies, indeed, the interaction is
dominated by the $s$-wave resonance, with higher angular momentum
channels safely ignored. However, an exception to this is the
scattering of {\em identical} fermions, corresponding to atoms in the
same hyperfine state in the present context. Because Pauli exclusion
principle forbids fermion scattering in the $s$-wave channel, the next
higher angular momentum channel, namely $p$-wave ($\ell=1$) scattering
dominates, with $s$-wave and $\ell > 1$ channels vanishing at low
energies~\cite{commentSvanish}.  Thus we see that $p$-wave Feshbach resonance is quite
special, being the dominant interaction channel for a single species
Fermi gas. With this motivation for our focus on a $p$-wave Feshbach
resonant superfluidity and in preparation for its study, we next
analyze a $p$-wave scattering amplitude.

Starting with \rfs{eq:scatamp} and expanding $F_p(k^2)$ similarly to
\rfs{eq:swaveexp} we find
\begin{equation}
\label{eq:exppp} F_p(k^2)=- v^{-1}+ \oh k_0  k^2.
\end{equation}
Here $v$ is the so-called scattering volume analogous to the
scattering length $a$ of the $s$-wave case, diverging and changing sign
when the system is taken through a $p$-wave Feshbach resonance. A
characteristic wavevector $k_0$ is everywhere negative and plays a
role similar to that of the effective range $r_0$ in the $s$-wave
channel, but has dimensions of an inverse length.

\begin{figure}[bt]
\includegraphics[height=2.5in]{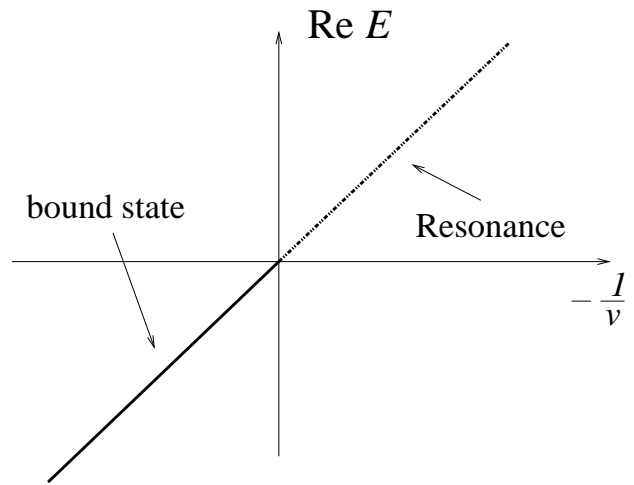}
\caption{\label{Fig:pwavepoles} The pole of a $p$-wave scattering amplitude
\rfs{eq:scatampp} as a function of $-1/v$ for $k_0<0$.}
\end{figure}

Hence at low energies the $p$-wave scattering amplitude takes the form
\begin{equation}
\label{eq:scatampp} f_p(k) = {k^2 \over -v^{-1}  + \oh k_0  k^2 - i
k^3}.
\end{equation}
Although the poles of the scattering amplitude \rfs{eq:scatampp} can
be found by solving a qubic equation, their exact positions are not
very illuminating and will not be pursued here. Instead, it will be
sufficient for our purpose to only consider an important low-energy
limit $|v^{-1}| \ll |k_0|^3$, in which the relevant pole of
\rfs{eq:scatampp} is close to zero and its position can be found by
neglecting (actually treating perturbatively in powers of $|v k_0^3|$)
$i k^3$ term in the scattering amplitude. To lowest order the pole is
then simply given by
\begin{equation}
E_{\rm pole}\approx{1 \over  m_r v k_0 }.
\end{equation}
This corresponds to a real bound state for $E_{\rm pole}<0$ (when
$v>0$) and a resonance for $E_{\rm pole}>0$ (when $v<0$), with a width
easily estimated to be
\begin{eqnarray}
\Gamma_p &\approx& {2 k_{\rm pole}^3\over m_r k_0}=E_{\rm pole}
\sqrt{32/|v k_0^3|},\\
&\ll& E_{\rm pole},
\end{eqnarray}
near a resonance, where $|v k_0^3|\rightarrow\infty$.  Thus, in
contrast to the $s$-wave case, where at sufficiently low energies ($E
< \Gamma_0=2/m_r r_0^2$) the width $\Gamma_s\approx \sqrt{\Gamma_0
E}\gg E$, here, because $\Gamma_p\sim E^{3/2}$, a $p$-wave resonance
becomes arbitrarily narrow at low energies. Consequently, as the
inverse scattering volume $v^{-1}$ is tuned through zero and the
relevant two-body energy $E_{\rm pole}=1/ (2\ m_r v k_0)$ vanishes, the
real bound state immediately turns into a resonance without going
through an intermediate virtual bound state (as it did in the $s$-wave
case). This is illustrated on Fig.~\ref{Fig:pwavepoles}. This resonant
pole behavior extends to all finite angular momentum ($\ell>0$)
channels.

The physical reason behind such a drastic difference between $s$-wave
and $p$-wave (and higher $\ell>0$ channels) resonances stems from the
centrifugal barrier that adds a long-ranged $1/r^2$ tail to the
effective scattering potential $U_{\rm eff}(r)=U(r)+{\hbar^2\ell
(\ell+1) \over 2 m_r r^2}$.  The width of a low lying resonant state
in such potential can be estimated by computing the decay rate through
$U_{\rm eff}(r)$, dominated by the long-ranged centrifugal barrier
$\sim 1/r^2$.  Employing the WKB approximation, at low energy $E$ the
decay rate is well approximated by
\begin{eqnarray}
\Gamma&\sim&e^{-{2\over\hbar}\int_d^{r_E}dr\sqrt{2m_r U_{\rm eff}}},\\
&\approx&e^{-2{\sqrt{\ell(\ell+1)}\int_d^{r_E}dr/r}},\\
&\approx&\left({r_E\over d}\right)^{-2\sqrt{\ell(\ell+1)}},
\label{Gamma_E}
\end{eqnarray}
In above $d$ and $r_E$ are the classical turning points of the
$-U_{\rm eff}(r)$, where $d$ can be taken as the microscopic range of
the potential (closed-channel molecular size), and more importantly
$r_E$ is determined by
\begin{equation}
E={\hbar^2\ell(\ell+1)\over 2m_r r_E^2}.
\label{r_E}
\end{equation}
Combining this with \rfs{Gamma_E} gives
\begin{equation}
\Gamma_\ell\sim E^{\sqrt{\ell(\ell+1)}}.
\end{equation}
Although WKB approximation does not recover the correct exponent of
$\ell + 1/2$, Eq.~\rf{eq:scatamp} (required by unitarity and
analyticity) except for the expected large $\ell$ limit (consistent
with the fact that for small $\ell$ the semiclassical criterion on
which it is based fails), it does correctly predict a narrowing of the
resonance at low energies and with increasing angular momentum $\ell$.

Of course, the expansion \rfs{eq:exppp} is only a good approximation
for small $k$. But in this regime it captures both low energy real
bound state (for $1/v > 0$) and narrow resonant state (for $1/v
<0$). Experimentally this regime is guaranteed to be accessible by
tuning the bound state and resonance energy $E_{\rm pole}=1/( m_r v
k_0)$ sufficiently close to zero so that $|v k_0^3| \ll 1$. In this
range the scattering amplitude \rfs{eq:scatampp} correctly captures
the physics of a resonant scattering potential and the related
Feshbach resonance without the need for higher order terms in the
expansion of $F_p(k^2)$.

\section{Resonant Scattering Theory: Microscopics}
\label{sec:MC}
\subsection{Potential scattering}
\label{Pscattering}
The next step in our program is to develop a model of a gas of
fermions interacting via a resonant pairwise potential $U(r)$ of the
type illustrated in Fig.~\ref{Fig-pot}, that exhibits a real bound
state or a resonance, controlled by tuning its shape (e.g., well
depth). It is of course possible to simply use a many-body theory with
a pairwise interactions literally taken to be $U(r)$ of
Fig.~\ref{Fig-pot}, with a (normal-ordered) Hamiltonian given by
\begin{eqnarray} \label{eq:explicitpotential}
\hat H&=&\int d^3 r~\sum_\sigma  \hat \psi^\dagger_\sigma \left(-{\nabla^2
\over 2 m} \right) \hat \psi_\sigma \\
&+&\oh\sum_{\sigma,\sigma'}\int d^3 r d^3 r' U(|\r-\r'|)
~\hat \psi^\dagger_\sigma(\r) \hat  \psi^\dagger_{\sigma'}(\r')
\hat \psi_{\sigma'}(\r') \hat \psi_\sigma(\r).
\nonumber
\end{eqnarray}
where $\hat \psi_\sigma(\r)$ ($\hat \psi^\dagger_\sigma(\r)$) is an annihilation
(creation) field operator of a fermion of flavor $\sigma$ at a point
$\r$. We would like first to discuss how a problem defined by the
Hamiltonian, \rf{eq:explicitpotential} leads directly to scattering
amplitudes \rfs{eq:scatamp}.

Motivated by experiments where studies are confined to gases of no
more than two fermion flavors (corresponding to a mixture of two
distinct hyperfine states) we will refer to $\sigma$ as simply spin,
designating a projection ($\sigma$) of the corresponding two-flavor
pseudo-spin along a quantization axis as a spin up, $\uparrow$, and
down, $\downarrow$.  In an equivalently and sometimes more convenient
momentum basis above Hamiltonian becomes
\begin{widetext}
\begin{eqnarray}
\label{eq:leoham} \hat H=\sum_\sigma\sum_\k
\frac{k^2}{2m}\hat a^\dagger_{\k,\sigma} \hat a_{\k,\sigma}
+\frac{1}{2V}\sum_{\sigma,\sigma'}\sum_{\k,\k',\p}
\tilde{U}(|\k-\k'|)
\hat a^\dagger_{\k'+\frac{\p}{2},\sigma} \hat a^\dagger_{-\k'+\frac{\p}{2},\sigma'}
\hat a_{-\k+\frac{\p}{2},\sigma'} \hat a_{\k+\frac{\p}{2},\sigma},
\end{eqnarray}
\end{widetext}
where $\hat  a_{\k,\sigma}$ ($\hat a^\dagger_{\k,\sigma}$) is an annihilation
(creation) operator of a fermionic atom of flavor $\sigma$ with
momentum $\k$, satisfying canonical anticommutation relations and
related to the field operator by $\hat \psi_\sigma(\r)=V^{-1/2}\sum_\k
\hat a_{\k,\sigma}e^{i\k\cdot\r}$. With our choice of momentum variables
above the relative center of mass momenta before (after) the collision
are $\pm\k$ ($\pm\k'$) and $\p$ is the conserved momentum of the
center of mass of the pair of scattering particles.

In the rest of this section, we would like to calculate the scattering
amplitudes $f_\ell$ given in \rfs{eq:partialwaves} in terms of the
interaction potential $\tilde{U}(|\k-\k'|)$.  With this goal in mind,
it is convenient to make the symmetry properties of the fermion
interaction $\hat H_{int}$ explicit, by taking advantage of the
rotational invariance of the two-body potential $U(|\r-\r'|)$ and the
anticommutation of the fermion operators. To this end we decompose the
angular dependence (arising through $\hat{\bf k}\cdot\hat{\bf k}'$,
where $\hat {\bf k}$ is a unit vector parallel to ${\bf k}$) of the
Fourier transform of the two-body potential, $\tilde{U}(|\k-\k'|)$
into spherical harmonics via
\begin{equation}
\tilde{U}(|\k-\k'|)\equiv U_{{\bf k},{\bf k}'}=
\sum_{\ell=0}^\infty (2\ell+1)u^{(\ell)}_{k,k'}
P_\ell(\hat{\bf k}\cdot\hat{\bf k}').
\label{eq:sphericalHarmonics}
\end{equation}
The $\ell$-th orbital angular momentum channel interaction amplitude
$u^{(\ell)}_{k,k'}$ can be straightforwardly shown to be given by
\begin{equation}
u^{(\ell)}_{k,k'}=4\pi\int_0^\infty dr ~r^2 U(r)j_\ell(k r)j_\ell(k'
r), \label{uell_kk}
\end{equation}
where $j_\ell(x)$ is the $\ell$-th spherical Bessel function.

Using anticommutativity of the fermion operators, it is possible to
decompose the interaction term in \rfs{eq:leoham} into the singlet and
triplet channels by introducing the two-body interaction vertex
$\tilde{U}_{\sigma_1\sigma_2}^{\sigma'_1\sigma'_2}(\k,\k')$ defined by
\begin{widetext}
\begin{eqnarray}
\hat H_{int}=\frac{1}{2V}\sum_{\sigma_1,\sigma_2,\sigma'_1,\sigma'_2}
\sum_{\k,\k',\p}\tilde{U}_{\sigma_1\sigma_2}^{\sigma'_1\sigma'_2}(\k,\k')
\hat a^\dagger_{\k'+\frac{\p}{2},\sigma'_1} \hat a^\dagger_{-\k'+\frac{\p}{2},\sigma'_2}
\hat a_{-\k+\frac{\p}{2},\sigma_2} \hat a_{\k+\frac{\p}{2}\sigma_1},
\end{eqnarray}
with
\begin{equation}
\tilde{U}_{\sigma_1\sigma_2}^{\sigma'_1\sigma'_2}(\k,\k')=
\frac{1}{2}\tilde{U}(|\k-\k'|)
\delta_{\sigma'_1\sigma_1}\delta_{\sigma'_2\sigma_2}
-\frac{1}{2}\tilde{U}(|\k+\k'|)
\delta_{\sigma'_1\sigma_2}\delta_{\sigma'_2\sigma_1}),
\end{equation}
\end{widetext}
that automatically reflects the antisymmetric (under exchange)
property of fermions, namely
\begin{eqnarray}
\tilde{U}_{\sigma_1\sigma_2}^{\sigma'_1\sigma'_2}(\k,\k')
&=&-\tilde{U}_{\sigma_2\sigma_1}^{\sigma'_1\sigma'_2}(-\k,\k'),\\
&=&-\tilde{U}_{\sigma_1\sigma_2}^{\sigma'_2\sigma'_1}(\k,-\k').
\end{eqnarray}
The vertex can be furthermore decomposed into spin singlet (s) and
triplet (t) channel eigenstates of the two-particle spin angular
momentum,
\begin{equation}
\tilde{U}_{\sigma_1\sigma_2}^{\sigma'_1\sigma'_2}(\k,\k')=
\tilde{U}^{(s)}_{\sigma_1\sigma_2,\sigma'_1\sigma'_2}(\k,\k')
+\tilde{U}^{(t)}_{\sigma_1\sigma_2,\sigma'_1\sigma'_2}(\k,\k').
\end{equation}
The singlet and triplet vertices
\begin{eqnarray} \label{eq:decomp}
\tilde{U}^{(s)}_{\sigma_1\sigma_2,\sigma'_1\sigma'_2}(\k,\k')&=&
U^{(e)}(\k,\k')\chi^{(s)}_{\sigma_1\sigma_2,\sigma'_1\sigma'_2}\\
\tilde{U}^{(t)}_{\sigma_1\sigma_2,\sigma'_1\sigma'_2}(\k,\k')&=&
U^{(o)}(\k,\k')\chi^{(t)}_{\sigma_1\sigma_2,\sigma'_1\sigma'_2}
\nonumber,
\end{eqnarray}
are expressed in terms of an orthonormal set of singlet and triplet
projection operators
\begin{eqnarray}
\chi^{(s)}_{\sigma_1\sigma_2,\sigma'_1\sigma'_2}&=&
\frac{1}{2}(\delta_{\sigma'_1\sigma_1}\delta_{\sigma'_2\sigma_2}
-\delta_{\sigma'_1\sigma_2}\delta_{\sigma'_2\sigma_1}),\\
\chi^{(t)}_{\sigma_1\sigma_2,\sigma'_1\sigma'_2}&=&
\frac{1}{2}(\delta_{\sigma'_1\sigma_1}\delta_{\sigma'_2\sigma_2}
+\delta_{\sigma'_1\sigma_2}\delta_{\sigma'_2\sigma_1}) \nonumber,
\end{eqnarray}
with coefficients
\begin{eqnarray} \label{eq:chandec}
U^{(e)}(\k,\k')&=&
\oh\left(\tilde{U}(|\k-\k'|)+\tilde{U}(|\k+\k'|)\right),\hspace{+1cm}\\
U^{(o)}(\k,\k')&=&\oh\left(\tilde{U}(|\k-\k'|)-\tilde{U}(|\k+\k'|)\right),
\nonumber
\end{eqnarray}
that, by virtue of decomposition, Eq.~\rf{eq:sphericalHarmonics} and
symmetry of Legendre polynomials, $P_\ell(-\hat{\bf k}\cdot\hat{\bf
k}')=(-1)^\ell P_\ell(\hat{\bf k}\cdot\hat{\bf k}')$ are vertices for
even and odd orbital angular momentum $\ell$ channels, respectively,
as required by the Pauli exclusion principle.  Physically, these
irreducible even and odd verticies make explicit the constructive and
distructive interference between scattering by angle $\theta$ and
$\pi-\theta$ of two fermions.

\begin{figure}[bt]
\includegraphics[height=.3in]{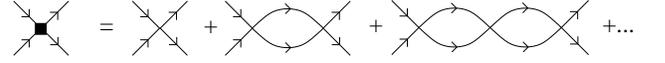}
\caption{\label{Fig:vertex} The renormalized 4-point vertex for
potential scattering, determining the $T$-matrix $T_{\k,\k'}$.}
\end{figure}

The two-body scattering amplitude $f(\k,\k')$ is proportional to the
$T$-matrix,
\begin{equation}
\label{eq:scatampT} f(\k,\k') = - {m\over 4 \pi} T_{\k,\k'}=-\frac
{m_r} {2 \pi} T_{\k,\k'},
\end{equation}
where $m_r=m/2$ is the reduced mass of two fermions.  The $T$-matrix
can be computed via standard methods. As illustrated on
Fig.~\ref{Fig:vertex}, it equals to a renormalized 4-point vertex
(1PI) $\Gamma^{(4)}(\k+\p/2,-\k+\p/2,\k'+\p/2,-\k'+\p/2;\varepsilon)$
for particles scattering with initial (final) momenta $\pm\k+\p/2$
($\pm\k'+\p/2$), and at a total energy in the center of mass frame
given by
\begin{eqnarray} \label{eq:some}
\varepsilon&=&\frac{(\k+\p/2)^2}{2m}+\frac{(-\k+\p/2)^2}{2m} -
\frac{p^2}{4m},\\
&=&\frac{k^2}{m}=\frac{k'^2}{m}=\frac{k^2}{2 m_r}, \nonumber
\end{eqnarray}
with the last relation valid due to energy conservation by a time
independent interaction.

Given the retarded Green's functions of fermions,
\begin{equation} \label{eq:greenferm}
G(\k, \omega) = \frac{1}{\omega-\frac {k^2}{2m} + i 0},
\end{equation} the main ingredient of the sequence of diagrams
from Fig.~\ref{Fig:vertex} is the polarization operator, denoted by a
bubble in the figure, and physically corresponding the Green's
function of the reduced fermion with momentum $\q$ and mass $m_r$,
\begin{eqnarray} \label{eq:defpi}
\Pi(\q,\ve) &=& \int \frac{d\omega}{2\pi i }
G\left( \frac{\p} 2+\q, \ve+ \frac{p^2}{4m}+\omega \right) G\left(
\frac{\p}2 -\q , -\omega\right)\nonumber\\
&&\nonumber\\
&=&  \frac{1}{\varepsilon -\frac{q^2}{m}+i0}.
 \end{eqnarray}
Although perhaps not immediately obvious, $\Pi(\q,\ve)$ as defined
above is independent of $\p$, the center of mass momentum of a pair of
fermions.


The sequence of diagrams in Fig.~\ref{Fig:vertex} then generates a
series for a $T$-matrix given by
\begin{equation}
\label{eq:Tmatrix} T_{\k,\k'} = U_{\k,\k'} + \sum_{\q} U_{\k, \q}
\Pi(\q,\ve) U_{\q, \k'} + \dots,
\end{equation}
that can formally be resummed into an integral equation
\begin{equation}
\label{eq:TmatrixI} T_{\k,\k'} = [(1- U \Pi)^{-1}U]_{\k,\k'},
\end{equation}
where a martix product over wavevectors inside the square brackets is
implied.

Utilizing the channel decomposition, Eq.~\rf{eq:chandec} of the vertex
$U_{\k,\k'}$ together with the closure-orthogonality relation
\begin{equation}
\int_{-1}^1P_\ell(\hat{\k}\cdot\hat{\q})
P_{\ell'}(\hat{\q}\cdot\hat{\k}')d\Omega_\q =
\frac{4\pi}{2\ell+1}\delta_{\ell\ell'}P_\ell(\hat{\k}\cdot\hat{\k}'),
\end{equation}
the $T$-matrix series separates into a partial waves sum
\begin{equation}
T_{\k,\k'}=\sum_{\ell=0}^\infty (2\ell+1) T^{(\ell)}_{k,k'} P_\ell(\hat{\k}\cdot\hat{\k}')
,
\label{Tpartialwaves}
\end{equation}
with
\begin{equation}
\label{eq:Tmatrix_ell} T^{(\ell)}_{k,k'} = u^{(\ell)}_{k,k'} +
{1\over V}\sum_{\q} u^{(\ell)}_{k,q} \Pi(q,\ve) u^{(\ell)}_{q, k'} + \dots,
\end{equation}
a $T$-matrix for scattering in an angular momentum channel $\ell$,
conserved by the spherical symmetry of the two-body interaction
potential. This demonstrates explicitly that the interaction vertices
in different $\ell$ channels do not mix, each contributing only to the
corresponding scattering amplitude channel $f_\ell(k)$ in
Eq.~\rf{eq:scatamp}.

Without specifying the interaction potential $U(r)$, a more explicit
expression for the $T$-matrix can only be obtained for the so-called
separable potential, discussed in detail in
Ref.~\cite{Nozieres1985}. Such separable interaction is a model that
captures well a low-energy behavior of a scattering amplitude of a
more generic short-range potential. To see this, we observe that a
generic short-range potential, with a range $d$, leads to a vertex
in the $\ell$-th channel, which at long scales, $kd \ll 1$, separates
into
\begin{eqnarray}
u^{(\ell)}_{k,k'}\approx \lambda k^\ell g^{(\ell)}_k k'^\ell
g^{(\ell)}_{k'} \approx \lambda k^\ell k'^\ell, \label{u_separates}
\end{eqnarray}
with
\begin{eqnarray}
\lambda&=&{4\pi U_0 d^{2\ell+3}\over[(2\ell+1)!!]^2},\\
U_0 d_0^{2\ell+3}&\equiv&\int_0^\infty d r ~ r^{2\ell+2} U(r).
\end{eqnarray}
Assuming that this separation holds at all $k$ (a definition of a
separable potential), we use this asymptotics inside
Eq.~\rf{eq:Tmatrix_ell}. This reduces the $T$-matrix to a geometric
series that resums to
\begin{equation}
T^{(\ell)}_{k,k'}={u^{(\ell)}_{k,k'}\over 1-\lambda\Pi^{(\ell)}(\ve)},
\label{DysonResummed}
\end{equation}
where $\Pi^{(\ell)}(\ve)$ is the trace over momentum of the atom
polarization ``bubble'' corresponding to the molecular self-energy at
energy $\ve$,
\begin{eqnarray}
\label{eq:polarization}
\hspace{-2cm}\Pi^{(\ell)}(\ve)&=&\frac{1}{V}\sum_{\bf q}
q^{\ell}g^{(\ell)}_q \Pi(q,\ve)q^\ell g^{(\ell)}_q,\\
&=&\int {d^3 q \over (2 \pi)^3 }{q^{2\ell}g^{(\ell)}(q)^2
\over\ve-{q^2 \over m}+i 0},\\
&=& -m\Lambda^{2 \ell+1} R\left( \frac {k^2} {\Lambda^2} \right) -
{i\over 4\pi} m^{\ell+3/2}\ \ve^{\ell+1/2}.
\end{eqnarray}
In above $R(x)$ is a Taylor-expandable function of its dimensionless
argument, the momentum cutoff $\Lambda\approx2\pi/d$ is set by the
potential range $d$, and, as before, $k^2/m=\ve$.  Putting this
together inside the $T$-matrix, we find the low-energy $\ell$-channel
scattering amplitude
\begin{equation} \label{eq:leo}
f^{(\ell)}(k)=-\frac 1 { \frac{4 \pi} {m k^{2 \ell}} \left( \frac 1
\lambda + m\Lambda^{2 \ell+1} R \left( \frac {k^2}{ \Lambda^2}
\right) \right) + i k}, \ k \ll \Lambda.
\end{equation}
This coincides with the general form, Eq.~\rf{eq:scatamp} arising from the
requirement of analyticity and unitarity of the scattering
matrix. However, we observe that in the $s$-wave case, for the full
range of accessible wavevectors up to ultraviolet cutoff, $k < \Lambda$
the scattering amplitude \rfs{eq:leo} is well approximated by the
non-resonant, scattering-length dominated form \rf{eq:scatlength},
with the scattering length given by $a^{-1}=4\pi/(m\lambda) +
4\pi\Lambda R(0)$.  The ``effective range" $r_0$ extracted from
\rfs{eq:leo} is $r_0 \sim 1/\Lambda$, namely microscopic, positive,
and is of the order of the spatial range of the potential $d$. Yet, as we
saw in Sec.~\ref{swave}, in order to capture possible resonances,
$r_0$ must be negative and much longer than the actual spatial range
of the potential. The fact that our calculation does not capture
possible resonances is an artifact of our choice of a separable
potential.

Although a more physical (nonseparable) potential $U(r)$, of a
resonant form depicted in Fig.~\ref{Fig-pot}, indeed exhibits
scattering via a resonant state (not just a bound and virtual bound
states), calculating the scattering amplitude $f_\ell(k)$ (beyond
\rfs{u_separates} approximation) is not really practical within the
second-quantized many-body approach formulated in
\rfs{eq:explicitpotential}.  In fact, the only way to derive the
scattering amplitude in that case is to go back to the Schr\"odinger
equation of a pair of fermions, reducing the problem to an effective
single-particle quantum mechanics. However, because we are ultimately
interested in condensed states of a finite density interacting atomic
gas, this two-particle simplification is of little value to our goals.

However, as we will show in Secs.~\ref{OneCM} and \ref{TwoCM}, a significant
progress can be made by formulating a much simpler pseudo-potential
model, that, on one hand reproduces the low-energy two-atom scattering
of the microscopic model \rf{eq:explicitpotential} in a vacuum
(thereby determining its parameters by dilute gas experiments), and on
the other hand is amenable to a standard many-body treatment even at
finite density.

Furthermore, as will see below, in cases of finite angular momentum
scattering, \rfs{eq:leo} can in principle describe scattering via
resonances as well as in the presence of bound states. Thus the
assumption of separability is no longer as restrictive as it is in the
$s$-wave case.

\subsection{Feshbach-resonant scattering}
\label{FRscattering}

As discussed in the Introduction, in fact, the physically most
relevant resonant scattering arising in the context of cold atoms is
microscopically due to a Feshbach resonance~\cite{Feshbach1959}. Generically it can be
described as a scattering, where the two-body potential,
$U_{\alpha,\alpha'}(|\r-\r'|)$ depends on internal quantum numbers
characterizing the two-atom state. These states, referred to as
channels, are not eigenstates of the interacting Hamiltonian and
therefore two atoms coming in one channel $\alpha$ in the process of
scattering will generically undergo a transition into a different
channel $\alpha'$.

The simplest and experimentally most relevant case is well
approximated by two channels $\alpha=o,c$ (often referred to as
``open'' and ``closed''), that approximately correspond to {\em
  electron} spin-triplet and {\em electron} spin-singlet states of two
scattering atoms; this is not to be confused with the {\em hyperfine}
singlet and triplet states discussed in the previous subsection.  Such
system admits an accessible Feshbach resonance when one of the
channels (usually the electron spin-singlet) admits a two-body bound state.
Furthermore, because pair of atoms in the two channels have very
different magnetic moments, their Zeeman splitting can be effectively
controlled with an external magnetic field. The corresponding
microscopic Hamiltonian is given by
\begin{widetext}
\begin{eqnarray} \label{eq:feshverygen}
\hspace{-1cm} \hat H&=&\int d^3 r~\sum_{\sigma,s} \hat \psi^\dagger_{\sigma,s}
\left(-{\nabla^2\over 2 m} \right)\hat \psi_{\sigma,s}
+\oh\sum_{\sigma,\sigma'\atop{s_1,s_2,s_1',s_2'}}\int d^3 r d^3 r'
U_{s_1s_2}^{s'_1s'_2}(|\r-\r'|)
~\hat \psi^\dagger_{\sigma, s_1'}(\r) \hat \psi^\dagger_{\sigma',s_2'}(\r')
\hat \psi_{\sigma', s_2}(\r') \hat \psi_{\sigma, s_1}(\r),
\end{eqnarray}
\end{widetext}
where $s$ labels the channel. The interaction
$U_{s_1s_2}^{s'_1s'_2}(|\r-\r'|)$ can be more conveniently reexpressed
in terms of the two-atom electron spin-singlet and spin-triplet
channels basis, $U_{\alpha,\alpha'}(|\r-\r'|)$, where
$U_{o,o}(|\r-\r'|)$, $U_{c,c}(|\r-\r'|)$ are the interaction for two
atoms in the open (triplet) and closed (singlet) channels,
respectively, and $U_{o,c}(|\r-\r'|)$, characterizes the interchannel
transition amplitudes, i.e., the strength
of o-c hybridization due to the hyperfine interactions.

The corresponding scattering problem would clearly be even more
involved than a single-channel model studied the previous subsection.
Yet, as the analysis of Section~\ref{TwoCM} will show, at low energies,
the scattering amplitude of two atoms, governed by
\rfs{eq:feshverygen}, is still of the same form, \rf{eq:scatamp}, as
that of a far simpler pseudo-potential two-channel model.  Indeed, the
form of a scattering amplitude is controlled by unitarity and
analyticity, not by precise details of realistic Hamiltonians.  Thus,
to capture either a microscopically potential- or a Feshbach resonant
scattering we will replace a realistic Hamiltonian, such as
\rfs{eq:feshverygen} with a simpler model, which, nevertheless
exhibits a low-energy scattering amplitude of the same form. To this
end, in the next two sections we examine two such effective models and
work out their scattering amplitudes. We will thereby determine and
justify our subsequent choice of a many-body model with the correct
low-energy two-body physics.

\section{One-Channel Model}
\label{OneCM}
\subsection{$S$-wave scattering}

The most drastic simplification of a resonant Fermi gas is to model
the two-body interaction by a featureless and short-ranged
single-channel pseudo-potential, that at long scales and low energies
is most commonly taken to simply be $U({\bf r})=\lambda
\delta^{(3)}({\bf r})$, with the corresponding many-body Hamiltonian
\begin{widetext}
\begin{eqnarray} \label{eq:singlechannelham}
\hspace{-0.5cm} \hat H_{s}^{1-ch}&=&\int d^3 r
\Big[\sum_\sigma \hat \psi^\dagger_\sigma \left(-{\nabla^2
\over 2 m} \right) \hat \psi_\sigma
+\lambda~\hat \psi^\dagger_\downarrow(\r) \hat \psi^\dagger_\uparrow(\r)
\hat \psi_\uparrow(\r) \hat \psi_\downarrow(\r)\Big].\nonumber\\
\end{eqnarray}
\end{widetext}

In analyzing the Hamiltonian like this one, one has to exercise a
certain amount of caution, as the repulsive $\delta$-function
potential is known to have a vanishing scattering amplitude in three
dimensions, and therefore does not make sense if understood
literally~\cite{commentVanish_delta}.

Hence $\delta$-function potential must be supplemented with a
short-scale cutoff $1/\Lambda$ (i.e., given a finite spatial extent),
that we will take to be much smaller than the wavelength of a
scattering particle, i.e., $k/\Lambda\ll 1$.  Furthermore, for
calculational convenience, but without modifying the properties on
scales longer than the cutoff, we will impose the cutoff $\Lambda$ on
each of the momenta $k$ and $k'$ independently, modeling the
interaction in \rfs{eq:singlechannelham} by a featureless separable
potential
\begin{equation}
\label{eq:sep} U_{\k,\k'}=u^{(0)}_{k,k'}=
\lambda ~\theta(\Lambda^2-\k^2)~
\theta(\Lambda^2-\k'^2),
\end{equation}
with $\theta(x)$ the usual step function, and interactions in all
finite angular momentum channels vanishing by contruction.  We note
that this separability of the potential is consistent with the general
long wavelength form of a generic short-scale potential found in
Eq.~\rf{u_separates}, although it does lead to some minor unphysical
features such as only a single bound state, independent of how strongly
attractive the potential (how negative $\lambda$)
is~\cite{Nozieres1985}.

As discussed in Sec.\ref{Pscattering}, the Dyson equation
\rf{eq:Tmatrix_ell} can be easily resummed into \rfs{DysonResummed},
with the $s$-wave polarization bubble
$\Pi_s(\ve)\equiv\Pi^{(\ell=0)}(\ve)$ (cf \rfs{eq:polarization})
easily computed to give
\begin{eqnarray}
\label{eq:polarization1}
\Pi_s(\ve)&=&\int {d^3 q\over (2 \pi)^3 }{\theta(\Lambda-q)
\over \ve-{q^2 \over m}+i 0}\nonumber\\
&=& -{m\over 2\pi^2} \Lambda - i {m^{3
/ 2}\over 4 \pi}\sqrt{\ve},
\end{eqnarray}
where we used $\ve \ll \Lambda^2/(2m)$. This then directly leads to
the $s$-wave scattering amplitude (vanishing in all other angular
momentum channels)
\begin{equation}
f_s(\k,\k') = -{1 \over {4 \pi \over m \lambda} + {2 \Lambda \over \pi}
+ i k},
\label{f1ch-model}
\end{equation}
which coincides with \rfs{eq:scatlength}, where the scattering length
is given by
\begin{eqnarray}
\label{eq:deltalength} a(\lambda)& =& \left({4\pi \over m \lambda} + {2 \Lambda
\over  \pi} \right)^{-1} \equiv \frac{m}{4\pi}\lambda_R,\\
&=&\frac{m}{4\pi}\frac{\lambda}{1-\lambda/\lambda_c},
\end{eqnarray}
where $\lambda_R$ can be called the renormalized coupling and
\begin{equation}
\lambda_c=-{2\pi^2 \over\Lambda m}
\end{equation}
is a critical value of coupling $\lambda$ at which the scattering
length diverges.

Hence we find that scattering off of a featureless potential of a
microscopic range $1/\Lambda$ (modeled by the cutoff
$\delta$-function separable potential), is indeed given by
\rfs{eq:scatlength}, with this form {\em exact} for $k/\Lambda \ll 1$,
i.e., for the particle wavelength $1/k$ longer than the range of the
potential. We also note that in the limit $\Lambda \rightarrow
\infty$, the scattering amplitude vanishes, in agreement with the
aforementioned fact that the ideal $\delta$-function potential does
not scatter quantum particles~\cite{commentVanish_delta}.

\begin{figure}[bt]
\includegraphics[height=2in]{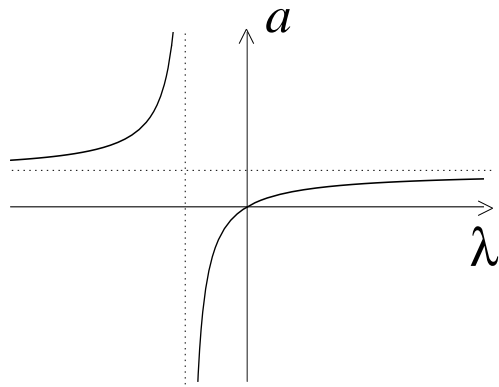}
\caption{\label{Fig-scatlength} The scattering length $a$ as a
function of strength $\lambda$ of the separable $\delta$-function
potential.}
\end{figure}

For finite cutoff, the scattering length $a$ as a function of
$\lambda$ is shown on Fig.~\ref{Fig-scatlength}. We note that in the
``hard ball'' limit of a strongly repulsive potential, $\lambda \gg {2
\over m \Lambda}$, the scattering length is given simply by its
spatial extent, $a=\pi/(2 \Lambda)\sim d$. For an attractive potential the
behavior is more interesting. For weak attraction, the scattering
length $a$ is negative.  However, for sufficiently strong attractive
potential, i.e., sufficiently negative $\lambda$, the scattering
length $a(\lambda)$ changes sign, diverging hyperbolically at the
critical value of $\lambda_c$, and becoming positive for $\lambda <
\lambda_c$. The critical value of $\lambda$ at which this takes place
corresponds to the threshold when the potential becomes sufficiently
attractive to admit a bound state. There is no more than one bound
state in a separable $\delta$-function potential, regardless of how
strongly attractive it is ~\cite{Nozieres1985}.

Finally, as above discussion (particularly, \rfs{f1ch-model})
indicates, although a one-channel $s$-wave model can successfully
reproduce the very low energy limit of the generic $s$-wave scattering
amplitude, such ultra-short range pseudo-potential models cannot
capture scattering via a resonance. The actual Feshbach resonance
experiments may or may not involve energies high enough (large enough
atom density) for the scattering to proceed via a resonance (most do
not, with the criteria for this derived in Subsection C,
below). However, our above findings show that the ones that do probe
the regime of scattering via a resonance must be described by a model
that goes beyond the one-channel $\delta$-function pseudo-potential
model. We will explore the simplest such two-channel model in
Sec.~\ref{tcswaveta}.

\subsection{Finite angular momentum scattering}
\label{oneCh_ell}

Unlike their $s$-wave counterpart, one-channel models for higher
angular momentum scattering can describe scattering via resonances.
This is already clear from the analysis after \rfs{eq:leo}. Let us
analyze this in more detail.

Above $s$-wave model \rf{eq:singlechannelham} can be straightforwardly
generalized to a pseudo-potential model at a finite angular momentum.
This is most easily formulated directly in momentum space by replacing
the two-body $\ell$-wave interaction in the microscopic model
\rf{eq:leoham} by a separable model potential
\begin{eqnarray}
u^{(\ell)}_{k,k'}&=&\lambda k^\ell g^{(\ell)}_{k} k'^\ell g^{(\ell)}_{k'}\\
&=&\lambda k^\ell k'^\ell~\theta(\Lambda^2-\k^2)~
\theta(\Lambda^2-\k'^2).
\label{uell_separates}
\end{eqnarray}
that simply extends the long wavelength asymptotics of a microscopic
interaction \rfs{u_separates} down to a microscopic length
scale $2\pi/\Lambda$.

Using results of the previous section, this model then immediately
leads to the scattering amplitude \rfs{eq:scatamp} with
\begin{equation}
F_{\ell}(k^2) =- \frac{ 4 \pi }{m \lambda} - \frac{2 \Lambda^{2
\ell+1} }{\pi (2 \ell+1)} - \frac {2 \Lambda^{2\ell-1}}{\pi \left(2
\ell-1 \right)} k^{2} +\dots .
\end{equation}
The corresponding scattering amplitude is given by
\begin{equation}
f_\ell(k)={k^{2\ell}\over -v^{-1}_\ell +
\frac{1}{2}k_0^{2\ell-1} k^2 - i k^{2\ell + 1}},
\label{f_ell}
\end{equation}
with the analogs of the scattering volume (of dimensions $2\ell +1$)
and effective range parameters given respectively by
\begin{eqnarray}
v_\ell &=&  \left(\frac{ 4 \pi }{m \lambda}+ \frac{2 \Lambda^{2
\ell+1} }{\pi (2 \ell+1)}\right)^{-1}
\equiv \frac{m}{4\pi}\lambda_\ell^R,\\
k_0^{2\ell-1}&=& -\frac {4 \Lambda^{2\ell-1}}{\pi \left(2
\ell-1 \right)}.
\label{vk0}
\end{eqnarray}
We note that $v_\ell$ diverges (hyperbollically) for a sufficiently
attractive interaction coupling, reaching a critical value
$$\lambda_c^{(\ell)}= -\frac{2 \pi^2 \left(2 \ell+1\right)}{m \Lambda^{2
\ell+1}},
$$

From the structure of $f_\ell(k)$ it is clear that at low energies
(length scales longer than $k_0^{-1}$), the imaginary term $i
k^{2\ell+1}$ is subdominant to the second $k^2$ term in the
denominator. Consequently, the pole is well-approximated by
\begin{eqnarray}
E_{\rm pole}&\approx& -\frac{2}{m k_0^{2\ell-1}v_\ell} -  i\Gamma_\ell/2,
\end{eqnarray}
where we defined
\begin{eqnarray}
\Gamma_\ell &\approx& \frac{4}{m}k_0^{-2\ell+1} k_{\rm pole}^{2\ell+1},\\
k_{\rm pole}^2&\approx&-2k_0^{-2\ell+1} v_\ell^{-1}.
\end{eqnarray}
For a positive detuning, $v_\ell < 0$, leading to the first term of $E_{\rm pole}$
real and positive, while the second one $-i\Gamma/2$ negative,
imaginary and at low energies ($k<k_0$) much smaller than
$Re[E_p]$. Thus, (in contrast to the s-wave case) for finite angular
momentum scattering, even a single-channel model with a separable
potential exhibits a resonance that is narrow for large, negative $v_\ell$.
For $v_\ell > 0$, the term $-i\Gamma/2$ becomes real and this resonance
directly turns into a true bound state, characterized by a pole $E_{\rm pole}$,
that is real and negative for $v_\ell > 0$.

\subsection{Model at finite density: small parameter}

Having established a model for two-particle scattering in a vacuum, a
generalization to a model at finite density $n$, that is of interest
to us, is straightforward. As usual this is easiest done by working
within a grand-canonical ensemble by introducing a chemical potential
$\mu$ that couples to a total number of particles operator $\hat{N}$
via
\begin{equation}
\hat{H}\rightarrow {\hat H}-\mu\hat{N}.
\label{muN}
\end{equation}
One thereby controls the average atom number and density by adjusting
$\mu$.

The single-channel models of the type \rfs{eq:singlechannelham} and
its corresponding finite angular momentum channel extensions have been
widely studied in many problems of condensed matter physics.  Although
(as most interacting many-body models) it cannot be solved exactly,
for sufficiently small renormalized coupling $\lambda_R(\lambda)$,
\rf{eq:deltalength},\rf{vk0}, (whether positive or negative), we
expect that one can analyze the system in a controlled perturbative
expansion about a mean-field solution in a dimensional measure of
$\lambda_R$, namely in the ratio of the interaction energy to a
typical kinetic energy $\epsilon_F$.

\subsubsection{Small parameter in an $s$-wave model}
\label{s-wave1CMparameter}

In the $s$-wave case this dimensionless ratio is just the gas
parameter
\begin{eqnarray}
\frac{|\lambda_R|\ n}{\epsilon_F}
&\sim& |a|n^{1/3}\propto k_F |a|.
\label{gasparam}
\end{eqnarray}
For weak repulsive $s$-wave interaction, $\lambda>0$, $a n^{1/3} \ll
1$, and the perturbation theory generically leads to a Fermi liquid
\cite{LandauFL}. For weak attractive interaction $\lambda<0$ and $|a|
n^{1/3} \ll 1$, it predicts a weak-coupling BCS superconductor.

However, as $\lambda$ is made more negative (increasing the strength
of the attractive interaction) $|a|$ increases according to
\rfs{eq:deltalength}, as illustrated in Fig.\ref{Fig-scatlength} and
eventually goes to infinity when $\lambda$ reaches the critical value
of $\lambda_c$. Near this (so-called) unitary point, the gas parameter
is clearly large, precluding a perturbative expansion within a
one-channel model.

On the other (BEC) side of the unitary point, a
molecular bound state appears and the phenomenology is that of
interacting bosonic molecules with a molecular scattering length
proportional to that of fermionic atoms, $a_m \approx 0.6
a$~\cite{Petrov2005,Brodsky2005,Levinsen2006}. Since on the BEC side $a$ also
diverges (this time from a {\em positive} side), the bosonic gas of these molecular
dimers is
strongly interacting near the unitary point and the situation is
as hopeless for quantitative analysis as it was on the BCS side of the unitary point.

Yet, at large negative $\lambda$, the bound state drops to a large
negative energy and $a$ becomes small again (this time positive). In
this deep-BEC regime, the resulting dilute repulsive gas of tightly
bound molecules then also exhibits the same small gas parameter as
that deep in the BCS regime. Hence its ground state is a weakly
interacting superfluid Bose-condensate~\cite{Leggett1980,Nozieres1985},
with properties that can be computed perturbatively in a small
parameter $a n^{1/3}$, although careful analysis of this sort was only
done recently~\cite{Levinsen2006}.

We note in passing, that, at a finite atom density the {\em effective}
measure of the strength of interaction is actually a dispersive
coupling $\lambda_{k_F}$, given by the $T$-matrix
$T_{k_F,k_F}=(4\pi/m)|f_s(k_F)|$
\bea
\label{lambdaWide_k}
\lambdah_{k_F}^{s} &\sim& {|T^s_{k_F}|\ n\over\epsilon_F},\\
&\sim& k_F |f_s(k_F)|,\\
&\sim&\frac{\kf}{|a^{-1} + i\kf|},\\
&\sim&
\cases{
{\kf |a|},& \text{for $\kf |a|\ll 1$,}\cr
1,& \text{for $\kf |a|\gg 1$,}\cr}.
\eea
Thus, in contrast to a two-body case, at finite density the growth of
this effective dimensionless coupling, $\sim k_F |a|$, actually
saturates at $1$ (i.e., at a large, nonperturbative, but noninfinite
value), due to a cutoff of the growing scattering length $a$ by atom
separation $\kf^{-1}$.

Hence, despite its many successes to predict qualitative behavior, the
Hamiltonian \rfs{eq:singlechannelham} has a limited ability to
describe a resonant interacting Fermi gas. First of all, as we just
saw, its two-body scattering amplitude, as given by
\rfs{eq:scatlength}, does not describe scattering via a resonant
state, capturing only a true bound state (for a sufficiently
attractive $\lambda$ and positive $a$), but not a resonance~\cite{Resonance} (possible
for negative $a$). Thus, if resonances (states at positive energy and
finite lifetime) are present, the model given by
\rfs{eq:singlechannelham} is insufficient. Even in the absence of such
resonant states, the perturbation theory about the mean-field state
commonly used to analyze \rfs{eq:singlechannelham} breaks down in the
course of the BCS to BEC crossover, where the scattering length
surpasses the inter-particle spacing and $|a| n^{1/3} \gtrsim 1$ is no
longer small.

However, it is quite common in literature to ignore these issues and
simply extend the mean-field analysis of \rfs{eq:singlechannelham}
into the nonperturbative unitary regime near $\lambda_c$. Given the
absence of a phase transition in the $s$-wave case, the prediction of
such mean-field theory is undoubtedly qualitatively correct even in
the strong coupling regime that smoothly interpolates between
Pauli-principle stabilized large Cooper pairs and a BEC of tightly
bound molecules. However, as we just discussed, such approach (all the
perturbative embellishments notwithstanding) cannot make any {\em
quantitatively} trustworthy predictions for $\lambda \approx
\lambda_c$, a regime where a bound state is about to, or just appeared
and $|a| n^{1 / 3}\gg1$. Since the question of the $s$-wave
BCS-BEC crossover is intrinsically a quantitative one, quantitatively
uncontrolled studies performed within above nonresonant model provide
little information about the details of such crossover, particularly
near the so-called unitary regime.

\subsubsection{Small parameter in a finite angular momentum model}
\label{p-wave1CMparameter}

As can be seen from the form of the scattering amplitude \rf{f_ell}
and its parameters $v_\ell$ and $k_0$, the case of a gas resonant at a
{\em finite} angular momentum is qualitatively quite different from that
of the $s$-wave model just considered. The reason is that, as
discussed in Sec.\ref{oneCh_ell}, on length scale longer than the
spatial range of the potential $\Lambda^{-1}$ (i.e., on effectively
all accessible scales) the $i k^{2\ell+1}$ in $f_\ell(k)$ is
subdominant and a one-channel finite angular momentum model exibits a
resonant state that continuously transforms into a bound state. As
discussed in Sec.\ref{sec:pwavescat}, physically this stems from the
existence of a finite $\ell$ centrifugal barrier
that strongly suppresses the molecular decay rate at low positive
energies.

Analogously (but distinctly) to the $s$-wave case, a dimensionless
parameter that measures the relative strength of interaction and
kinetic energy in the $\ell$-wave case is given by
\begin{eqnarray}
\lambdah_{k_F}{(\ell)}&\sim&\frac{|T^{(\ell)}_{\kf}|\ k_F^{2\ell}\ n}
{\epsilon_F},\\
&\sim&\frac{k_F^{2\ell+1}}{||v_\ell^{-1}|+k_0^{2\ell-1} k_F^2|},\\
&\sim&
\cases{
\kf^{2\ell+1}|v_\ell|,& \text{for $k_0^{2\ell-1}k_F^2|v_\ell|\ll 1$,}\cr
\left(\frac{k_F}{k_0}\right)^{2\ell-1},
& \text{for $k_0^{2\ell-1}k_F^2|v_\ell|\gg 1$,}}.\nonumber\\
\label{lambda_kell}
\end{eqnarray}
Since, (as found above) $k_0\sim\Lambda$, we find that for a finite
angular momentum resonance, although the effective coupling
$\lambdah_{k_F}^{(\ell)}$ grows with $v_\ell$ as a resonance is
approached, it saturates at a value $\ll 1$, cutoff by a finite density.
Thus, this heuristic argument suggests that
in principle a controlled perturbative treatment of full BCS-BEC
crossover is possible for finite angular-momentum Feshbach resonances,
even within a one-channel model. Such analysis has not yet been done, and it is an interesting
research problem for future work.

\section{Two-Channel Model}
\label{TwoCM}

As we have seen, there are considerable shortcomings of a local
one-channel model, particularly for the $s$-wave case, as it does not
exhibit a resonant state, nor does it have a dimensionless parameter
that can be taken to be small throughout the BCS-BEC crossover. Thus,
we now consider a more involved fermion-boson two-channel model that
is free of these deficiencies. Furthermore, the appeal of this
two-channel model is that it is inspired by and more accurately
reflects the microscopics of the Feshbach resonance physics discussed
in the Introduction and above, but applies more universally to any
system where a resonant interaction (e.g., a shape resonance of the type
illustrated in Fig.~\ref{Fig-pot}) is at work. A general two-channel
model Hamiltonian, that in cold-atom context for the special $s$-wave
case was first introduced by Timmermans \cite{Timmermans1999}, is
given by
\begin{widetext}
\begin{eqnarray}
\label{eq:Hgeneral} &\hat H^{2-ch} &=
\sum_{\k,\sigma} {k^2 \over 2m} ~\hat a^\dagger_{\k,\sigma} \hat
a_{\k,\sigma} + \sum_{\p,\ell,m\atop{\mu}}
\left(\epsilon_0^{(\ell,m)} +{p^2 \over 4m} \right) \hat {\bf
b}_{\p,\ell,m}^{\mu\;\dagger}
\hat {\bf b}_{\p,\ell,m}^\mu
+\sum_{\k,\p,\ell\atop{\sigma,\sigma'}} ~{g_k^{(\ell)}\over \sqrt{V}}\;
k^\ell\left(\hat{b}_{\p,\ell\atop{\sigma,\sigma'}}(\hat{\k})\hat
a^\dagger_{\k+{\p\over 2},\sigma} \hat a^\dagger_{-\k+{\p \over
2},\sigma'} + h.c.\right).\nonumber\\
\end{eqnarray}
\end{widetext}
In above model Hamiltonian $\hat a_{\k,\sigma}$ ($\hat a^\dagger_{\k,\sigma}$)
is a fermionic annihilation (creation) operator of an atom of flavor
$\sigma$ with momentum $\k$, representing atoms in the open-channel
(typically corresponding to the electron [physical, as opposed to
flavor] spin-triplet state of two atoms) continuum. The annihilation
operator $\hat b_{\p,\ell\atop{\sigma,\sigma'}}(\hat{\k})$ destroys a
bosonic diatomic molecule of mass $2m$, with a center of mass momentum
$\p$, internal (atoms') momenta $\pm\k$.  It is a cartesian
spin-tensor that transforms as a tensor-product of two spin-1/2
representations and an orbital angular momentum $\ell$ representation.
It is convenient to decompose it into $2\ell+1$ components
$b_{\p,\ell,m\atop{\sigma,\sigma'}}$ corresponding to its projections
along an orbital quantization axis, according to:
\begin{eqnarray} \label{eq:defbone}
\hat b_{\p,\ell\atop{\sigma,\sigma'}}(\hat{\k})&=&
\sum_{m=-\ell}^{\ell}\hat b_{\p,\ell,m\atop{\sigma,\sigma'}}
Y^*_{\ell,m}(\hat\k) \nonumber \\
&\equiv& {\bf b}_{\p,\ell\atop{\sigma,\sigma'}}
\cdot{\bf Y}^*_{\ell}(\hat\k),
\label{blm}
\end{eqnarray}
where $Y_{\ell,m}(\hat{\k})$ are the spherical harmonics, $\hat{\k}$
is a unit vector along $\k$, and in the last line the scalar product
is over $2\ell+1$ components labeled by $-\ell\leq m\leq \ell$.
These bosonic orbital components can be further decomposed into a
singlet ($\mu=s$) and a triplet (with three spin projection components
$s_z=0,\pm1$ linear combinations of $\mu=(x,y,z)$ cartesian
components) spinor representations according to:
\begin{equation} \label{eq:defbtwo}
\hat {\bf b}_{\p,\ell\atop{\sigma,\sigma'}}=
\frac{1}{\sqrt{2}}\sum_{\mu=s,x,y,z}\hat {\bf b}_{\p,\ell}^\mu\cdot
i(\sigma_y, \sigma_y\vec{\sigma})^\mu_{\sigma,\sigma'}
\end{equation}
with $\vec\sigma$ a vector of Pauli spin matrices; notice that
$i\sigma_y$ is a fully antisymmetric (and thus a singlet) spin tensor
and the components of $\sigma_y\vec{\sigma}$ are linear combinations of
the spin-triplet projections $s_z=0,\pm1$, represented by $2\times 2$
symmetric matrices, with the relations
\begin{eqnarray}
\hat {\bf b}_{\p,\ell}^{(0)}&=& -\hat {\bf b}_{\p,\ell}^z\ ,\\
\hat {\bf b}_{\p,\ell}^{(\pm1)}&=& \pm\frac{1}{\sqrt{2}}\left(\hat {\bf b}_{\p,\ell}^x
\pm i \hat {\bf b}_{\p,\ell}^y\right) .
\end{eqnarray}
Within a Feshbach resonant system context the molecule $\hat {\bf
  b}_{\p,\ell}^{\mu}$ represents a (quasi-) bound state of two atoms in a
closed channel (usually electronic spin-singlet state of two atoms), a
true bound state in the limit of a vanishing coupling $g_k^{(\ell)}$
(proportional to o-c channels hybridization energy $U_{oc}$) for the
decay of a closed-channel molecule into an open-channel pair of atoms.
As discussed in the Introduction, in this case the 'bare' molecular
rest energy $\epsilon_0^{(\ell,m)}$ (the detuning relative to the
bottom of the open-channel continuum) corresponds to the Zeeman
energies that can be readily tuned with an external magnetic field.
For generality we allowed this detuning to have a nontrivial $m$
dependence, encoding an explicit breaking of orbital rotational
invariance seen in the experimental systems \cite{Ticknor2004}. This
ingredient will be central to our analysis in Sec.\ref{PWaveChapter},
for a determination of the correct ground state of a $p$-wave
paired superfluid.  Focusing on the closed-channel bound state, the model
clearly ignores the continuum (with respect to relative coordinate) of
closed-channel states.  Because for the experimentally interesting
regime of a resonance tuned to low energies, these states are at a
finite energy, they can be adiabatically eliminated (thereby only
slightly modifying model parameters) and can therefore be safely
omitted. In the context of a shape resonance the molecule $\hat {\bf
  b}^\mu_{\p,\ell}$ represents a resonance that is long-lived in the limit
of large potential barrier.

We would like to emphasize that for a nonzero Feshbach resonance
coupling $g_k^{(\ell)}$ it would be incorrect to consider
$b$-particles to be the true bound states (diatomic molecules) of
$a$-particles (atoms). Indeed, freely propagating $b$-particles are
not even eigenstates of the Hamiltonian \rfs{eq:Hgeneral}. The true,
physical molecule is a linear combination of $b$-particles and a
surrounding cloud of $a$-particles. They can be found by studying the
scattering problem posed by \rfs{eq:Hgeneral}.  In particular, the
true bound states of \rfs{eq:Hgeneral} can be spatially quite large
with their spatial extent set by a scattering length $a$ and at finite
atom density can easily overlap. In contrast, the $b$-particles
(related to the true bound states only in the limit of vanishing
Feshbach resonance couplings $g_k^{(\ell)}$) are point-like, with
their size set by a microscopic length scale corresponding to the
range $d=2\pi/\Lambda$ of the interatomic atomic potential,
$U_{s_1s_2}^{s'_1s'_2}(|\r-\r'|)$.

With this in mind, we now turn to the analysis of the two-channel
model, considering separately the $s$-wave and $p$-wave cases.

\subsection{$S$-wave}
\label{Sec:SWaveTCM}
\subsubsection{Two-atom scattering}
\label{tcswaveta}

As discussed in the Introduction, in the case of a two-flavor atomic
gas, at low energies it is appropriate to focus on the dominant
$s$-wave channel, which by virtue of Pauli principle automatically
also selects the singlet two-atom states.  Ignoring all other
scattering channels in the model \rf{eq:Hgeneral}, the $s$-wave
two-channel model Hamiltonian reduces to
\begin{widetext}
\begin{eqnarray}
\label{eq:ham_s}\hat H_{s}^{2-ch} &=& \sum_{\k, \sigma} {k^2 \over 2m} ~\hat
a^\dagger_{\k,\sigma} \hat a_{\k,\sigma} + \sum_\p \left(\epsilon_0
+{p^2 \over 4m} \right) \hat b_\p^\dagger \hat b_\p
+\sum_{\k,\p}
~{g_s \over \sqrt{V}}
\left(\hat b_\p ~ \hat a^\dagger_{\k+\frac{\p}{2},\uparrow}
\hat a^\dagger_{-\k+\frac{\p}{2}, \downarrow} +
\hat b^\dagger_\p ~ \hat a_{-\k+\frac{\p}{2},\downarrow}
\hat a_{\k+\frac{\p}{2},\uparrow}\right),
\end{eqnarray}
\end{widetext}
where to simplify notation we defined
\begin{eqnarray}
\hat b_\p &\equiv& \hat b_{\p,0,0}^{(0)},\\
g_s\theta(\Lambda-k) &\equiv& \frac{1}{\sqrt{2\pi}}g_k^{(0)},
\end{eqnarray}
incorporating the short-scale (shorter than the atomic interaction
range $d=2\pi/\Lambda$) falloff of the Feshbach resonant coupling
$g_k^{(0)}$ as an implicit sharp cutoff at $\Lambda$ on the momentum
sums.

\begin{figure}[bt]
\includegraphics[height=.5in]{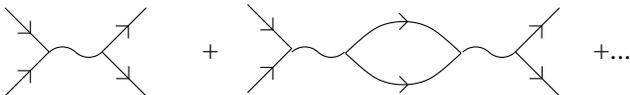}
\caption{\label{Fig-twochannel} The diagrams contributing to the
$T$-matrix of the two-channel model \rfs{eq:ham_s}. The straight and
wavy lines represent fermionic and bosonic Green's functions,
respectively.}
\end{figure}

Within this model, the fermions of the same spin do not interact at
all, and the scattering amplitude of two fermions of opposite spin can
be calculated exactly. In addition to the free fermion Green's
function \rfs{eq:greenferm}, the Green's function for a free boson is
given by
\begin{equation}
D_0(\p,\omega) = \frac{1}{\omega-\frac{p^2}{4m}-\epsilon_0+i0}.
\end{equation}
The $T$-matrix is then given by a geometric series depicted in
Fig.~\ref{Fig-twochannel}, and written algebraically as
\begin{equation} \label{eq:twochantmat}
T(\k, \k') = g_s D_0 g_s  + g_s D_0 g_s \Pi_s g_s D_0 g_s + \dots = \frac 1
{\frac 1 {g_s^2} D_0^{-1} - \Pi_s}.
\end{equation}
Above,
\begin{eqnarray}
D_0&=&D_0\bigg(\p,\ve+\frac{p^2}{4m}\bigg)=D_0({\bf 0},k^2/m),\\
\Pi_s&=&\Pi_s(k^2/m),
\end{eqnarray}
latter defined in \rfs{eq:polarization1} and $\ve=k^2/m$ is the
two-atom center of mass energy.  The scattering amplitude is then
given exactly by the low-to-intermediate energy form in
\rfs{eq:swaveintro}, with the $s$-wave scattering length and effective
range given by
\begin{eqnarray}
\label{eq:fanolengthrange1}
a&=&-\frac{m g_s^2}{4\pi \left( \epsilon_0 -
\frac{g_s^2 m \Lambda }{2\pi^2}\right)}=\frac{2}{m r_0 \omega_0},\\
\label{eq:r02Ch}
r_0 &=& - \frac{8\pi}{m^2 g_s^2}.
\end{eqnarray}
We note that as expected from general considerations of~Sec.~\ref{swaveMidE},
the effective range parameter $r_0$ is indeed
{\em negative}. Proportional to $1/g_s^2$, it controls the lifetime of
the $b$-bosons to decay into two atoms, corresponding to the inverse
width of the Feshbach resonance. $r_0$ therefore becomes arbitrarily
{\em long} (compared to the microscopic range $1/\Lambda$) of the
potential with decreasing $g_s$. We recall from Sec.\ref{swaveMidE} that
these two conditions are precisely those required for the resonance to
exhibit a positive energy, finite lifetime resonant state.

From Eqs.(\ref{eq:fanolengthrange1},\ref{eq:r02Ch}) we also identify
the characteristic crossover energy scale $\Gamma_0=4/(m r_0^2)$ and
the parameter $\omega_0=2/(m r_0 a)$ appearing in the $s$-wave
scattering amplitude \rf{fsE}, given by
\begin{eqnarray}
\label{eq:Gamma0}
\Gamma_0&=&\frac{m^3 g_s^4}{16\pi^2},\\
\label{eq:omega0}
\omega_0 &=&\epsilon_0 - \frac{g_s^2 m \Lambda}{2\pi^2},
\end{eqnarray}
In terms of these derived quantities, all the scattering phenomenology
discussed in Sec.\ref{swaveMidE} follows immediately.

We observe that the scattering length diverges at a critical value of
the bare detuning $\epsilon_0^c = \frac{g^2 m}{2\pi^2}\Lambda$,
corresponding to the point at which the bound state
appears.  This should be contrasted with the naive expectation that the
bound state, being a $b$-particle, appears when $\epsilon_0$ goes
through zero. We refer to this shift as a renormalization of detuning
(``mass renormalization'' of the closed-channel boson $b$ in the field theory
parlance). The origin of the shift from $\epsilon_0$ to $\omega_0$
lies in the fact that the $b$-particle is, of course, not the bound
state (physical molecule) of the two $a$-particles (atoms). Rather, an
actual bound state is a superposition of a $b$-particle and a cloud of
$a$-particles. The $b$-particle only corresponds to the part of the
physical bound state (molecule) which lies within the closed-channel.
We emphasize that while the $b$-particle can be safely treated as a
point particle, whose size is related to the detuning-independent
cutoff $1/\Lambda$, the size of the actual bound state (physical
molecule) can get arbitrarily large, with its size diverging with $a
\rightarrow \infty$, as is further discussed in Appendix~\ref{Ap:BVP}.

Since $a$ diverges where the parameter $\omega_0$ vanishes, we
identify this additively-renormalized detuning parameter $\omega_0$
with the physical detuning corresponding to Zeeman energy splitting
between closed- and open channels, controlled by the magnetic field
$H$ and vanishing at field $H_0$. Comparing the prediction
\rf{eq:fanolengthrange1} for $a$ with its empirical form,
\rfs{as}, \cite{commentExpWidth}, allows us to identify parameters of the
two-channel model with the experimental parameters according to
\begin{eqnarray}
\omega_0&\approx& 2\mu_B (H-H_0),\\
\Gamma_0&\approx& 4m\mu_B^2a_{bg}^2 H_w^2/\hbar^2,
\end{eqnarray}
where we estimated the magnetic moment responsible for the Zeeman
splitting between the open- and closed-channels (corresponding
respectively to electron spin triplet and singlet, respectively) to be
$2\mu_B$.

Hence, the conclusion is that indeed the two-channel model faithfully
describes a scattering in the presence of a resonant state, as well as
a bound and virtual bound states, depending on the value of the
detuning parameter. It is thus a sufficient model to capture all the
generic features of a Feshbach-resonant atomic gas, without resorting
to a fully microscopic (and therefore typically intractable)
description. This should be contrasted with the one-channel model
\rfs{eq:singlechannelham}, which is only able to capture scattering in
the absence of a resonant state, i.e., only in presence of either
bound or virtual-bound states and as such insufficient to capture an
intermediate energy behavior of a Feshbach-resonant atomic gas.

We close this section with a comment. In the literature it is common
to study models that in addition to the two-channel Feshbach resonant
interaction considered above, a featureless nonresonant four-Fermi
atomic interaction is also included. It is simple to show that in
three-dimensions, doing so does not add any new physics to the pure
two-channel model considered here. Instead it just amounts to
redefining the relation between model's parameters ($\epsilon_0$, $g_s$,
$\lambda$) and the experimentally determined parameters $\omega_0$ and
$r_0$. Please see Appendix~\ref{AppendixHybrid} for details.

\subsubsection{Model at finite density: small parameter}
\label{s-wave2CMparameter}

As already discussed in the Introduction, a finite density $s$-wave
resonant Fermi gas, described by a two-channel model, \rfs{eq:ham_s} is characterized by an average atom spacing
$n^{-1/3}\sim 1/k_F$ in addition to the scattering length $a$ and the
effective range $r_0$, derived above and discussed phenomenologically
in Sec.~\ref{sec:RRST}. Hence, in addition to the dimensionless gas
parameter $k_F a$ discussed in Sec.~\ref{s-wave1CMparameter}, a
two-channel model admits another key dimensionless parameter $\gamma_s
\propto 1/(\kf |r_0|)$ that is the ratio of the average atom spacing
$\kf^{-1}$ to the effective range length $|r_0|$.  Equivalently,
$\gamma_s$ is related to the square-root of the ratio of the Feshbach
resonance width $\Gamma_0$ (controlled by the Feshbach
resonance coupling $g_s$ and defined by \rfs{fsE}) to the Fermi energy,
and equivalently to the ratio of the resonance width (at the Fermi energy)
$\sqrt{\Gamma_0\ef}$ to the Fermi energy:
\be
\gamma_s \equiv  \frac{g^2 N(\ef)}{\ef}
= \frac{\sqrt{8}}{\pi} \sqrt{\frac{\width}{\ef}}
=\frac{g^2c}{\sqrt{\ef}} = \frac{8}{\pi} \frac{1}{\kf |r_0|}.
\label{eq:gammadef}
\ee

The two-channel model, \rfs{eq:ham_s} is described by an interacting
Hamiltonian, whose interaction strength is controlled by a coupling
$g_s$. The corresponding dimensionless parameter $\gamma_s \propto g_s^2$
controls a perturbative expansion in the Feshbach resonant interaction
(about an exactly solvable non-interacting $g_s=0$ limit) of any
physical quantity.  The key observation is that $\gamma_s$ is
independent of the scattering length $\as$ and detuning $\omega_0$,
and as such, if indeed small, remains small throughout the crossover,
even for a Feshbach resonance tuned through zero energy, where the
scattering length $a$ and the gas parameter $|a|n^{1/3}$ diverge.

Hence, we arrive an important conclusion: the two-channel model
predictions for a {\em narrow} Feshbach resonance, (defined by
$\gamma_s \ll 1$, i.e., width of the resonance $\Gamma_0$ much smaller
than the Fermi energy, or equivalently effective range $r_0$ much
longer than atom spacing $n^{-1/3}$) are {\em quantitatively } accurate
throughout the BCS-BEC crossover, no matter how large the value of the
gas parameter $|a|n^{1/3}$ gets.

As discussed in Sec.\ref{s-wave1CMparameter}, this availability of
small parameter in the two-channel model contrasts strongly with the
one-channel model, characterized by a dimensionless gas parameter
$|a|n^{1/3}$ that diverges for a Feshbach resonance tuned to zero (i.e.,
\lq\lq on resonance\rq\rq), that therefore does {\em not} admit a
small perturbative expansion parameter throughout the entire crossover (with the exception of
deep BCS and deep BEC regimes).

In contrast, for   the broad-resonance $\gamma_s \gg 1$ system, the two-channel model is
no more solvable than the one-channel model; in fact, as we will show
in the next subsection, in this limit the two models become
identical. The perturbatively accessible and nonperturbative regions
of the two-channel model in the $k_F$ and $a^{-1}$ parameter space
are illustrated in Fig.\ref{fig:perturbative}. In terms of the
Fig.\ref{Fig-polesEr}, the broad and narrow resonance limits
respectively correspond to $k_F$ falling inside and outside the
virtual bound state regime, defined by $1/|a| < 1/(2|r_0|)$.

The dimensionless parameter $\gamma_s$ naturally emerges in a
perturbative expansion in atom-molecule coupling. More physically, it
can also be deduced by estimating the ratio of the energy associated
with the atom-molecule Feshbach-resonance interaction to the kinetic
energy, i.e, the non-interacting part of the Hamiltonian
Eq.~(\ref{eq:ham_s}). To see this, note that the atom-molecule
coupling energy $E^s_{FR}$ per atom scales like
\be
E^s_{FR} \sim g_s n^{1/2},
\ee
where we estimated the value of $\bh(\r)$ by $\bh(\r)\sim \sqrt{n}$. This
interaction energy is to be compared to the non-interacting part of
the Hamiltonian, i.e., the kinetic energy per atom
\be
E_0 \sim \epsilon_F,
\ee
with the square of the ratio
\bea
\gamma_s&\sim& (E^s_{FR}/E_0)^2,
\\
&\sim& g_s^2 n/\ef^2\sim {m^2 g_s^2\over k_F},
\eea
giving the scale of the dimensionless parameter $\gamma_s$ in
Eq.~(\ref{eq:gammadef}).

In the spirit of the discussion in Sec.~\ref{s-wave1CMparameter},
another instructive way to estimate the interaction strength and to
derive the dimensionless coupling that controls the two-channel
model's perturbation theory is to integrate out (in a coherent-state
path-integral sense) the closed-channel molecular field $b(\r)$ from
the action. As $b(\r)$ couples to atoms only linearly this can be done
exactly by a simple Gaussian integration. The resulting action only
involves fermionic atoms that interact via an effective four-Fermi
{\it dispersive\/} vertex.  After incorporating fermion-bubble
self-energy corrections of the $T$-matrix the latter is given by $T_{k_F}
\approx (4\pi/m)f_s(\kf)$, with a key factor that is the
finite-density analog of the scattering amplitude, $f_s(k)$,
Eq.~(\ref{eq:amplitudeswaver0}). To gauge the strength of the molecule-mediated
interaction energy we compare the interaction per atom
$(4\pi/m)f_s({k_F})n$ to the kinetic energy per atom $\ef$.  Hence,
dropping numerical prefactors, the dimensionless coupling that is a
measure of the atomic interaction, is given by
\bea
\lambdah^s_{k_F} &\equiv&\frac{4\pi n}{m\ef}|f_s(\kf)| , \\
&\sim & k_F |f_s(k_F)|.
\eea
At large detuning (i.e., deep in the BCS regime) $\lambdah^s_{k_F}\sim \kf
|\as|\ll 1$ and the theory is perturbative in $\lambdah^s_{k_F}$.  However,
as detuning is reduced $|\as(\omega_0)|$ and $\lambdah^s_{k_F}(\omega_0)$
grow, and close to the resonance $\as^{-1}$ may be neglected in the
denominator of Eq.~(\ref{fsk}).  In this regime, the coupling
saturates at $\lambdah_{k_F}^\infty$:
\be
\lambdah_{k_F}^\infty\sim \frac{\kf}{|r_0 \kf^2/2 - i\kf|},
\ee
whose magnitude crucially depends on the dimensionless ratio
$\gamma_s \propto 1/(k_F |r_0|)$, with
\bea
\label{six}
\hspace{-0.7cm}
\lambdah_{k_F}^\infty
&\sim&
\cases{\frac{1}{r_0\kf},
& \text{for $|r_0| \kf \gg 1$,}\cr
1,
& \text{for $|r_0| \kf \ll 1$.}\cr}
\eea
Hence, in contrast to a two-particle vacuum scattering, in which the
cross section diverges when the Feshbach resonance is tuned to zero
energy, at finite density, for sufficiently large $\as$, the effective
coupling $\lambdah^s_{k_F}$ ceases to grow and saturates at
$\lambdah_{k_F}^\infty$, with the saturation value depending on whether
this growth is cut off by the atom spacing $1/\kf$ or the effective
range $r_0$. The former case corresponds to a narrow resonance
[$\gamma_s \propto (|r_0|\kf)^{-1} \ll 1$], with the interaction
remaining weak (and therefore perturbative) throughout the BCS-BEC
crossover, right through the strong-scattering $1/(\kf |\as|)=0$
point. In contrast, in the latter wide-resonance case [$\gamma_s \propto
(|r_0|\kf)^{-1} \gg 1$], discussed in Sec.\ref{s-wave1CMparameter},
sufficiently close to the unitary point $1/\as=0$ the effective
coupling $\lambdah_{k_F}^\infty$, Eq.~(\ref{six}), grows to $\curO(1)$
precluding a perturbative expansion in atom interaction near the
unitary point.

\subsubsection{Relation to one-channel model}
\label{relation1CM}

In this section we would like to demonstrate that in the  broad-resonance limit, of relevance to most
experimentally-realized Feshbach resonances to date, the $r_0 k^2$
contribution to the dispersion (arising from the molecular kinetic
energy) of the effective coupling $\lambdah^s_k$ can be neglected and
one obtains an effective single (open-) channel description. Thus the one and two channel models
are equivalent in the limit $\gamma_s \rightarrow \infty$.

The reduction to a single-channel model in the broad resonance limit
can be executed in an operator formalism, with the derivation becoming
exact in the infinite Feshbach resonance width
($\gamma_s\rightarrow\infty$) limit. (For this same reduction in the functional integral formalism,
see Appendix A of Ref.~\cite{Levinsen2006}.)
The expression for the scattering length, \rfs{eq:fanolengthrange1}
\be
\frac{1}{\as} = - \frac{4\pi}{m g_s^2}\big(\epsilon_0
- \frac{g_s^2m\Lambda}{2\pi^2}
\big),
\label{eq:as2}
\ee
dictates that a proper transition to the broad resonance limit
corresponds to $g_s\to\infty$ while adjusting the bare detuning
according to
\be
\epsilon_0 =  -\frac{g_s^2}{\lambda},
\label{eq:connection}
\ee
such that the physical scattering length $\as$ remains fixed. This
allows us to trade the bare detuning $\epsilon_0$ and coupling $g_s$ for
a new coupling $\lambda$ that physically corresponds to a non-resonant
attractive interaction depth, that can be used to tune the scattering
length.  The Heisenberg equation of motion governing the molecular
field $\bh_\p$ dynamics under Hamiltonian \rf{eq:ham_s}, with
condition Eq.~(\ref{eq:connection}), is given by:
%
\bea
&&\hspace{-1.2cm} \dot{\bh}_{\bf p} = -i\big[\bh_\p,\hat H_{s}^{2-ch}\big],
\\
&&\hspace{-.8cm}\,=
-i\Big[\bigg(\frac{p^2}{4m} - \frac{g_s^2}{\lambda}\bigg) \bh_\p +
{g_s\over V^{1/2}} \sum_\k
\ah_{-\k+\frac{\p}{2}\downarrow}^{\phdag}\ah_{\k+\frac{\p}{2}\uparrow}^{\phdag}
\Big].
\label{moleom}
\eea
%
Now, in the large $g_s \to\infty$ limit (keeping $\lambda$ fixed) the
molecular kinetic energy term $\propto p^2/4m$ on the right and the
$\dot{\bh}_{\bf p}$ term on the left are clearly subdominant, reducing the
Heisenberg equation to a simple constraint relation
\be
\bh_\p = \frac{\lambda}{g_s V^{1/2}} \sum_\k
\ah_{-\k+\frac{\p}{2}\downarrow}^{\phdag}
\ah_{\k+\frac{\p}{2}\uparrow}^{\phdag}.
\label{318}
\ee
Hence, we see that in the extreme broad-resonance limit the molecular
field's dynamics is \lq\lq slaved\rq\rq\ to that of the pair of atoms,
according to Eq.~(\ref{318}). Substituting this constraint into the
Hamiltonian, \rf{eq:ham_s} allows us to eliminate the closed-channel
molecular field in favor of a purely open-channel atomic model with
the Hamiltonian
\begin{widetext}
\begin{eqnarray}
\label{eq:ham_s1ch}
\hat H_{s}^{1-ch} &=& \sum_{\k, \sigma} {k^2 \over 2m} ~\hat
a^\dagger_{\k,\sigma} \hat a_{\k,\sigma}
+ {\lambda\over V}\sum_{\k,\k',\p}
\ah_{\k'+\frac{\p}{2}\uparrow}^{\dagger}
\ah_{-\k'+\frac{\p}{2}\downarrow}^{\dagger}
\ah_{-\k+\frac{\p}{2}\downarrow}
\ah_{\k+\frac{\p}{2}\uparrow}^{\phdag}
\end{eqnarray}
\end{widetext}
a momentum space version of the one-channel model,
\rfs{eq:singlechannelham} discussed in Sec.\rf{OneCM}.

A clear advantage of the one-channel model is that, as shown above, it
naturally emerges as the correct Hamiltonian in the
experimentally-relevant case of a wide resonance, $\gamma_s \gg 1$.
However, as discussed in Sec.~\ref{s-wave1CMparameter}, a notable
disadvantage is that, in the most interesting regime of a Feshbach
resonance tuned to zero energy, its dimensionless gas parameter $\kf
|\as|\to \infty$  precluding a controlled perturbative calculation
throughout the BCS-BEC crossover.

\subsection{$P$-wave}
\label{Sec:PWaveTwoChannel}
\subsubsection{Two-atom scattering}

As discussed in the Introduction, for a single component Fermi gas
Pauli principle forbids interaction in the $s$-wave channel, and,
consequently the dominant interaction is in the $p$-wave
channel.  In addition to this motivation, a study
of a $p$-wave resonance is attractive because, as we will see below,
(and is already clear from scattering phenomenology discussion in
Sec.~\ref{Pscattering}) they can in principle be made arbitrarily
narrow by simply decreasing the particle density (as opposed to
increasing $n$ in the $s$-wave case) and therefore are amanable to a
quantitatively accurate description possibly in experimentally
accessible regimes. Finally, as we will see, $p$-wave superfluids
exhibit richer set of possibilities and thereby allow genuine phase
transitions (some quite exotic), not just crossover as a function of
detuning.

With this motivation in mind, in this section we focus on the dominant
$p$-wave, $s_z=+1$ triplet channel (a gas of atoms in a single
hyperfine state $\uparrow$) in the model \rf{eq:Hgeneral}, described
by the following $p$-wave Hamiltonian
\begin{widetext}
\begin{eqnarray}
\label{eq:ham_p} \hat H_{p}^{2-ch} &=& \sum_{\k} {k^2 \over 2m} ~\hat
a^\dagger_{\k,\sigma} \hat a_{\k,\sigma}
+ \sum_{\p,\alpha} \left(\epsilon_\alpha+{p^2 \over 4m}\right)
\hat b_{\p,\alpha}^\dagger \hat b_{\p,\alpha}
+\sum_{\k,\p}~{g_p \over \sqrt{V}}\
k_\alpha\left(\hat b_{\p,\alpha} ~ \hat a^\dagger_{\k+\frac{\p}{2}}
\hat a^\dagger_{-\k+\frac{\p}{2}} +
\hat b^\dagger_{\p,\alpha} ~ \hat a_{-\k+\frac{\p}{2}}
\hat a_{\k+\frac{\p}{2}}\right).\nonumber\\
\end{eqnarray}
\end{widetext}
Here, as before, we defined the $p$-wave coupling to be $g_p$, where the subscript $p$ refers to the ``$p$-wave",
not to be confused with momentum.
To simplify notation, we defined three (cartesian tensor
components) $p$-wave bosonic operators ($\alpha=x,y,z$) in terms of
the three bosonic (closed-channel) operators $\hat
b_{\p,1,m}^{(s_z=+1)}$ with definite projections of orbital angular
momentum, $m=(\pm 1,0)$ (and $s_z=+1$), defined in
Sec.\ref{TwoCM}
\begin{eqnarray} \label{relationsbetweenmuandalpha1}
\hat b_{\p,x} &\equiv& \frac{1}{\sqrt{2}}
\bigg(\hat b_{\p,1,1}^{(+1)}+\hat b_{\p,1,-1}^{(+1)}\bigg)\ ,\\
\label{relationsbetweenmuandalpha2}
\hat b_{\p,y} &\equiv& -\frac{i}{\sqrt{2}}
\bigg(\hat b_{\p,1,1}^{(+1)}-\hat b_{\p,1,-1}^{(+1)}\bigg)\ ,\\
\label{relationsbetweenmuandalpha3}
\hat b_{\p,z} &\equiv& \hat b_{\p,1,0}^{(+1)}\ ,
\end{eqnarray}
and we have dropped the hyperfine subscript $\uparrow$ on these molecular
operators.  We also defined the corresponding Feshbach resonance
coupling and bare detunings
\begin{eqnarray}
g_p\theta(\Lambda-k) &\equiv& \sqrt{\frac{3}{4\pi}}g_k^{(1)},\\
\epsilon_z&=&\epsilon_0^{(1,0)},\\
\epsilon_{x,y}=\epsilon_\perp
&=&\frac{1}{2}\big(\epsilon_0^{(1,1)}+\epsilon_0^{(1,-1)}\big),
\end{eqnarray}
incorporating the short-scale (shorter than the atomic interaction
range $d=2\pi/\Lambda$) falloff of the Feshbach resonant coupling
$g_k^{(1)}$ as an implicit sharp cutoff at $\Lambda$ on the momentum
sums. The coupling $g_p$ is the amplitude for the transition between a
pair of identical fermionic atoms with one unit of orbital (relative)
angular momentum into a closed-channel molecule with an internal
angular momentum $\ell=1$.

In $H_p$, \rfs{eq:ham_p}, we have specialized to the experimentally
relevant time-reversal invariant Hamiltonian~\cite{Ticknor2004} for
degenerate $m=\pm1$ resonances and thereby omitted a contribution
\begin{equation}
\hat H_{t-break}=\frac{i}{2}\big(\epsilon_0^{(1,+1)}-\epsilon_0^{(1,-1)}\big)
\sum_{\p}(\hat b_{\p,y}^\dagger \hat b_{\p,x} -
\hat b_{\p,x}^\dagger \hat b_{\p,y}),
\end{equation}
that vanishes in the case $\epsilon_0^{(1,+1)}=\epsilon_0^{(1,-1)}$ of
interest to us here.

By construction, the fermionic atoms ($a$-particles) scatter only in
the $p$-wave channel. The scattering amplitude can be easily
calculated in the same $T$-matrix formalism, as in the $s$-wave case,
\rfs{eq:twochantmat}.

The propagator of the $b_\alpha$-particles is given by
\begin{equation}
D_{\alpha \beta}(\p,\omega) = \frac{ \delta_{\alpha \beta}}{
\omega-\frac{p^2}{4m}-\epsilon_\alpha+i0}
\equiv D_\alpha(\p,\omega)\delta_{\alpha,\beta},
\end{equation}
Graphically, the $T$-matrix is represented by the geometric series in
Fig.~\ref{Fig-twochannel}, with vertices proportional to $k_\alpha$. It is given by
\begin{widetext}
\begin{equation} \label{eq:pwaveT1}
T_{{\bf k}, {\bf k'}} =  2g_p^2\sum_{\alpha} k_{\alpha} D_\alpha k'_{\alpha}
+ 2 g_p^4 V^{-1}\sum_{{\bf q},\alpha,\beta}k_\alpha D_\alpha q_{\alpha}\ 2\Pi\
q_{\beta} D_\beta k'_{\beta} + \dots = \sum_{\alpha} {2g_p^2 k_\alpha
k'_{\alpha} \over D_\alpha^{-1}-\frac{2}{3}g_p^2V^{-1}\sum_\q q^2 \Pi},
\end{equation}
\end{widetext}
where $D_\alpha$ stands for $D_\alpha\left({\bf
    0},\frac{k^2}{m}\right)$, $\Pi$ stands for the polarization bubble
$\Pi\left(\q,\frac{k^2}{m}\right)$, defined in \rfs{eq:defpi},
$\alpha=x,y,z$, and overall factor of $2$ comes from the definition of
the $T$-matrix in this many-body language (see factor of $1/2$ in the
definition of the interaction term in \rfs{eq:leoham}). A related
symmetry factor of $2$ appearing in front of $\Pi$ in \rfs{eq:pwaveT1}
is also a consequence of identical fermions appearing the diagrams in
Fig.~\ref{Fig-twochannel}, that allows two possible contractions of
atomic lines inside $\Pi$, which contrasts to one such contraction for
$s$-wave scattering of atoms distinguished by (hyperfine-) spin.

Calculating the momentum $\q$ sum in the $p$-wave polarization bubble in
the denominator of \rfs{eq:pwaveT1}, we find
\begin{eqnarray}
\label{eq:polap}
\Pi_p(\ve)&\equiv&\frac{1}{V}\sum_{\q} q^2\Pi(\q,\ve),\nonumber\\
&=&\int {d^3 q \over (2 \pi)^3} {q^2\over \ve -{q^2 \over m}+i0}\nonumber\\
&=& - {m \Lambda^3 \over 6 \pi^2} - { m^2 \Lambda \over  2 \pi^2}\ \ve
- i { m^{5 / 2}\over 4\pi}\ \ve^{3/ 2} .
\end{eqnarray}
where as before $\ve=k^2/m$ is the molecule's internal energy in the
center of mass frame.  Just as in \rfs{eq:polarization1} in
$\Pi_p(\ve)$ we have cut off the (otherwise ultra-violently divergent)
integral at high momentum $\Lambda$ corresponding to the inverse
(closed-channel) molecular size, with the calculation (and the whole
approach of treating $b$ as a point particle) valid only as long as
$\ve \ll \Lambda^2/m$. However, in contrast to the $s$-wave,
\rfs{eq:polarization1}, here the integral for the $p$-wave case scales
as $p^3$ at large momenta. As a result, in addition to the constant
contribution (first $\Lambda^3$ term that is analogous to linear
$\Lambda$ term, Eqs.\rf{eq:polarization1}, \rf{eq:omega0}) that leads
to the detuning shift, the polarization bubble shows a second
$\Lambda$-dependent contribution that multiplicatively renormalizes
the molecular dispersion. For a future reference, we introduce two
cutoff-dependent parameters related to these two terms
\begin{eqnarray}
\label{eq:c1}
c_1 &=& { m \over 9 \pi^2} {\Lambda^3 g_p^2}, \ \\
\label{eq:c2}
c_2 &=& { m^2 \over 3 \pi^2} g_p^2 \Lambda\ ,\nonumber\\
&\equiv&\frac{\Lambda}{k_g}\ ,
\end{eqnarray}
where $c_1$ is a constant with dimensions of energy, $c_2$ is
an important dimensionless constant and we defined a new momentum scale
\begin{equation}
k_g =\frac{3\pi^2}{m^2 g_p^2}.
\end{equation}

The two-body scattering amplitude is obtained through its relation
$f(\k,\k') = - {m\over 4 \pi} T_{\k,\k'}$, \rfs{eq:scatampT} to the
$T$-matrix. Combining this with \rfs{eq:pwaveT1}, \rfs{eq:polap} and
Eqs.~\rf{eq:c1}, \rf{eq:c2}, we thereby obtain
\begin{equation}
\label{eq:fkaka}
f(\k,\k') =
\sum_\alpha{3 k_\alpha k'_\alpha\over {6 \pi \over m g_p^2}
\left(\epsilon_\alpha-c_1 \right) - {6 \pi \over m^2 g_p^2}(1 + c_2)k^2- i k^3}.
\end{equation}

For an isotropic interaction, $m=0,\pm 1$, Feshbach resonances are
degenerate, $\ve_\alpha=\ve_0$, and the scattering amplitude is (not
surprisingly) entirely in the $p$-wave channel
\begin{equation}
f({\bf k},{\bf k'})=3 f_p(k) \cos(\theta),
\end{equation}
where $\theta$ is the angle between momenta $\k$ and $\k'$ before and
after the scattering event. The partial wave scattering amplitude
$f_p(k)$ in the $p$-wave channel, as follows from
\rfs{eq:partialwaves} and \rf{Tpartialwaves}, is given by
\begin{equation}
\label{eq:fp2ch} f_p(k) = {k^2 \over {6 \pi \over m g_p^2}
\left(\epsilon_0-c_1 \right) - {6 \pi \over m^2 g_p^2} \left(1 + c_2
\right)k^2- i k^3}.
\end{equation}
Therefore, as argued on general grounds, the scattering of identical
fermionic atoms is indeed exactly of the form \rfs{eq:scatampp}, with
\begin{eqnarray}
\label{eq:v}
v^{-1}&=&-{6 \pi \over m g_p^2} \left( \epsilon_0-c_1
\right)\ , \\
\label{eq:k0}
k_0 &=& -{12 \pi \over m^2 g_p^2} \left(1 + c_2 \right) ,\nonumber\\
&=&-{4\over\pi} k_g \left(1 + \frac{\Lambda}{k_g} \right) ,
\end{eqnarray}
and this result essentially {\em exact}, valid on all momentum scales
up to the cutoff $\Lambda$. As required for a resonant state, indeed $k_0
< 0$ is negative definite.

We note that $k_0$ is the characteristic momentum scale beyond which
the width of the resonance $k^3$ becomes larger than its energy $k_0
k^2/2$, i.e., a crossover scale beyond which the resonant state
disappears. It is clear from its form, \rfs{eq:k0} that $k_0$ is given
by the following limits
\begin{eqnarray}
\label{k0asymp}
k_0&=&
-\frac{4}{\pi}\times \cases{
k_g, & \text{for $k_g \gg \Lambda$,}\cr
\Lambda, & \text{for $k_g \ll \Lambda$,}\cr
}
\end{eqnarray}
depending on the ratio $\Lambda/k_g$, but with $k_0 \ge \Lambda$ for
all $k_g$, set by Feshbach resonance coupling $g_p$.

It is useful to introduce the physical detuning $\omega_0$
\begin{equation} \label{eq:omegap}
\omega_0 = {\epsilon_0-c_1 \over 1+c_2},
\end{equation}
that corresponds to the energy of the pole in $f_p(k)$, \rfs{eq:fp2ch}
when this pole is tuned to low energy.  In terms of the detuning
$\omega_0$, the $p$-wave scattering amplitude is given by
\begin{equation}
\label{eq:scatTrenorm} f_p(k) ={k^2 \over {6 \pi \over m g_p^2}
\left(1+c_2 \right) \left( \omega_0 - {k^2 \over  m} \right) - i
k^3}.
\end{equation}
Adjusting $\omega_0$ from negative to positive, turns the scattering
in the presence of a low-lying bound state at $-|\omega_0|$ into the
scattering in the presence of a resonance at $\omega_0$.

Thus, the $p$-wave two-channel model \rfs{eq:ham_p} captures the most
general low-energy scattering in almost exactly the same way as its
$s$-wave counterpart does. The most obvious difference from the
$s$-wave model lies in how the cutoff $\Lambda$ enters the scattering
amplitude. In the $s$-wave case $\Lambda$ could be eliminated via a
redefinition of the detuning energy from $\epsilon_0$ to $\omega_0$,
as in \rfs{eq:omega0}, and thereby disappears from all other
computations.  In the $p$-wave case, however, $\Lambda$ enters the
scattering amplitude not only additively but also {\em
  multiplicatively}, and therefore explicitly appears in the
scattering amplitude $f_p(k)$, \rfs{eq:scatTrenorm} through the
dimensionless parameter $c_2$, even after the shift to physical
detuning $\omega_0$.

Interestingly, appearance of $c_2$ persists when we calculate the
phases of the $p$-wave condensate later in this paper. While the
parameter $c_1$ drops out of all predictions when written in terms of
physical parameters, the dimensionless parameter $c_2$, controlled by
the closed-channel cutoff $\Lambda$ and coupling $g_p$ continues to
appear explicitly.  Unfortunately, it is not easy to extract $c_2$
from experimental measurements of the scattering amplitude, as it
enters the amplitude in the combination $(1+c_2)/(m^2 g_p^2)$. We note
that if $c_2\ll 1$ then it and the uv-cutoff $\Lambda$ indeed drop out
from all physical quantities with $k_0\sim -k_g$. However, if $c_2 \gg
1$, then $k_0\sim -(1+c_2)/(m^2 g_p^2) \sim -\Lambda$ reduces to a
quantity that is completely independent of $g$. Indeed in this limit,
the bare dispersion of the closed-channel $b$-field can be ignored
(in comparison to the polarization bubble $\Lambda$-dependent
corrections) and the field $b$ can be integrated out, just
like in the strong coupling $s$-wave two-channel model, and leads to a
$p$-wave single-channel model analog of \rf{eq:singlechannelham}.

\subsubsection{Model at finite density: small parameter}
\label{sec:pwave2Cfinitedensity}

A finite density $p$-wave resonant Fermi gas is characterized by the
following three length scales: the average atom spacing (Fermi
wavelength) $1/k_F$, the analog of effective range (characterizing
resonance intrinsic width $\Gamma_0$) $1/k_0$, and the scattering
length ${\left| v \right|}^{1/3}$. Consequently, we can form two dimensionless
constants. One is the $p$-wave gas parameter $k_F {\left| v \right|}^{1/3}$, that,
although small for large positive and negative detuning, diverges near
the resonance (tuned to low energy), and thereby precludes a
controlled perturbative expansion in $k_F {\left| v \right|}^{1/3}$ (or equivalently in
$n v$) throughout the phase diagram. However, in the two-channel
$p$-wave model the second dimensionless parameter, defined in \rfs{gamma_p} and approximately
given by
\begin{equation}
\label{gamma_p-approx}
\gamma_p\sim \frac{k_F}{k_0},
\end{equation}
( that is approximately independent of $\omega_0$ and $v$), offers such a
controlled expansion even in the region where $v$ diverges.

As in the $s$-wave case, we can get a better physical sense of the
dimensionless parameter that controls perturbation theory by looking
at the ratio of the $p$-wave Feshbach resonance interaction energy
\be
E^p_{FR} \sim g_p n^{1/2} k_F,
\ee
to the typical kinetic energy per atom
\be
E_0 \sim \epsilon_F,
\ee
where we estimated the value of $\bh(\r)$ by $\bh(\r)\sim \sqrt{n}$.
and took the value of a typical internal momentum $k$ (appearing in
the $p$-wave vertex) to be $k_F$.  We find
\begin{eqnarray}
(E^p_{FR}/E_0)^2&\sim& g_p^2 n k_F^2/\ef^2,\nonumber\\
&\sim& m^2 g_p^2 k_F\sim m^2 g_p^2 n^{1/3},\nonumber\\
&\sim&\frac{k_F}{k_g},
\end{eqnarray}
that, as expected is indeed controlled by $\gamma_p$ in the limit of
large $k_g$, i.e., small $g_p$.

As in the $s$-wave case above we can more carefully gauge the
interaction strength and the corresponding dimensionless coupling of
the $p$-wave two-channel model by formally integrating out the
closed-channel molecular field ${\bf b}(\r)$. This leads to an
effective one-channel model, with atoms that interact via an effective
four-Fermi {\it dispersive\/} vertex.  After incorporating the
fermion-bubble self-energy corrections of the $T$-matrix the latter is
given by $T_{k_F} \approx (4\pi/m)f_p(\kf)$, with a key factor that is the
finite-density analog of the scattering amplitude, $f_p(k)$,
Eq.~(\ref{eq:scatampp}). Following the finite angular momentum one-channel
model analysis of Sec.~\ref{p-wave1CMparameter} we can gauge the
strength of the molecule-mediated interaction energy by comparing the
interaction per atom $(4\pi/m)f_p({ k_F})n$ to the kinetic energy per atom
$\ef$.  Hence, dropping numerical prefactors, the dimensionless
$p$-wave coupling that is a measure of the atomic interaction, is
given by
\bea
\lambdah_{k_F}^{(p)} &\equiv&\frac{4\pi n}{m\ef}|f_p(\kf)| , \\
&\sim & k_F |f_p(k_F)|,\nonumber\\
&\sim& \frac{\kf^3}{|v^{-1}+\frac{1}{2} k_0 \kf^2/2 - i\kf^3|},
\eea
We first note that, in principle, at high densities (energies) $k_F >
k_0$, the $k_0 k_F^2$ is subdominant and the dimensionless coupling is
given by the gas parameter $v k_F^3\sim v n$ that saturates at $1$ as
$v^{1/3}$ grows beyond atom spacing $1/k_F$. However, given the $k_0$
asymptotics, \rfs{k0asymp}, we have $k_F \ll k_0 \ge\Lambda$ and in
contrast to the $s$-wave case this nonperturbative regime is {\em
  never} accessible in the $p$-wave case. Namely, at all physically
accessible densities the width term $k_F^3$ is subdominant and the
effective coupling is given by
\begin{eqnarray}
\hspace{-1.8cm}
\lambdah_{k_F}^{(p)}
&\sim&\frac{k_F^3}{||v_\ell^{-1}|+\oh k_0 k_F^2|},\\
&\sim&
\cases{
\kf^3|v|=\frac{k_F}{k_0}(k_0 k_F^2|v|), & \text{for $k_0 k_F^2|v|\ll 1$,}\cr
\frac{k_F}{k_0}, & \text{for $k_0 k_F^2|v|\gg 1$.}\nonumber
}
\label{lambda_kp}
\end{eqnarray}

Since, as emphasized above $k_F\ll k_0$ for all densities of physical
interest (see \rfs{k0asymp}), we conclude that a $p$-wave resonant gas
is {\em always} in a perturbative regime with the perturbation theory
controlled via a small dimensionless coupling $\gamma_p$ given in
\rfs{gamma_p-approx}. It thereby allows quantitatively accurate
predictions given in powers of $\gamma_p$ (However, if $c_2 \gg 1$, this
description might go beyond mean field theory~\cite{c2largeComment}. We are grateful
to Y. Castin~\cite{Castin2006} for pointing this out to us.)

\section{$S$-Wave BCS-BEC Crossover}
\label{SWaveChapter}

In Sec.~\ref{Sec:SWaveTCM} we developed and justified the
proper two-channel model of an $s$-wave resonantly interacting atomic
gas and related its parameters to a two-body scattering experiment. We
now turn to the main goal of our work, namely a study of this model at
a fixed chemical potential, with the aim to establish the
thermodynamics of an $s$-wave resonant Fermi gas as a function of
temperature $T$, density $n$, and detuning $\omega_0$.

The thermodynamics is encoded in the partition function $Z={\rm Tr}~
e^{-\beta\hat{H}}$ and the corresponding free energy $F=-T\ln Z$.  The
partition function can be conveniently formulated in terms the
imaginary-time path-integral over coherent states labeled by
commuting closed-channel fields $\bphi(\r)$, $\bphib(\r)$ (bosonic
molecules) and anticommuting open-channel fields $\psi(\r)$,
$\psib(\r)$ (fermionic atoms), and their complex conjugates
\begin{equation}
\label{eq:partfunc}
Z_s=\int {\cal D} {\psi} {\cal D} {\psib} {\cal D} \bphi {\cal D}{\bphib}
~e^{-S_s[\phi,\psi]},
\end{equation}
with the action $S_s[\phi,\psi]$ corresponding to the Hamiltonian
$H_s^{2-ch}$, \rfs{eq:ham_s}, given by
\begin{widetext}
\begin{equation}
\label{eq:Ss}
S_s\left[\phi,\psi \right]= \int_0^\beta d\tau \int d^3 r~\left[
 \sum_{\sigma=\uparrow,\downarrow}
\bar \psi_{\sigma} \left( \d_\tau -{\grsq \over 2 m}-\mu \right)
\psi_{\sigma} +   \bar \phi
\left( \d_\tau+\epsilon_0 - 2 \mu -{\grsq \over 4 m} \right) \phi
+ g_s \left( \phi ~\bar \psi_{\uparrow} \bar \psi_{\downarrow} +
\bar \phi ~\psi_{\downarrow} \psi_{\uparrow} \right) \right],
\end{equation}
\end{widetext}
where $\beta=1/T$ is the inverse temperature.  In above, because
fermionic atoms and bosonic molecules are in equilibrium, able
interconvert into each other via the Feshbach resonant coupling $g_s$,
the chemical potential $\mu$ couples to the {\em total} conserved
particle number, that is the sum of the number of free atoms
(fermions) and twice the total number of bosons.

Given that fermions $\psi$ appear quadratically in \rfs{eq:partfunc},
they can be formally integrated out, giving an effective purely bosonic
action
\begin{equation}
\label{eq:partfunceff} Z_s = \int {\cal D} \phi {\cal D} \bar
\phi~e^{- S_{s}\left[\phi \right]},
\end{equation}
where the effective action $S_s \left[\phi \right]$ is given by
\begin{widetext}
\begin{equation}
\label{Seff} S_s\left[\phi \right]=\int_0^\beta d \tau\int d^3 r~\bar
\phi \left(\d_\tau+\epsilon_0 - 2 \mu -{\grsq \over 4 m} \right) \phi -
\tr \ln \left( \matrix { i \omega -{\grsq \over 2m} - \mu & g_s
\phi \cr g_s \bar \phi & i \omega +{\grsq \over 2m} + \mu } \right).
\end{equation}
\end{widetext}
The bosonic action $S_s[\phi]$ completely characterizes our system.
However, it is nonlinear in $\phi$, describing effective bosonic
interactions controlled by $g_s$ and therefore cannot be solved
exactly. Nevertheless it can be studied via standard many-body methods
as we will describe below.

\subsection{Infinitely-narrow resonance limit}
\label{sec:inrl}

It is enlightening to first consider the limit of a vanishingly
narrow resonance, $\gamma_s \sim {g_s^2 m^{3/2}}{\epsilon_F^{-{1/2}}}
\rightarrow 0$.  As can be seen most clearly from the original
action, \rfs{eq:Ss} or the corresponding Hamiltonian, \rfs{eq:ham_s},
in this $g_s\rightarrow 0$ limit the system breaks into two
non-interacting parts: fermionic atoms of mass $m$ and bosonic
molecules of mass $2m$. Despite of a vanishing interaction, we
emphasize an implicit order of limits here. Namely, the vanishing
interaction is still sufficiently finite so that on experimental times
scales the resulting fermion-boson mixture is nevertheless in
equilibrium, with only total number of particles (but not the separate
fermion and boson number)
\begin{equation}
\label{eq:pne} n_f+2 n_b = n,
\end{equation}
that is conserved, with $n_f$, $n_b$ the atom and molecule densities,
respectively. This key feature is captured by a common chemical
potential $\mu$ even in the $g_s \rightarrow 0$ limit.  In this limit
the boson lifetime, even for $\epsilon_0>0$ when the boson is actually
a resonance~\cite{Resonance}, becomes infinitely long (but still short
enough to establish equilibrium with fermions), and therefore the
bosons can be considered as stable particles on equal footing with
fermions. Also in this limit the distinction between parameter
$\epsilon_0$ and the renormalized physical detuning $\omega_0$,
\rfs{eq:omega0}, disappears
\begin{equation}
\epsilon_0=\omega_0.
\end{equation}

\begin{figure}[bt]
\includegraphics[height=4in]{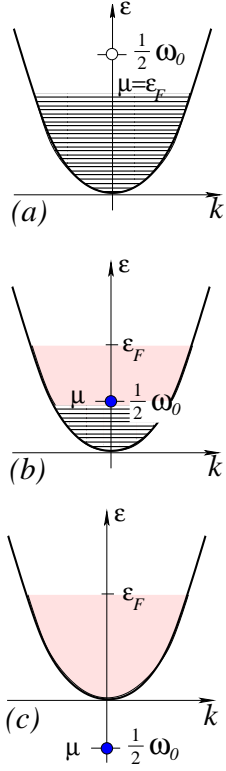}
\caption{\label{g0BCSBEC} An illustration of the BCS-BEC crossover
  in the limit of a vanishing resonance width $\gamma_s \rightarrow
  0$. The evolution with detuning $\omega_0$ is illustrated, with (a)
  the BCS regime of $\omega_0 > 2\epsilon_F$, where particles are all
  free atoms forming a Fermi sea, (b) the crossover regime of $0 <
  \omega_0 < 2\epsilon_F$, where a fraction of atoms between
  $\omega_0$ and $\epsilon_F$ have converted into BEC of bosonic
  molecules, and (c) the BEC regime of $\omega_0 < 0$, where only
  Bose-condensed molecules are present.}
\end{figure}

\begin{figure}[bt]
\includegraphics[height=2in]{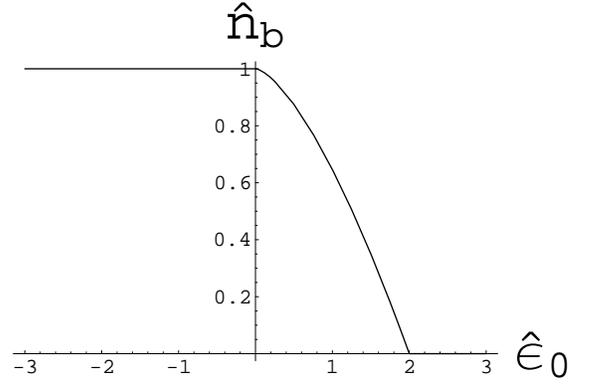}
\caption{\label{Fig-FreeCond} The normalized density of bosonic molecules $\hat n_b = 2 n_b/n$ vs
  the normalized detuning $\hat \omega_0 = \omega_0/\epsilon_F$ in
  the limit of a vanishing resonance width, $\gamma_s \rightarrow 0$
  .}
\end{figure}

At $T=0$, the condition $2\mu \le \omega_0$ holds, since the lowest
energy level of bosons cannot be negative. We therefore arrive at the
following picture illustrated in Fig.~\ref{g0BCSBEC}.
For a large positive detuning $\omega_0>2 \epsilon_F$, molecules are too
energetically costly to be produced in equilibrium, and all particles
are fermionic atoms forming a Fermi sea, with the chemical potential
locked to the Fermi energy $\mu=\epsilon_F$, with
\begin{equation}
\label{eq:fermi} \epsilon_F={ \left(3 \pi^2 n \right)^{2/3}
\over 2 m}.
\end{equation}
set completely by the total particle density $n$.  However, as
illustrated in Fig.~\ref{g0BCSBEC}, for an intermediate range of
detuning $\omega_0<2 \epsilon_F$, it becomes energetically
advantageous to convert a fraction of fermions in the Fermi sea
between $\omega_0$ and $2\epsilon_F$ into Bose-condensed molecules,
thereby keeping the effective bosonic chemical potential
$2\mu-\omega_0$ at its lowest value of zero. This atom-to-molecule
conversion regime continues as detuning is reduced, with $\mu$ locked
to $\omega_0/2$. It terminates when $\omega_0$ reaches $0$, at which
point atom-to-molecule conversion is complete and the system enters
into the BEC regime of a pure molecular condensate for $\omega_0<0$.
The full range of behavior can be summarized by the evolution of the
molecular boson density, $n_b(\omega_0)$ with detuning $\omega_0$,
that can be easily found to be
\begin{eqnarray}
\label{eq:narrowcrossover}
\hspace{-1cm}
n_b &=&
\cases{
0, \
&\text{for $\omega_0>2 \epsilon_F$} \cr
{n\over2}\left(1-\left({\omega_0\over2\epsilon_F}\right)^{3/2}\right), \
&\text{for $0 \le \omega_0 \le 2\epsilon_F$} \cr
{n \over 2}, \
&\text{for $\omega_0 <0$}.
}
\end{eqnarray}
and is displayed in Fig.~\ref{Fig-FreeCond}.

For finite temperature the chemical potential is no longer
locked to the detuning and is determined by the particle number
equation \rfs{eq:pne}, together with the noninteracting expressions
for the fermionic density
\begin{equation}
n_f = 2 \int {d^3 k \over (2 \pi)^3} {1 \over e^{{k^2\over2m T}
- {\mu \over T}}+1},
\end{equation}
and the bosonic density
\begin{equation}
n_b = \int {d^3 k \over (2 \pi)^3} {1 \over
e^{{k^2 \over 4 m T} - {2 \mu - \omega_0 \over T}} - 1}+n_0,
\end{equation}
where $n_0=|B|^2$ is, as usual, the density of the bosonic condensate.
This total number constraint must be supplemented by the free-energy
minimization rule that $n_0>0$ only if $2 \mu = \omega_0$ and
vanishes otherwise.

\begin{figure}[bt]
\includegraphics[height=2in]{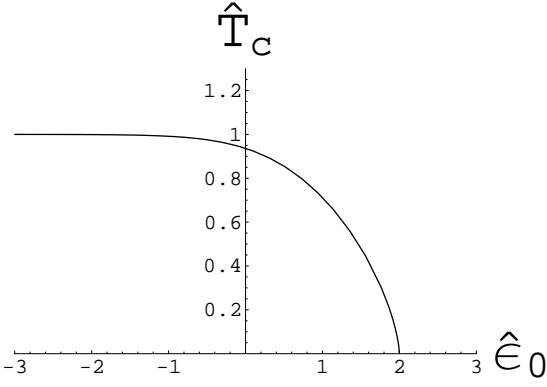}
\caption{\label{Fig-CritTempZero} The normalized critical temperature $\hat T_c=T_c/T_{\rm BEC}$ as a function
  of the normalized detuning $\hat\omega_0 = \omega_0/\epsilon_F$
  in the limit of a vanishing resonance width $\gamma_s \rightarrow
  0$.}
\end{figure}

These equations can then be used to determine the normal-superfluid
transition temperature $T_c(\omega_0)$, defined as a temperature at a
given detuning $\omega_0$ at which $n_0$ first vanishes.
Setting $2 \mu=\omega_0$ and $n_0=0$, we find an implicit equation
\begin{equation}
\label{eq:bcsbecn}
\int {d^3 k \over (2 \pi)^3} {1 \over e^{{k^2
\over  2 m T_c} - {\omega_0 \over 2 T_c}}+1} +\int {d^3 k \over
(2 \pi)^3} {1 \over e^{{k^2 \over 4 m T_c} } - 1}={n \over 2},
\end{equation}
that uniquely gives $T_c(\omega_0)$. The numerical solution of
\rfs{eq:bcsbecn} is presented in Fig.~\ref{Fig-CritTempZero}.

The limiting behavior of $T_c(\omega_0)$ is easy to deduce. Deep in
the BEC regime, for $\omega_0 \ll -\epsilon_F$, the first integral is
exponentially small, reflecting the fact that in this regime the
fermion chemical potential $\mu$ is large and negative and a number of
thermally created fermionic atoms is strongly suppressed.  The second
integral then gives the critical temperature, that in this regime
coincides with the BEC transition temperature
\begin{equation}
\label{eq:BECtt} T_{\rm c}(\omega_0\ll-\epsilon_F)
\approx T_{\rm BEC} = {\pi \over m} \left( {n \over 2 \zeta\left(
{3 / 2} \right)} \right)^{2 / 3},
\end{equation}
that is indeed on the order of $\epsilon_F$. As $\omega_0$ is
increased through the BEC and crossover regimes, $T_c(\omega_0)$
decreases, as the contribution of thermally created free atoms from
the first integral increases.  When detuning reaches $\omega_0=2
\epsilon_F$, the solution of \rfs{eq:bcsbecn} drops down to
$T_c(2\epsilon_F)=0$. Beyond this point, for $\omega_0>2 \epsilon_F$
in the BCS regime, the bosons are completely converted into free
fermions forming a Fermi sea and $T_c(\omega_0)$ sticks at $0$.

\subsection{Narrow-resonance limit}

We extend our study of the $s$-wave resonant Fermi gas, described by
the Bose-Fermi mixture action to the limit where $\gamma_s$ is small
but nonzero. The overall qualitative picture is quite similar to the
$g_s\rightarrow 0$ limit discussed in the previous subsection and
summarized in Fig.\ref{g0BCSBEC}, with only a few new features.

Because of the $\phi$ nonlinearities in $S_s[\phi]$, \rfs{Seff}, the
functional integral \rfs{eq:partfunceff} in general cannot be
evaluated exactly.  However, as discussed in the Introduction and in
Sec.~\ref{Sec:SWaveTCM}, for small $g_s$ ($\gamma_s\ll 1$) the theory
can be analyzed by a controlled perturbative expansion in powers of
$\gamma_s$ around the saddle-point (mean-field) approximation of
$Z_s$. To this end, we look for the spatially uniform field
configuration $\phi({\bf r})=B$ that minimizes the action
$S_s\left[\phi \right]$. We find the following saddle-point equation
$$
 {1 \over B} \left. { \delta S_s\left[\phi \right] \over  ~\delta \bar
\phi}  \right|_{\phi=B}
  =  $$
\begin{equation}
\epsilon_0 - 2 \mu - g_s^2 T \sum_{\omega_n}   \int  {d^3 k \over (2
\pi)^3} {1 \over \omega_n^2 + \left({k^2 \over 2 m} - \mu\right)^2
+ g_s^2 \bar B B}=0.
\end{equation}
where $\omega_n=\pi T (2 n + 1)$ are the fermion Matsubara
frequencies.  The sum over the frequencies can be done in a closed
form, leading to the so-called BCS-BEC gap equation for the mean field
$B(T,\mu,\epsilon_0)$ (and the corresponding condensate density
$|B|^2$)
\begin{equation}
\label{eq:BECBCS} \epsilon_0-2 \mu = {g_s^2 \over 2} \int {d^3 k
\over (2 \pi)^3} { \tanh  \frac{ E_k }{  2 T }  \over E_k},
\end{equation}
where $E_k$ is given by
\begin{equation}
\label{eq:Ek} E_k =\sqrt{ \left({k^2 \over 2 m} - \mu \right)^2 +
g_s^2 \bar B B}
\end{equation}
The integral on the right hand side of the BCS-BEC gap equation is
formally divergent, scaling linearly with the uv momentum cutoff
$\Lambda$. However, expressing the bare detuning parameter
$\epsilon_0$ in terms of the physical, renormalized detuning
$\omega_0$ using \rfs{eq:omega0}, we can completely eliminate the
appearance of the microscopic uv scale $\Lambda$ in all physical
quantities, and thereby obtain a uv-convergent form of the BCS-BEC gap
equation
\begin{equation}
\label{eq:BECBCS1} \omega_0 - 2 \mu = {g_s^2 \over 2} \int {d^3 k
\over (2 \pi)^3} \left[ {\tanh {E_k \over 2 T} \over E_k} -{2 m
\over k^2} \right].
\end{equation}

To calculate $Z_s$ in the saddle-point approximation, we write the
bosonic field
\begin{equation}
\phi({\bf r})=B+\varphi({\bf r}),
\end{equation}
in terms of a fluctuation field $\varphi(\r)$ about the saddle point
$B$.  Expanding the action, we find
\begin{equation} \label{eq:Seff}
S_{s}\left[B,\varphi\right]\approx S_0[B]+S_{\rm fluct}[B,\varphi],
\end{equation}
where
\begin{widetext}
\begin{equation}
\label{eq:Seff1} S_0[B]=\left. S_{s} \right|_{\phi=B}={V \over
T} \left(
\epsilon_0 - 2 \mu \right) \bar B B  -
V \sum_{\omega_n} \int {d^3 k \over (2 \pi)^3} \ln \left[
\omega_n^2 + \left( {k^2 \over 2m}-\mu \right)^2 + g_s^2 \bar B B,
\right]
\end{equation}
and
\begin{equation} \label{eq:Sfluct}
S_{\rm fluct}[B,\varphi] =  \sum_{p,\Omega_n} \left( \matrix { \bar
\varphi_{p,\Omega_n} & \varphi_{-p,-\Omega_n}} \right) \left(
\matrix { \oh s_B(p,\Omega_n)+\Sigma_{11} \left(p,\Omega_n \right)
& \Sigma_{20} \left( p,\Omega_n \right) \cr \Sigma_{20}
(p,\Omega_n) & \oh s_B(p,-\Omega_n)+\Sigma_{11} \left(p,-\Omega_n
\right) } \right) \left( \matrix {\varphi_{p,\Omega_n} \cr \bar
\varphi_{-p,-\Omega_n}} \right).
\end{equation}
\end{widetext}
Here $s_B(p,\Omega_n)=i \Omega_n +\epsilon_0-2 \mu +{p^2 \over 4m}$
represents the free part of the bosonic action,
$\Sigma_{11}(p,\omega)$ and $\Sigma_{02}(p,\omega)$ are the normal and
anomalous fermion polarization operators, and $\Omega_n=2 \pi T n$ are
the bosonic Matsubara frequencies.  We used here a fact that
$\Sigma_{20}$ is symmetric under the sign change of its arguments. The
explicit expressions for the polarization operators are not very
illuminating at this stage and will be discussed later (see
Eqs.~\rf{eq:sigma11} and \rf{eq:sigma20}). Within this saddle-point
approximation, the partition function $Z_s$ is given by
\begin{equation}
Z_s \approx  \exp \left(-S_0 \right) \int {\cal D} \varphi {\cal D}
\bar \varphi \exp \left( - S_{\rm fluct} \right).
\end{equation}

In the applications of \rfs{eq:BECBCS} to atomic gases, it is the
total particle number $N$, rather than the chemical potential that is
controlled experimentally. Of course, as usual, in the thermodynamic
limit there is no distinction between the two ensembles and it is
sufficient to work in the grand-canonical ensemble (as we have done
above), and then eliminate $\mu$ in favor of $N$ through the particle
number equation
\begin{equation}
\label{eq:partnumbdef} N= T \pp{\mu} \ln Z_s.
\end{equation}
Solving \rf{eq:partnumbdef} simultaneously with the BCS-BEC gap
equation \rf{eq:BECBCS1} determines the condensate density $B$ and
chemical potential $\mu$ as a function of experimentally controlled
parameters, the detuning $\omega_0$, temperature $T$, and particle
number $N$.

\subsubsection{Zero temperature BCS-BEC crossover}
\label{ztbcsbeccrossover}

At zero temperature the BCS-BEC gap equation reduces to
\begin{equation}
\label{eq:BECBCSzero} \omega_0-2 \mu = {g_s^2 \over 2} \int {d^3 k
\over (2 \pi)^3} \left[ { 1  \over E_k} - {2 m \over k^2} \right].
\end{equation}
The particle number equation can also be evaluated noting that at
$T=0$ and small $\gamma_s$, most of the weakly-interacting bosons
remain in the condensate $B$ (with only a small interaction-driven
depletion set by $\gamma_s$) and therefore the fluctuations
$\varphi(\r)$ are small and can be safely neglected.  Omitting $S_{\rm
  fluct}$ from $Z_s$ above and using \rfs{eq:Seff1}, the particle
number-density ($n=N/V$) equation is then given by
\begin{eqnarray}
\label{eq:zeropart}
n &=& -{T\over V} \pbyp{S_0}{\mu},\nonumber\\
  &=& \int {d^3 k \over (2 \pi)^3} \left[ 1 -
{ \xi_k \over E_k} \right] +2 \bar B B,
\end{eqnarray}
where $E_k$ is still given by \rfs{eq:Ek}, and
\begin{equation}
\label{eq:xik} \xi_k={k^2 \over 2m}-\mu.
\end{equation}

Before solving this equation together with the BCS-BEC gap equation,
\rf{eq:BECBCSzero} for the condensate $B$ and the chemical potential
$\mu$ as a function of $\omega_0$, let us comment on the nature of the
ground state. The bosons, within the narrow resonance approximation,
where the condensate depletion has been neglected, are all located in
the condensate. The fermions, on the other hand, form a paired
superfluid described by the BCS-like Hamiltonian
\begin{eqnarray}
\label{eq:hamferm} H -\mu~N_f = \sum_{\k, \sigma}\xi_k ~\hat
a^\dagger_{\k \sigma} \hat a_{\k \sigma}  + \nonumber  \\
\sum_{\k}~g_s \left(B ~ \hat a^\dagger_{\k \uparrow} \hat a^\dagger_{-\k
\downarrow} +
 \bar B~ \hat a_{-\k \downarrow} \hat a_{\k \uparrow} \right),
\end{eqnarray}
with the condensate $B$ appearing as pairing parameter to be
self-consistently determined. In practice, $B$ can always be chosen to
be real due to the symmetry $B \rightarrow e^{i\varphi} B$, so that
$B=\bar B$.  The role of the BCS gap is played by the bosonic density,
via
\begin{equation}
\label{eq:gapdelta}
\Delta=g_s B.
\end{equation}
The ground state of this Hamiltonian is the BCS wavefunction~\cite{Schrieffer}
\begin{equation}
\label{eq:gr} \left| {\rm BCS} \right\rangle = \prod_k \left( u_k +
v_k ~a^\dagger_{ k \uparrow} a^\dagger_{-k \downarrow} \right)
\left|0 \right\rangle,
\end{equation}
with $u_k$, $v_k$ given by the standard expressions
\begin{equation}
\label{eq:vkuk} v_k=\sqrt{\oh \left( 1-{\xi_k \over E_k} \right)},
\ u_k=\sqrt{\oh \left( 1+{\xi_k \over E_k} \right)}.
\end{equation}

When $\omega_0$ is large and positive, $\omega_0 \gtrsim 2
\epsilon_F$, the closed-channel molecules (bosons $\hat{b}_\p$) are
energetically suppressed and \rfs{eq:hamferm} predicts phenomenology of
a BCS superconductor~\cite{Schrieffer}. Namely, most of the particles will be fermionic
atoms with a weak attraction due to exchange of virtual bosons
(resonances) with energy detuned much higher than the chemical
potential. Such degenerate fermions will therefore form a BCS ground
state \rf{eq:gr}. Due to fermionic pairing, the bosons will still be
present, albeit in the exponentially small numbers since the BCS gap
\rfs{eq:gapdelta} and the associated Cooper-pair density are
exponentially small.

In a qualitative picture similar to the $g_s\rightarrow0$ limit (see Fig.~\ref{g0BCSBEC}), as
$\omega_0$ is decreased the number of boson condensate will grow,
while the number of fermions will decrease, with the substantial
change approximately (within a window $\gamma_s \epsilon_F$) taking place when $\omega_0$ drops below
$2\epsilon_F$. As $\omega_0$ crosses $0$, the chemical potential that
tracks it for $\omega_0<2\epsilon_F$ will also change sign and become
negative. At this point the remaining fermions will form a
``strongly-coupled superconductor'' (in the notation of
Ref.~\cite{Read2000}) that exhibits pairing in the absence of a Fermi
surface, driven by the existence of a true two-body bound state. Such
situation is not typically encountered nor experimentally accessible
in ordinary, solid-state superconductors. As discussed in
Ref.~\cite{Read2000}, a strongly coupled $s$-wave superconductor is
not qualitatively different from the more standard one with $\mu>0$.
This contrasts with the $p$-wave case, where $\mu>0$ and $\mu<0$
regimes are separated by a (possibly topological) quantum phase
transition, that we will discussed in Sec.~\ref{PWaveChapter} devoted
to $p$-wave resonances.  Because throughout the entire range of
detuning $\omega_0$ (particularly for $0<\omega_0<2\epsilon_F$) the
system will be a superposition of Bose-condensed bosons and Cooper-paired
fermions, we refer to it as a BCS-BEC condensate.

It is useful to contrast this small $\gamma_s$ picture of BCS-BEC
condensate with the earlier studies of wide resonances (corresponding
to the large $\gamma_s$ limit in our setting). When the resonance is
wide, most of the particles are fermions regardless of the value of
$\omega_0$ (open-channel fermions in the atomic physics parlance) and
only a small fraction of the condensate will be bosons (closed-channel
fermions). As a result, for a wide resonance no sharp features exist in the
BCS-BEC crossover.

Now we are in the position to solve the BCS-BEC gap and the particle
number equations. The former can be written as
\begin{equation}
\label{eq:saddleren1} \omega_0 - 2 \mu = {g_s^2 \left(2 m \right)^{3
\over 2} \sqrt{|\mu|} \over 4 \pi^2} ~I \left(u \right),
\end{equation}
where $u={g_s B/\mu}$ (with $B=\bar B$), and the integral $I(u)$ is
given by
\begin{equation}
\label{eq:I} I(u)=\int_0^\infty dx~ \left[{x^2 \over
\sqrt{\left(x^2-\sign~u  \right)^2 + u^2 }} - 1 \right],
\end{equation}
with $\sign~u=\sign~\mu$. Similarly, the particle number equation
reduces to
\begin{equation}
\label{eq:part1} {(2 m)^{3 \over 2} |\mu|^{3 \over 2} \over 3
\pi^2} K(u) + 2 B^2 =N,
\end{equation}
where $K(u)$ is defined as
\begin{equation}
K(u) = {3 \over 2} \int_0^\infty dx~x^2\left[1-{ x^2-\sign~u \over
\sqrt{(x^2-\sign~u)^2+u^2}} \right].
\end{equation}
$K(u)$ essentially measures the deviation of the particle number from
the usual Fermi distribution in the absence of pairing, with
$K(u\rightarrow 0^+)\rightarrow 1$.

In the BCS regime we expect a small condensate with $u \ll 1$, for
which
\begin{equation}
I(u ) \approx \ln \left(8 \over e^2 u \right).
\end{equation}
The logarithmic divergence of $I(u)$ as $u$ goes to zero is the
standard Fermi surface contribution to the gap equation. In the same
regime we can replace $K(u)$ by $K(0^+)=1$. We thus find
\begin{equation}
\label{eq:sadr1} B \approx g_s^{-1}8 e^{-2}\mu \exp\left(-{ 4 \pi^2
\left( \omega_0-2 \mu \right) \over (2 m)^{3 / 2} g_s^2
\sqrt{\mu}}\right),
\end{equation}
and
\begin{equation}
\label{eq:parr1}
{(2 m \mu)^{3 / 2} \over 3 \pi^2} + 2 B^2 = n.
\end{equation}
Since $B$ is exponentially small, \rfs{eq:parr1} gives $\mu \approx
\epsilon_F$. Combined, these two equations give us $B(\omega_0)$ as a
function of detuning. The quantitative validity of this regime, $u \ll
1$ is given by
\begin{equation}
\omega_0 \gg 2 \epsilon_F + \epsilon_F\gamma_s,
\end{equation}
where $\gamma_s$ is the small parameter given by \rfs{eq:gammadef}
characterizing an $s$-wave resonant gas at density $n$.

As $\omega_0$ is decreased below $2\epsilon_F$, a crossover regime is
entered as fermions begin to be converted into bosons, and the
chemical potential $\mu$ tracks $\omega_0/2$ to accuracy ${\cal
  O}(\gamma_s)$. As is clear from the infinitely narrow resonance
analysis, \rfs{eq:narrowcrossover} and Fig.\ref{g0BCSBEC}, most of
the atoms will pair up into bosons that become true bound states
inside the BEC regime of $\omega_0 < 0$.  In this BEC regime, for a
sufficiently negative $\omega_0$, so that $u<0$ and $|u| \ll 1$, we
can use the following simple asymptotics $I(-0)=-{\pi \over 2}$, and
$K(u) \approx {3 \pi \over 16} u^2$, to reduce the gap and particle
number equations deep in the BEC regime to be
\begin{eqnarray}
\omega_0-2 \mu & \approx & -{g_s^2 (2 m)^{3 / 2} \sqrt{|\mu|} \over 8
\pi},\\
n& \approx &{(2 m)^{3 / 2} g_s^2 B^2 \over 16 \pi \sqrt{|\mu |}}  + 2 B^2.
\end{eqnarray}
These give a BCS-BEC condensate, in which most particles are molecular
(true bound states) bosons, and only a small fraction of the total
number are the Cooper-paired fermions.

The full solution of Eqs.~\rf{eq:saddleren1} and \rf{eq:part1} can
only be found numerically and is displayed in Fig.~\ref{Fig:cond-det}
for $\gamma=0.1$. We can see that in contrast of the infinitely narrow
resonance case presented earlier on Fig.~\ref{Fig-FreeCond}, the
molecular condensate density $n_b=B^2$ extends to $\epsilon_0 \ge 2
\epsilon_F$, although only as an exponentially small tail. This
represents the BCS condensate absent in the limit of an infinitely narrow
resonance, $\gamma_s \rightarrow 0$.

\subsubsection{Ground state energy across BCS-BEC crossover}
\label{subsec:EG}

It is instructive to also calculate the zero temperature grand
canonical ground-state energy density $\ve_{GS} (\mu,\omega_0,B)$, that is given by
the $T \rightarrow 0$ limit of $T S_0/V$, where $S_0$ is given by
\rfs{eq:Seff1}. Calculating it at arbitrary $B$ and then minimizing it
with respect to $B$ will be of course be equivalent to solving the gap
equation \rfs{eq:BECBCSzero}.  As in the previous subsection, to
lowest order in $\gamma_s$, ignoring quantum fluctuations of $\phi$,
we find
\be
\ve_{GS}
= (\omega_0-2\mu) B^2 +
\sum_k ( \xi_k - E_k +g_s^2 \frac{B^2}{2\epsilon_k}).
\label{eq:gsefirst}
\ee
Here we again chose $B$ to be real, traded $\epsilon_0$ for $\omega_0$, and used the notations Eqs.~\rf{eq:xik}, \rf{eq:Ek} with
the additional notation
\begin{equation}
\epsilon_k = \frac{k^2}{2m}.
\end{equation}

Setting $B = 0$ the normal state energy is easily computed as
%
\bea
\ve_{GS}(B = 0)&=& \sum_k (\xi_k - |\xi_k|),
\label{eq:normalstategsefirst}
\\
&=&  -\frac{8}{15} c\mu^{5/2}\Theta(\mu),
\label{eq:normalstategsefirstp}
\eea
%
where we converted the sum to an integral and used the three-dimensional
density of states $N(E) = c \sqrt{E}$ with
\be
\label{eq:cdefappendix}
c\equiv \frac{m^{3/2}}{\sqrt{2}{\pi^2}}.
\ee
Combining this with Eq.~(\ref{eq:gsefirst}) then gives:
\be
\ve_{GS} = (\omega_0-2\mu)B^2
- \frac{8}{15}c\mu^{5/2}\Theta(\mu) + J(\mu,B),
\label{eq:cfive}
\ee
where
\be
J(\mu,B) \equiv
\int \frac{d^3 k}{(2\pi)^3} \big(|\xi_k|-E_k + g_s^2 \frac{B^2}{2\epsilon_k}\big),
\label{eq:gseintsecond}
\ee
and we have converted the momentum sum to an integral.

The gap equation discussed previously, \rfs{eq:BECBCSzero}, and the
particle number equation \rfs{eq:zeropart}
obviously follow from 
\be
\label{eq:eomzeroh}
0 = \frac{\partial \ve_{GS}}{\partial B},
\ee
and
\be
n = -\frac{\partial \ve_{GS}}{\partial \mu}.
\label{eq:numzeroh}
\ee

For a narrow Feshbach resonance ($\gamma_s \ll 1$), we can find an
accurate analytic approximations to $\ve_{GS}$ in Eq.~(\ref{eq:cfive})
in all relevant regimes.  The first step is to find an appropriate
approximation to Eq.~(\ref{eq:gseintsecond}), which has drastically
different properties depending on whether $\mu>0$ (so that the
low-energy states are near the Fermi surface) or $\mu<0$ (so that
there is no Fermi surface and excitations are gapped with energy
bounded from below by $|\mu|$). We proceed by first evaluating the
derivative $\frac{\partial J}{\partial B}$ and then integrating the
expression with a constant of integration $J(\mu,0)=0$.
\bea
&&\hspace{-1cm}\frac{\partial J}{\partial B} = -g_s^2 B  \int \frac{d^3 k}{(2\pi)^3} \big(
\frac{1}{E_k} - \frac{1}{\epsilon_k}
\big),
\\
 &&\hspace{-1cm}\quad\,\,\,\,\,= -g_s^2 B  \int_0^{\infty} \sqrt{\epsilon} d\epsilon
 \big(
\frac{1}{\sqrt{(\epsilon-\mu)^2+\Delta^2}} - \frac{1}{\epsilon}
\big),
\\
&&\hspace{-1cm}\quad\,\,\,\,\,\simeq -2g_s^2   N(\mu) B\ln \frac{8{\rm e}^{-2} \mu}{g_s B},
\hskip 0.1cm \mu>0; \mu \gg g_s B,
\label{eq:iderbcs}
\\
&&\hspace{-1cm}\quad\,\,\,\,\,\simeq  g_s^2  N(\mu)B
\Big[ \pi + \frac{\pi}{16} \big(\frac{g_s B}{\mu}\big)^2\Big],
 \mu \ll- g_s B,
\label{eq:ibec}
\eea
%
This calculation proceeds through the evaluation of the integral for
$I(u)$, \rfs{eq:I}.  Integrating back up with respect to $B$, we thus
have
\bea
\label{eq:i}
&&\hspace{-0.5cm}J \simeq
\cases{
 - N(\mu)\big(\frac{g_s^2 B^2}{2}
+ g_s^2 B^2 \ln \frac{8{\rm e}^{-2}\mu}{g_s B}\big),\
 & \text{$\mu>0$; $\mu \gg g_s B$},
\cr
    N(\mu) \frac{g_s^2 B^2}{2}
\Big[ \pi + \frac{\pi}{32} \big(\frac{g_s B}{\mu}\big)^2\Big],\
&\text{$\mu<0$; $|\mu| \gg g_s B$},
\cr}\nonumber\\
\eea
%
Having computed $\ve_{GS}(\mu,B)$ in the regimes of interest, the
phase diagram is easily deduced by finding $B$ that minimizes
$\ve_{GS}(\mu,B)$, subject to the total atom number constraint
Eq.~(\ref{eq:numzeroh}).

\paragraph{BCS regime:}
\label{par:BCSregime}

The BCS regime is defined by $\omega_0 \gg 2\ef$, where $g_s B \ll
\mu$ and $\mu\simeq \ef>0$, with pairing taking place in a thin shell
around the well-formed Fermi surface.  In this regime, $\ve_{GS}$ is
given by
\bea
&& \ve_{GS} \simeq -c\frac{\sqrt{\mu}}{2}\Delta^2 +\frac{\Delta^2}{g_s^2} (\delta - 2\mu)  +
c\sqrt{\mu} \Delta^2 \ln\frac{\Delta} {8 {\rm e}^{-2} \mu}
\nonumber \\
&& \qquad \qquad  -\frac{8}{15} c\mu^{5/2}
\label{eq:eg0fin}
\eea
with $\Delta \equiv g_s B$, see \rfs{eq:gapdelta}.

It is convenient to work with the dimensionless variables defined by
\begin{equation}
\hat \mu = \frac{\mu}{\epsilon_F}, \ \hat \Delta = \frac{\Delta}{\epsilon_F}, \
\hat \omega_0 = \frac{\omega_0}{\epsilon_F}.
\end{equation}

The normalized ground-state energy $e_{GS}$ in the BCS regime is then
given by
\bea
&&e_{GS}\equiv \frac{\ve_{GS}}{c\ef^{5/2}} \simeq -\frac{\sqrt{\muh}}{2}\Deltah^2 +\Deltah^2 (\deltah - 2\muh) \gamma_s^{-1}
\nonumber \\
&& \qquad +
\sqrt{\muh} \Deltah^2 \ln\frac{\Deltah} {8 {\rm e}^{-2} \muh} - \frac{8}{15} \muh^{5/2},
\eea
where, $\gamma_s$, defined in Eq.~(\ref{eq:gammadef}), is a
dimensionless measure of the Feshbach resonance width $\width$ to the
Fermi energy.  With this, Eqs.~(\ref{eq:eomzeroh}) and \rf{eq:numzeroh} become
%
\bea
&&\hspace{-1.2cm}0 = \frac{\partial e_{GS}}{\partial\Deltah},
\\
&&\hspace{-1.2cm}\quad\simeq   2\Deltah (\deltah - 2\muh )\gamma_s^{-1} +
2\sqrt{\muh} \Deltah \ln \frac{\Deltah} {8 {\rm e}^{-2} \muh},
\label{eq:gapzeroh2}
\\
&&\hspace{-1.2cm}\frac{4}{3} = -\frac{\partial e_{GS}}{\partial\muh},
\\
&&\hspace{-1.2cm}\quad\simeq\frac{5}{4} \frac{\Deltah^2}{\sqrt{\muh}} + \frac{4}{3} \muh^{3/2} + 2\Deltah^2 \gamma_s^{-1}
- \frac{\Deltah^2}{2\sqrt{\muh}}\ln \frac{\Deltah}{8{\rm e}^{-2} \muh},
\label{eq:numzeroh2}
\eea
%
that admits the normal state ($\Deltah = 0$, $\muh = 1$) and the BCS
SF state
%
%
\bea
\label{eq:baredeltanought}
\Deltah&\simeq&\Deltahbcs(\muh) \equiv 8{\rm e}^{-2} \muh{\rm e}^{-\gamma_s^{-1}(\deltah - 2\muh)/\sqrt{\muh}},
\\
\frac{4}{3} &\simeq&  \frac{4}{3} \muh^{3/2} + 2\Deltah^2 \gamma_s^{-1}.
\label{eq:weakcouplingbcsnum}
\eea
%
where in the second line we approximately neglected the first term on
the right side of Eq.~(\ref{eq:numzeroh2}), valid since $\Deltahbcs
\ll 1$ (and $\gamma_s\ll 1$).  It is easy to show that the BCS solution
is always a minimum of $\ve_{GS}(B)$.

The meaning of the two terms on the right side of
Eq.~(\ref{eq:weakcouplingbcsnum}) is clear once we recall its form in
terms of dimensionful quantities:
\be
n \simeq \frac{4}{3}c\mu^{3/2} + 2|B|^2,
\ee
i.e., the first term simply represents the total unpaired atom
density, reduced below $n$ since $\mu<\ef$, while the second term
represents the density of atoms bound into molecules, i.e., twice the
molecular condensate density $|B|^2$.  Qualitatively, we see that at large
$\deltah$, $\Deltah \ll 1$, implying from the number equation that
$\muh \approx 1$.

\paragraph{BEC regime:}
\label{par:BECregime}
We next consider the BEC regime defined by $\omega_0 < 0$.  As we
shall see, in this regime $\mu < 0$ and $|\mu| \gg \Delta$, so that
Eq.~(\ref{eq:i}), $I(\mu,\Delta)$, applies.  This yields, for the
normalized ground-state energy,
\be
e_{G} \simeq (\deltah - 2\muh ) \Deltah^2 \gamma_s^{-1}
 + \sqrt{|\muh|} \frac{\Deltah^2}{2}  \Big[ \pi + \frac{\pi}{32}
\big(\frac{\Deltah}{\muh}\big)^2\Big],
\ee
and, for the gap and number equations (dividing by an overall factor
of $\Deltah$ in the former)
%
\bea
\label{becgap}
0 &\simeq& 2\gamma_s^{-1} (\deltah - 2\muh ) + \sqrt{|\muh|}\Big[ \pi + \frac{\pi}{16} \big(\frac{\Deltah}{\muh}\big)^2\Big],
\\
\frac{4}{3}  &\simeq& 2\gamma_s^{-1} \Deltah^2 + \frac{\Deltah^2 \pi }{ 4\sqrt{|\muh|}}.
\label{becnum}
\eea
%
In the BEC regime the roles of the two equations are reversed, with
$\muh$ approximately determined by the gap equation and $\Deltah$
approximately determined by the number equation.
Thus, $\muh$ is well-approximated by neglecting the term proportional
to $\Deltah^2$ in Eq.~(\ref{becgap}), giving
\bea
\muh &\approx& \frac{\deltah}{2} \Big[
\sqrt{1+
\frac{\gamma_s^2 \pi^2}{32 |\deltah|}} -
\frac{\gamma_s \pi}{\sqrt{32 |\deltah|}}
\Big]^2.
\label{eq:eomzeroh2p}
\eea
At large negative detuning, $|\deltah| \gg 1$, in other words in
the BEC regime, Eq.~(\ref{eq:eomzeroh2p}) reduces to $\muh\approx
\deltah/2$, with the chemical potential tracking the detuning.

Inserting Eq.~(\ref{eq:eomzeroh2p}) into Eq.~(\ref{becnum}) yields
\be
\label{eq:deltazeroh}
\Deltah^2 = \frac{2\gamma_s}{3}\Big[
1- \frac{\gamma_s \pi}{\sqrt{(\gamma_s \pi)^2 + 32|\deltah|}}
\Big].
\ee
Using $\Deltah = \Delta/\ef$ and the relation $\Delta^2 = g_s^2n_b$
between $\Delta$ and the molecular density, we have
\bea
n_b &=& \frac{3}{4} \gamma_s^{-1} \Deltah^2 n,
\\
&\simeq&\frac{n}{2}\Big[
1- \frac{\gamma_s \pi}{\sqrt{(\gamma_s \pi)^2 + 32|\deltah|}}
\Big],
\eea
which, as expected (given the fermions are nearly absent for $\mu<0$)
simply yields $n_b \approx n/2$ in the asymptotic (large $|\deltah|$)
BEC regime.

These results of course match those derived purely on the basis of the
gap and particle number equations in Sec.~\ref{ztbcsbeccrossover}.

\subsubsection{Zero temperature collective excitations and condensate
depletion}

We would now like to calculate the spectrum of collective excitations
of the BEC-BCS condensate, which is contained in the $S_{\rm fluct}$
part of the effective action. In order to do that, we need expressions
for the self-energies $\Sigma_{11}$ and $\Sigma_{20}$ appearing in
\rfs{eq:Sfluct}. At zero temperature these are given by
\begin{widetext}
\begin{equation}
\label{eq:sigma11}
 \Sigma_{11}\left( q,\Omega \right) = {g_s^2 \over 2} \int {d
\omega \over 2 \pi} {d^3 k \over (2 \pi)^3} \frac{ \left[ i \left(
{ \Omega \over 2} + \omega \right) - \xi_+ \right] \left[i \left(
\omega -{ \Omega \over 2} \right) + \xi_- \right]} { \left[ \left(
{ \Omega \over 2} + \omega \right)^2 + \xi_+^2 + g_s^2 B^2 \right]
\left[ \left(\omega -{ \Omega \over 2}\right)^2 + \xi_-^2 + g_s^2
B^2 \right]},
\end{equation}
and
\begin{equation}
\label{eq:sigma20} \Sigma_{20} \left(q,\Omega, \right) = {g_s^4 B^2
\over 2} \int {d \omega \over 2 \pi} {d^3 k \over (2 \pi)^3}
\frac{1}{{ \left[ \left( { \Omega \over 2} + \omega \right)^2 +
\xi_+^2 + g_s^2 B^2 \right] \left[ \left(\omega -{ \Omega \over
2}\right)^2 + \xi_-^2 + g_s^2 B^2 \right]}},
\end{equation}
\end{widetext}
where
\begin{equation}
\label{eq:xis} \xi_+ ={1 \over 2 m} \left( {\bf k} + {{\bf q} \over
2} \right)^2  - \mu, \ \xi_- ={1 \over 2 m} \left( {\bf k} - {{\bf
q} \over 2} \right)^2  - \mu.
\end{equation}
The self-energy $\Sigma_{11}$ involves an IR divergent integral
over $k$. This divergence can be regularized if one notices that
$\Sigma_{11}$ enters the effective action $S_{\rm fluct}$ in the
combination $\epsilon_0-2 \mu + 2 \Sigma_{11}$. It is
straightforward to check, however, that
\begin{equation}
\label{eq:Gold} \epsilon_0 - 2 \mu + 2\Sigma_{11}(0,0)=2
\Sigma_{20}(0,0),
\end{equation}
by virtue of the saddle-point equation \rfs{eq:BECBCSzero}. This
situation is typical in the interacting Bose gas, and \rfs{eq:Gold} is
nothing but the Goldstone theorem ensuring that the collective
excitations remain massless (also referred to as Hugenholtz-Pines relation~\cite{Hugenholtz1959}
in the interacting Bose gas
literature). Therefore, we are really interested not
in $\Sigma_{11}(q,\Omega)$, but rather in the linear combination
$\Sigma_{11}(q,\Omega)-\Sigma_{11}(0,0)$, which remains finite.

The spectrum of collective excitations is given by the condition
that the propagator computed with the help of \rfs{eq:Sfluct} has
a pole. To simply the calculations, we will only compute the
spectrum at low momentum and energy. Following
\cite{LevitovUnpublished}, in anticipation that the collective
excitations are sound waves and so, $\Omega_q \sim q$, we expand the
self-energies in powers of energy and momentum according to
\begin{eqnarray}
\epsilon_0-2 \mu+2 \Sigma_{11}(q,\Omega) &\approx&  2 \Sigma_0 +
\Sigma i \Omega+\Sigma_1 \Omega^2+\Pi_1 q^2 \nonumber \\
\Sigma_{20}(q,\Omega)&\approx&\Sigma_0 + \Sigma_2 \Omega^2+ \Pi_2 q^2
\end{eqnarray}
where we have used \rfs{eq:Gold}. Therefore, the spectrum is given
by the condition that the determinant of the matrix in
\rfs{eq:Sfluct} vanishes. That matrix in our case takes the form
\begin{widetext}
\begin{equation}
\det \left( \matrix { \oh \left( i\Omega \left(1+\Sigma \right) +
\frac{q^2}{4m} + 2 \Sigma_0 + \Sigma_1 \Omega^2+\Pi_1 q^2 \right) &
\Sigma_0 + \Sigma_2 \Omega^2 + \Pi_2 \Omega^2 \cr \Sigma_0 + \Sigma_2
\Omega^2 + \Pi_2 \Omega^2 & \oh \left( -i\Omega \left(1+\Sigma
\right) + \frac{q^2}{4m} + 2 \Sigma_0 + \Sigma_1 \Omega^2+\Pi_1 q^2
\right) } \right)=0.
\end{equation}
\end{widetext}
This gives for the spectrum $\Omega_q$ of excitations
\begin{equation}
\label{eq:Omegaq} \Omega_q^2=q^2  \frac{4 \Sigma_0 \left(
\frac{1}{4m}+\t \Pi\right) }{\left( 1+\Sigma \right)^2 + 4 \Sigma_0
\t \Sigma},
\end{equation}
where we introduced the notation
\begin{equation}
\t \Pi=\Pi_1-2 \Pi_2, \ \t \Sigma=\Sigma_1-2\Sigma_2.
\end{equation}
In other words, the excitations are indeed sound modes, with the speed
of sound
\begin{equation} \label{eq:SOS}
c= \sqrt{\frac{4 \Sigma_0 \left( \frac{1}{4m}+\t \Pi\right) }{\left(
1+\Sigma \right)^2 + 4 \Sigma_0 \t \Sigma}}.
\end{equation}

We now evaluate $\Sigma_0$, $\Sigma$, $\t \Pi$, and $\t \Sigma$.
Doing the frequency integral in $\Sigma_{2,0}(0,0)$ gives
\begin{equation}
\label{eq:Delta} \Sigma_0= {g_s^4 B^2 \over 8} \int_0^\infty {k^2 dk \over 2
\pi^2} {1 \over \left[ \left( {k^2 \over 2 m} - \mu \right)^2 +
g_s^2 B^2 \right]^{3 / 2}},
\end{equation}
Differentiating $\Sigma_{11}(0,\Omega)$ with respect to $\Omega$
at $\Omega=0$ gives
\begin{equation}
\label{eq:Sigma} \Sigma=  {g_s^2 \over 4} \int_0^\infty {k^2 dk \over 2
\pi^2} {{k^2 \over 2m} - \mu \over \left[ \left( {k^2 \over 2 m} -
\mu\right)^2 + g_s^2 B^2 \right]^{3 / 2}}.
\end{equation}
Finally, we also find after some algebra
\begin{equation}
\label{eq:tsigma} \t \Sigma = \frac{g_s^2}{8} \int_0^\infty {k^2 dk \over 2
\pi^2} {1 \over  \left[ \left( {k^2 \over 2 m} - \mu\right)^2 +
g_s^2 B^2 \right]^{3  / 2}},
\end{equation}
\begin{equation}
\label{eq:Pi} \t \Pi = \frac{\Sigma}{4m}+\frac{g_s^4 B^2}{8m} \int_0^\infty
{k^2 dk \over 2 \pi^2} \frac{\frac{k^2}{2m}}{\left(
\left(\frac{k^2}{2m}-\mu \right)^2 + g_s^2 B^2 \right)^{5 / 2}}.
\end{equation}

In general, evaluation of the integrals in Eqs.~\rf{eq:Delta},
\rf{eq:Sigma}, \rf{eq:tsigma}, and \rf{eq:Pi} is straightforward
but cumbersome. We will present results only in the deep  BEC
and  BCS regimes (that is, $\omega_0 \lesssim - \epsilon_F$
and $\omega_0 \gtrsim \epsilon_F$).

First, consider the BEC side of the crossover. There $\omega_0$ and
$\mu$ are negative, and it is clear that at small $g_s$, $\Sigma \ll
1$, $\Sigma_0 \t \Sigma \ll 1$, and $m \t \Pi \ll 1$, and thus they can be
neglected. The speed of sound is then simply given by
\begin{equation}
c_{\rm BEC} = \sqrt{\frac{\Sigma_0}{m}}.
\end{equation}
Since $|\mu| \gg g_s^2 B^2$ in this BEC regime, we can neglect $g_s^2
B^2$ in the denominator of \rfs{eq:Delta} to find
\begin{equation}
\Sigma_0 \approx {g_s^2 B^2 \over 8} \int {k^2 dk \over 2 \pi^2} {1
\over \left(\frac{k^2}{2m}-\mu\right)^3}={g_s^4 B^2 m^{3 / 2}
\over 32 \pi |\omega_0|^{3 / 2}},
\end{equation}
where we have used $2 \mu=\omega_0$ in this BEC regime. Therefore,
the square of the speed of sound is simply
\begin{equation}
c^2_{\rm BEC}={g_s^4 B^2 m^{1 /  2} \over 32 \pi
|\omega_0|^{3 / 2}}.
\end{equation}
where $B^2 \approx {n/2}$. We compare this expression for $c^2_{\rm
  BEC}$ with a standard expression for a BEC condensates of point
bosons (see, for example, Ref.~\cite{LL9})
\begin{equation}
c^2={4 \pi a_b B^2 \over (2 m)^2},
\end{equation}
where $a_b$ is the boson scattering length (not to be confused with
the scattering length of fermions, given by $a$ in
\rfs{eq:fanolengthrange1}). By inspection, we therefore conclude that
on the BEC side of the crossover, the BCS-BEC paired condensates
behaves as an effective gas of weakly repulsive bosons with a
scattering length
\begin{equation}
\label{eq:ab} a_b = {g_s^4 m^{5/2} \over 32 \pi^2
|\omega_0|^{3/2}}.
\end{equation}
We note that the scattering length $a_b$, together with the speed of
sound $c$, decreases as the detuning $\omega_0$ is made more negative,
deeper into the BEC regime.  This is of course to be expected, as
paired-bosons interaction arises due to their polarization into their
constituent fermions, followed by a fermions exchange \--- the process
of the order of $g_s^4$. Since this
virtual fermion creation process costs a molecular binding energy,
\rfs{eq:poleR_E}, deep in the BEC regime approximately given by
$|\omega_0|$, it is suppressed with increasing $|\omega_0|$, as is the
effective bosonic interaction and $a_b$.

Here an important remark is in order. Our narrow resonance result for
$a_b$, \rfs{eq:ab}, contrasts sharply  with the well-known ($g_s$ and
$m$ independent) result for the molecular scattering length deep in
the BEC regime, namely $a_b\approx 0.6 ~a$, where $a$ is the fermion
scattering length \cite{Petrov2005}.  The short answer explaining this
difference is that the $a_b\approx 0.6 ~a$ prediction is for the wide
resonance BEC regime, corresponding to
$\gamma_s\rightarrow\infty$ instead of the limit of narrow
resonance considered here.  In more detail, the results of
Ref.~\cite{Petrov2005} apply only in the regime where $a \gg |r_0|$.
In our narrow resonance problem, this regime is realized only in a
very narrow range of $\omega_0$, satisfying $-\Gamma_0 \lesssim \omega_0 < 0$ ($\Gamma_0$ is the resonance width given in
\rfs{eq:Gamma0}).
This is not what one should call the BEC regime of the
narrow resonance crossover, which should be defined as $\omega_0 \lesssim -\epsilon_F$, with $\epsilon_F$,
in turn, being much bigger than $\Gamma_0$.

Moreover, if one does tune $\omega_0$  to this narrow window, the Fermi energy
of the gas under study here will be much bigger than the binding energy of the bosonic
molecules, and the condensate cannot be treated at all as weakly interacting bosons. So even though
the scattering length of bosons within this window of $\omega_0$ is indeed $0.6~a$, this
will not get reflected in the speed of sound in the condensate.

Contrast this with the BEC regime of the {\em broad} resonance BCS-BEC superfluid,
where $a \gg r_0$ for a wide range of the detuning $\omega_0<0$, and where Fermi energy is small
compared to $\Gamma_0$.  For further
details, including the calculation of the speed of sound in the broad
resonance BEC regime of \rfs{eq:ham_s}, see Ref.~\cite{Levinsen2006}.

Thus we conclude that \rfs{eq:ab} is the correct scattering length of molecules in the narrow resonance
problem. In fact, \rfs{eq:ab} can also be derived independently by studying the scattering of bosons in
vacuum perturbatively. We will not do it here.

Let us now turn to the BCS regime $\omega_0 >2 \epsilon_F$, where $\mu
\approx \epsilon_F$. We evaluate $\Sigma_0$, $\Sigma$, $\t \Sigma$, and
$\t \Pi$, from Eqs.~\rf{eq:Delta}, \rf{eq:Sigma}, and \rf{eq:tsigma},
\rf{eq:Pi}. These integrals are easiest to compute if we change
variables $k^2/(2m)-\mu=\xi$, and notice that only small $\xi$
essentially contribute to the integrals. We find
\begin{eqnarray}
\Sigma_0 &\approx& {g_s^2 (2 m)^{3/2} \over 16 \pi^2} \mu^{1/2},
\ \Sigma \approx {\omega_0 - 2 \mu \over 4 \mu},\\
\t \Pi &\approx&   \frac{\mu \sqrt{(2m)^3 \mu}}{24 m B^2 \pi^2}, \ \t \Sigma \approx  \frac{\sqrt{(2m)^3 \mu}}{16 B^2 \pi^2}.
\end{eqnarray}
The speed of sound, \rfs{eq:SOS}, is now dominated by $\t \Sigma$, $\t
\Pi$, and give
\begin{equation}
c_{\rm BCS} = \sqrt{\frac{\t \Pi}{\t
      \Sigma}} = \sqrt{\frac{2}{3}\frac{\mu}{m}}=\frac{v_F}{\sqrt{3}}.
\end{equation}
that reassuringly recovers the well-known result for the speed of
sound in a neutral BCS superconductor \cite{Nambu1960,Bogoliubov}.

In the intermediate crossover regime between BEC and BCS, where $0
\lesssim \omega_0 \lesssim 2 \epsilon_F$, the integrals in $\Sigma_0$,
$\Sigma$, $\t \Sigma$, and $\t \Pi$ should be evaluated numerically to
give the speed of sound which interpolates between its BEC and BCS
values.

Using our understanding of the collective excitations, we can now
compute the interaction-driven depletion of the condensate, namely the
number of bosons that are not Bose-condensed into a single-particle
$\k=0$ state, even at zero temperature.  As we will show, the
depletion number turns out to be much smaller than the number of
particles in the condensate (with the ratio controlled by the
smallness of $\gamma_s\sim g_s^2 m^{3/2} \epsilon_F^{-{1/2}}$), which
justifies our neglecting it in the analysis of the crossover, above.
We also note that smallness of depletion justifies the expansion in
powers of fluctuations (controlled by $\gamma_s$) across the whole
range of the BCS-BEC crossover in a narrow resonance atomic system.

The number of excited bosons can be simply computed from the Green's
function of the fluctuations,
\begin{equation}
n_{\rm exc} = \lim_{\tau \rightarrow 0^+}  \VEV{\bar \varphi(0)
\varphi(\tau)}.
\end{equation}
Evaluating the Green's function gives
\begin{equation}
\VEV{\bar \varphi(\Omega,q) ~\varphi(-\Omega,-q)} =
\frac{s_B(q,-\Omega) + 2 \Sigma_{11}(q,-\Omega)}{\det M},
\end{equation}
where $M$ is the matrix in \rfs{eq:Sfluct}. Evaluating this expression
analytically in general is difficult, so we concentrate on limiting
BEC regime.

In the BEC regime, where we can approximate $\Sigma_{20}\approx
\Sigma_0$, $\epsilon_0-2 \mu+\Sigma_{11}\approx 2 \Sigma_0$ and sum over
frequency, we find
\begin{equation}
\label{eq:ntbos}
 n_{\rm exc} =  \int {d^3 q \over (2 \pi)^3}
  {{q^2 \over 4 m} + 2 \Sigma_0 - \Omega_q \over 2 \Omega_q},
\end{equation}
where $\Omega_q$ is the spectrum, $\Omega_q = 2 \sqrt{\left({q^2 \over
      8 m} + \Sigma_0\right)^2 -\Sigma_0^2}$ (\rfs{eq:Omegaq} is the small
momentum version of it). Doing the integral gives
\begin{equation} \label{eq:pnwng}
n_{\rm exc}=\frac{8}{3 \pi^2} \left( m \Sigma_0
\right)^{3/2} = \frac{8}{3} n \sqrt{\frac{n a_b^3}{\pi}},
\end{equation}
which coincides with the standard expressions of the condensate
depletion in a weakly interacting Bose gas. As advertised above, the
depletion is small and vanishes in the limit of a vanishingly narrow
resonance, $\gamma_s\ll 1$.

As $\omega_0$ is increased from negative towards positive values,
\rfs{eq:pnwng} is no longer applicable and analysis is best performed
numerically.

\subsubsection{Critical temperature}
\label{Sec:CritTemp}

We expect the condensate to be reduced with increasing temperature,
vanishing at a critical temperature $T_c(\omega_0)$, that we compute below.  In
contrast to superfluids of point bosons, in paired superfluids there
are two physically distinct effects that contribute to the condensate
reduction with temperature\cite{Nozieres1985}.  One is the
dissociation of Cooper pairs (and closed-channel molecules hybridized
with them), and, simultaneously, thermal bosonic excitations. One of
them is captured by the finite temperature gap equation
\rfs{eq:BECBCS1} and is responsible for $T_c$ in the BCS regime, while
the other must be included in the finite-temperature particle number
equation and is at work in the BEC regime.


For simplicity we focus on the Bose-Fermi mixture at the critical
temperature, where the condensate density vanishes. This allows us to
take advantage of technique of Ref.~\cite{Nozieres1985} to find the
number of particles. This method ignores the interactions between the
bosons and concentrates solely on the bosonic propagator modified by
the presence of fermions. This amounts to approximation of
$S_{s}[\phi]$, \rfs{Seff} by a quadratic expansion in
$\phi(\r)=\varphi(\r)$, reducing it to $S_{\rm fluct}[\phi]$,
\rfs{eq:Sfluct}. In the case of a broad resonance, considered in
Ref.~\cite{Nozieres1985}, this is not quantitatively justified for
$\omega_0$ sufficiently close to zero so that $a \gg n^{-1/3}$. In
contrast, in the case of a narrow resonance system studied here, this
expansion is justified, since the strength of interactions, governed
by $\gamma_s$, is weak.

\begin{figure}[bt]
\includegraphics[height=2in]{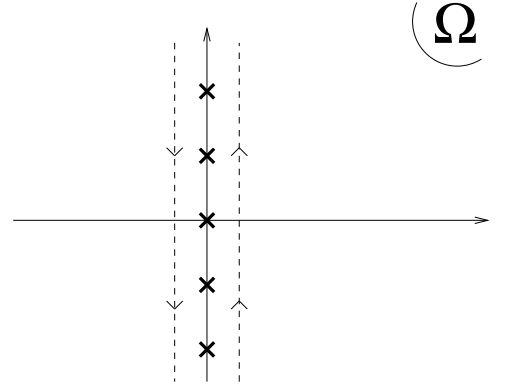}
\caption{\label{Fig-contour1} The contour of integration in
\rfs{eq:intoverom}. The crosses depict the positions of the poles
of the Bose-Einstein distribution $1/(e^{\Omega/T}-1)$.}
\end{figure}

At $T=T_c$ the condensate vanishes, $B=0$, and the gap equation
reduces to
\begin{equation}
\label{eq:BECBCS2} \omega_0 - 2 \mu = {g_s^2 \over 2} \int {d^3 k
\over (2 \pi)^3} \left[ {\tanh {\xi_k \over 2 T_c} \over \xi_k}
-{2 m \over k^2} \right],
\end{equation}
with $\xi_k$ given by \rfs{eq:xik}. To find the particle number
equation, we need to evaluate the contribution to the partition
function due to fluctuations of $\phi$ in \rfs{Seff}. This can be
expressed in terms of polarization operators $\Sigma_{11}$ and
$\Sigma_{20}$. In fact, for $B=0$, $\Sigma_{20}=0$, and only
$\Sigma_{11}$ survives. This gives for \rfs{eq:Seff1}
\begin{equation}
\label{eq:SeffTc} S_0[B]= -  V \sum_n \int {d^3 k \over (2 \pi)^3}
\ln \left[ \omega_n^2 + \left( {k^2 \over 2m}-\mu \right)^2\right],
\end{equation}
where $\omega_n=\pi T (2n+1)$ are fermionic Matsubara frequencies,
with fluctuation corrections to $S_0[B]$ given by
\begin{equation} \label{eq:sumoverfreq}
S_0^{\rm fluct}[B] = V \sum_n \int {d^3 q \over (2\pi)^3} \ln s_n,
\end{equation}
where
\begin{equation}
s_n(q,\Omega_n)=i\Omega_n + {q^2 \over 4m} - 2 \mu + \epsilon_0+2
\Sigma_{11}(q,\Omega_n),
\end{equation}
and $\Omega_n=2 \pi T n$ are bosonic Matsubara frequencies. To
simplify this expression further, we can use the technique discussed
in Ref.~\cite{Nozieres1985}. To that end, we introduce the many-body
finite temperature phase-shift
\begin{equation}
\delta(q,\Omega) = {\rm Im} \ln \left(  \Omega  - {q^2 \over 4m}
+ 2 \mu - \epsilon_0-2 \Sigma_{11}(q,i \Omega) \right),
\end{equation}
that is a generalization of the vacuum phase-shift, which can be
deduced from \rfs{eq:swaveintro} and the relation $f(p)=\left(e^{2 i \delta}
  -1\right)/(2 i p)$. We can now transform the sum over frequencies in
\rfs{eq:sumoverfreq} into the integral
\begin{equation} \label{eq:intoverom}
S_0^{\rm fluct}[B] ={V \over T} \int {d^3 q \over 2 \pi^3} \oint {d
\Omega \over 2 \pi i}~ {\delta(q,\Omega) \over e^{\Omega/T}-1},
\end{equation}
with the integration over $\Omega$ done along the contour depicted in
Fig.~\ref{Fig-contour1}.

Combining this all together, the particle number equation
\rf{eq:partnumbdef}
\begin{equation}
N=-T \pbyp{S_0}{\mu}-T\pbyp{S_0^{\rm fluct}}{\mu},
\end{equation}
takes the form
\begin{equation} \label{eq:pneft}
n=\int {d^3 q \over (2 \pi)^3} \left[ {1 \over
e^{\frac{q^2}{2mT}-\frac{\mu}{T}}+1} - \oint \frac{d\Omega}{2 \pi
i}{\pbyp{\delta (q,\Omega)}{\mu} \over e^{\Omega /
T}-1}\right],
\end{equation}
with the first and second terms giving the number of fermion and
bosons, respectively.

To make further progress, we need to know $\Sigma_{11}$, which, when
evaluated at $T=T_c$ is given by
\begin{equation}
\Sigma_{11}(q,\Omega) = - \frac{g_s^2 T}{2} \sum_n \int {d^3 k \over
(2\pi)^3} {1 \over \left[ \xi_+ - i \left( \omega_n - \Omega
\right) \right] \left[ \xi_- + i \omega_n \right]},
\end{equation}
where $\xi_+$, $\xi_-$ are given by \rfs{eq:xis}, above. The sum over
the frequencies is elementary, with the result
\begin{equation}\label{eq:Sigma11ft} \Sigma_{11} (q,\Omega)=-{g_s^2 \over 4} \int {d^3 k \over (2 \pi)^3}
{\tanh \left[ {\xi_+ \over 2 T_c} \right] +\tanh \left[ {\xi_-
\over 2 T_c} \right] \over \xi_+ +\xi_- + i \Omega }.
\end{equation}
Unfortunately the remaining integral can only be done numerically, and
we will not evaluate it in this paper, except deep in the BEC regime.
Fortunately, however, we do not need to know it in the narrow
resonance limit of $g_s \rightarrow 0$.  In this limit, $\Sigma_{11}$
becomes small and the phase-shift reduces to
\begin{equation}
\delta(q,\Omega) = {\rm Im} \ln \left(  \Omega  - {q^2 \over 4m}
+ 2 \mu - \omega_0 \right).
\end{equation}
In this limit, we can use $\omega_0$ and $\epsilon_0$ interchangeably
since they now coincide.

Substituting into \rfs{eq:pneft}, we can transform the contour
integral over $\Omega$ to the form
\begin{equation}
-2 \oint \frac{d\Omega}{2 \pi i}
\frac{1}{\Omega-\frac{q^2}{4m}-\omega_0+2 \mu}
\frac{1}{e^{\Omega/T}-1}.
\end{equation}
The contour in this integral can now be transformed to enclose the
pole at $\Omega=\frac{q^2}{4m}$
going in the clockwise direction, finally giving for
the total particle density
\begin{equation} \label{eq:hTpne}
\frac {n}{2}=\int {d^3 q \over (2 \pi)^3} \left[ {1 \over
e^{\frac{q^2}{2mT_c}-\frac{\mu}{T}}+1} + {1 \over e^{ \frac{q^2}{4
mT_c} }-1}\right],
\end{equation}
where we used $\omega_0=2\mu$ valid the small $g_s$ limit.

This equation coincides with \rfs{eq:bcsbecn} which we derived in the
$g_s=0$ limit, as could have been guessed from the outset.  However,
here we are in principle in the position to compute corrections to
this equation if $\Sigma_{11}$ is evaluated and included in
\rfs{eq:pneft}.

Let us now use the gap equation \rfs{eq:BECBCS2} and the particle
number equation \rfs{eq:hTpne} to compute the critical temperature as
a function of detuning $\omega_0$.

In the BCS regime, $\omega_0 \gtrsim 2 \epsilon_F$, we expect the
transition temperature to be exponentially small. As a result, the
particle number equation forces $\mu$ to be very close to $\epsilon_F$
(slightly below it). Indeed, the number of excited bosons at a low
temperature is expected to be small, and the particle number is
saturated by fermions, whose chemical potential must therefore be in
the vicinity of $\epsilon_F$. We recall that for $T_c=0$,
\rfs{eq:hTpne} would be solved simply by setting $\mu=\epsilon_F$.

We then need to use \rfs{eq:BECBCS2}, with $\epsilon_F$ substituted
for $\mu$ with sufficient accuracy to determine $T_c$. The actual
calculations are identical to the ones employed by the BCS theory. One
technique for solving \rfs{eq:BECBCS2} in this regime is described in
Ref.~\cite{LL9}.  Evaluating the integral in \rfs{eq:BECBCS2} we find
\begin{equation}
\label{eq:BCStt}
T_c = \frac{8 e^{C-2}}{ \pi}~\epsilon_F \exp \left[{-4 \pi^2
\frac{ \omega_0 - 2 \epsilon_F}{g_s^2 \left( 2 m
\right)^{3/2} \sqrt{\epsilon_F}}} \right],
\end{equation}
where $C$ is the Euler constant, $\ln C \approx 0.577$. We see that
indeed, the critical temperature is exponentially small in the ratio
$(\omega_0-2\epsilon_F)/g_s^2$. This could have been guessed without
any calculation as this simply coincides with the standard BCS result
in the same way as \rfs{eq:sadr1} coincides with the appropriate BCS
result, with $T_c/\Delta = e^C/\pi$.

In the deep BEC regime, where $\omega_0$ is negative, we expect the
chemical potential $\mu$ to roughly follow $\omega_0$, in the way
quite similar to the infinitely narrow resonance limit described in
section \ref{sec:inrl}. The critical temperature will then be given by
solving \rfs{eq:hTpne} and noting that the fermion part of the
particle number is going to be very small.  Therefore, it will reach
its asymptotics coinciding with the critical temperature of a
non-interacting Bose gas, given by \rfs{eq:BECtt}.

Between the BEC and BCS regime through the crossover the temperature
will interpolate between the BEC \rfs{eq:BECtt} and the BCS
\rfs{eq:BCStt} values, in the precise way that can be obtained through
a numerical solution.

An interesting question is whether the critical temperature decreases
monotonously as the detuning is increased or perhaps has a maximum at
some intermediate value of the detuning. Recall that Nozi\`eres and
Schmitt-Rink observed a maximum in the $T_c$ vs $\omega_0$ diagram,
see Ref.~\cite{Nozieres1985}, and so did subsequent papers which
followed their techniques. However, as these authors themselves
observed, their calculations were done in the case of a broad
resonance, where their approach was an uncontrolled approximation that
could not guarantee that the maximum was not an unphysical artifact of
their approximation. In contrast, in our case of narrow resonance, we
can actually calculate the entire curve $T_0(\omega_0)$ perturbatively
in powers of $g_s$, and predict the behavior of $T_c$ in a trustworthy
way, at least for a $\gamma_s < 1$ system.

For our purpose it is sufficient to concentrate on the deep BEC regime
where $\omega_0 \ll -2 \epsilon_F$. In this regime we expect $T_c(\omega_0)$ to
approach the limiting value \rf{eq:BECtt} of order $\epsilon_F$ from
either above or below.  Given that the high $\omega_0$ BCS
asymptotics, \rfs{eq:BCStt}, is exponentially small compared to
$\epsilon_F$, the approach of the asymptotic BEC value (at large
negative $\omega_0$) from {\em above} implies unambiguously that the
curve $T_c(\omega_0)$ must have a {\em maximum} somewhere.

In the infinitely narrow resonance case we observe that the transition
temperature decreases with increasing $\omega_0$, since the fermion
number, suppressed in the BEC regime as $e^{\omega_0/2T}$, would start
increasing in accordance with \rfs{eq:bcsbecn}. However, for a narrow
but finite width resonance, fluctuations must also be taken into
account.

Let us first evaluate the contribution of the fluctuations to the
particle number equation \rfs{eq:hTpne}. First, we compute
\rfs{eq:Sigma11ft}, which in the BEC regime can be evaluated and leads
to a correction to the particle number equation which we now discuss.
At $\omega_0 \lesssim -2\epsilon_F$ (and consequently, $\mu \lesssim
-\epsilon_F$), we can safely neglect the hyperbolic tangents in the
numerators of \rfs{eq:Sigma11ft} to arrive at
\begin{equation}
\label{eq:Sigma11ft2}
\Sigma_{11} (q,i \Omega)=-{g^2 \over 2} \int {d^3 k \over (2 \pi)^3}
{1 \over  \frac{k^2}{m}+\frac{q^2}{4m}-2\mu- \Omega }.
\end{equation}
This expression basically coincides with the corresponding expression
for the polarization operator in a vacuum, \rfs{eq:polarization}.
Physically this is expected since deep in the BEC regime there are
only exponentially small number of fermions, so from the point of view
of bosons, the situation is indistinguishable from a vacuum.

\begin{figure}[bt]
\includegraphics[height=2in]{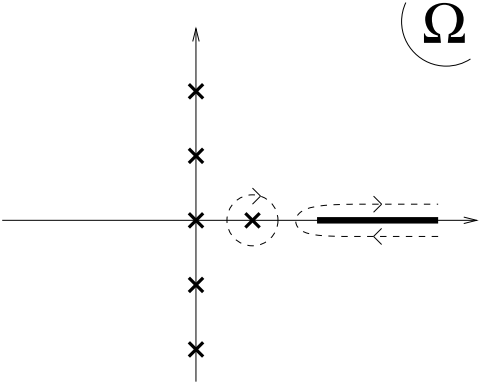}
\caption{\label{Fig-contour2} In the BEC regime we can deform the
  contour from Fig.~\ref{Fig-contour1} to this one, which encloses the
  pole corresponding to the bound state and goes around the cut
  corresponding to the scattering states. }
\end{figure}

We now observe that the phase-shift $\delta(q,\Omega)$ has a
singularity at
\begin{equation} \label{eq:qdeq}
\Omega-\frac{q^2}{4m}+2\mu-\epsilon_0-2\Sigma_{11}(q,i \Omega)=0.
\end{equation}
The value of $\Omega$ that solves this equation is given by
\begin{equation} \label{eq:qdeqq}
\Omega_q=\frac{q^2}{4m}.
\end{equation}
To see this we observe that $\Sigma$ only depends on $\Omega$ and $q$
through the combination $\Omega-\frac{q^2}{4m}$, and therefore the $q$
dependence of the solution to \rfs{eq:qdeq} is simply
$\frac{q^2}{4m}$. At the same time, at $q=0$, the solution to
\rfs{eq:qdeq} must be $\Omega=0$, owing to the Goldstone theorem
\rfs{eq:Gold}, giving the result \rf{eq:qdeqq}.

In addition to this pole, the phase-shift $\delta(q,\Omega)$ will also
have a cut along the real axis of $\Omega$, corresponding to the
scattering fermionic states. This cut goes from
$\Omega=\frac{q^2}{4m}-2\mu$ to infinity (notice that $\mu<0$).  Using
this information, we can transform the contour in the integral over
$\Omega$ in \rfs{eq:pneft} to the one depicted in
Fig.~\ref{Fig-contour2}. The integral around the pole gives back the
atom number confined inside thermally-excited bosons,
\begin{equation}
\label{eq:hTpne25}
2 \int {d^3 q \over (2 \pi)^3}  {1 \over e^{ \frac{q^2}{4 mT_c}}-1},
\end{equation}
while the remaining integral along the cut gives terms suppressed
exponentially as $e^{\frac{\omega_0}{2T_c}}$. These terms represent
corrections to the particle number equation \rfs{eq:hTpne}.  They must
be combined with properly evaluated fermion number in \rfs{eq:hTpne}
and with the additional terms given by the expansion of the hyperbolic
tangent in \rfs{eq:Sigma11ft} to give corrections to the critical
temperature in the deep BEC regime $\omega_0 \ll - 2 \epsilon_F$. The
key observation is that all these contributions are exponentially
small as $e^{\frac{\omega_0}{2T_c}}$.

However, all this ignores interactions between the bosons (in other
words, higher terms in the expansion in powers of $\phi$ in
\rfs{Seff}). It turns out that the interactions change the critical
temperature in a way which is {\em not} exponentially suppressed in
the deep BEC regime.  A weakly interacting Bose gas with a given
scattering length (for our case given by \rfs{eq:ab}) has been
extensively studied in the literature.  Although the correction to the
critical temperature due to interactions is still a controversial
subject, there is a reasonable agreement in the literature that this
correction is positive and is given by a bosonic gas parameter (see
Ref.~\cite{Andersen2004})
\begin{equation}
\frac{T_c -T_{c0}}{T_{c0}} \sim \left(\frac{n}{2}
\right)^{1/3} a_b = \left(\frac{n}{2}
\right)^{1/3} {g_s^4 m^{5/2} \over 16 \pi^2
|\omega_0|^{3/2}}
\end{equation}
Here $T_{c0}$ represents the critical temperature in the
noninteracting limit \rfs{eq:BECtt} and $n/2$ is the density of
bosons. This expression is clearly much bigger than the exponentially
small corrections due to fermion number and $\Sigma_{11}$ and hence
those other corrections can be neglected.

Therefore, we conclude that as $\omega_0$ is increased from large
negative values, $T_c(\omega_0)$ actually {\em increases}, according
to
\begin{equation}
T_c = {\pi \over m} \left( {n \over 2 \zeta\left( {3 \over 2}
\right)} \right)^{2/3} \left[ 1 + \alpha \left(\frac{n}{2}
\right)^{1/3} {g_s^4 m^{5/2} \over 16 \pi^2
|\omega_0|^{3/2}} + \dots \right],
\end{equation}
where $\alpha$ is an unknown constant of the order of $1$. At the same
time, in the BCS regime, at large positive $\omega_0$, it drops off
exponentially, according to \rfs{eq:BCStt}. Thus, $T_c(\omega_0)$ must
exhibit a {\em maximum} somewhere for the intermediate values of $\omega_0$.

Although this conclusion about the existence of a maximum in
$T_c(\omega_0)$ agrees with those appearing in a number of papers
devoted to broad resonances, beginning from Ref.~\cite{Nozieres1985},
here, in contrast to those studies our arguments for a maximum are
robust and quite general, being based on a quantitatively trustworthy
(in a narrow resonance case) calculation, rather than on an uncontrolled
approximations.

A schematic diagram depicting $T_c(\omega_0)$ is illustrated in
Fig.~\ref{Fig-crit}. To set the proper scale on the vertical axis, we
notice that the ratio of $\epsilon_F$ given by \rfs{eq:fermi} to
$T_{BEC}$, \rfs{eq:BECtt}, is approximately $5$. Compare this with the
critical temperature of an infinitely narrow resonance, $\gamma_s
\rightarrow 0$, shown on Fig.~\ref{Fig-CritTempZero}. For a finite
resonance width, the first qualitative difference is that the critical
temperature is nonzero even for $\omega_0 \ge 2 \epsilon_F$,
representing the BCS regime, absent in the limit of infinitely narrow
resonances. Secondly, a $T_c(\omega_0)$ for a finite-width resonance
exhibits a maximum at intermediate values of $\omega_0$, reflecting
the boson-boson interaction correction to the critical temperature in
the BEC phase.

\section{$P$-Wave BCS-BEC Crossover and Phase Transitions}
\label{PWaveChapter}

\subsection{Coherent-state formulation and saddle-point approximation}
\label{csPwave}

In Sec.~\ref{Sec:PWaveTwoChannel} we established the proper two-channel model for a
$p$-wave resonantly interacting atomic gas and determined its
parameters in terms of results of a two-body scattering experiment. We
now turn to the study of this model at a fixed chemical potential,
with the aim to establish the phases and phase transitions of such a
Fermi gas at finite density.

As usual, the thermodynamics is encoded in the partition function
$Z=\tr e^{-\beta H}$ and the corresponding free energy $F=-T\ln
Z$. The partition function can be conveniently formulated in terms the
imaginary-time path-integral over coherent states labelled by
commuting closed-channel fields $\bphi(\r)$, $\bphib(\r)$ (bosonic
molecules) and anticommuting open-channel fields $\psi(\r)$,
$\psib(\r)$ (fermionic atoms), and their complex conjugates
\begin{equation}
\label{eq:partfunc1}
 Z_p=\int {\cal D} {\psi} {\cal D} {\psib} {\cal D} \bphi {\cal D}{\bphib}
~e^{-S_p},
\end{equation}
with the action $S_p$ corresponding to the Hamiltonian $H_p^{2-ch}$,
\rfs{eq:ham_p}, given by
\begin{widetext}
\begin{eqnarray}
\label{Sp}
S_p[\bphi,\psi]= \int_0^\beta d\tau
\int d^3 r~\left[\psib\left(\d_\tau
-{\grsq \over 2 m}-\mu \right)\psi
+\sum_\alpha\phib_\alpha\left(\d_\tau+\epsilon_\alpha
- 2\mu -{\grsq \over 4 m} \right) \phi_\alpha
+ g_p\sum_\alpha\left(\phi_\alpha\psib\ i \nabla_\alpha\psib
                   +\phib_\alpha \psi\ i\nabla_\alpha
                   \psi\right)\right]
\nonumber\\
\end{eqnarray}
As with the $s$-wave case in \rfs{eq:partfunceff} the fermionic atoms can be
formally integrated out exactly, to give the effective bosonic action
\begin{eqnarray}
\label{SeffP}
S_p[\phi]=-\oh \tr \ln \left(
\matrix { \oh \left( i\omega_n -{\grsq \over 2m} - \mu  \right)
& i g \bphi\cdot\nabla \cr
  i g \bphib\cdot\nabla
& \oh \left(i \omega_n + {\grsq \over 2m} + \mu \right) } \right)
+ \int_0^\beta d \tau d^3 r ~\phib _\alpha
\left( \d_\tau+\epsilon_\alpha - 2 \mu -{\grsq \over 4 m} \right)
\phi_\alpha.
\end{eqnarray}
\end{widetext}
where the trace is over the $2\times2$ matrix structure, space $\r$ and the
fermionic Matsubara frequencies $\omega_n=\pi(2n+1)/\beta$.

The field theory $S_p[\phi]$ is nonlinear in $\phi$ and therefore
cannot be solved exactly. However, as discussed in
Sec.~\ref{sec:pwave2Cfinitedensity} for a narrow resonance it is
characterized by a dimensionless {\em detuning-independent} parameter
$\gamma_p$, \rfs{gamma_p}, and can therefore be systematically
analyzed as a perturbative expansion in $\gamma_p\ll 1$.

A lowest order in this expansion in $\gamma_p$ corresponds to a
computation of the function integral over $\phi$ via a saddle-point
method. The dominant saddle-point configuration is a constant
${ \phi_\alpha}(\r)=B_\alpha$, that is proportional to the condensate of the
zero-momentum bosonic operator according to
\begin{equation}
\bB\delta_{\p,{\bf 0}}={1\over\sqrt{V}}{\bf b}_{\bp=0}.
\label{condensateB}
\end{equation}
The resulting saddle-point action then becomes quadratic in the
fermionic fields and can therefore be easily computed. Within this
approximation it gives the free-energy density $f_p=S_p[\bB]/(\beta
V)$
\begin{equation}
f_p[\bB]=\sum_\alpha(\epsilon_\alpha - 2\mu)\Bb_\alpha B_\alpha
-\frac{T}{2V}\sum_{\k,\omega_n}\ln(\omega_n^2+E_\k^2),
\label{Sp_sp}
\end{equation}
where
\begin{equation} \label{eq:Ekp}
E_\k = \sqrt{\left({k^2 \over 2 m} - \mu\right)^2  + 4 g_p^2
|\bB\cdot\k|^2}
\end{equation}
and is also the spectrum of the Bogoliubov quasiparticles discussed
below. In above, $\bB$ is determined by the minimum of $f_p[\bB]$,
given by the saddle-point equation $\delta f_p[\bB]/\delta\Bb_\alpha=0$:
\begin{equation}
\label{eq:saddlepT}
\left( \epsilon_\alpha - 2 \mu \right) B_\alpha
= \sum_\beta I^{(T)}_{\alpha \beta}\left[ {\bf B} \right] B_\beta,
\end{equation}
where
\begin{equation} \label{eq:Iab}
I^{(T)}_{\alpha \beta}\left[ {\bf B} \right] =g_p^2 \int {d^3 k \over (2 \pi)^3} { k_\alpha
k_\beta\tanh\left({E_\k\over 2 T}\right)\over  E_\k},
\end{equation}
obtained by simple contour integration over $z\equiv i\omega_n$

Above expressions can also be equally easily obtained by working
within the operator (rather than functional integral) formalism,
approximating the Hamiltonian by a fermionic quadratic form with a
variational parameter $\bB$, performing a standard Bogoliubov
transformation, followed by a trace over decoupled Bogoliubov
quasiparticles and minimizing the resulting free energy over $\bB$.

The complex vector ``order parameter'' $\bB$ can be uniquely and
conveniently decomposed according to
\begin{equation}
\label{eq:Buv} \bB={\bf u} + i {\bf v},
\end{equation}
where $\bu$ and $\bv$ are two real vectors.

For latter use it is important to establish a relation between vectors
$\bu$ and $\bv$ ($6$ real components) and states with a definite
angular momentum, characterized by $3$ complex wavefunctions
$B^{(m=0,\pm 1)}$.  This connection is contained in
Eqs.~\rf{relationsbetweenmuandalpha1}, \rf{relationsbetweenmuandalpha2}, \rf{relationsbetweenmuandalpha3}.  Firstly, we note that under a global gauge
transformation $\bB\rightarrow e^{i\varphi}\bB$,
\begin{equation} \label{eq:barBBp}
\bB\cdot\bB = (u^2-v^2) + i 2\bu\cdot\bv,
\label{BB}
\end{equation}
transforms as a two-dimensional rank-2 tensor, with real and
imaginary components rotated into each other by an angle $2\varphi$,
while
\begin{equation}
\bBb\cdot\bB = u^2+v^2,
\label{B*B}
\end{equation}
is a gauge-invariant scalar. Using these transformations, it can be
shown that as long as $\bu$ and $\bv$ are {\em not} parallel, a phase
$\varphi$ can always be chosen to make them perpendicular.  If $\bu$
and $\bv$ are parallel, then they remain parallel, and $\bB$ can be
made real by a choice of $\varphi$; hence, a $\bu || \bv$ state is
equivalent to a state with $\bv=0$.  We also note that a state
characterized by $u=v$ and $\bu\cdot\bv=0$ retains these properties.

Using Eqs.~\rf{relationsbetweenmuandalpha1}, \rf{relationsbetweenmuandalpha2}, \rf{relationsbetweenmuandalpha3}, and \rf{eq:Buv} we find
\begin{eqnarray}
B^{(0)}&=&u_z + i v_z,\\
B^{(\pm1)}&=&(u_x \mp v_y) + i (v_x\pm u_y),
\end{eqnarray}
which shows that the $m=0$ $p$-wave superfluid corresponds to $\bu
||\bv$ (equivalently $\bv = 0$) pointing along the $m=0$ quantization
axis, and $m=\pm1$ superfluids are characterized by states with $\bu
\perp \bv$, $u=v$ with the projection of the angular momentum onto
$\bu \times \bv$ equal to $\pm 1$, respectively. All other $\bu$,
$\bv$ states are related to a linear combination of above three
eigenstates by a gauge transformation.

It is also important to summarize symmetries of the free energy
$f_p[\bB]$, \rfs{Sp_sp},\rf{eq:Ekp}. Firstly, quite clearly $f_p[\bB]$
is invariant under gauge transformations. Secondly, in a symmetric
case of degenerate $m=0,\pm 1$ Feshbach resonances with
$\epsilon_\alpha=\epsilon_0$, the free energy is also rotationally
invariant. Thus, at a quadratic level $f_p[\bB]$ must be a function of
the only rotationally, gauge-invariant quadratic form, \rf{B*B}. At a
higher order in $\bB$, all terms can be expressed as powers of this
quadratic invariant and an independent quartic term $|\bB\cdot\bB|^2$,
a magnitude-squared of the quadratic form in \rf{BB}.  In the
physically interesting case where the rotationally symmetry is
explicitly broken by distinct $\epsilon_\alpha$'s, generically
$f_p[\bB]$ will not exhibit rotational symmetry. However, within the
saddle-point approximation, it is easy to see that the first,
quadratic term in $f_p[\bB]$ is the only one that breaks rotational
symmetry, with higher order terms a function of the two independent
gauge- and rotationally-invariant combinations $\bBb\cdot\bB$ and
$|\bB\cdot\bB|^2$.

\subsection{Zero-temperature: ground state of a $p$-wave resonant Fermi gas}
\label{groundstatePwave}

\subsubsection{Saddle-point equation and ground-state energy}
\label{EgsPwave}

We focus on the case of zero temperature, for which the free-energy
density reduces to the ground-state energy density
$f^{T=0}_p[\bB]=\ve_{GS}[\bB]$
\begin{equation}
\ve_{GS}[\bB]=\sum_\alpha(\epsilon_\alpha - 2\mu)\Bb_\alpha B_\alpha
-\oh\int{d\omega d^3 k\over(2\pi)^4}\ln(\omega^2+E_{\bf k}^2),
\label{Egs_sp}
\end{equation}
with the saddle-point (gap) equation given by \rf{eq:saddlepT} and
\begin{equation} \label{eq:Iabzero}
I_{\alpha \beta}\left[ {\bf B} \right] =g_p^2 \int {d^3 k \over (2 \pi)^3}
{ k_\alpha k_\beta\over  E_\k}.
\end{equation}

It is advantageous at this stage to trade the parameter
$\epsilon_\beta$ for a physical detuning $\omega_\beta$, according to
\begin{equation} \omega_\beta = \frac{\epsilon_\beta-c_1}{1+c_2},
\end{equation}
introduced in \rfs{eq:omegap}. This gives a renormalized saddle-point equation
\begin{equation} \label{eq:saddlepren}
\left( \omega_\alpha (1+c_2) - 2 \mu \right) B_\alpha = \sum_\beta
\left( I_{\alpha \beta} - c_1 \delta_{\alpha \beta} \right)
B_\beta.
\end{equation}

To proceed further, we need to calculate $I_{\alpha \beta}$, that we
do in detail in Appendix \ref{appendixIab}. Since the integral is
formally divergent as $k^3$, the leading contribution to $I_{\alpha
  \beta}$ comes from short scales (high energies), cut off by
$\Lambda$ corresponding to the inverse size of the closed-channel
molecule. This leading $\Lambda^3$ contribution is given by
\begin{equation}
I_{\alpha \beta}^{\left( \Lambda^3 \right)}=c_1\ \delta_{\alpha
  \beta},
\label{IabLambda3}
\end{equation}
with $c_1$ defined in \rfs{eq:c1} by the two-atom $p$-wave scattering
calculation, \rfs{eq:polap}, that led to the definition of
$\omega_\alpha$.  Hence, as in the $s$-wave case, this leading
uv-cutoff dependent contribution identically cancels the $c_1$ term in
\rfs{eq:saddlepren}, and therefore does not contribute to any physical
quantity expressed in terms of a physical detuning $\omega_\alpha$.

However, $I_{\alpha \beta}$ also has a subleading cutoff-dependent
contributions that scale linearly with $\Lambda$, and are given by
\begin{eqnarray}
I_{\alpha \beta}^{\left( \Lambda^1 \right)}&=&
2 \mu c_2 \delta_{\alpha\beta}
-\frac{8}{5} m g_p^2 c_2 \left(\delta_{\alpha \beta} |B|^2 +
\Bb_\alpha B_\beta + \Bb_\beta B_\alpha\right),\nonumber\\
\label{IabLambda1}
\end{eqnarray}
with the dimensionless constant $c_2$ identical to that defined by the
two-atom scattering theory, \rfs{eq:c2}.

A tensor $I_{\alpha\beta}$ also contains uv-cutoff independent
low-energy contributions coming from momenta around Fermi surface.
Because these are infrared divergent at $\bB=0$, they are nonanalytic
in $B=|\bB|$, and therefore (as usual) are in fact dominant at small $B$,
relevant to the positive detuning BCS regime.  As detailed in Appendix~\ref{appendixIab} these contributions are easiest to evaluate in the
gauge where $\bu$ and $\bv$ are perpendicular, and together with
Eqs.\rf{IabLambda3}, \rf{IabLambda1} in the $\bu\cdot\bv=0$ gauge
finally give the explicit gap equation
\begin{widetext}
\begin{eqnarray}
\label{sp_final}
 \left(1 + c_2 \right)(\omega_\alpha -2 \mu) B_\alpha &=&
 -\gamma_p c_2\frac{8\epsilon_F}{5n}
\sum_\beta\left[\delta_{\alpha \beta} \left(
u^2 + v^2 \right) + 2 u_\alpha u_\beta + 2
v_\alpha v_\beta\right]B_\beta\nonumber\\
&&+ \gamma_p \mu\sqrt{\frac{\mu}{\epsilon_F}}
\sum_\beta\left\{\delta_{\alpha\beta}\ln \left[ {8 e^{-8/3}\mu \over
m g_p^2 (u+v)^2} \right] - \frac{2u_\alpha
u_\beta}{u(u+v)} - \frac{2v_\alpha
v_\beta}{v (u+v)} \right\} B_\beta\ .
\end{eqnarray}
\end{widetext}

Integrating these saddle-point equations over $\Bb_\alpha$ we obtain
the ground state energy density
\begin{widetext}
\begin{eqnarray}
\label{Egs}
\frac{\ve_{GS}(\bu,\bv)}{1+c_2}= \sum_\alpha
\left(u_\alpha^2+v_\alpha^2 \right) \left[ \omega_\alpha - 2 \mu
+a_1 \ln\left\{ a_0 \left(u+v \right)\right\} \right]
+ a_1 {u^3+v^3\over u+v} + a_2\left[ \left(u^2+v^2 \right)^2 +
\oh \left( u^2-v^2 \right)^2 \right],
\end{eqnarray}
\end{widetext}
where
\begin{eqnarray}
\label{eq:a1} a_1&=&{2\gamma_p\over 1+c_2}\mu \sqrt{\frac{\mu}
{\epsilon_F}}~\theta(\mu),\\
\label{eq:a2} a_2&=&{8\over 5} {c_2\gamma_p\over 1+c_2}
\frac{\epsilon_F}{n},\\
\label{eq:a0} a_0&=& e^{5/6} (\epsilon_F/\mu)^{1/2}(\gamma_p/8n)^{1/2},
\end{eqnarray}
and
\begin{eqnarray} \gamma_p &=& \frac{\sqrt{2}}{3 \pi^2} g_p^2
\epsilon_F^{1/2} m^{5 / 2},\\
&=&\frac{m^2 g_p^2}{3 \pi^2}k_F=k_F/k_g
\end{eqnarray}
is the dimensionless $p$-wave Feshbach resonance coupling discussed
previously. It is straightforward to check that $\delta\ve_{GS}/\delta
\Bb_\alpha=0$ gives back \rfs{sp_final}.

We emphasize that Eqs.\rf{sp_final}, \rf{Egs} are written in the
$\bu\cdot\bv=0$ gauge. However, once obtained we can utilize the
gauge-invariance of $\ve_{GS}$ to reexpress it in an arbitrary gauge.
To this end we note that $u^2+v^2=\bBb\cdot\bB$ is already invariant.
However, while $u^2-v^2$ is not gauge invariant (being a real part of
$\bB\cdot\bB$, \rfs{eq:barBBp}), its square is a gauge-invariant operator
written in $\bu\cdot\bv=0$ gauge, i.e., in the $\bu\cdot\bv=0$ gauge
$(u^2-v^2)^2 = |\bB\cdot\bB|^2$.

Thus, a gauge invariant form of $\ve_{GS}[\bB]$ is given by
\begin{widetext}
\begin{eqnarray}
\label{EgsGaugeInvnt}
\frac{\ve_{GS}(\bu,\bv)}{1+c_2}= \sum_\alpha
\left(u_\alpha^2+v_\alpha^2 \right) \left[ \omega_\alpha - 2 \mu
+a_1 \ln\left\{ a_0 \left(u+v \right)\right\} \right]
+ a_1 {u^3+v^3\over u+v} + a_2\left[ \left(\bBb\cdot\bB\right)^2 +
\oh |\bB\cdot\bB|^2\right],
\end{eqnarray}
\end{widetext}
where
\begin{eqnarray}
\label{uGaugeInvnt}
u &\rightarrow&\frac{1}{\sqrt{2}}\sqrt{\bBb\cdot\bB + |\bB\cdot\bB|},\\
\label{vGaugeInvnt}
v &\rightarrow&\frac{1}{\sqrt{2}}\sqrt{\bBb\cdot\bB - |\bB\cdot\bB|},
\end{eqnarray}
A global minimum of the energy density function $\ve_{GS}[\bB]$,
\rfs{EgsGaugeInvnt} then determines the ground state of a $p$-wave
paired superfluid at fixed chemical potential, and possible quantum phase
transitions as a function of detuning and chemical potential as the
nature of the minimum changes.

\subsubsection{Particle number equation}

As discussed earlier in the context of an $s$-wave superfluid, for
atomic gas experiments of interest to us, it is more relevant to
determine the ground state at a fixed total atom number $N$, rather
than a chemical potential.  As usual, however, this problem is related
to the fixed $\mu$ result by supplementing a minimization of
$\ve_{GS}$ (the gap equation, \rfs{sp_final}) with the total atom number
equation. The latter is given
\begin{eqnarray}
\label{NeqnGeneral1}
n&=&\frac 1 V \langle\bB|\hat{N}|\bB\rangle,\\
\label{NeqnGeneral2}
&=&-{\partial\ve_{GS}\over\partial\mu},
\end{eqnarray}
where the right-hand side is the expectation value of the total atom
number computed in the grand-canonical ensemble, i.e., at fixed $\mu$,
in the ground state $|\bB\rangle$ (a BCS-type variational one,
labelled by $\bB$ in the case of the saddle-point approximation).
This gives a relation between $N$ and $\mu$, thereby allowing one to
eliminate the latter in favor of the former.  We thus turn to the
computation of the atom number equation.

Within the above saddle-point approximation (that ignores molecular
field fluctuations) valid at a small $\gamma_p$, the atom number
density equation is given by
\begin{equation}
\label{nEqn_pwave}
n=2 |\bB|^2 + n_f,
\end{equation}
where the fermion density is given by
\begin{equation} \label{eq:pnp}
n_f = \oh \int {d^3 k \over (2 \pi)^3}
\left[ 1-\frac{k^2/2 m - \mu } {E_\k} \right].
\end{equation}
The coefficient $1/2$ in front of the integral, absent in the $s$-wave
case, \rfs{eq:zeropart} is due to the fact that here there is only a
single species of fermions (``polarized'' isospin).  Clearly,
according to \rfs{NeqnGeneral2}, result \rf{nEqn_pwave},\rf{eq:pnp}
can be equivalently obtained by differentiating $\ve_{GS}$ with
respect to $\mu$.

It is essential to note a crucial qualitative difference between
\rfs{eq:pnp} and its $s$-wave counterpart \rfs{eq:zeropart}. For $g_s^2
B^2 < \mu$, the $s$-wave fermion density \rfs{eq:zeropart} at nonzero
$B$ can be estimated by the density $p_\mu^3/(3 \pi^2)$
($p_\mu=\sqrt{2 m \mu}$) of a degenerate noninteractive fermion gas at
the same chemical potential. However, because in the $p$-wave case,
for $\bB\neq 0$ the occupation number $n_f(k)$ (integrand in
\rfs{eq:pnp}) exhibits a long tail, the integral in \rfs{eq:pnp} is
formally linearly divergent at large momenta, cutoff only by the
inverse closed-channel molecular size
$\Lambda$~\cite{Yip2005,Gurarie2005}. To compute the fermion number, we
separate out this large $\bB$-dependent short-scale contribution,
finding
\begin{equation}
\label{nf_uvPart}
n_f =  n_{0f}+ 2 c_2 |\bB|^2,
\end{equation}
where $c_2=m^2 g_p^2\Lambda/3\pi^2$ is the dimensionless parameter that
already appeared in the two-body study of the $p$-wave two-channel
model, Sec.~\ref{Sec:PWaveTwoChannel}, see \cite{c2largeComment}, and
\begin{equation}
\label{eq:redpartp}
n_{0f}=\oh \int {d^3 k \over (2 \pi)^3}
\left[ 1-\frac{k^2/2 m - \mu } {E_\k}
-\frac{8m^2 g_p^2}{3k^2}|\bB|^2 \right]
\end{equation}
is a remaining contribution to $n_f$ that is uv-convergent, i.e., is
not dominated by large momenta, and as a result for $g_p^2 |\bB|^2 \ll
\mu$, can be estimated by its $g_p=0$ value
\begin{equation}
\label{n0fNormal}
n_{0f} \approx \frac{\left( 2 m \mu \right)^{3/2}}{6 \pi^2}\theta(\mu),
\ \ \text{for}\ g^2|\bB|^2\ll \mu,
\end{equation}
where $\theta(\mu)$ is the usual theta-function, equal to $1$ for
positive argument and to $0$ for negative argument.

In the range of detuning where $\mu \lesssim g_p^2 |\bB|^2$, the full
integral in \rfs{eq:redpartp} must be computed more precisely, but
this is a very narrow range of the chemical potential and can (and
will) be ignored.

Thus we find that the atom number-density equation is given by
\begin{equation}
\label{eq:pnepr}
n=2 (1+c_2)|\bB|^2 + n_{0f},
\end{equation}
to be contrasted with its $s$-wave analog, \rfs{eq:zeropart}. As noted
above the number equation can be directly obtained from $\ve_{GS}$,
\rfs{Egs}, via \rfs{NeqnGeneral2} and in particular the key
enhancement factor $(1+c_2)$ above arises from the same factor in
$\ve_{GS}$. Its implication depends on $c_2$. If $c_2 \ll 1$, then the
number equation is no different than its $s$-wave counterpart and for
example in the BEC regime, where $\mu < 0$ the total atom number is
``carried'' by the bosons. If, however, $c_2 \gg
1$,\cite{c2largeComment} then it shows that even deep in the BEC
regime, where $\mu$ is large and {\em negative} and correspondingly
$n_{0f}$ is vanishingly small, the density of bosons is given by $n/(2
c_2)$ and is a small fraction of the total atom density, $n$. In this
case the total atom number is in a form of free atoms with density
given by $2c_2|\bB|^2$, the last term in \rfs{nf_uvPart}. This is a
reflection of the fact that the $p$-wave interactions (proportional to
$k^2$, due to a centrifugal barrier diverging at short scales) are
strong at large momenta and therefore for large $c_2$ lead to a large
depletion of the molecular condensate, even in the BEC regime where
fermions are at a negative chemical potential. This is a phenomenon
not previously discussed in the literature.

\subsubsection{Phases and phase transitions of the $p$-wave BCS-BEC superfluid}
\label{PPWC}
Zero-temperature phases, crossover and transitions as a function of
detuning in a $p$-wave resonant Fermi gas are completely encoded
inside the ground state energy function $\ve_{GS}[\bB,\mu]$, \rf{Egs}
or, equivalently the associated gap and number equations,
Eqs.\rf{sp_final},\rf{eq:pnepr}.  From our earlier analysis of the
$s$-wave BCS-BEC crossover for a narrow resonance in
Sec.\ref{sec:inrl}, we can already anticipate some of the qualitative
phenomenology associated with changing of the detuning.  At zero
temperature the gas will condense into a $p$-wave superfluid that at
large positive $\omega_0$ will be of a BCS type with weakly-paired,
strongly-overlapping Cooper pairs and correspondingly an exponentially
small boson number. As $\omega_0$ is lowered past $2\epsilon_F$, the
number of bosons in the condensate will grow as a power of
$2\epsilon_F-\omega_0$, while the number of fermions will diminish,
reflected in the tracking of the chemical potential with detuning,
$\mu\approx\omega_0/2$. This intermediate crossover regime will
thereby consist of a superposition of small (of size $\Lambda^{-1}$)
closed-channel molecular bosons and much larger Cooper pairs.
Finally, for $\omega_0$ lowered below zero, the tracking chemical
potential will change sign to $\mu < 0$ and (for small $c_2$) the
condensate will transform into a purely molecular Bose-Einstein
condensate.

Although very generally this picture remains correct, there are a
number of qualitatively important differences in evolution with the
detuning between $s$-wave and $p$-wave superfluids.  Firstly, $p$-wave
superfluid is characterized by a richer complex {\em vector} order
parameter $\bB$, associated with $\ell=1$ angular momentum of the
condensing boson, and therefore admits a possibility of a variety of
distinct $p$-wave superfluid ground states and associated quantum
phase transitions between them. Possible superfluid ground states are
distinguished by a projection of condensate's angular momentum along a
quantization axis. This allows for a possibilities of a time-reversal
breaking $m=1$ states (and its rotated and time-reversed versions)
referred to as a $p_x + i p_y$-superfluid with a projection of the
angular momentum of the condensed bosons onto the $z$-axis equal to
$+1$, or a $p_z$-superfluid (and its rotated analog), with a
projection of the condensate's angular momentum onto the $z$-axis
equal to 0.  As discussed in Sec.\ref{csPwave} these two phases are
characterized by $\bu$ and $\bv$, defined in \rfs{eq:Buv}, with the
$u=v$, $\bu\perp\bv$ state corresponding to $p_x+i p_y$ phase, and
$v=0$ state the $p_z$ phase (which we will often refer to as $p_x$ phase as well,
~\cite{pwaveNames}), respectively. It is also possible to have
a ``superposition'' phase, where $\bu\perp\bv$, but with unequal
lengths, $u\neq v$, corresponding to a time-reversal breaking state in
which all bosons condense into a linear combination of $p_z$ and
$p_x+ip_y$ orbitals.

Secondly, and related to above, a $p$-wave gas is characterized by
(potentially) three distinct detunings, $\omega_\alpha$, one for each
component of the $\ell=1$-field $b_\alpha$. As discussed by Ticknor,
et al. \cite{Ticknor2004}, in systems of interest to us, this resonance
splitting, $\delta$ arises due to the interatomic dipolar interaction
predominately due to electron spin.  Although it is rotationally
invariant in the spin-singlet closed-channel, the source of anisotropy
is a small admixture of the spin-triplet channel, with a result that,
with the quantization axis along the external magnetic field ${\bf H}$
(that we take to be along $\hat{\bf x}$), the $m=0$ resonance is lower
by energy $\delta > 0$ than the degenerate $m=\pm 1$ doublet. Thus we
will take
\begin{eqnarray}
\label{splitOmega}
\omega_x&=&\omega_0,\nonumber\\
\omega_{y,z}&=&\omega_0+\delta,
\end{eqnarray}
This feature will be key to a nontrivial phase diagram possibilities
illustrated in Figs.~\ref{Fig-phasediagno}, \ref{Fig-phasediaglow}, \ref{Fig-phasediaginter},
\ref{Fig-phasediaghigh}.

Finally, another important difference that has already been noted in
the previous subsection is the large $c_2$ limit of the $p$-wave
number equation, \rf{eq:pnepr}, in which even for $\mu<0$ (in what one would
normally call the BEC regime) the fermion density is large and
correspondingly the boson density $|\bB|^2\approx n/(2 c_2)$ is
vanishingly small for $c_2\gg 1$~\cite{c2largeComment}.

To determine which of the $p$-wave superfluid phases is realized by
the BCS-BEC condensate, we minimize the ground-state energy
$\ve_{GS}$, \rfs{Egs} with respect to $\bu$ and $\bv$ for
$\omega_\alpha$ of interest, while enforcing the total atom
number-density constraint \rfs{eq:pnepr}.

\paragraph{Isotropic $p$-wave Feshbach resonance}
\label{isotropicPwave}

We first consider a simpler isotropic case, where
$\omega_\alpha=\omega_0$ for all $\alpha$.  Utilizing the rotational
invariance of $\ve_{GS}$, it sufficient to minimize it over magnitudes
$u$ and $v$. Analogous to other isotropic problems with a vector order
parameter (e.g., a Heisenberg magnet), the actual global (as opposed
to their relative) direction of vectors $\bf u$, $\bf v$ in the
ordered phase will be chosen {\em spontaneously}.

Although ultimately we need to minimize $\ve_{GS}[u,v]$ at fixed total
atom number, i.e., subject to the atom number equation contraint
\begin{equation} \label{eq:pneuv}
2 (1+c_2) (u^2+v^2) + n_{0f} = n.
\end{equation}
it is important to first study $\ve_{GS}[u,v]$ at fixed $\mu$.
Standard analysis of $\ve_{GS}$, \rfs{Egs} shows that there are four
extrema: (i) $u=v=0$ (normal state), (ii) $u\neq 0$, $v=0$
($p_x$-superfluid state), (iii) $u=0$, $v\neq0$ ($p_x$-superfluid
state), and (iv) $u=v\neq0$ ($p_x+i p_y$-superfluid state), where
clearly (ii) and (iii) correspond to the same superfluid
state.\cite{pwaveNames} After some standard algebra, one can show that
the normal ($u=v=0$) state is always a maximum with energy
$\ve_{GS}[0,0]\equiv\ve_{GS}^{N}=0$.

\begin{figure}[bt]
\includegraphics[height=3in]{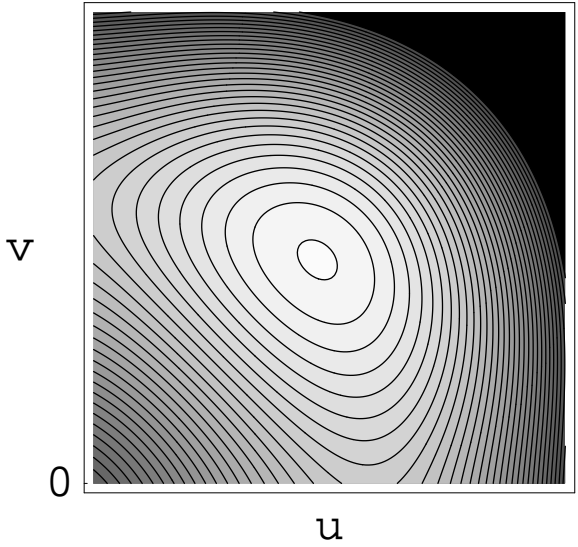}
\caption{A contour plot of $\ve_{GS}(u,v)$ in the absence of
splitting, $\delta=0$. The global minimum at $u=v\neq0$, saddle-points at
$u=0, v\not=0$ and $u\not =0$, $v=0$, and a maximum at $u=v=0$ can
clearly be seen.}
\label{Egspxpy}
\end{figure}

The nature and relative stability of the other extrema is decided by
the parts of $\ve_{GS}[u,v]$ that do not depend on the $u^2+v^2$
combination, namely by terms
$$
a_1 \left( (u^2+v^2)\ln(u+v) + \frac{u^3+v^3}{u+v} \right) +
\frac{a_2}{2} \left(u^2-v^2\right)^2.
$$ As illustrated in Fig.~\ref{Egspxpy} and standard analysis shows
that $u\neq 0, v=0$ and $u=0, v\neq 0$ extrema are degenerate
(guaranteed by $u\leftrightarrow v$ symmetry) saddle-points and
$u=v\neq0$ is a global minimum, independent of the actual values of
$a_1$ and $a_2$, as long as they are positive. In the BEC regime,
where $a_1=0$, this is clear since the last $a_2$ term prefers the
$u=v$ state, but also remains true throughout the BCS and the
crossover regimes.

Because there is only one local minimum, this fixed chemical potential
result automatically applies to the minimization of $\ve_{GS}$ at
fixed total atom density $n$, with $\mu$ eliminated through
\rfs{eq:pneuv}.  Thus we conclude that at $T=0$, the ground state of a
Fermi gas interacting with an {\em isotropic} Feshbach resonance is a
$p_x + i p_y$-wave superfluid throughout the BEC-BCS crossover.

In the BCS regime, this results agrees with the well-known prediction
by Anderson and Morel \cite{Anderson1961}, who showed (in the context
of A$_1$-phase of $^3$He) that a polarized $p$-wave BCS superconductor
at $T=0$ is always in the $p_x+ip_y$ state. Thus, our above conclusion
extends their result to the BEC ($\mu < 0$) and crossover ($0 < \mu <
\epsilon_F$) regimes of a resonantly-paired superfluid.

We now compute this $p_x + i p_y$ ground state explicitly.  To this
end, we substitute $u=v$ into the ground state energy, \rfs{Egs} to
obtain $\ve_{GS}^{p_x + i p_y}[u]\equiv\ve_{GS}[u,u]$
\begin{equation} \label{EgsPx+iPy}
\frac{\ve_{GS}^{p_x + i p_y}}{1+c_2}=
2 u^2 \left[\omega_0 - 2 \mu + a_1 \ln(2 a_0 u)\right]
+ a_1 u^2 + 4 a_2  u^4,
\end{equation}
and minimize it with the constraint \rfs{eq:pneuv}. At fixed $\mu$,
the saddle-point equation $\partial\ve_{GS}^{p_x + i p_y}[u]/\partial
u=0$ is given by
\begin{equation}
\label{spPx+iPy}
u[\omega_0 - 2\mu + a_1 + a_1\ln(2a_0 u)] + 4 a_2 u^2 = 0.
\end{equation}
As in the $s$-wave case, once the atom number constraint is
implemented, the detailed behavior is quite different in three
regimes, depending on the range of detuning $\omega_0$.

{\em BCS regime:}

For $\omega_0 > 2\epsilon_F$, closed-channel molecules (${\bf b}$) and
the corresponding condensate are energetically costly leading to a
small $u$. This justifies us to neglect the molecular contribution
(first term) inside the number equation \rf{eq:pneuv}. Then, with
approximation of $n_{0f}(\mu)$ by the normal state atom density (i.e.,
also ignoring the small condensate density there) \rf{n0fNormal}
immediately gives $\mu\approx\epsilon_F$.  Furthermore, similarly
neglecting the subdominant quartic term, $a_2 u^4$ inside
$\ve_{GS}^{p_x + i p_y}(\delta=0)$, \rfs{EgsPx+iPy}, the corresponding
saddle-point equation can then be solved analytically, giving
\begin{eqnarray}
\label{eq:upx+ipy}
u^{\delta=0}_{p_x+i p_y} &=& \frac{1}{2 a_0 e} ~e^{-(\omega_0 - 2 \epsilon_F)/a_1},\\
&=&e^{-11/6}\sqrt{2n\over\gamma_p}~e^{-(\omega_0 - 2
  \epsilon_F)(1+\gamma_p\Lambda/k_F)/(2\gamma_p\epsilon_F)},\nonumber\\
&&\hspace{2cm}\ \ \text{for}\ \ \omega_0 > 2\epsilon_F,\ \delta=0,\nonumber
\end{eqnarray}
that is indeed exponentially small in this BCS regime. The
corresponding condensation energy density $ \ve_{GS}^{p_x + i
  p_y} (\delta) \equiv \ve_{GS}^{p_x + i
  p_y}[u^\delta_{p_x + i p_y},\delta] $ is given by
\begin{eqnarray}
\label{eq:Egspx+ipy}
\ve_{GS}^{p_x + i p_y}(0)&=&-(1+c_2)a_1 (u^0_{p_x+i p_y})^2,\\
&=&-4e^{-11/3}\ \epsilon_F n ~e^{-(\omega_0 - 2 \epsilon_F)
(1+\gamma_p\Lambda/k_F)/(\gamma_p\epsilon_F)}\nonumber\\
&&\hspace{2cm}\ \ \text{for}\ \ \omega_0 > 2\epsilon_F,\ \delta=0,\nonumber
\end{eqnarray}

Within the same set of approximations it is also straightforward to
compute the corresponding quantities for the $p_x$-state, obtaining ($u^0
\equiv u^{\delta=0}$)
\begin{eqnarray}
\label{eq:upxEgs}
u^0_{p_x} &=& \frac{1}{a_0 e^{3/2}} ~e^{-(\omega_0 - 2 \epsilon_F)/a_1},\\
&=&e^{-14/6}\sqrt{8n\over\gamma_p}~e^{-(\omega_0 - 2
  \epsilon_F)(1+\gamma_p\Lambda/k_F)/(2\gamma_p\epsilon_F)},\nonumber\\
\ve_{GS}^{p_x}(0)&=&-8e^{-14/3}\ \epsilon_F n~e^{-(\omega_0 - 2
  \epsilon_F)(1+\gamma_p\Lambda/k_F)/(\gamma_p\epsilon_F)},\nonumber\\
&&\hspace{2cm}\ \ \text{for}\ \ \omega_0 > 2\epsilon_F,\ \delta=0,
\end{eqnarray}
that gives a ratio ${\cal R}(\delta=0)=\ve_{GS}^{p_x+i
  p_y}(0)/\ve_{GS}^{p_x}(0)=e/2$ of condensation energies for the two
states, consistent with the numerical value reported in
Ref.~\cite{Anderson1961} and thereby confirms that $p_x+ip_y$ state is
energetically more favorable.

{\em Crossover and BEC regimes:}

For $\omega_0<2 \epsilon_F$, it becomes favorable (even in
$g_p \rightarrow0$ limit) to convert a finite fraction of the Fermi sea
(between $\omega_0$ and $2\epsilon_F$) into a BEC of closed-channel
molecules.  Consistent with this, the $\log$ contribution in
\rfs{EgsPx+iPy} is no longer large, with $\ve_{GS}^{p_x + i p_y}$
immediately giving a chemical potential that tracks the detuning
according to $\mu\approx \omega_0/2$ with accuracy of $O(\gamma_p)$.

As previously noted \cite{Andreev2004,Sheehy2006a}, we observe that the
roles of number and gap equations interchange in the $\omega_0 <
2\epsilon_F$ regime, with the former determining the molecular
condensate density and the latter giving the chemical potential.
Consistent with this, the number equation, \rfs{eq:pneuv} then gives
the growth of the bosonic condensate according to
\begin{equation}
u^2+v^2 \approx \frac{n}{2 (1+c_2)}\left[1-
\left({\omega_0\over2\epsilon_F}\right)^{3/2}\theta(\omega_0)\right],
\end{equation}
reaching a maximum value of
\begin{equation}
u^2+v^2=\frac{n}{2 (1+c_2)},
\end{equation}
for a negative detuning.  As already noted above, it is remarkable
that even for a large negative detuning the boson density never
reaches its (ideal, $\gamma_p\ll 1$) maximum value of $n/2$
corresponding to the total atom density $n$.  Instead, due to $p$-wave
interaction that is strong at short scales, $c_2 > 0$, and a $p$-wave
molecular condensate is depleted into open-channel atoms. We will
nevertheless continue to refer to this range of detuning as the BEC
regime.

Since as shown above, in the crossover and BEC regimes the ground
state remains a $p_x+i p_y$-wave superfluid the $p$-wave order
parameter is given by
\begin{eqnarray} \label{eq:upx+ipyBEC}
u^0_{p_x+i p_y}\approx \frac{n^{1/2}}{2(1+c_2)^{1/2}}\bigg[1&-&
\left({\omega_0\over2\epsilon_F}\right)^{3/2}\theta(\omega_0)\bigg]^{1/2},
\nonumber\\
&&\ \ \text{for}\ \ \omega_0 < 2\epsilon_F.
\end{eqnarray}

\paragraph{Anisotropic $p$-wave Feshbach resonance}
\label{anisotropicPwave}

We now analyze the more experimentally relevant anisotropic case
\cite{Ticknor2004}, where the triplet Feshbach resonance is split by
dipolar interactions into a $m=\pm 1$ degenerate doublet resonance and
an $m=0$ resonance, with $\omega_\alpha$ given by \rfs{splitOmega}.
With the magnetic field ${\bf H}$ picking out a special direction
(that we take to be $\hat{\bf x}$), the ground-state energy function
$\ve_{GS}$ is no longer rotationally invariant. Within our
saddle-point approximation this uniaxial anisotropy only enters
through the detuning part
\begin{equation}
\label{eq:noniso}
\ve_{GS}^{\text{anisot.}}[\bu,\bv]
=\omega_0 (u^2+v^2)+\delta(u_y^2+u_z^2+v_y^2+v_z^2).
\end{equation}
With $\delta > 0$, this uniaxial single-particle energy is clearly
minimized by $u_y=u_z=v_y=v_z=0$, i.e., when $\bu$ and $\bv$ are
parallel and point along ${\bf H}=H\hat{\bf x}$, corresponding to the
$p_x$-wave ground state. In our more convenient $\bu\cdot\bv=0$ gauge
choice, this $p_x$ state is equivalent to either $\bu$ or $\bv$
pointing along ${\bf H}$ and with the other vanishing. Furthermore, in
this transverse gauge for a $p_x+i p_y$ state \cite{pwaveNames} (that,
as we saw above is preferred by the interactions) in which neither
$\bu$ nor $\bv$ vanish, $\ve_{GS}^{\text{anisot.}}$ is clearly
minimized by choosing the {\em longer} of the $\bu$ and $\bv$ to be along
${\bf H}=H\hat{\bf x}$, while the shorter one spontaneously selects a
direction anywhere in the (yz-) plane perpendicular to
${\bf H}$.  For $u=v$, their overall orientation is chosen
spontaneously.

An explicit minimization over the {\em direction} of $\bu$-$\bv$
orthogonal set confirms these arguments, giving
\begin{equation}
\label{eq:nonisoMin}
\ve_{GS}^{\text{anisot.}}[u,v]
=\omega_0 (u^2+v^2)+\delta~\text{Min}[u^2,v^2].
\end{equation}
It is convenient to take advantage of the exchange symmetry $\bu
\leftrightarrow \bv$, $\ve_{GS}[u,v]=\ve_{GS}[v,u]$, and for $u\neq v$
(without loss of generality) always choose $\bu$ to be the longer
vector, with the other state physically equivalent. With this choice
and \rfs{eq:nonisoMin} the ground-state energy is minimized by $\bu$
directed along ${\bf H}$.  The resulting ground state energy as a
function of magnitudes $u$ and $v$, with $u>v$ and $\bu=u\hat{\bf x}$
takes the form
\begin{widetext}
\begin{eqnarray}
\label{EgsAnisot}
\hat{\ve}_{GS}[\uhh,\vhh]&=&
\w(\uhh^2+\vhh^2)+(\uhh^2+\vhh^2)^2+\oh(\uhh^2-\vhh^2)^2
+\dl\vhh^2
+ \a\left[(\uhh^2+\vhh^2)\ln(\uhh+\vhh)+ {\uhh^3+\vhh^3\over \uhh+\vhh}\right],\\
&&\hspace{5cm}\qquad\qquad\qquad\qquad\qquad
\text{for}\ \ u>v,\ \bu=u\hat{\bf x}, \bu\perp\bv.\nonumber
\end{eqnarray}
\end{widetext}
where to simplify notation we introduced dimensionless variables
\begin{eqnarray}
\w&\equiv&\big(\omega_0-2\mu+a_1\ln(a_0\sqrt{n})\big)/(a_2 n),\\
\dl&\equiv&\frac{\delta}{a_2 n},\\
\a&\equiv&\frac{a_1}{a_2 n},\\
\hat{\ve}_{GS}&\equiv&{\ve_{GS}\over(1+c_2)a_2
  n^2}\equiv{\ve_{GS}\over\ve_0},\\
\uhh&\equiv&\frac{u}{\sqrt{n}},\ \ \vhh\equiv\frac{v}{\sqrt{n}}.
\end{eqnarray}

\begin{figure}[bt]
\includegraphics[height=3in]{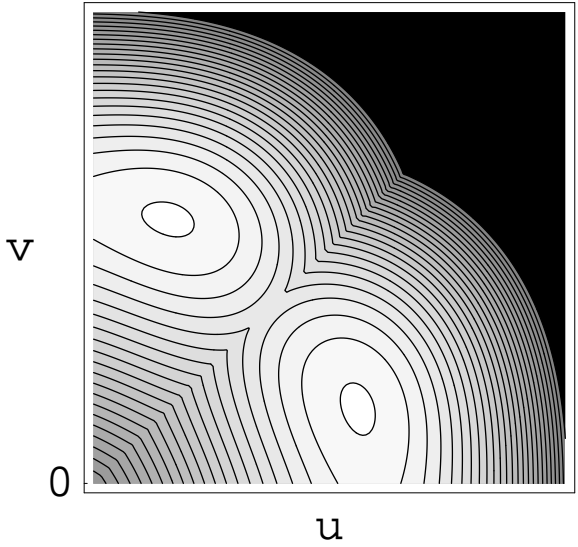}
\caption{A contour plot of $\ve_{GS}(u,v)$ in the presence of a small
splitting $\delta$. The global minimum at $u\not =v\neq 0$ and saddle-points
at $u=0, v\not=0$ and $u\not =0$, $v=0$, can clearly be seen.}
\label{Egspxpluspy}
\end{figure}

\begin{figure}[bt]
\includegraphics[height=3in]{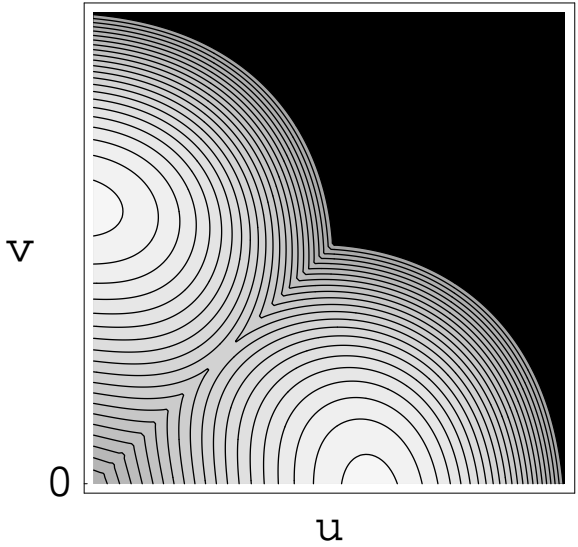}
\caption{A contour plot of $\ve_{GS}(u,v)$ in the presence of a large
splitting $\delta$. Only the global minima at $u=0, v\not=0$ and
$u\not =0$, $v=0$ are present}
\label{Egspx}
\end{figure}

As in the isotropic case above, the ground state as a function of
detuning $\omega_0$ and dipolar splitting $\delta$ is found by
minimizing $\hat{\ve}_{GS}[\u,\v]$ over magnitudes $\u$ and $\v$ with
the constraint of the total atom density equation, \rfs{eq:pneuv}
\begin{equation}
\label{numberCrosshat}
\uhh^2+\vhh^2=\frac{1}{2(1+c_2)} \left( 1-\left(\frac{\mu}{ \epsilon_F}
\right)^{3 \over 2} \theta(\mu)\right).
\end{equation}
Standard analysis shows that $\hat{\ve}_{GS}[\uhh,\vhh]$ generically
has three physically distinct\cite{pwaveNames} (confined to $\uhh >
\vhh > 0$ quadrant) extrema: (i) $\uhh=\vhh=0$ (normal state), (ii)
$\uhh > 0$, $\vhh=0$ ($p_x$-superfluid state), and (iii) $\uhh > \vhh
> 0$ ($p_x + i p_y$-superfluid state \cite{pwaveNames}). At zero
temperature for a fixed atom density the normal state is always a
maximum with energy $\ve_{GS}[0,0]\equiv\ve_{GS}^{N}=0$. However, in
contrast to the isotropic case, here the relative stability of the
$p_x$ and $p_x+i p_y$ states crucially depends on the detuning
$\omega_0$ and dipolar splitting $\delta$. This is summarized by the
contour plots of $\ve_{GS}(u,v)$ for $\delta=0$ (Fig.~\ref{Egspxpy}),
$\delta$ small (Fig.~\ref{Egspxpluspy}) and $\delta$ large
(Fig.~\ref{Egspx}).  We now study these in detail.

{\em BCS regime:}

As a generic property of the BCS regime, for $\omega_0 > 2\epsilon_F$,
the molecules are energetically suppressed, and only exponentially
small condensate is expected. The number equation then leads to
$\mu\approx\epsilon_F$. It also allows us to neglect the subdominant
quartic in $\uhh$ and $\vhh$ contributions inside $\ve_{GS}$,
\rfs{EgsAnisot}, allowing the corresponding saddle-point equation to
be solved analytically. For the two candidate $p$-wave superfluid
states we find:
\begin{eqnarray}
\label{px+ipyEgsAni}
u_{p_x+i p_y} &=&
(1+\delta/a_1)~e^{-\delta/2a_1} u^0_{p_x+i p_y},\\
&=&\frac{1}{2a_0 e}(1+\delta/a_1)~e^{-(\omega_0-2\epsilon_F+\delta/2)/a_1},
\nonumber\\
v_{p_x+i p_y} &=&
(1-\delta/a_1)~e^{-\delta/2a_1} v^0_{p_x+i p_y},\nonumber\\
&=&\frac{1}{2a_0 e}(1-\delta/a_1)~e^{-(\omega_0-2\epsilon_F+\delta/2)/a_1},
\nonumber\\
\ve_{GS}^{p_x + i p_y}(\delta)&=&-(1+\delta^2/a_1^2)e^{-\delta/a_1}
\ve_{GS}^{p_x+i p_y}(0),\nonumber\\
&&\hspace{3cm}\text{for}\ \ \omega_0 > 2\epsilon_F,\nonumber
\end{eqnarray}
and
\begin{eqnarray}
\label{pxAni}
u_{p_x} &=& u^0_{p_x},\\
&=&\frac{1}{a_0 e^{3/2}} ~e^{-(\omega_0 - 2 \epsilon_F)/a_1},\nonumber\\
v_{p_x} &=& 0,\nonumber\\
\ve_{GS}^{p_x}(\delta)&=& \ve_{GS}^{p_x}(0),\nonumber\\
\hspace{2cm}&&\ \ \text{for}\ \ \omega_0 > 2\epsilon_F,\nonumber
\end{eqnarray}
where the corresponding ground-state energies at finite splitting
$\delta$ have been expressed in terms of $\delta=0$ energies,
Eqs.\rf{eq:Egspx+ipy},\rf{eq:upxEgs}. The ratio of $p_x+i p_y$ and
$p_x$ ground-state energies is then given by
\begin{eqnarray}
{\cal R}(\delta)&=&{\ve_{GS}^{p_x+i
    p_y}\over\ve_{GS}^{p_x}},\nonumber\\
&=&{e\over 2}\left(1+{\delta^2\over a_1^2}\right)e^{-\delta/a_1},
\end{eqnarray}
and reduces to the previously found result of $e/2$ for $\delta=0$.

Consistent with the analysis of the ``{\em isotropic} resonance''
subsection, for low dipolar splitting $\delta$, ${\cal R}(\delta) > 1$
and in the BCS regime the $p_x + i p_y$ superfluid \cite{pwaveNames}
is the ground state, as seen on Fig.~\ref{Egspxpluspy}. However, ${\cal R}(\delta)$ reaches $1$ at
$\delta_c^{BCS}$, given by
\begin{eqnarray}
\label{delta_cBCS}
\delta_c^{BCS}&=&a_1,\\
&=&{2\gamma_p\over 1+c_2}~\epsilon_F,\nonumber\\
\end{eqnarray}
signaling a quantum phase transition from the $p_x+i p_y$ to $p_x$
ground state for $\delta > \delta_c$ \cite{pwaveNames}. This is
consistent with intrinsically positive quantity $v_{p_x+i
  p_y}(\delta)$, in \rfs{px+ipyEgsAni} turning negative (unphysical)
for $\delta > \delta_c^{BCS}$.

{\em BEC regime:}

We can similarly evaluate the order parameters, ground state energies
and the $p_x$-$p_x+i p_y$ quantum phase transition boundary in the
opposite, BEC regime of a large negative detuning $\omega_0$ and $\mu
< 0$, which reduces the number equation, \rfs{numberCrosshat} to
\begin{equation}
\label{numberBEChat}
\uhh^2+\vhh^2=\frac{1}{2(1+c_2)}\equiv \hat{n}_B.
\end{equation}
As discussed for the isotropic resonance case, in the BEC regime the
condensate is no longer exponentially small (given by a finite
fraction of total atom density, as seen above), and as a result we can
neglect the $\a$ terms in \rfs{EgsAnisot} for small $\a$. Standard
minimization of the resulting ground-state energy function, together
with the number equation \rf{numberBEChat} gives for two extrema,
one corresponding to a $p_x + i p_y$ superfluid \cite{pwaveNames}
\begin{eqnarray}
\label{px+ipyBECani}
u_{p_x+i p_y}&=&\frac{1}{\sqrt{2}}~n^{1/2}
\left(\hat{n}_B + \dl/2\right)^{1/2},\\
&=& \frac{1}{2}~n^{1/2}
\left(\frac{1}{1+c_2} + \frac{5(1+c_2)}{8 c_2\gamma_p}
\frac{\delta}{\epsilon_F}\right)^{1/2},\nonumber\\
v_{p_x+i p_y}&=&\frac{1}{\sqrt{2}}~n^{1/2}
\left(\hat{n}_B - \dl/2\right)^{1/2},\nonumber\\
&=& \frac{1}{2}~n^{1/2}
\left(\frac{1}{1+c_2} - \frac{5(1+c_2)}{8 c_2\gamma_p}
\frac{\delta}{\epsilon_F}\right)^{1/2},\nonumber\\
\ve_{GS}^{p_x+i p_y}(\delta)&=&-\ve_0(\hat{n}_B^2 +
\dl^2/8),\nonumber\\
&=&-\left({2c_2\gamma_p\over5(1+c_2)^2} +
{5(1+c_2)^2\over
  8c_2\gamma_p}\frac{\delta^2}{\epsilon_F^2}\right)\epsilon_F n,\nonumber
\end{eqnarray}
and one corresponding to a $p_x$ superfluid
\begin{eqnarray}
\label{pxBECani}
u_{p_x}&=& n^{1/2}\hat{n}_B^{1/2},\\
&=& \frac{1}{\sqrt{2(1+c_2)}}~n^{1/2},\nonumber\\
v_{p_x}&=& 0,\nonumber\\
\ve_{GS}^{p_x}(\delta)&=&-\frac{3}{2}\ve_0~\hat{n}_B^2,\nonumber\\
&=&-\frac{3}{5}{c_2\gamma_p\over(1+c_2)^2}~\epsilon_F n.\nonumber
\end{eqnarray}

As argued earlier for the isotropic case, for low dipolar splitting
$\delta$ the ground state is a $p_x+i p_y$ superfluid, with order
parameter and ground-state energy given in \rfs{px+ipyBECani}, illustrated on Fig.~\ref{Egspxpluspy}. As we
can see from the form of $v_{p_x+i p_y}$ this minimum and the
corresponding state disappears for $\dl > 2\hat{n}_B\equiv\dl_c^{BEC}$, which
gives the critical splitting
\begin{eqnarray}
\label{delta_cBEC}
\delta^{BEC}_c&=&2 a_2 n\hat{n}_B,\\
&=&\frac{8}{5}{c_2\gamma_p\over(1+c_2)^2}~\epsilon_F\nonumber
\end{eqnarray}
for the quantum phase transition from $p_x+i p_y$ to $p_x$ superfluid
\cite{pwaveNames}.

The behavior of the $p$-wave superfluid order parameters and ground
state energy as a function of splitting $\delta$ and for full range of
detuning $\omega_0$ is best mapped out numerically and gives a smooth
interpolation between above extreme (BCS and BEC) limits derived
above. However, the phase boundary $\delta_c(\omega_0)$ for the
quantum phase transition between $p_x+i p_y$ and $p_x$ superfluids can
in fact be obtained analytically.

To this end we start at a large dipolar splitting, for which the
$p_x$-superfluid ($u > v=0$) is a stable ground state and therefore
the eigenvalues of the curvature matrix of $\ve_{GS}[u,v]$ are
positive in this state.  We then locate the critical phase boundary
$\delta_c(\omega_0)$ by a point where the eigenvalue along $v$
direction changes sign, becoming negative and therefore signaling an
instability toward development of a finite value of $v$ characteristic
of the $p_x+i p_y$-superfluid.

\begin{figure}[bt]
\includegraphics[height=1.5in]{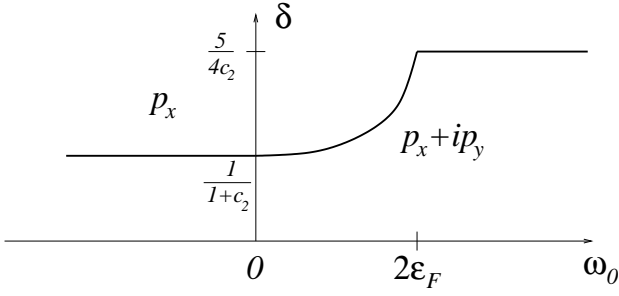}
\caption{The phase diagram of an anisotropic $p$-wave superfluid at zero
  temperature, illustrating a phase boundary of (dimensionless)
  dipolar splitting $\dl_c(\omega_0)$ as function of detuning, that
  marks a phase boundary of a continuous quantum phase transition
  between a $p_x$- and $p_x+i p_y$-superfluid.}
\label{Fig-zerotphasediagram}
\end{figure}

To carry this out, we first minimize $\ve_{GS}[u,v]$ to implicitly
determine the value of $u_{p_x}$ (with $v_{p_x}=0$), that is given by:
\begin{equation}
\label{pxSPcrossover}
(2\w + 3\a)+ 6 \uhh_{p_x}^2 + 2\a\ln\uhh_{p_x}=0.
\end{equation}
Although above saddle-point equation cannot be explicitly solved for
$u_{p_x}$, it can be used to evaluate the eigenvalues of the curvature
matrix at the $p_x$ minimum, and thereby determine the transition
boundary $\delta_c(\omega_0)$.  Computing the eigenvalues of the
curvature matrix of the ground-state energy at the $p_x$ minimum we find
that $p_x$ superfluid is stable for
\begin{equation}
\label{eigenvalue}
\dl-\a-2\uhh_{p_x}^2>0,
\end{equation}
which when combined with the atom number equation
\rf{numberCrosshat} gives (to lowest order in $\gamma_p$)
\begin{widetext}
\begin{eqnarray}
\label{delta_cHat}
\dl(\omega_0)&\approx&
\a(\mu) + 2\hat{n}_B
\left(1-\left({\mu\over\epsilon_F}\right)^{3/2}\theta(\mu)\right),\cr
&\approx&
\cases{\dl_c^{BCS}=\frac{5}{4c_2}\ ,& \text{for $\omega_0 > 2\epsilon_F$}\cr
\dl_c^{cross.+ BEC}(\omega_0)
=\left(\frac{5}{4c_2}-\frac{1}{1+c_2}\right)
\left({\omega_0\over2\epsilon_F}\right)^{3/2}\theta(\omega_0) +
\frac{1}{1+c_2}\ ,& \text{for $\omega_0 < 2\epsilon_F$},\cr}
\end{eqnarray}
\end{widetext}
for the (dimensionless) critical boundary illustrated in
Fig.~\ref{Fig-zerotphasediagram}, with the system transitioning into
the $p_x + i p_y$-superfluid \cite{pwaveNames} for $\delta <
\delta_c(\omega_0)$. In above we used small $\gamma_p$ (narrow
Feshbach resonance) approximation for the chemical potential
$\mu(\omega_0)$ (derived above) appropriate for different regimes. As
anticipated the phase boundary $\delta_c(\omega_0)$ smoothly
interpolates as a function of detuning between the BCS and BEC results
found in Eqs.\rf{delta_cBCS}, \rf{delta_cBEC}. Since for all values of
the dimensionless coupling $c_2=\gamma_p\Lambda/k_F$,
$\dl_c^{BEC}<\dl_c^{BCS}$, for $\dl$ falling between these two values
we predict a continuous quantum phase transition at a critical value
of detuning, given by (to ${\cal{O}}(\gamma_p)$)
\begin{equation}
\label{omega0c}
\omega_0^c(\delta)\approx 2\epsilon_F
\left({\delta-\delta_c^{BEC}
\over\delta_c^{BCS}-\delta_c^{BEC}}\right)^{2/3}.
\end{equation}

\subsection{Finite temperature: phases and transitions in a $p$-wave
resonant Fermi gas}

We now extend our study of the phase behavior of a $p$-wave resonant
gas to finite temperature. This involves a calculation of the
free-energy density, $f_p[\bB]$, \rfs{Sp_sp}, and its minimization
along the lines similar to the above $T=0$ analysis of the
ground-state energy density $\ve_{GS}[\bB]$. The former amounts to a
computation of the polarization tensor $I^{(T)}_{\alpha\beta}[\bB]$,
\rfs{eq:Iabzero}, details of which we relegate to Appendix~\ref{appendixIab}.

The upshot of detailed calculations, presented in the Appendix~\ref{appendixIab},
is that (as usual) at finite $T$
the low-energy singularities arising from Fermi-surface low-energy
contributions to $I_{\alpha\beta}[\bB]$ are cutoff by $T$.
Consequently, (in contrast to the $T=0$ case, above), the free energy,
$f_p[\bB]$ is an analytic function of $\bB$, that at high
temperatures, where $|\bB|$ is small is Taylor-expandable in powers of
the gauge-invariant tensor $\Bb_\alpha B_\beta$.  Naturally, in the
isotropic case $\omega_\alpha=\omega_0$ and $f_p[\bB]$ only involves
rotationally invariant traces of the powers of a tensor
$Q_{\alpha\beta}=\Bb_\alpha B_\beta$ and its transpose
$Q_{\beta\alpha}$.

Within the saddle-point approximation only the quadratic contribution
is anisotropic, and the resulting free-energy density is given by a
standard Landau form
\begin{eqnarray}
\label{f_sp}
f_p[\bB]=\sum_{\alpha=x,y,z}t_\alpha|B_\alpha|^2+
\lambda_1\left(\bBb\cdot\bB\right)^2 +
\lambda_2|\bB\cdot\bB|^2,\nonumber\\
\end{eqnarray}
where, because of the dipolar-anisotropy splitting $t_x(T,\omega_0) <
t_y(T,\omega_0) = t_z(T,\omega_0)\equiv t_\perp(T,\omega_0)$,
reflecting uniaxial symmetry of the system, and these parameters
vanish linearly at respective $T_c's$, with
\begin{eqnarray}
\label{t}
t_x(T,\omega_0)&\sim& T-T_c^{x}(\omega_0),\\
t_\perp(T,\omega_0)&\sim& T-T_c^{\perp}(\omega_0),\nonumber
\end{eqnarray}
and
\begin{equation}
T_c^x(\omega_0) > T_c^\perp(\omega_0).
\end{equation}
The parameters $\lambda_{1,2}$ are only weakly temperature dependent.

Beyond the saddle-point approximation, we expect that generically only
the gauge-invariance is preserved by all the terms in $f_p[\bB]$ and
lack of rotational symmetry for finite $\delta$ will be reflected by
all terms. However, for our purposes it will be sufficient to keep
only the dominant non-rotational invariant contribution entering
through the quadratic term as reflected in $f_p[\bB]$ above. In terms
of $\bu$ and $\bv$ parametrization the free-energy density is given
by
\begin{widetext}
\begin{eqnarray}
\label{fuv_sp}
f_p[\bu,\bv]=\sum_{\alpha=x,y,z}t_\alpha (u_\alpha^2 + v_\alpha^2)
+ \lambda_1(|\bu|^2+|\bv|^2)^2
+ \lambda_2\left((|\bu|^2-|\bv|^2)^2+4(\bu\cdot\bv)^2\right),
\end{eqnarray}
\end{widetext}
where the ratio of $4$ between the two $\lambda_2$ terms is a generic
feature that is a reflection of the underlying gauge-invariance.

\subsubsection{Isotropic}
\label{IsotropicT}

In the isotropic case ($\delta=0$), $t_x=t_\perp\equiv t(T)$, and the
free-energy density is fully rotationally invariant, given by
\begin{equation}
\label{f_spBiso}
f_p^{iso}[\bB]=t|\bB|^2 + \lambda_1\left(\bBb\cdot\bB\right)^2 +
\lambda_2|\bB\cdot\bB|^2.
\end{equation}
For $t > 0$ ($T>T_c$), $f_p^{iso}[\bB]$ is minimized by $\bB=0$ and
the gas is in its normal (nonsuperfluid) phase. Upon lowering $T$
below $T_c$, a minimum develops at a finite value of $\bB$. As can be
seen from the its expression in terms of $\bu$ and $\bv$, \rfs{fuv_spAni}
the minimum is at $u=v$ and $\bu\cdot\bv=0$ (or any of its
gauge-equivalent states corresponding to unequal and nontransverse
$\bu$ and $\bv$). Thus, the finite-$T$ normal-to-superfluid transition
is to a $p_x+i p_y$-superfluid (SF$_{p_x+i p_y}$), consistent with our
earlier finding \cite{Anderson1961} that the ground state is a
$p_x+i p_y$-superfluid, for all detunings. At this transition the global
$U(1)$ gauge-symmetry is spontaneously broken, corresponding to a
choice of a phase of $\bB$ or equivalently the relative orientation
and magnitudes of $\bu$ and $\bv$ (as long as they are not parallel
or one of them does not vanish, since this would correspond to a
$p_x$ state that is not connected to the $p_x + i p_y$ state by a
gauge transformation). In addition, an arbitrary choice of an
overall orientation of $\bB$ (i.e., of the $\bu-\bv$ frame, that by
gauge-choice can be taken to be orthogonal) spontaneously breaks
$O(3)$ rotational symmetry. Clearly time-reversal symmetry is also
spontaneously broken in the $p_x+i p_y$-superfluid state.

This finite-temperature transition is in the {\em complex} $O(3)$
universality class, which can be thought of as a well-explored real
$O(6)$ model\cite{ZinnJustin}, explicitly broken by $\lambda_2$
crystal symmetry-like breaking fields, analogous to $O(3)$ ferromagnet
in a crystal-fields due to spin-orbit coupling to a lattice.  Its
critical behavior has been extensively explored by Vicari, {\em et
  al.}~\cite{Vicari2004}.

\subsubsection{Anisotropic}
\label{anisotropicT}

We now turn to the more experimentally relevant uniaxially {\em
  anisotropic} case, of a Feshbach-resonance triplet split by $\delta
> 0$ (as described above) by dipolar interactions in the presence of
an external magnetic field ${\bf H}=H\hat{\bf x}$. The dipolar
splitting considerably enriches the phase diagram, allowing for three
possible phase diagram topologies, illustrated in
Figs~\ref{Fig-phasediaglow}, \ref{Fig-phasediaginter} and
\ref{Fig-phasediaghigh}.  In terms of the complex $O(3)$ model
dipolar-splitting leads to an easy-axis (Ising) anisotropy, with the
free-energy density given by
\begin{widetext}
\begin{eqnarray}
\label{fIsing_sp}
f^{anisot}_p \left[ \bB \right] &=&t_x|B_x|^2 + t_\perp|\bB_\perp|^2+
\lambda_1\left(\bBb\cdot\bB\right)^2 +
\lambda_2|\bB\cdot\bB|^2,\\
\label{fuv_spAni}
&=&t_x (u_x^2 + v_x^2) + t_\perp (u_\perp^2 + v_\perp^2) +
\lambda_1(|\bu|^2+|\bv|^2)^2 +
\lambda_2\left((|\bu|^2-|\bv|^2)^2+4(\bu\cdot\bv)^2\right),
\end{eqnarray}
\end{widetext}
where $\perp$ indicates two components in the plane perpendicular to
the external magnetic field ${\bf H}$ axis that we have taken to be
$\hat{\bf x}$. For $t_x < t_\perp$ it is clear that $B_x$ part of
$\bB$ will order first, with $\bB_\perp=0$. Namely, since $T_c^x >
T_c^\perp$, $\bu$ and $\bv$ will always both order parallel to the
$x$-axis, showing that for arbitrary small splitting $\delta > 0$ and
arbitrary detuning $\omega_0$, the finite temperature normal to
$p$-wave superfluid transition is always to the $p_x$-superfluid SF$_{p_x}$.  We
designate this upper-critical temperature by $T_{c2}(\omega_0)$ and
expect it to be set (up to renormalization by fluctuations) by
$T_c^x(\omega_0)$.  Clearly from the structure of $f^{anisot}_p[\bB]$,
\rfs{fIsing_sp}, the noncritical (``massive'') $\bB_\perp$ component
can be safely integrated out at the N-SF$_{p_x}$ transition, leaving a
Landau model of a single complex order parameter $B_x$. Hence the
finite-$T$ N-SF$_{p_x}$ classical transition is in 3D XY universality
class, at which only a global $U(1)$ gauge symmetry is broken.

What follows upon further lower the temperature qualitatively depends
on the strength of the dipolar splitting $\delta$. This follows from
the zero-temperature analysis of Sec. \ref{PPWC} and is summarized by
phase diagrams in Figs~\ref{Fig-phasediaglow},
\ref{Fig-phasediaginter} and \ref{Fig-phasediaghigh}.

\begin{figure}[bt]
\includegraphics[height=3in]{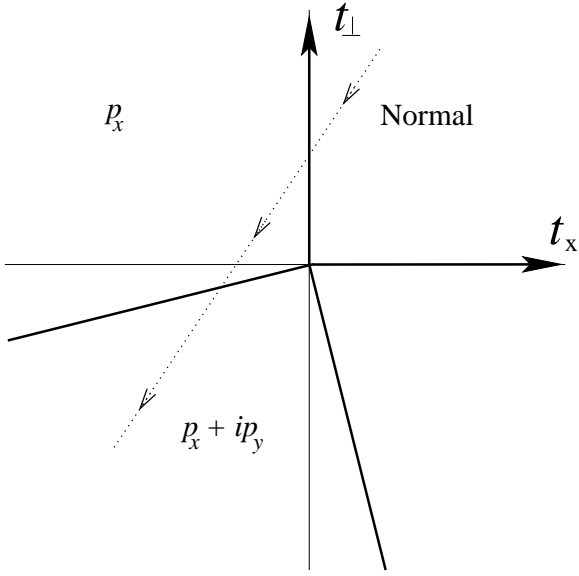}
\caption{Finite-temperature phase diagram illustrating continuous
  transitions between normal (N), $p_x$-superfluid (SF$_{p_x}$) and
  ${p_x+i p_y}$-superfluid (SF$_{p_x+i p_y}$). The parameters
  $t_x(T)< t_\perp(T)$ are reduced temperatures split by
  $\delta$. Only $t_x < t_\perp$ part of the figure is physically
  relevant.}
\label{finiteTtx-tperp}
\end{figure}

For {\em weak} (normalized) Feshbach resonance dipolar splitting
$0<\dl < \dl_c^{BCS}$, upon further lowering temperature from a
$p_x$-superfluid phase, the system always undergoes a transition to a
$p_x + i p_y$-superfluid for all detuning $\omega_0$; we designate
this critical temperature by $T_{c1}(\omega_0)$. To see this, we
observe that for this low range of $\delta$, the parameter $t_\perp$
becomes negatives with reduced $T$ and thereby leads to another
critical temperature at which the $\bB_\perp$ component also orders.
This ordering takes place in the presence of a finite $p_x$ order
parameter $B_{x0}$, within mean-field theory given by
\begin{equation}
B_{x0}=\sqrt{-t_x\over2(\lambda_1+\lambda_2)}.
\end{equation}
The resulting Landau theory for $\bB_\perp$ is then given by
\begin{widetext}
\begin{eqnarray}
\label{fpx-pxpy}
f_{p_x\rightarrow p_x+i p_y}&=&(t_\perp + 2\lambda_1|B_{x0}|^2)|\bB_\perp|^2
+(\lambda_1+\lambda_2)|\bB_\perp|^4
+\lambda_2\left(\Bb_{x0}^2 \bB_\perp\cdot\bB_\perp
                + B_{x0}^2 \bBb_\perp\cdot\bBb_\perp\right),\\
&=& \left(\tilde{t}_\perp + 2\tilde{\lambda}
\cos(2\varphi_\perp-2\varphi_{x,0})\right)
|\bB_\perp|^2 + \lambda|\bB_\perp|^4
\end{eqnarray}
\end{widetext}
where $\lambda\equiv\lambda_1+\lambda_2$ and
\begin{eqnarray}
\tilde{\lambda}&\equiv&\lambda_2|B_{x0}|^2=
{\lambda_2\over2(\lambda_1+\lambda_2)}|t_x|,\\
\tilde{t}_\perp&\equiv& t_\perp+2\lambda_1|B_{x0}|^2
=t_\perp+{\lambda_1\over\lambda_1+\lambda_2}|t_x|.
\end{eqnarray}
We note that at this transition the phase $\varphi_\perp$ of $\bB_\perp$
locks to the phase $\varphi_{x,0}$ of $B_{x,0}$ so that the relative phase is
$\pm\pi/2$.  This is exactly what is expected upon ordering into one
of the two degenerate $p_x \pm i p_y$ states.  We thus find that the
SF$_{p_x}$ to SF$_{p_x+i p_y}$ transition is modified by the presence
of $p_x$ order and takes place at
\begin{eqnarray}
\label{px-pxpyTc}
t_\perp(T)&=& 2(\lambda_2-\lambda_1)|B_{x0}|^2,\nonumber\\
&=& - t_x(T){\lambda_2-\lambda_1\over\lambda_2+\lambda_1},
\end{eqnarray}
which then in turn determines $T_{c1}(\omega_0)$.  Since the $U(1)$
gauge-symmetry is already broken in the $p_x$-superfluid phase and
since, as seen above $\varphi_\perp$ is automatically locked to
$\varphi_{x,0}$, the remaining symmetries that are broken at this
transition are the $O(2)$ rotations of $\bB_\perp$ about the $x$-axis (set by the magnetic
field $H$)
and the time-reversal symmetry associated with a choice of one of the
locking angles $\pm\pi/2$, corresponding to angular momentum
projection $m=\pm 1$. Thus the SF$_{p_x}$ to SF$_{p_x+i p_y}$
transition is also in the well-studied 3D XY universality class. Above
results are summarized by a finite temperature part of the phase
diagram, illustrated in Fig.~\ref{finiteTtx-tperp}

At {\em intermediate} dipolar splitting $\dl_c^{BEC}< \dl <
\dl_c^{BCS}$, the N-SF$_{p_x}$ transition can also be followed by the
SF$_{p_x}$ to SF$_{p_x+i p_y}$ transition, but only for detuning
$\omega_0 > \omega_{0c}(T)$, as illustrated in
Fig~.\ref{Fig-phasediaginter}. The zero-temperature critical frequency
$\omega_{0c}$ (with limits
$\omega_{0c}(\delta\rightarrow\delta_c^{BCS})\rightarrow +\infty$ and
$\omega_{0c}(\delta\rightarrow\delta_c^{BEC})\rightarrow -\infty$) is
given by \rfs{omega0c} and $\omega_{0c}(T)$ is its finite-$T$ extension.
Hence, for this intermediate range of $\delta$, we predict on general
grounds that this SF$_{p_x}$-- SF$_{p_x+i p_y}$ phase boundary
$T_{c1}(\omega_0)$ has a maximum. Thus at fixed $T$ the gas for this
range of parameters should exhibit a {\em reentrant}
$SF_{p_x}\rightarrow SF_{p_x+i p_y}\rightarrow SF_{p_x}$ transition
with detuning $\omega_0$.

Finally, for {\em large} dipolar splitting $\dl_c^{BCS} < \dl$,
$p_x$-superfluid is stable below $T_{c2}(\omega_0)$ throughout, as
illustrated in Fig.\ref{Fig-phasediaghigh}.  We note, however, that
for a Feshbach resonance splitting much larger than the Fermi energy,
we expect that on sufficiently short time scales (set by time scale
for energy relaxation in the system) the two ($m=0$ and $m=\pm1$)
split Feshbach resonances will act independently, so that one can come
in resonance with each of them separately. If so, either
Fig.~\ref{Fig-phasediaghigh} or \ref{Fig-phasediaglow} will be
experimentally observed, depending on to which of the two resonances,
$m=0$ or $m=\pm1$, respectively the system has been tuned.

All of the above discussed transitions are quite conventional and
should be experimentally identifiable through (among other signatures)
their standard universal thermodynamic singularities (e.g., in heat
capacity). Furthermore, the phases should be distinguishable through
their spectroscopic properties, with the normal state gapless
throughout, and for positive chemical potential, the $p_x$-superfluid
displaying a gap with an equatorial line of nodes, i.e., for
$\k_{nodes}^{p_x}=\k^\perp_F$, and $p_x + i p_y$-superfluid exhibiting
a gap with nodes at the north and south pole, i.e., at
$\k_{nodes}^{p_x+i p_y}=\pm k_F\hat{\bf z}$.

In addition to above transitions that are characterizable by an order
parameter, we expect the $p$-wave superfluid to exhibit a number of
non-Landau type of (the so-called) topological transitions at $\mu=0$.
The simplest argument for the existence of such transitions is the
fact that a $p$-wave superfluid exhibits the aforementioned {\em
  gapless} excitations around Fermi surface for $\mu > 0$, and is
 gapped to single-particle excitations for $\mu < 0$, as
clearly seen in $E_\k$, \rfs{eq:Ekp}. Thus we predict continuous
$SF_{p_x}^{gapless}\rightarrow SF_{p_x}^{gapped}$ and $SF_{p_x+ i
  p_y}^{gapless}\rightarrow SF_{p_x+i p_y}^{gapped}$ transitions at
$\mu=0$. One might expect a clear signature of such transitions from
the change in the low-$T$ thermodynamic behavior, e.g., with the heat
capacity changing from a power-law in $T$ to an activated form, with
gap for $\mu < 0$ set by the molecular binding energy. However, a
collective sound mode, present in any superfluid, that contributes
power-law in $T$ contributions, might obscure the distinction between
the $\mu > 0$ and $\mu < 0$ $p$-wave superfluid phases. Local
spectroscopic experimental probes (some atomic gas analog of tunneling
experiments) should prove useful for detection of these transitions.
Despite a lack of local Landau order parameter, these weakly- and
strongly-paired $p$-wave superfluids are distinguishable by their
topological properties~\cite{VolovikBook,VolovikBook1,VolovikReview,Volovik2004,Read2000}, as we
discuss in the next section.

\section{Topological phase transitions and non-Abelian statistics}
\label{Sec:TopPhaseTrans}

In addition to the rich (but conventional) phenomenology of $p$-wave
resonant gases obtained in previous subsections, as we will show next,
they can also exhibit a more subtle (in some cases topological) order
and associated phase transitions, that cannot be classified by a local
order parameter, nor associated Landau theory~\cite{ChaikinLubensky}. The existence of such continuous non-Landau
type phase transitions have long been appreciated in the
literature. Examples range from Anderson's metal-insulator transition
and transitions between different quantum Hall ground states to the
well-understood Kosterlitz-Thouless transition (e.g., in superfluid
films) and two-dimensional melting, all separating two disordered
states undistinguished by any local order parameter or conventional
symmetry operation.

\subsection{$P$-wave superfluid in $3$ dimensions}
\label{sec:top3d}

It is remarkable that $p_x$- and $p_x+i p_y$-superfluids are examples
of a system, that can undergo such non-Landau type phase transition
when the chemical potential changes sign. In three dimensions this can
be simply seen from the qualitative change in the spectrum $E_\k$,
\rfs{eq:Ekp} of single-particle fermionic excitations, that for the
$p_x$ and $p_x + i p_y$ states
\begin{eqnarray}
\label{eq:pxsub}
(B_x, B_y, B_z )_{p_x} &=& \frac{\Delta}{2 g_p}~(1, 0, 0),\\
\label{eq:pxpysub}
(B_x, B_y, B_z )_{p_x+i p_y} &=& \frac{\Delta}{2 g_p}~(1, i, 0),\
\end{eqnarray}
are respectively given by
\begin{eqnarray}
\label{eq:Ekpx}
E^{p_x}_{\bf k} &=&
\sqrt{ \left( \frac{k^2}{2m}-\mu \right)^2+|\Delta|^2 k_x^2},\\ \label{eq:Ekpxpy}
E^{p_x + i p_y}_{\bf k} &=&
\sqrt{\left( \frac{k^2}{2m}-\mu \right)^2+|\Delta|^2(k_x^2+k_y^2)},
\end{eqnarray}
with $\Delta$ the maximum gap of each state, related to the
corresponding order parameter. Clearly, in a $p_x$-superfluid, for
$\mu>0$, $E^{p_x}_{\k}=0$ (i.e., gapless) for $k_x=0$ and $k_y$, $k_z$
arbitrary, and for $\mu < 0$, $E^{p_x}_{\k} > 0$ (i.e., gapped) for
all $\k$. Similarly, in a $p_x+ i p_y$-superfluid, for $\mu>0$,
$E^{p_x+i p_y}_{\k}=0$ (i.e., gapless) for $k_z=\sqrt{2 m \mu}$ and
$k_x=k_y=0$,  and for $\mu < 0$, $E^{p_x+i p_y}_{\k} >
0$ (i.e., gapped) for all $\k$.  Physically these spectral distinctions
arise because for $\mu >0$, a phase that we refer to as $SF_{p}^{\text
  weak}$, the pairing is a collective Fermi surface phenomenon and
finite angular momentum forces the gap to vanish on some subspace of
the Fermi surface. On the other hand, for $\mu < 0$, in the
$SF_{p}^{\text strong}$ the gap is single-pair of fermions phenomenon
and is simply set by the molecular binding energy, independent of the
angular momentum state of the molecule.

These changes in the spectrum lead to qualitatively distinct
single-particle correlation functions and therefore require a genuine
quantum phase transition (illustrated in Figs.~\ref{Fig-phasediaglow},
\ref{Fig-phasediaginter}, \ref{Fig-phasediaghigh}) separating two
distinct (weakly- and strongly-paired) $p_x$- and two distinct
(weakly- and strongly-paired) $p_x + i p_y$-superfluids, as $\mu$
changes from positive to negative, respectively.

In a classic BCS $p$-wave paired superfluid, such as He$^3$, $\mu$ is
always positive and these transitions are not experimentally
accessible. However, in $p$-wave resonant atomic gases they should be
easily realizable (if a $p$-wave superfluid is produced) by changing
the detuning parameter $\omega_0$ (controlled by an external magnetic
field), that is, as we have shown in Sec.~\ref{PWaveChapter}, is closely tracked
(up to terms of the order of $g_p$) by $\mu$.

In addition to the above quasiparticle spectrum and correlation
function argument for the transition at $\mu=0$, the existence of the
$SF_{p_x+i p_y}^{\text weak}\rightarrow SF_{p_x+i p_y}^{\text strong}$
transition can be also seen by noting that the two types of $p_x+i
p_y$-superfluids can be distinguished by {\em topological} order as
discussed in detail in
Refs.~\cite{VolovikBook1,Volovik2004,VolovikReview}.  Although, as
argued above a spectral distinction between $\mu > 0$ and $\mu <0$
$p_x$-superfluids exists, and therefore we expect a corresponding
$SF_{p_x}^{\text weak}\rightarrow SF_{p_x}^{\text strong}$ transition
at $\mu=0$, we are not aware of any topological distinction between
these two phases similar to the $p_x+i p_y$ classification.  For the
rest of the section, below, we will focus on the analogous transition
in two dimensions, where for a $p_x+i p_y$-superfluid clearly no
spectral distinction exists, with both $SF_{p_x+i p_y}^{\text weak}$
and $SF_{p_x+i p_y}^{\text strong}$ gapped, but still distinguishable
by topological order.

\subsection{$P$-wave superfluid in $2$ dimensions}

The three-dimensional calculations of this paper, arguing for the
existence of a $p$-wave superfluid can be easily extended to two
dimensions, with only minor quantitative distinctions (e.g., the
dimensionless parameter $c_2$ in $2D$ scales logarithmically with the
uv cutoff $\Lambda$). Thus, we expect the existence of a fully-gapped
two-dimensional $p_x+i p_y$-superfluid for $\mu>0$ and $\mu<0$.  As we
will see, a plethora of especially interesting phenomena takes place in such
a system, that we expect to be realizable by confining the degenerate
atomic gas to a highly oblate magnetic trap.

Although much of our discussion of this system  follows an
excellent paper by N. Read and D. Green, Ref.~\cite{Read2000}, as well
as Refs.~\cite{VolovikBook1,VolovikReview,Gurarie2006}, we elaborate
on details of the analysis (particularly on the existence of the zero
modes), and thereby hope to elucidate a number of points discussed
there.  Furthermore, while above papers are well-known and appreciated
in the quantum Hall community, they are less familiar to the atomic
community and thus, their main results are worth 
elaborating on here.

Following Ref.~\cite{Read2000}, let us first construct a ground state
wave function of a two-dimensional $p_x+ip_y$ superfluid.  As discussed
in Sec.~\ref{Sec:PWaveTwoChannel} its mean-field Hamiltonian (valid in the
narrow-resonance limit) follows directly from \rfs{eq:ham_p}, with the
substitution $\hat b_{{\bf p},\alpha} \rightarrow \delta_{{\bf p},0}
B_{\alpha}$ and \rfs{eq:pxpysub}, and is given by
\begin{eqnarray}
\label{eq:hamfermp} \hat H -\mu ~\hat N_f &=& \sum_{\k}\xi_k ~\hat a^\dagger_{\k
} \hat a_{\k }  - \oh \sum_{\k} ~\bigg[\Delta \left(k_x+ik_y \right)~
\hat a^\dagger_{\k } \hat a^\dagger_{-\k }  \cr &+&
\Delta \left(k_x-ik_y\right) ~ \hat a_{-\k } \hat a_{\k }\bigg].
\end{eqnarray}
Here we fix the phase of $\Delta$ by choosing $\Delta=\bar \Delta$
and, as before, $\xi_k$ is given by \rfs{eq:xik},
$$\xi_k=\frac{k^2}{2m}-\mu.$$
This Hamiltonian is diagonalized by a
unitary transformation to the Bogoliubov quasiparticles
\begin{eqnarray} \label{eq:bogquasi}
\hat \gamma_\k &= &u_\k^* \hat a_\k + v_\k^* \hat a^\dagger_{-\k}, \cr
\hat \gamma^\dagger_{\k} &=& v_\k \hat a_{-\k} + u_\k \hat a^\dagger_{\k},
\end{eqnarray}
taking the form
\begin{equation}
\hat H = E_{GS}+\sum_k E_\k \hat \gamma^\dagger_\k \hat \gamma_\k.
\end{equation}
The ground state of this Hamiltonian is similar to its $s$-wave
counterpart \rfs{eq:gr}, and is given by
\begin{equation} \label{eq:grp}
\left| {\rm BCS} \right\rangle = \prod_\k \left( u^*_\k + v^*_\k
~a^\dagger_{-\k} a^\dagger_{\k} \right) \left| 0 \right\rangle.
\end{equation}
with each pair $\k$, $-\k$ in the product taken only once. Here $u_\k$,
$v_\k$ are $p$-wave analogs of \rfs{eq:xik}, and satisfy the
Bogoliubov-de-Gennes equations
\begin{equation} \label{eq:BdGnoVor}
\left( \matrix {  \xi_k & - \Delta \left(k_x+i k_y \right) \cr
-\Delta \left(k_x-i k_y \right) & -\xi_k } \right) \left( \matrix
{ u_\k \cr v_{\k} } \right)= E_k  \left( \matrix { u_\k \cr v_{\k} }
\right),
\end{equation}
The solution of these equations is straightforward with the result
\rfs{eq:Ekpxpy} and with normalized $u_\k$, $v_\k$ being
\begin{eqnarray}
u_\k &=& -\sqrt{ \frac{E_\k+\xi_k}{2 E_\k}},
\cr
v_\k &=&
\frac{ \left(k_x-i k_y \right) \Delta}{\sqrt{ 2 E_\k
\left(E_\k+\xi_k \right)} }
\end{eqnarray}
We note that unlike the $s$-wave case \rfs{eq:vkuk}, the relative
phase of $u_\k$ and $v_\k$ is nonzero. Let us construct a real space
version of \rfs{eq:grp}. It is given by
\begin{equation} \label{eq:wf}
\Psi( \r_1, \r_2,\dots) = \sum_P (-1)^P g(\r_{P_1}- \r_{P_2})
~g( \r_{P_3}-\r_{P_4}) \dots.
\end{equation}
Here $\r_i$ are two-dimensional vectors denoting the position of the
$i$-th fermion, and $g(\r)$ is a Cooper-pair (or molecular)
wavefunction given by
\begin{equation} \label{eq:grwave}
g(\r) = \int {d^2 k \over ( 2 \pi)^2}~ e^{i \k \cdot \r}
~\frac{v_\k}{u_\k}.
\end{equation}
$P$ stands for a permutation of numbers $1, 2, \dots, N_f$, where
$N_f$ is the total number of fermions, and $(-1)^P$ is the sign of the
permutation, thereby enforcing the antisymmetrization of the many-atom
ground-state wavefunction. Notice that $g(\r)=-g(-\r)$ due to the
$p$-wave symmetry of the superfluid (since $u_\k=u_{-\k}$,
$v_{\k}=-v_{-\k}$).

Now suppose $\mu>0$. Then at small $k \ll \sqrt{2 m \mu}$ or
equivalently $|\r| \gg n^{-1/3}$, we can estimate the function to be
integrated in \rfs{eq:grwave} to go as
\begin{equation}
\frac{v_\k}{u_\k} =  \frac{ k_x - i k_y}{E_\k+\xi_k} \Delta \sim
\frac{k_x-i k_y}{k^2}.
\end{equation}
It immediately follows that for $|\r| \gg l$,
\begin{equation}
  g(\r) \sim \frac{1}{z},
\end{equation}
where $z$ is the complex number representing the two-dimensional
vector $\r = x \hat {\bf x} + y \hat{\bf y}$ as $z=x+iy$. Therefore,
for $\mu>0$ the wave function takes the form
\begin{equation} \label{eq:mr}
\Psi(z_1,z_2,\dots) = \sum_P (-1)^P \frac{1}{z_{P_1}-z_{P_2}}
~\frac{1}{z_{P_3}-z_{P_4}} \dots.
\end{equation}
This wave function occurs in the context of the quantum Hall effect
(modulo the Gaussian and Jastrow factors not essential for the present
discussion) and is called the Pfaffian or Moore-Read state \cite{MR}.

To understand the connection with the quantum Hall effect, we recall
that for the last few years attempts have been made to realize quantum
Hall states~\cite{Gunn2000,Gunn2001}  out of Bose-Einstein condensates
by rotating them~\cite{Cornell2004}.
In the $p_x+ip_y$ condensate, thanks to the relative angular momentum
$\ell=1$ of each Cooper pair (or closed-channel molecule), the
(fermionic) condensate already automatically rotates and therefore
does not require any externally imposed rotation to be in the quantum
Hall ground state.

A key observation is that for $\mu<0$  $u_\k/v_\k \sim k_x-ik_y$
at small $k$ and $g(\r)$ no longer has the power-law fall off
characteristic of the quantum Hall-like ground state in \rfs{eq:mr}.
Instead, the integral in \rfs{eq:grwave} is then dominated by large
$k$, and generally we expect that $g(\r)$ will be an exponentially
decaying function. The authors of Ref.~\cite{Read2000} referred to the
$\mu<0$ as the strongly-coupled phase. For BCS-BEC condensates studied
here, $\mu<0$ corresponds to the BEC regime obtained for negative
detuning $\omega_0$.

As mentioned above, despite the qualitative distinction in the
ground-state wavefunctions, in a two dimensions $E_k>0$, i.e., gapped
for all $\k$, for both $\mu>0$ and $\mu<0$. The only special point
where there are gapless excitations is $\mu=0$. Nevertheless given the
qualitative distinction between the ground states at $\mu>0$ and
$\mu<0$, they much be separated by a quantum phase transition at the
gapless point $\mu=0$.  The situation is again reminiscent of quantum
Hall transitions where gapless points separates gapped quantum Hall
states.

Although this transition at $\mu=0$ is not of Landau type (not
exhibiting any obvious local order parameter) the weakly- ($\mu >0$)
and strongly-paired ($\mu<0$) $p_x+i p_y$ states are topologically
distinct and therefore the transition is topological. The topological
distinction lies in the properties of the two complex functions $u_\k$
and $v_\k$, constrained by $|u_\k|^2+|v_\k|^2=1$. Since their overall
phase is unimportant, they are parametrized by two real parameters.
Thus, $u_\k$ and $v_\k$ represent a map from a two-dimensional space
of $\k$ (which can be thought of topologically as a sphere $S_2$, if
the point $k=\infty$ is added) to the two-dimensional space of $u_\k$,
$v_\k$.  Such maps are characterized by the winding numbers, called
the homotopy classes corresponding to the homotopy group
$\pi_2(S_2)={\mathbb Z}$. Roughly speaking, these winding numbers are
the number of times $u_\k$, $v_\k$ wraps around a sphere as $\k$
varies.  Quite remarkably, one can see that these numbers are
different for $\mu>0$ and $\mu<0$.

To see this explicitly we construct a unit vector $\vec n$ which
points in the direction of the spinor $(u_\k, v_\k)$. To this end,
recall the standard relation between a spinor $\psi_\alpha$ and a
vector $n_\mu$, $n_\mu = \sigma^\mu_{\alpha\beta} \psi^*_\alpha
\psi_\beta$, which gives
\begin{eqnarray} \label{eq:vector}
n^x_\k &=& u_\k^* v_\k + v_\k^* u_\k = -\frac{k_x \Delta}{E_\k}, \cr
n^y_\k &=& i \left( u_\k v_\k^*-v_\k u_\k^* \right)
=\frac{k_y \Delta}{E_\k},\cr
n^z_\k &=& u_\k u_\k^* - v_\k v_\k^*=\frac{\xi_k}{E_\k}.
\end{eqnarray}
It is important to keep in mind that $\Delta$ is in fact a function of
$k^2$, being a constant for $k \ll \Lambda$, but quickly dropping off
to zero at $k \gg \Lambda$, where $\Lambda$ is the ultraviolet cutoff
associated with the interatomic potential range.  The winding number
associated with $\pi_2(S_2)$ is given by the well known topological
invariant (discussed in our context in Ref.~\cite{VolovikBook1}, whose notations we borrow
here)
\begin{equation}
\label{Ninvariant}
\t N_3=\frac 1 {8 \pi} \int d^2 k
~\left[ \vec n \cdot \partial_\alpha \vec n \times
\partial_\beta \vec n ~ \epsilon_{\alpha \beta} \right].
\end{equation}
Substituting $\vec{n}_\k$, \rfs{eq:vector} into the expression for the
topological invariant we find, after an appropriate rescaling of $k$
and with $\Delta=\hat{\Delta}/\sqrt{2m}$
\begin{equation}
\t N_3=\frac{1}{2} \int_0^\infty k dk~ \hat{\Delta}~ \frac{(k^2+\mu)
\hat\Delta-2 k^2 \left(k^2-\mu \right)
\pbyp{\hat\Delta}{k^2}}{\left( \left(k^2-\mu
\right)^2 + k^2\hat{\Delta}^2 \right)^{\frac 32}}.
\end{equation}
As required by the general form of $\t N_3$, \rfs{Ninvariant}, this
expression is a total derivative, and the integral can be computed
directly with the result
\begin{equation}
\t N_3 = \oh \left. \frac{k^2-\mu
}{\sqrt{\left(k^2-\mu \right)^2 + k^2\hat\Delta^2}}
\right|_{k=0}^{k=\infty}=\oh \left( 1+{\rm sign}~\mu \right).
\end{equation}
Thus, for $\mu<0$, $\t N_3=0$, while for $\mu>0$ $\t N_3=1$, and
indeed $u_\k$, $v_\k$ define a topologically nontrivial map only for
$\mu>0$.  Hence, a $p_x+ip_y$-superfluid ground state exhibits
topological order only for $\mu>0$.

It is interesting to observe that the topological invariant $\t N_3$ for
the $p_x$-state gives $\t N_3=0$ independent of $\mu$, since its $u_\k$,
$v_\k$ are real and therefore define a trivial map.  The same is true
for an $s$-wave condensate. Thus, at least based on this topological
invariant, neither of these states are topological, nor is the
transition between weakly- (BCS) and strongly-paired (BEC) states in
these systems.

Finally, we remark that the topological invariant $\t N_3$ constructed here
constitutes a particular case of more general topological invariants
studied in Ref.~\cite{VolovikBook1}.

\subsection{Vortices  and  zero modes of
a two-dimensional $p_x+ip_y$ superfluid: non-Abelian statistics and
``index theorem''}

We can further elucidate the nature of the $p_x+i p_y$ condensates if
we study the solutions to the Bogoliubov-de-Gennes (BdG) equation in
the presence of vortices in the condensate $\Delta$. In fact as we
will see below, a nontrivial topological order exhibited by the
weakly-paired ($\mu > 0$) $p_x+i p_y$-superfluid will reflect itself
in the nature of the spectrum in the presence of vortices. Recall that
a phase of the condensate wavefunction changes by an integer number
times $2\pi$ every time one goes around the vortex. Thus, in the
presence of collection of vortices at positions $z_i$, the gap
function $\Delta(\r)$, proportional to the condensate wavefunction can
generally be written as
\begin{equation}
\Delta(\r) = \prod_i \left( \frac{z-z_i}{\bar z - \bar z_i }
\right)^{m_i/2} D(\r),
\end{equation}
where $D(\r)$ is a function of position whose phase is single
valued. Since its square is the condensate density, and $D(\r)$ is
expected to vanish inside vortex cores.

Generically, in the presence of vortices, one expects solutions
localized on them. It has been appreciated for some time, based on a
variety of arguments~\cite{Read2000,VolovikBook1} (without an explicit solution of
the Bogoliubov-de-Gennes equation), that a $p_x+i p_y$-superfluid is
special in that its fundamental $2\pi$ vortex in thermodynamic limit
is guaranteed to carry a state (referred to as ``zero mode'') at
exactly zero energy.

Recently, we have studied the question of existence and robustness of
such zero modes for the more general problem of a collection of
vortices~\cite{Gurarie2006}. As we will show below, we found that
for a macroscopic sample (i.e., ignoring the boundary physics),
without fine-tuning, strictly speaking there is only {\em one} or {\em
zero} Majorana-fermion mode depending only on whether the total
vorticity of the order parameter (in elementary vortex units of
$2\pi$) is {\em odd} or {\em even}, respectively.  For a collection of
well-separated vortices, within an exponential accuracy one zero mode
per an isolated odd-vorticity vortex persists. As two of such vortices
are brought closer together the corresponding pair of ``zero'' modes
splits away to finite $\pm E$ (vortex-separation dependent) energies.
Generically, even-vorticity vortices do not carry any zero modes.

Before we proceed to construct these solutions explicitly, let us
discuss in general what we expect from the solutions of these BdG
equations. A generic Bogoliubov-de-Gennes Hamiltonian can always be
represented in the form
\begin{equation}
\hat H = \sum_{ij} \left(\hat{a}^\dagger_i h_{ij} \hat{a}_j
-\hat{a}_j h_{ij} \hat{a}^\dagger_i+
\hat{a}_i \Delta_{ij} \hat{a}^\dagger_j + \hat{a}^\dagger_j
\Delta^*_{ij} \hat{a}^\dagger_i
\right).
\end{equation}
Here the indices $i$, $j$ represent a way to enumerate fermion
creation and annihilation operators, being for example, points in
space and/or spin, if the fermions also carry spin. $h_{ij}$ is  a
hermitian operator, while $\Delta_{ij}$ is an antisymmetric
operator. The study of this Hamiltonian is then equivalent to the
study of a matrix
\begin{equation} \label{eq:classD}
{\cal H}  = \left( \matrix { h & \Delta \cr \Delta^\dagger & - h^T}
\right).
\end{equation}
This matrix possesses the following important symmetry property
\begin{equation} \label{eq:symmetryD}
\sigma_1 {\cal H} \sigma_1 = - {\cal H}^*.
\end{equation}
Here $\sigma_1$ is the first Pauli matrix acting in the 2 by 2 space
of the matrix \rfs{eq:classD}. In the terminology of
Ref.~\cite{Altland1997}, we say that this matrix belongs to symmetry
class $D$.  As a result of this property, if $\psi$ is an
eigenvector of this matrix with the eigenvalue $E$, then $\sigma_1
\psi^*$ has to be an eigenvector with the eigenvalue $-E$. Indeed,
\begin{equation}
{\cal H} \sigma_1 \psi^* = - \sigma_1 {\cal H}^* \psi^*=-E \sigma_1
\psi^*.
\end{equation}
As a result, all nonzero eigenvalues of ${\cal H}$ come in pairs, $\pm
E$. A special role is played by the zero eigenvectors of this matrix,
namely the zero modes discussed above. If $\psi$ is a zero mode,
$\sigma_1 \psi^*$ is also a zero mode. Taking linear combinations
$\psi+\sigma_1 \psi^*$, $i \left(\psi-\sigma_1 \psi^* \right)$ of
these modes, we can always ensure the relation
\begin{equation}
\label{eq:rela}
\sigma_1 \psi^*=\psi
\end{equation}
for every zero mode.  In the absence of other symmetries of ${\cal H}$
it is quite clear that generically there is nothing that protects the
total number $N_z$ of its zero modes under smooth changes of the
Hamiltonian matrix that preserve its BdG form, namely retain the
properties in Eqs.~\rf{eq:classD} and \rf{eq:symmetryD} . However,
since non-zero modes have to always appear and disappear in $\pm E$ pairs, as long
as the symmetry property \rf{eq:symmetryD} is preserved by the
perturbation, the number of zero modes can only change by multiples of
$2$.  Thus, while the number $N_z$ of zero modes of the Hamiltonian
\rf{eq:classD} may change, this number will always remain either odd
or even, with $(-1)^{N_z}$ a ``topological invariant''
\cite{ZirnbauerPrivate,NickReadUnpublished}.

The value of this invariant is easy to establish if one observes that
${\cal H}$ is an even-sized matrix, with an even number of
eigenvalues. Since the number of non-zero modes must be even, this
implies that the number of zero modes is also even.  Thus
$(-1)^{N_z}=0$, and generally the BdG problem does not have any
topologically protected zero modes.  Furthermore, since, as
demonstrated above, zero modes must appear in pairs, there can only be
an even number of accidental zero modes, which will nevertheless be
generally destroyed by any perturbation of ${\cal H}$ (preserving its
BdG structure \rfs{eq:classD}).  We believe this observation was first
made by N. Read~\cite{NickReadUnpublished}.

The situation should be contrasted with that of the Dirac operators
${\cal D}$. Those operators, being generally of one of chiral
classes in the terminology of Ref.~\cite{Altland1997}, obey the
symmetry
$$ \sigma_3 {\cal D} \sigma_3 = -{\cal D}.$$ Thus if $\psi$ is an
eigenvector of ${\cal D}$ with the eigenvalue $E$, $\sigma_3 \psi$ is
an eigenvector with the eigenvalue $-E$. The zero modes of ${\cal D}$
must obey the relation $$ \sigma_3 \psi_{L,R} = \pm \psi_{L,R}.$$ Here
``left" zero modes $\psi_L$ come with the eigenvalue $+1$, while
``right" zero modes $\psi_R$ have the eigenvalue $-1$ of the operator
$\sigma_3$. As the operator ${\cal D}$ is deformed, the number of zero
modes changes, but the non-zero modes always appear in pairs where one
of the pair has to be ``left" and the other ``right". Therefore, while
the number of zero modes is not an invariant, the difference between
the number of left and right zero modes has to be a topological
invariant, determined (through the index theorem) by the monopole
charge of the background gauge-field.

Contrast this with zero modes of ${\cal H}$, which obey the relation
\rfs{eq:rela}. Because of complex conjugation of $\psi$, these zero
modes cannot be split into ``left" and ``right". Indeed, even if we
tried to impose $\sigma_1 \psi^* = - \psi$, a simple redefinition of
$\psi \rightarrow i \psi$ brings this relation back to
\rfs{eq:rela}. Thus, the most an ``index theorem" could demonstrate in
case of the Bogoliubov-de-Gennes problem, is whether there is 0 or
exactly 1 zero mode. Moreover, since the Bogoliubov-de-Gennes problem
is defined by an even dimensional Hamiltonian, generically there will
not be any topologically protected zero
modes~\cite{NickReadUnpublished}.

Yet it is quite remarkable that in case of an isolated vortex of odd
vorticity in a macroscopic sample (i.e., ignoring the sample
boundaries) of a $p_x+ip_y$-superfluid of spinless fermions, there is
exactly one zero mode localized on this vortex
\cite{Kopnin1991,Read2000,Nayak2006a,Tewari2006}.  To be consistent
with above general property of the BdG Hamiltonian (namely, that the
total number of zero modes must be even) another vortex is situated at
the boundary of the system \cite{Read2000,NickReadUnpublished},
preserving the overall parity of the number of zero modes. Hence,
although even in this odd-vorticity case the one zero mode is not
protected topologically, able to hybridize with a vortex at a boundary
of the sample, it survives (up to exponentially small corrections)
only by virtue of being far away from the boundary (and from other
odd-vorticity vortices).

To see this explicitly we now consider a $p_x + i p_y$-superfluid in the
presence of a single rotationally symmetry vortex, characterized by
\begin{equation}
\Delta(\r)=\frac{i}{2} e^{i l \varphi} f^2( r),
\end{equation}
where $f(r)$ is a real function of $r$ (vanishing at small $r$), $l$
is the vorticity of the vortex, $r$, $\varphi$ are the polar
coordinates centered on the vortex, and the factor of $i/2$ is chosen
to simplify subsequent calculations. In this case the
Bogoliubov-de-Gennes equations take the form
\begin{eqnarray}
\label{eq:BdGvortex}
\left(- \frac{\nabla^2}{2m} - \mu \right) u(\r)- f(r)
e^{\frac{il\varphi}{2}} \pp{\bar z} \left[  e^{\frac{il\varphi}{2}}
f(r) v(\r) \right] = E u(\r), \cr
\left( \frac{\nabla^2}{2m} + \mu \right)v(\r)
- f(r)  e^{- \frac{il\varphi}{2}} \pp{z}
\left[e^{-\frac{il\varphi}{2}}  f(r)  u(\r) \right] = E v(\r), \cr
\end{eqnarray}
We remark that once solutions to these equations $u_n(\r)$, $v_n(\r)$,
corresponding to energies $E_n$, are known, the Bogoliubov
quasiparticle creation and annihilation operators are given by
\begin{eqnarray}
\label{eq:nonuniuv}
\hat{\gamma}_n &=& \int d^2 r \left[ u_n^*(\r) \hat a(\r) +
v_n^*(\r) \hat a^\dagger(\r) \right] \cr
\hat{\gamma}_n^\dagger &=& \int d^2 r
\left[ u_n(\r) \hat a^\dagger(\r) + v_n(\r) \hat a(\r) \right].
\end{eqnarray}
If the condensate was uniform, then the solutions to the
Bogoliubov-de-Gennes equations would be plane waves, immediately
leading to \rfs{eq:bogquasi}. The inverse to \rfs{eq:nonuniuv} reads
\begin{eqnarray}
\hat{a}(\r) = \sum_n \hat \gamma_n u_n(\r) + \hat \gamma^\dagger_n v^*_n(\r),
\cr \hat a^\dagger(\r) = \sum_n \hat \gamma_n^\dagger u_n^*(\r) + \hat \gamma_n
v_n(\r).
\end{eqnarray}

Next we observe that for the case of a vortex of {\em even} vorticity,
$l=2n$, we can eliminate the phase dependence of \rfs{eq:BdGvortex}
entirely. Indeed, making a transformation
\begin{equation} \label{eq:cltr}
u \rightarrow u e^{i n\varphi}, \  v \rightarrow v e^{-i n \varphi}.
\end{equation}
leads to equations
\begin{eqnarray} \label{eq:BdGvortex1}
\left(- \frac{\nabla^2}{2m} +\frac{n^2}{2m r^2}- \mu \right) u-
\frac{ i n}{m r^2} \pbyp{u}{\varphi}- f(r)  \pp{\bar z} \left[  f(r)
v \right] = E u, \cr \left( \frac{\nabla^2}{2m} -\frac{n^2}{2m r^2}+
\mu \right) v- \frac{ i n}{m r^2} \pbyp{v}{\varphi}-f(r)  \pp{z}
\left[  f(r)  u \right] = E v. \cr
\end{eqnarray}
Now we note that these equations are topologically equivalent to the
BdG equation without any vortices. Indeed, the only difference between
these equations and those for a uniform condensate is the presence of
the terms $2 i n/r^2[\partial/\partial \varphi]$, $n^2/r^2$, and
$f(r)$ that is a constant at large $r$ and vanishes in the core of the
vortex for $r<r_{\rm core}$.  We can imagine smoothly deforming these
equations to get rid of the first two terms (for example, by replacing
them with $\alpha \left(n^2/r^2 - 2 i n/r^2[\partial/\partial \varphi]
\right) u$ and taking $\alpha$ from $1$ to $0$), and smoothly
deforming $f(r)$ into a constant equal to its asymptotic value at
large $r$; in order to be smooth, the deformation must preserve the
BdG structure \rfs{eq:classD} and the vorticity of the condensate, if
there is any.  These equations then become equivalent to
\rfs{eq:BdGnoVor} for a constant, vortex-free order parameter with an
exact spectrum \rfs{eq:Ekpxpy}, that for $\mu\neq 0$ in two
dimensional space clearly does not exhibit any zero modes.

As Eqs.~\rf{eq:BdGvortex1} are smoothly deformed to get rid of the vortex,
in principle it is possible that its solutions will change and that it
will develop zero modes (although, as demonstrated above, this can
only happen in $\pm E$ pairs, leading to an even number of
these). However, these modes will not be topologically protected, and
even a small deformation of, say, the shape of the order parameter
shape $f(r)$ will destroy these modes.  We note that this argument
easily accommodates vortices that are not symmetric, as those can be
smoothly deformed into symmetric ones without changing the
topologically protected parity of $N_z$.  The conclusion is that
generically there are no zero modes in the presence of an isolated
vortex of even vorticity.

The situation is dractically different if the vorticity of the vortex
is odd, i.e., if $l=2n-1$. In this case the transformation
\rfs{eq:cltr} cannot entirely eliminate the vortex from the equations
(even with the help of a smooth deformation), leaving at least one
fundamental unit of vorticity. This thereby guarantees at least one
one zero mode localized on the odd-vorticity vortex.  To see this,
recall that due to the condition \rfs{eq:rela}, the zero mode
satisfies
\begin{equation}
\label{uv}
u=v^*.
\end{equation}
Combining this with the transformation \rfs{eq:cltr}, we find the
equation for the zero mode
\begin{eqnarray} \label{eq:shroddd}
-  f(r) e^{-\frac{i\varphi}{2}} \pp{\bar z}
\left[  e^{-\frac{i \varphi}{2}}   f(r) u^* \right]  &=& \cr
\left(\frac{\nabla^2}{2m} - \frac{n^2}{2m r^2}+ \mu \right) u+\frac{
i n}{m r^2} \pbyp{u}{\varphi}.
&&
\end{eqnarray}
We look for the solution to this equation in terms of a spherically
symmetric real function $u(r)$. This gives
\begin{equation}
\label{eq:shrspsym}
-\frac{1}{2m} u'' - \left(  \frac{f^2}{2} + \frac{1}{2m r} \right)
u' - \left( \frac{f^2}{4r} + \frac{f f'}{2} -\frac{n^2}{2mr^2}
\right) u = \mu u.
\end{equation}
A transformation
\begin{equation} \label{eq:mcs}
u(r) = \chi(r) \exp \left( - \frac m 2 \int_0^r dr'~ f^2(r') \right)
\end{equation}
brings this equation to a more familiar form
\begin{equation}
-\frac {\chi'' } {2m} - \frac{\chi' }{2mr} + \left( m
\frac{f^4(r)}{8} + \frac{n^2}{2mr^2} \right) \chi = \mu \chi.
\end{equation}
This is a Schr\"odinger equation for a particle of mass $m$ which
moves with angular momentum $n$ in the potential $m f^4(r)/8$, that is
everywhere positive. We observe that this potential vanishes at the
origin, and quickly reaches its asymptotic bulk value $mf_0^4/8$ at large
$r$.  Then for $\mu>mf_0^4/8$, there always exist a solution to this
equation finite at the origin and at infinity. Moreover, if
$\mu<mf_0^4/8$, then the solution finite at the origin will diverge at
infinity as
\begin{equation}
\chi \sim e^{r\sqrt{m^2 \frac{f_0^4}{4}-2m\mu}}.
\end{equation}
Combining this with \rfs{eq:mcs}, we see that $u(r)$ will still be a
bounded function at infinity as long as $\mu>0$. Thus the conclusion
is, there exist zero mode as long as $\mu>0$. For a special case of
the $n=0$ vortex of vorticity $-1$, the small and large $r$
asymptotics of the solution we found here was discussed recently in
Ref.\cite{Nayak2006a}.

In the simplest London approximation of a spatially uniform condensate
when $f(r)=f_0$ for all $r$ except inside an infinitesimal small core,
the zero mode localized on an isolated odd-vorticity vortex is simply given by
\begin{equation}
u(r) =\left\{ \matrix{  J_n\left(r \sqrt{2\mu m -m^2 \frac{f_0^4}{4}}
\right) e^{-\frac{m}{2} f_0^2 r}, \ {\rm for}~\mu>m \frac{f_0^4}{8},
\cr
I_n\left(r \sqrt{m^2 \frac{f_0^4}{4}-2m \mu} \right)
e^{-\frac{m}{2} f_0^2 r},  \ {\rm for}~0<\mu<m \frac{f_0^4}{8} ,}
\right.
\end{equation}
where $J_n(x)$, $I_n(x)$ are Bessel and modified Bessel functions.

We note that it may seem possible to construct additional zero modes
in the following way. Instead of the ansatz of a rotationally
invariant $u(r)$ just after \rfs{eq:shroddd}, we could have chosen
an ansatz
\begin{equation}
u(\r) = u_\alpha(r) e^{i\alpha\varphi} + u_{-\alpha}(r) e^{-i\alpha\varphi}.
\end{equation}
Then two second order differential equations follow, relating these
two functions.  These are
\begin{widetext}
\begin{eqnarray}
-\frac{1}{2m} u''_\alpha
-\frac{1}{2mr} u'_\alpha+\frac{(n+\alpha)^2}{2mr^2}u_{\alpha} &=&
\frac{f^2}{r} \left(\frac 1 4 - \frac{\alpha}{2} \right) u_{-\alpha}
+ \frac{f}{2} \left( f' u_{-\alpha} + f u'_{-\alpha} \right) +\mu u_\alpha. \\
-\frac{1}{2m} u''_{-\alpha} -\frac{1}{2mr}
u'_{-\alpha}+\frac{(n-\alpha)^2}{2mr^2}u_{-\alpha} &=&
\frac{f^2}{r} \left(\frac 1 4 + \frac{\alpha}{2} \right) u_{\alpha}
+ \frac{f}{2} \left( f' u_{\alpha} + f u'_{\alpha} \right) + \mu u_{-\alpha}.
\end{eqnarray}
\end{widetext}
Generally there are going to be four solutions to these equations which
go as $r^{|n-\alpha|}$ or $r^{|n+\alpha|}$ at small $r$.  The other
two will diverge as $r^{-|n-\alpha|}$ or as $r^{-|n+\alpha|}$.
%
At infinity the four solutions of these equations go as $\exp \left[ r
\left( \pm \frac m 2 f_0^2 \pm \sqrt{f_0^4 m^2 - 8 \mu m}
\right)\right]$.  Obviously, only two of these solutions are finite at
infinity.  However, barring a coincidence, none of those solutions
finite at $r=0$ are also finite at infinity.  Even if they are for
some special value of $\mu$, by the above arguments the additional
zero modes must appear in topologically unprotected pairs, that will
be split to finite $\pm E$ energies by a slight generic deformation of
the potential (order parameter distortion). Hence we conclude that
generically there will be no additional zero modes (except the one
found above) for an odd-vorticity vortex.

Thus we indeed find that the number of zero modes in a symmetric
odd-vorticity vortex must be one. Since a smooth deformations of the
order parameter can only change the zero mode number by multiples of
two, an arbitrarily shaped odd-vorticity vortex must also have an odd
number of zero modes . However, any number of zero modes other than
one is not generic and will revert to one under an arbitrary
deformation of the order parameter.

Now for a collection of well-separated $r \gg 1/(m \Delta)$ vortices
of odd vorticity, each of them will have one zero mode localized on
it. However, as they are brought closer to each other, these zero
modes will actually split into a band of low lying $\pm E$
modes~\cite{Read2000}.  However, since other excited modes are
separated from the zero modes by a gap 
\cite{Kopnin1991}, the narrow band will only mix very weakly with
other states of the system.

It is this band of nearly degenerate zero modes that exhibit
non-Abelian statistics. Following Ref.~\cite{Ivanov2001}, we briefly
describe how it is realized here. Each of the zero modes is in fact a
Majorana fermion, as follows directly from \rfs{eq:nonuniuv} and
condition \rf{uv}, given by
\begin{equation}
\hat\gamma = \int d^2 r \left(u^*(\r) \hat a(\r)  + u(\r) \hat
  a^\dagger(\r)\right).
\end{equation}
It is straightforward to check that
\begin{eqnarray}
\hat\gamma^\dagger&=&\hat\gamma,\\
\hat\gamma^2&=&1,
\end{eqnarray}
when $u(\r)$ is properly normalized. A Majorana fermion is essentially
half of a real fermion, thus they must always come in pairs (in case
of an odd number of odd vorticity vortices, the last remaining
Majorana fermion is located at the boundary). Given $2n$ Majorana
fermions, we can construct creation and annihilation operators of $n$
real fermions, according to
\begin{equation}
\hat c_j=\hat\gamma_{2j-1}+i \hat\gamma_{2j}, \ \
\hat c_j^\dagger=\hat\gamma_{2j-1}-i\hat\gamma_{2j}.
\end{equation}
with $\hat\gamma_j$ a Majorana annihilation operator of a fermion
localized on a vortex at position $\r_j$. Thus clearly a real fermion
$\hat c_j$ is actually split between two vortices at $\r_{2j-1}$ and
$\r_{2j}$.

In the presence of $2n$ vortices, there are $2^n$ states corresponding
to $n$ pair of vortices being either occupied or empty. Now it is
possible to show that if two vortices are adiabatically exchanged \---
moved around each other \--- these nearly degenerate zero states mix
with each other. More precisely,
\begin{equation}
\label{eq:mix}
\left( \matrix{ \psi_1 \cr \psi_2 \cr \dots \cr \psi_{2^n}} \right)
\rightarrow
U\left( \matrix{ \psi_1 \cr \psi_2 \cr \dots \cr \psi_{2^n}}\right),
\end{equation}
where $U$ is a $2^n$ by $2^n$ unitary matrix, representing the unitary
transformation of the $2^n$ ground states $\psi_j$. The matrix $U$
depends on which two vortices are exchanged (and on the direction of
the exchange). It does not, however, depend on the path along which
the vortices are moved, and is thus topological.

The matrix $U$ is not a general unitary matrix. In fact, all the $2^n$
states should be split into two subsets of size $2^{n/2}$, one with
even, and the other with odd number of fermions. $U$ only mixes states
within each of these subsets. The reason for this is that the fermions
which occupy or vacate the zero modes must come in pairs, since they
are produced from a Cooper pair which is being split into two
fermions, or being assembled back from two fermions.

Matrices $U$ can be constructed by considering the change in $\psi_j$
as one vortex is slowly moved around another, while others are kept
fixed.  For such adiabatic change, the effect on the
Bogoliubov-de-Gennes equation written in the vicinity of the first
vortex is simply through $\Delta$ slowly changing its phase.  This can
be incorporated into a change of the phase of $u$ and $v$ by absorbing
half of the phase into $u$ and the other half (with the opposite sign)
into $v$. As a result, when one vortex moves all the way around
another vortex, each of its Majorana fermions changes sign.  A change
in sign of the Majorana fermions can be translated into the change in
the states $\psi_j$, by constructing an appropriate operator such that
$U^\dagger \hat\gamma_j U \rightarrow - \hat\gamma_j$.  Then the
action of $U$ on states $\psi_j$ constitutes a transformation as in
\rfs{eq:mix}.  For a more detailed discussion and an explicit
construction of $U$ for a $p_x+ i p_y$-superfluid we refer the reader
to Ref.~\cite{Ivanov2001}.

The transformation (by $U$) upon exchange of two vortices in a $p_x +
i p_y$-superfluid is a generalization of a standard quantum statistics
of bosons and fermions familiar from standard quantum mechanics. This
exchange transformaton also generalizes the two-dimensional anyonic
quantum statistics (familiar from Abelian quantum Hall states), where,
upon a two-particle exchange a many-particle wavefunction gets
multiplied by a phase factor $e^{i\theta}$ (with a phase $\theta$ not
necessarily just $0$ or $\pi$).  Since generically unitary matrices
$U$, corresponding to different pairs of vortex exchanges do not
commute, the resulting quantum statistics is termed
non-Abelian\cite{MR}.  Thus odd-vorticity vortices in a $p_x+i
p_y$-superfluid at positive detuning ($\mu > 0$) are excitations
(``particles'') with non-Abelian statistics.

Now, in addition to a basic interest, recent excitement about states
that exhibit such non-Abelian statistics is the observation that they
can form a basis for building a fault-tolerant ``topological" quantum
computer \cite{Kitaev2003}. More conventional quantum bit (q-bit)
schemes, such as the Josephson-junction charge, flux and phase q-bits, ions in
an electrostatic trap, or spin q-bits suffer from decoherence due to
interaction with the environment. In contrast, a q-bit based on
non-Abelian statics, as e.g., a state of $2n$ vortices is
topologically protected because to change it requires a global
operation on vortices such as one encircling another, something that
environmental noise will not generically do.

Based on the analysis presented here we propose~\cite{Gurarie2005}
that a $p_x + i p_y$-superfluid, that is likely to be realized in a
resonant Fermi gas interacting via a $p$-wave Feshbach resonance is
a viable candidate for an implementation of such a non-Abelian q-bit
and associated topological quantum computation.  One advantage of the
realization of such a q-bit in degenerate atomic systems (as opposed
to solid state superconductors) is the tunability of their interaction
via an external magnetic field, that allows a tuning of the chemical
potential closer to the $\mu=0$ transition, while taking care to
remain in the topological phase $\mu>0$. This in turn will allow a
more energetically stable $BEC$ superfluid, whose transition
temperature and the size of the gap are set by the Fermi energy
$\epsilon_F$, as opposed to a tiny fraction of it as in conventional
superconductors stuck in the exponentially weak BCS regime.

Of course, even if such topological $p$-wave superfluid state and the
associated non-Abelian q-bit are realized in atomic resonantly-paired
condensates, many challenges remain, such as a scheme for addressing
the q-bits by manipulation of vortices and reading off the state
$\psi_j$ of their zero modes \cite{Nayak2006a}.

\section{Comparison with experiment}
\label{sec:CWE}

\subsection{$S$-wave}
\label{sec:CWEs}

An important remaining question which must be addressed is whether
current experiments are characterized by a narrow or a broad
resonance. While it is generally believed that most $s$-wave Feshbach
resonances realized in current experiments are wide, we will show
below that some are indeed narrow in the sense that a relevant
dimensionless parameter $\gamma_s$ controlling the quantitative
validity of our theory is small.

As discussed throughout the paper, theoretically, an absolute
characterization of a width of an $s$-wave resonance is through the
value of a dimensionless parameter $\gamma_s \sim g_s^2 m^{3/2}
\epsilon_F^{-1/2}$, \rfs{eq:gamma_s1}, that is set by the ratio of the
resonance energy width to the Fermi energy. Although this parameter is
never measured directly, it can be related to experimentally
determined quantities through the atomic scattering length $a(H)$ as a
function of magnetic field $H$, that is either measured or calculated
(see Fig.~\ref{fig:as}).  From that data, magnetic field width $H_{w}$
can be extracted, as the range of the magnetic field change between
the resonance (where $a \rightarrow \infty$) and the point where
$a=0$. Alternatively, one can look at the range of $H$ where $a$
deviates significantly (e.g., by a factor $2$) from the background
scattering length $a_{bg}$. Both methods produce similar definitions
of $H_w$.

By itself, $H_w$ carries little information about the dimensionless
many-body resonance width $\gamma_s$, which, as we recall, is not only
a properties of the resonance, but depends also on the particle
density. That is, to assess how wide the resonance is, its width,
controlling strength of interactions must be compared to another
energy scale, which in this case is the typical kinetic energy
$\epsilon_F$.  To establish a relation between $\gamma_s$ and $H_w$,
we recall \rfs{eq:fanolengthrange1}
$$ a= \frac{2}{m r_0 \omega_0},$$ where $\omega_0$ is a detuning, that
measures the deviation of the Zeeman splitting (between the open and
closed channel of Feshbach resonance) from its value at the resonance,
where $a \rightarrow \infty$ (see Fig.~\ref{Fig-Feshbach}). By
matching our results for the scattering length with its experimental
dependence on the magnetic field, we determined that $\omega_0 \approx
2 \mu_B \left(H-H_0 \right)$, with $H_0$ the field at which the
resonance is tuned to zero energy and $\mu_B$ the Bohr magneton. This
allows us to express the effective-range length, $r_0$ (entering the
expression for the energy width of the resonance; see below) purely in
terms of experimentally measured quantities, namely:
\begin{equation}
\label{eq:critwidthexp}
|r_0| \simeq \frac{ \hbar^2}{m a_{bg} \mu_B H_w}.
\end{equation}
Here $\hbar$ was restored to facilitate calculations below.  Once
 $r_0$ is found, we can compute $\gamma_s$ by using
\begin{equation}
\label{gamma_sExp}
\gamma_s = \frac{l}{|r_0|} \frac{8}{\left( 3 \pi^5 \right)^{1/3}}
\approx 0.8 \frac{l}{|r_0|}.
\end{equation}
From $r_0$ we can also estimate the intrinsic (density independent)
energy width of an $s$-wave resonance,
\begin{eqnarray}
\Gamma_0&\sim& \frac{\hbar^2}{m r_0^2},\cr
&\sim&{(\mu_B H_w)^2\over \hbar^2/(m a_{bg}^2)},\\
&\sim&\frac{(\mu_B H_w)^2}{\epsilon_F}\left({a_{bg}\over l}\right)^2.
\end{eqnarray}
We note that in contrast to a naive guess, this energy width of the
resonance is {\em not} simply the Zeeman energy (converted with a Bohr
magneton) associated with the width-field $H_w$. From $\Gamma_0$ the
dimensionless parameter $\gamma_s$ is then found to be
\begin{eqnarray}
\gamma_s&\sim&\sqrt{\Gamma_0\over\epsilon_F},\cr
&\approx&\frac{\mu_B H_w}{\epsilon_F}{a_{bg}\over l}.
\end{eqnarray}

Equation \rf{gamma_sExp} can now be used as a criterion on whether a
resonance is narrow or wide (which is of course, equivalent to the one
discussed in Sec.~\ref{validity}) . We also remark that the use of
$a_{bg}$ to find $H_w$ is completely arbitrary; we could have instead
define $H_w$ as a range of the magnetic field where $a(H)$ exceeds
some given value $|a|>a_w$ as a reference point, but this would still
lead to an exactly the same $|r_0|$, \rfs{eq:critwidthexp} (with
$a_{bg}$ replaced by $a_w$)
\begin{equation}
\label{eq:critwidthexpw}
|r_0| \simeq \frac{ \hbar^2}{m a_{w} \mu_B H_w}.
\end{equation}
and the same criterion for the narrowness of the resonance; only the
product $a_w H_w$ enters $r_0$.

Physically, $\gamma_s$ can be though of as the ratio of the (energy)
range of $\omega_0$, where $a$ exceeds inter-particle spacing $l$, to
the Fermi energy of the gas.  Now, of course, it is in principle
possible to make a resonance narrow by increasing the atom density $n$
(reducing spacing $l$). However, it is our understanding that due to
experimental limitations, the Fermi energy is typically in the 1$\mu$K
or less range and cannot be significantly increased above this value.

We now apply this dimensionless criterion to the experiment reported
in Ref.~\cite{Hulet2003}. The $s$-wave resonance studied there is in
$^6$Li at $H_0\approx 543.25$ G and is probably the most narrow one
discussed in the literature. These authors report the density of their
condensate to be $3 \times 10^{12}$ cm$^{-3}$, corresponding to the
inter-particle separation of $l\approx 7\times 10^{-5}$ cm $\approx
1.3 \cdot 10^4$ au. We also note that the size of a closed channel
molecule is set by the range of the van der Waals interaction, which
is about $50$ au. This length $d$ plays the role of the inverse uv
cutoff $1/\Lambda$ of our theory. We note that the ratio $d/l\approx
1/250$ thus justifying our assumption throughout the paper that
$\Lambda^2/(2m)$ ($\Lambda$) can be treated as the largest energy
(momentum) scale of the system.

To estimate $\gamma_s$ for the resonance in Ref.~\cite{Hulet2003}, we
use their Fig.~1 for the scattering length $a(H)$ to extract $H_w$ and
$a_w$. Although from this figure it is difficult to deduce the range
of the magnetic field where $a(H)$ is larger than the background
length $a_{bg}$, we use the arbitrariness of $a_w$ to pick
$a_w=500$au. This corresponds to $|H-H_0| \le 0.015$ G, that can be
converted into an energy by multiplying by $\mu_B$.
Using \rfs{eq:critwidthexpw} then gives
\begin{equation}
|r_0| \simeq 6\cdot 10^4 \ {\rm au},
\end{equation}
and
\begin{equation}
\gamma_s \simeq \frac{l}{|r_0|} \approx 0.2.
\end{equation}
We thus conclude that the $543.25$ G $s$-wave resonance in $^6$Li is
in fact quite narrow in the absolute, dimensionless sense. It is
therefore a good candidate for a quantitative comparison with our
predictions for a narrow $s$-wave BCS-BEC crossover.

However, we note that, in contrast to above estimate of $\gamma_s$
based on Fig.~1, according to Fig.~4 of the same Ref.~\cite{Hulet2003}
the BCS-BEC crossover occurs over the range of magnetic fields of the
order of $1$ G or so, corresponding to the range of the detuning of
$100 \mu$K $ >> \epsilon_F\approx 1.5\mu$ K. Thus based on this
Fig.~4, using the narrow-resonance theory we would instead conclude
that the resonance is wide.  The reason for this discrepancy is
currently unclear to us. One should also notice that the authors of
Ref.~\cite{Hulet2003} were unable to convert all the atoms into the
molecules, so perhaps there were other factors in their experiment
which made its direct comparison with the narrow-resonance theory
difficult.

Although the resonance in $^6$Li at 543.25 G is unusually narrow, more
typical resonances have magnetic width $H_w \simeq 10$ G
with $a_w \simeq 50$ au~\cite{Ketterle1998}. For such a resonance
(assuming $^6$Li for the mass $m$ of the atom) ,
\begin{equation}
|r_0| \simeq 10^3 \ {\rm au},
\end{equation}
with
\begin{equation}
\gamma_s \simeq \frac{l}{|r_0|} \approx 10.
\end{equation}
Thus a more typical $s$-wave Feshbach resonance experiments lie in the
class of wide resonances.

\subsection{$P$-wave}
\label{sec:CWEp}

Although so far no atomic $p$-wave BCS-BEC superfluid has been
realized, $p$-wave Feshbach resonances have been demonstrated and
explored experimentally. To get a sense of future $p$-wave superfluid
possibilities, it is useful to look at the
Ref.~\cite{Ticknor2004}. Unfortunately, as we will see below from the
data reported there it is not possible to extract $g_p$.  Nevertheless
some conclusions can be made about which phases of a $p$-wave
condensate may be realized with the $p$-wave Feshbach resonance in
$^{40}$K.

We first look the data concerning the value of parameter $k_0$ from
\rfs{eq:scatampp}. The parameter $c$, as given in that paper (Eq.~(8)
of Ref.~\cite{Ticknor2004}) is magnetic field dependent, but in the
relevant range of the magnetic field it is roughly $k_0 \approx -0.04$
au$^{-1}$. As in the previous $s$-wave analysis, we estimate the uv
cutoff $\Lambda$ to be roughly of the inverse size of the
closed-channel molecule, i.e., $\Lambda\approx 0.02$ au$^{-1}$.  This
indicates that in the expression for $k_0$, \rfs{eq:k0},
$$
k_0=-\frac{12 \pi}{m^2 g_p^2} \left(1+\frac{m^2 }{3 \pi^2} g_p^2 \Lambda
\right)
$$
most likely the dimensionless uv parameter
$c_2=g_p^2 m^2 \Lambda/(3\pi^2) \gg 1$, which gives
\begin{equation}
k_0\approx -\frac{4}{\pi}\Lambda.
\end{equation}
Otherwise, $|k_0|$ would have been much bigger than $\Lambda$.  Thus
we deduced that $c_2 \gg 1$ for the experiment of
Ref.~\cite{Ticknor2004}.

This implies that this experiment is done in the regime where the
mean-field theory considered in this paper might become quantitatively
unreliable~\cite{c2largeComment}. Assuming that it
does not, large $c_2$ indicates that even in the BEC regime of this
system, most of the particles will be in the form of free atoms, not
bosonic molecules, as indicated in our analysis above.

Since the experiment reported in Ref.~\cite{Ticknor2004} is likely to
be in the regime where $c_2 \gg 1$, it is impossible to extract the
Feshbach resonance coupling $g_p$. We expect that $g_p$ gets
renormalized in the regime of large $c_2$, so that its bare value
simply drops out. However, since the complete theory of $p$-wave
superfluids at large $c_2$ is yet to be constructed, we cannot tell
what this implies for the experiment~\cite{c2largeComment}.

Next, we note that the dipolar-interaction splitting $\delta$ between
the $m_z=0$ and $m_z=\pm 1$ Feshbach resonances is quoted in this
paper as approximately $4 \mu$K. This is presumably several times
bigger than currently experimentally achievable $\epsilon_F$, that are
typically in the range of $0.5$ to $1\mu$K.  Thus we conclude that
under conditions described in Ref.~\cite{Ticknor2004}, at low
temperatures the gas will be in the $p_x$-superfluid state. However,
because the splitting is considerably larger than $\epsilon_F$ it
might be possible to bring $m_z=\pm 1$ molecules in resonance with the
atoms independently of $m_z=0$ molecules. If so, a $p_x+ip_y$-superfluid
state should be realizable for tuning near the $m_z=\pm1$ resonance.

\section{Conclusions}
\label{conclusion}

In this paper we presented a study of a degenerate Fermi gases
interacting through a tunable narrow Feshbach resonance, as recently
demonstrated experimentally. Starting with an analysis of the two-body
scattering physics we developed and justified generic models for
description of such systems. We paid a particular attention to regimes
of validity for a perturbative analysis of such systems at finite
density, and showed the existence of a small dimensionless parameter,
the ratio of the Feshbach resonance width to the Fermi energy. It  allows perturbative description
throughout the full BEC-BCS range of detuning, within the framework of two channel
model. Focussing on the most
interesting cases of the $s$- and $p$-wave resonances, we analyzed in
detail the corresponding systems. For the $s$-wave resonance, we
obtained predictions for the behavior of the system across the BEC-BCS
crossover, that we expect to be {\em quantitatively} accurate for the
case of a narrow resonance. For the far richer $p$-wave resonance,
dominant for a single hyperfine species of atoms, we showed the
existence of and analyzed a number of classical, quantum and
topological phase transitions exhibited by this system as a function
of temperature, Feshbach resonance detuning and resonance dipolar
splitting, and calculated the corresponding phase diagrams,
illustrated in Figs.~\ref{Fig-phasediaglow}, \ref{Fig-phasediaginter},
and \ref{Fig-phasediaghigh}. Finally, we studied topological
properties of the weakly-paired $p_x + i p_y$-superfluid, as well as
zero-modes inside vortices of such a topologically-ordered superfluid
in two dimensions. We hope that our analysis will be useful for
probing the associated non-Abelian quantum statistics of such
vortices, and more generally, for experimental realization and studies
of a  resonant $p$-wave superfluidity in degenerate atomic gases.

\acknowledgements

This paper would not have been possible if it were not for our earlier 
collaboration with  A. V. Andreev. We are also grateful to many people for discussions which helped us shape our understanding
of the material
discussed here, including D. Sheehy, M. Veillette, J. Levinsen, D. Jin, E. Cornell, J. Bohn, M. Holland, C. Greene, L. Levitov, R. Barankov, T.-L. Ho, G. Shlyapnikov, D. Petrov, Y. Castin, C. A. R. S\'a de Melo, N. Read, M. Zirnbauer, C. Nayak, and many others. This work was supported by the NSF grants DMR-0449521 (V.G.) and DMR-0321848 (L.R.).

\appendix
\section{Bosonic Vacuum Propagator}
\label{Ap:BVP} It is instructive to analyze the renormalized vacuum
propagator of the $b$-particles of the two-channel model. We will do
it in the $s$-wave case. Since the renormalized propagator is
nothing but the sequence of diagrams depicted on
Fig.~\ref{Fig-twochannel} with external legs amputated, its
calculation parallels that of the two fermion scattering amplitude.
The answer is given by
\begin{equation}
D(\k,\omega)=\frac{1}{\omega-\frac{k^2}{4m}-\omega_0 + i \frac{g_s^2
m^{3/ 2} }{4 \pi} \sqrt{\omega-\frac{k^2}{4m}}}.
\end{equation}
The renormalized propagator is of course simply proportional to the
two fermion scattering amplitude. Thus the {\sl poles} of this
propagators, which describe the physical bound states (or
resonances) of two fermions, coincide with the poles of the
scattering amplitude. They are given by
\begin{equation} \label{eq:polephys1}
\omega = \frac{k^2}{4m}+\omega_0 + \left[\sqrt{1-\frac{64 \pi^2
\omega_0 }{ m^3 g_s^4}}-1 \right] \frac{m^3 g_s^4 }{ 32 \pi^2}.
\end{equation}


It is now straightforward to calculate the residue of the propagator
$D$ at its poles. The result is
$$ Z=\left[ 1+i \frac{g_s^2 m^{\frac 3 2}}{8 \pi} \left(\omega_0 +
\left[\sqrt{1-\frac{64 \pi^2 \omega_0 }{ m^3 g_s^4}}-1 \right]
\frac{m^3 g_s^4 }{ 32 \pi^2} \right)^{-1} \right]^{-1}.$$ Although the
result is rather cumbersome, it is straightforward to see that the
residue goes to 1 if $g_s$ goes to zero, and it goes to $0$ if $g_s$
goes to infinity.  Thus, the contribution of $b$ to the actual bound
state of two fermions (physical molecule) reduces to zero in the limit
of large $g_s$ (wide resonance limit).  Notice that while the size of
the closed channel molecules $b$ is of the order of $d=2\pi/\Lambda$,
the actual size of the molecule, which is a superposition of $b$ and a
cloud of open channel fermions, could be quite large (and is in fact
of the order of the scattering length $a$). Thus one common perception
that since the molecules are large $b$ cannot be a point particle is
incorrect. $b$ is not a molecule, but only its closed channel part,
whose contribution to the actual molecule may in fact reduce to zero
in a wide resonance regime.

\section{Scattering matrix via single-body Hamiltonian}
\label{appendixFano}

In this appendix we compute scattering amplitudes of a number of
models relevant to the problem of resonantly interacting Fermi
gases. The results that we find here have already been obtained in
Sec.~\ref{sec:MC}, Sec.~\ref{OneCM} and Sec.~\ref{TwoCM} of the main
text, working directly with many-body Hamiltonians. However, a problem
of two particles interacting with a potential $U(\r_1-\r_2)$ can
always be reduced to a decoupled evolution of their center of mass and
that of their relative coordinate $\r=\r_1-\r_2$, whose dynamics is
governed by a Hamiltonian for a single effective particle with a
reduced mass $m_r=m_1m_2/(m_1+m_2)$, moving in a single-body potential
$U(r)$. Hence, a many-body Hamiltonian, when restricted to act on a
two-particle Hilbert subspace (as in the computation of two-particle
scattering amplitude) has an equivalent single-particle Hamiltonian
with which scattering physics can be equivalently straightforwardly
analyzed.  Thus the analysis in this appendix will complement the main
text in that we will compute scattering amplitudes for many-body
models studied there using equivalent relative coordinate
single-particle Hamiltonian.

\subsection{Fano-Anderson model}

This is a model of a particle, created by
$\hat a^\dagger_\k$, moving freely in space (representing the ``open
channel''), which when it hits the origin can turn into another
particle, created by $\hat b^\dagger$. The $b$-particle (representing the
``closed channel") cannot move at all and has a fixed energy
$\epsilon_0$. The Hamiltonian of this problem can be written as
\begin{equation}
\label{eq:FanoAnderson}\hat H =\sum_\k {k^2 \over 2 m_r} \hat a^\dagger_\k \hat a_\k
+\epsilon_0 \hat b^\dagger \hat b+ g_s ~\left( \hat b^\dagger \hat
a(0) + \hat a^\dagger(0) \hat b\right),
\end{equation}
where $g_s$ is the interconversion rate between $a$- and
$b$-particles, and $m_r$ is the mass of the $a$-particle. This is
called the Fano-Anderson model \cite{Fano1961}, or a model of a
localized state in the continuum~\cite{Mahan}. $\epsilon_0$ is a
parameter which plays the role of ``detuning". It is the energy of the
$b$-particle if it was left alone and did not interact with the
$a$-particle. The model represents the two-body version of the
$s$-wave two-channel model given by \rfs{eq:ham_s} (where $m_r$
corresponds to the reduced mass of fermions in \rfs{eq:ham_s}, hence
the notation).

The scattering amplitude of the $a$ particles can be easily
evaluated using the $T$-matrix formalism. The $T$-matrix is given
by
\begin{eqnarray}
\label{eq:pT}
T_{\k, \k'} &=& g_s D(E)g_s  + g_s D(E) g_s \Pi(E) g_s D(E)g_s + \dots \nonumber \\
&=& {g_s^2 \over D^{-1}(E) - g_s^2 \Pi(E)},
\end{eqnarray}
where $D(E)$ is the Green's function of the $b$-particle,
\begin{equation} \label{eq:D} D(E)={1 \over  E-\epsilon_0 +i
0 },
\end{equation}
and $\Pi(E)$ is the trace of the $a$-particle Green's function
\begin{equation}
\Pi(E) = \int \frac{d^3 q}{(2\pi)^3} \frac{\theta\left(\lambda-q\right)}{E- \frac{q^2}{2m_r} + i0},
\end{equation}
and $E=k^2/(2m_r)$. The value of $\Pi(E)$ was already computed in
\rfs{eq:polarization1}.  Doing the algebra, we arrive at
\begin{equation}
\label{eq:Fano} f(\k,\k') = - {1 \over {\pi \over m_r^2 g_s^2 } k^2 -
{2 \pi \over m_r g_s^2} \omega_0 + i k},
\end{equation}
where $\omega_0$ is the ``renormalized" energy of the
$b$-particle, \begin{equation} \label{eq:shift} \omega_0 =
\epsilon_0 - g_s^2 m_r \Lambda/\pi^2.\end{equation} We see that
the scattering length and effective range extracted out of \rfs{eq:Fano}
\begin{equation}
\label{eq:fanolengthrange} a^{-1} = -{2 \pi \over m_r g_s^2} \omega_0,
\ r_0 =-{2 \pi \over m_r^2 g_s^2}.
\end{equation}
coincides with Eqs.~\rf{eq:fanolengthrange1}, \rf{eq:r02Ch}.

\subsection{Hybrid model}
\label{AppendixHybrid}

In the literature it is popular to consider Feshbach-resonant 
interactions together with the interactions via
a short range potential. The single-particle Hamiltonian which captures 
a combination of
these interactions takes the form
\begin{equation} \label{eq:hybridd}
\hat H=\sum_\k {k^2 \over 2m_r} \hat  a^\dagger_\k \hat a_\k + \epsilon_0 \hat b^\dagger \hat b +
\lambda ~\hat a^\dagger(0) \hat a(0)+ g_s \left( \hat  b^\dagger \hat a(0) + \hat a^\dagger(0)
\hat b\right).
\end{equation}
The many-body version would then be a combination of a two-channel
model with a direct four-fermion point interaction scattering term
\begin{widetext}
\begin{eqnarray}
\label{eq:hybrid} \hat H &=& \sum_{\k, \sigma} {k^2 \over 2m} ~\hat
a^\dagger_{\k,\sigma} \hat a_{\k,\sigma} + \sum_\p \left(\epsilon_0
+{p^2 \over 4m} \right) \hat b_\p^\dagger \hat b_\p
+\sum_{\k,\p}
~{g_s \over \sqrt{V}}
\left(\hat b_\p ~ \hat a^\dagger_{\k+\frac{\p}{2},\uparrow}
\hat a^\dagger_{-\k+\frac{\p}{2}, \downarrow} +
\hat b^\dagger_\p ~ \hat a_{-\k+\frac{\p}{2},\downarrow}
\hat a_{\k+\frac{\p}{2},\uparrow}\right)  \cr
&&+ \frac{\lambda} V ~\sum_{\k, \k', \p}\hat a^\dagger_{\k'+\frac{\p}2, \downarrow} \hat a^\dagger_{-\k'+\frac{\p}2, \uparrow}
\hat a_{-\k-\frac{\p}2, \uparrow} \hat a_{\k-\frac{\p}2, \downarrow} \Big].
\end{eqnarray}
\end{widetext}
It is instructive to calculate the scattering amplitude of the
$a$-particles which follow from \rfs{eq:hybridd}. The calculation
largely parallels that given in \rfs{eq:pT}, except $g_s D g_s$ gets
replaced by $g_s D g_s +\lambda$.  After some algebra we obtain
\begin{equation}
f_0(k)=-{1 \over {2 \pi \over m_r} {{k^2 \over 2m_r} - \epsilon_0
\over \lambda \left( {k^2 \over 2m_r} - \epsilon_0 \right) + g_s^2 }
+{2 \Lambda \over \pi}  + i k}
\end{equation}
We see that the $a$-particles scatter in a rather complicated
fashion. If we are only interested in low-energy scattering, we
can expand the denominator of the scattering amplitude and read
off the scattering length and the effective range as
\begin{equation}
a^{-1} =  {2 \pi \epsilon_0 \over m_r ( \epsilon_0 \lambda-g_s^2)} +{2
\Lambda \over \pi}, \ r_0=- {2 \pi g_s^2 \over m_r^2 \left( \epsilon_0
\lambda-g_s^2\right)^2}.
\end{equation}
In principle, we can now redefine the parameters $\epsilon_0$,
$g$, and $\lambda$ in such a way that the scattering length and
the effective range computed here would coincide with the ones
produced by the pure Fano-Anderson model.  We see that
ultimately including both the $\delta$-like potential and the
Fano-Anderson term in the Hamiltonian does not produce new low
energy physics compared to the pure Fano-Anderson case, and
amounts to just redefining the parameters of the Fano-Anderson
model \--- and thus, of the two-channel model \rfs{eq:ham_s}. Its only 
physical effect is to accommodate for $a_{\rm bg} \sim d$ absent 
in the pure two-channel model.  

We also remark that this may sometimes not be true in lower
dimensions. In 1D the inclusion of the contact interaction term may
change the physics described by the two-channel model
qualitatively~\cite{Gurarie2006a}.

\subsection{$P$-wave Fano-Anderson Model}
The $p$-wave version of the Fano-Anderson model
\rfs{eq:FanoAnderson} is given by
\begin{equation}
\label{eq:FanoAndersonP} \hat H =\sum_{\k} {k^2 \over 2m_r} \hat a^\dagger_\k \hat a_\k
+\epsilon_0 \sum_{\alpha=1}^3 \hat b^\dagger_\alpha \hat b_\alpha +  {g_p
\over \sqrt{V}} \sum_{\k,\alpha} k_\alpha \left(\hat  b^\dagger_\alpha
\hat a_{\k} + \hat a^\dagger_{\k} \hat b_\alpha \right).
\end{equation}
Here the $a$ particle scatters in the $p$-wave channel and can
convert into a $b$ particle which carries internal angular
momentum $\ell=1$, since the angular momentum is conserved. The
angular momentum is represented by the vector index $\alpha$. It
is related to the states with definite projections of angular
momentum $\hat b_{m=1}$, $\hat b_{m=0}$, and $\hat b_{m=-1}$ via the standard
formulae (Ref.~\cite{LL}) discussed in Eqs.~\rf{relationsbetweenmuandalpha1}, \rf{relationsbetweenmuandalpha2}, \rf{relationsbetweenmuandalpha3}, which we repeat
here for convenience
\begin{eqnarray} \label{eq:vecspin}
\hat b_z &=& \hat b_{m=0}, \cr  -{1 \over \sqrt{2}} \left( \hat b_x+i \hat b_y \right)& = &
\hat b_{m=1}, \cr {1 \over \sqrt{2}} \left( \hat b_x - i \hat b_y \right)&=& \hat b_{m=-1}.
\end{eqnarray}

By construction, $a$-particles scatter only in the $p$-wave
channel. The scattering amplitude can be easily calculated in the
same $T$-matrix formalism, as in the $s$-wave case,
Eqs.~\rf{eq:pT}, \rf{eq:Fano}.

The propagator of the $b$-particles is now
\begin{equation}
D_{\alpha \beta}(E) = \delta_{\alpha \beta} D(E)
\end{equation}
with $D(E)$ given by \rfs{eq:D}. The $T$-matrix is given by the
$p$-wave version of \rfs{eq:pT},
\begin{widetext}
\begin{equation} \label{eq:pwaveT}
T_{{\bf k}, {\bf k'}} =  \sum_{\alpha} g_p k_{\alpha} D k'_{\alpha}
g_p + \sum_{{\bf k''},\alpha,\beta} g_p k_\alpha D  g_p k''_{\alpha}
G(k'',E) g_p k''_{\beta} D g_p k'_{\beta} + \dots = \sum_{\alpha} {g_p^2
k_\alpha k'_{\alpha} \over D^{-1}(E)-g_p^2 \Pi(E)},
\end{equation}
\end{widetext} where $G(E,k)$ is the Green's
function of the $a$-particles and $\Pi(E)$ is now
\begin{eqnarray}
\label{eq:polapApp} \Pi(E) = {1 \over 3} \int {d^3 q \over (2 \pi)^3}
{ q^2 \over E-{q^2 \over 2 m_r}+i0}\nonumber  \\ = - {m_r \Lambda^3
\over 9 \pi^2} -{2 m_r^2 \Lambda \over 3 \pi^2} E - i {\sqrt{2} m_r^{5
/ 2} E^{3 / 2} \over 3 \pi  }.
\end{eqnarray}
Here $E=k^2/(2m_r)$.  Just like everywhere else throughout the paper,
we cut off the divergent integral at $q\sim \Lambda$. Unlike
\rfs{eq:polarization}, the integral is divergent as $q^3$ and produces
two cutoff dependent terms. With the help of the notations
\rfs{eq:c1}, \rfs{eq:c2}, introduced earlier, we find the $p$-wave
scattering amplitude
\begin{eqnarray} f_p &=& -\frac{m_r}{6\pi} \frac{g_p^2 k^2}{D^{-1}(E)-g_p^2 \Pi(E)}\cr
&=&\frac{k^2}{\frac{6 \pi}{m g^2} \left(\epsilon_0-c_1 \right) - \frac{3 \pi}{m^2 g_p^2}
\left( 1+2 c_2 \right) k^2 - i k^3}.\end{eqnarray}
This coincides with the result of the many-body calculations reported in \rfs{eq:fkaka}
with the exception of a numerical coefficient. This difference is the result
of the indistinguishability of identical particles which was important in \rfs{eq:fkaka}
but played no role here, in the one-body scattering calculation.

\section{Details of the $p$-wave saddle-point equation and
  free energy}
\label{appendixIab}

The thermodynamics of a $p$-wave superfluid is completely determined
by the free energy, \rfs{Sp_sp} and the corresponding saddle-point
equation, \rfs{eq:saddlepT}, derived from it. These were expressed in
terms of one key tensor $I_{\alpha\beta}^{(T)}[\bB]$, defined in
\rfs{eq:Iab}.  Here we compute $I_{\alpha\beta}^{(T)}[\bB]$ at zero and
finite temperatures and thereby obtain the corresponding ground-state
energy and the free energy.
\subsection{Zero temperature}
At zero temperature $I_{\alpha\beta}^{(T)}[\bB]$ reduces to
\begin{eqnarray}
\label{eq:Iab0App}
\hspace{-1cm}
I_{\alpha \beta} &=&g_p^2 \int {d^3 k \over (2 \pi)^3}
{ k_\alpha k_\beta\over E_\k},\nonumber\\
&=&g_p^2 \int {d^3 k \over (2 \pi)^3}
{ k_\alpha k_\beta\over \left[(\epsilon_k-\mu)^2 +
4g_p^2|\bB\cdot\bk|^2\right]^{1/2}},
\end{eqnarray}
where we used the spectrum $E_\k$, \rfs{eq:Ekp}.  The integral is
naturally computed in spherical coordinates, with the radial part over
$k$ conveniently expressed as an integral over the free spectrum
$\epsilon_k=k^2/2m$
\begin{eqnarray}
\label{eq:Iab0App2}
I_{\alpha \beta} &=& g_p^2~ 2m|\mu| N(|\mu|)
\int {d\Omega_\k\over 4\pi}\kh_\alpha\kh_\beta~I(Q_{\bf\kh}),
\end{eqnarray}
where we defined a function
\begin{eqnarray}
\label{eq:IQ}
I(Q)&=& \int_0^{\El} d\eh {\eh^{3/2}\over
\left[(\eh-\mh)^2 + Q\eh\right]^{1/2}},
\end{eqnarray}
that arises from an integral over $\epsilon$ scaled by the chemical
potential $\eh=\epsilon/|\mu|$,
$N(\mu)=m^{3/2}|\mu|^{1/2}/(2^{1/2}\pi^2)\equiv c|\mu|^{1/2}$ is the
density of states, $\mh\equiv\mu/|\mu|=\pm 1$,
$\El\equiv(\Lambda^2/2m)/|\mu|$, and
\begin{eqnarray}
\label{eq:Q}
Q&=& {8m g_p^2\over\mu}|\bB\cdot{\bf\kh}|^2,\\
&=& {8m g_p^2\over\mu}\left((\bu\cdot{\bf\kh})^2+(\bv\cdot{\bf\kh})^2\right).
\end{eqnarray}

Because at large $\eh$, $I(Q)$ scales as $\eh^{3/2}$, its one set of
dominant contributions comes from the region of integration near the
uv cutoff $\El$. We isolate these $I_\Lambda$ contributions by writing
\begin{eqnarray}
\label{eq:IQsplit}
I(Q)&=& I_\Lambda + \delta I,
\end{eqnarray}
with
\begin{eqnarray}
\label{eq:IQ_Lambda}
I_\Lambda&=&\int_0^{\El} d x\left[x^{1/2} + {\mh-Q/2\over
    x^{1/2}}\right],\\
&=&\frac{2}{3}\El^{3/2}+2(\mh-Q/2)\El^{1/2},
\end{eqnarray}
and
\begin{eqnarray}
\label{eq:IQfs}
\hspace{-0.5cm}
\delta I&=& \int_0^{\infty} d\eh\left\{{\eh^{3/2}\over
\left[(\eh-\mh)^2 + Q\eh\right]^{1/2}}-\eh^{1/2} - {\mh-Q/2\over
    \eh^{1/2}}\right\}.\nonumber\\
&&
\end{eqnarray}
Because the remaining contribution $\delta I$ is uv-convergent, we
have extended its uv cutoff $\El$ to infinity, thereby only neglecting
insignificant terms that are down by a factor of order ${\cal
  O}(\mh/\El,Q/\El)\ll 1$.

Combining the uv contribution $I_\Lambda(Q)$ inside $I_{\alpha\beta}$,
\rfs{eq:Iab0App2}, and doing the angular integrals
we obtain the uv contribution
\begin{widetext}
\begin{eqnarray}
\label{eq:IabLambda}
I^\Lambda_{\alpha \beta} &=& g_p^2~ 2m|\mu| N(|\mu|)
\int {d\Omega_\k\over
  4\pi^2}\kh_\alpha\kh_\beta~I_\Lambda(Q_{\bf\kh}),\\
&=& g_p^2~ 2m|\mu| N(|\mu|)
\int {d\Omega_\k\over 4\pi}\kh_\alpha\kh_\beta
\left[\frac{2}{3}\El^{3/2}+2\mh\El^{1/2}
-\frac{8m g_p^2}{|\mu|}\El^{1/2}(\bBb\cdot{\bf\kh})(\bB\cdot{\bf\kh})\right],
\nonumber\\
&=& \frac{2}{3}g_p^2~\frac{(2m)^{5/2}}{4\pi^2}
\left[\left(\frac{1}{3}E_\Lambda^{3/2}
+\mu E_\Lambda^{1/2}\right)\delta_{\alpha\beta}
-\frac{4m g_p^2}{5}E_\Lambda^{1/2}
\left(\delta_{\alpha\beta}\bBb\cdot\bB
+ \Bb_\alpha B_\beta + \Bb_\beta B_\alpha\right)\right],\nonumber\\
&=& (c_1 + 2\mu c_2)\delta_{\alpha\beta}
-\frac{8}{5}m g_p^2 c_2
\left(\delta_{\alpha\beta}\bBb\cdot\bB
+ \Bb_\alpha B_\beta + \Bb_\beta B_\alpha\right),\nonumber
\end{eqnarray}
\end{widetext}
where $c_1$ and $c_2$ constants were defined in Eqs.\rf{eq:c1},
\rf{eq:c2}, and we used three-dimensional spherical averages
\begin{eqnarray}
\int \frac{d\Omega_{\bf\kh}}{4\pi}\kh_\alpha \kh_\beta
&=&\frac{1}{3}\delta_{\alpha\beta},\\
\int \frac{d\Omega_{\bf\kh}}{4\pi}
\kh_\alpha \kh_\beta\kh_\gamma \kh_\delta
&=&\frac{1}{15}\left(\delta_{\alpha\beta}\delta_{\gamma\delta}+
\delta_{\alpha\gamma}\delta_{\beta\delta} +
\delta_{\alpha\delta}\delta_{\beta\gamma}\right).\nonumber
\end{eqnarray}
This confirms the uv contribution to $I_{\alpha\beta}[\bB]$ used in
the main text, Eqs.\rf{IabLambda3} and \rf{IabLambda1}.

The value of the second contribution, $\delta I(Q)$ in \rfs{eq:IQsplit}
critically depends on the sign of $\mu$. In the BEC regime of $\mu
<0$, $\mh=-1$ and the integral in $\delta I(Q)$ is convergent
everywhere, making only a strongly subdominant ${\cal
  O}(\mh/\El,Q/\El)\ll 1$ contribution to $I_\Lambda(Q)$, that can be
safely neglected.

In contrast, in the BCS regime of $\mu > 0$ and $Q\ll 1$, the integral
in $\delta I(Q)$, while uv convergent, makes a large
contribution to $I_\Lambda(Q)$ that is, in fact, logarithmically
divergent with a vanishing $Q$. This large contribution arises from a
region around $\eh=1$, physically corresponding to low-energy
excitations near the Fermi surface.

We focus on the integration above and below the Fermi-surface, $\delta
I=\theta(\mu)(\delta I^- + \delta I^+$), with
\begin{eqnarray}
\label{eq:IQfs2}
\delta I^- &=&\int_0^1 d\eh g(\eh)=\int_0^1 dx g(1-x),\nonumber\\
&=&\delta\tilde{I}^- + J^-\\
\delta I^+ &=&\int_1^\infty d\eh g(\eh)=\int_0^1 dx g(1+x)+
\int_1^\infty dx g(1+x),\nonumber\\
&=&\delta\tilde{I}^+ + J^+,
\end{eqnarray}
and the integrand $g(\eh)$ given by \rfs{eq:IQfs} with $\mh=1$. Above
we separated the dominant logarithmic contribution
$\delta\tilde{I}^{\pm}$ out of $\delta I^{\pm}$,
\begin{eqnarray}
\label{eq:IQfs3}
\delta\tilde{I}^{\pm}(Q) &=&\int_0^1{dx\over[x^2+Q(1\pm x)]^{1/2}},\\
&=&\ln{2\over\sqrt{Q}} + f^{\pm}(Q)\nonumber\\
\end{eqnarray}
with
\begin{eqnarray}
\label{eq:fQ}
f^{-}(Q) &=&\ln\left[{1+Q/2+\sqrt{1+2Q}\over2+Q^{1/2}}\right],\\
&\approx&-\oh Q^{1/2} +\frac{7}{8}Q-{\cal O}(Q^{3/2}),\nonumber\\
f^{+}(Q) &=&\ln\left[{2-Q/2\over2-Q^{1/2}}\right],\\
&\approx&\oh Q^{1/2} -\frac{1}{8}Q+{\cal O}(Q^{3/2}),\nonumber
\end{eqnarray}
and the subdominant in small $Q$ contributions
\begin{eqnarray}
\label{eq:Js}
\hspace*{-1cm}
J^{-}(Q) &=&\int_0^1 dx\left[g(1-x)-{1\over[x^2+Q(1-x)]^{1/2}}\right],
\quad\quad\quad\\
J^{+}(Q) &=&\int_0^\infty dx\left[g(1+x)-{1\over[x^2+Q(1+ x)]^{1/2}}\right],
\quad\quad\quad
\end{eqnarray}
that are finite for $Q\rightarrow0$. Taylor-expanding these
subdominant contributions to lowest order in $Q$ and combining
everything together, we obtain:
\begin{eqnarray}
\label{eq:dIfinal}
\delta I(Q)=\theta(\mu)\ln\left({64e^{-16/3}\over Q}\right)
+\theta(\mu){\cal O}(Q\ln Q).
\end{eqnarray}
Combining this with $I_\Lambda(Q)$ inside $I_{\alpha\beta}(Q)$,
\rfs{eq:Iab0App2}, we obtain $I_{\alpha\beta}=I^\Lambda_{\alpha\beta}-
\delta\hat{I}_{\alpha\beta}~\theta(\mu)~g_p^2 \sqrt{2 m^{5}\mu^{3}}/\pi^2$, where
\begin{eqnarray}
\label{IabSph}
\delta\hat{I}_{\alpha\beta}=
\int {d\Omega_\k\over 4\pi}\kh_\alpha\kh_\beta
~\ln\left[(\uh\cdot{\bf\kh})^2+(\vh\cdot{\bf\kh})^2\right],
\end{eqnarray}
and
\begin{eqnarray}
\label{Bhat}
\hat{\bB}=\uh + i\vh\equiv \frac{1}{8}e^{8/3}\sqrt{8 m g_p^2\over\mu}\bB.
\end{eqnarray}
The spherical average in \rfs{IabSph} is easiest to evaluate in the
transverse gauge $\uh\cdot\vh=0$, taking $\uh$ and $\vh$ to be the
$k_x\equiv k_u$ and $k_y\equiv k_v$ axes. This reduces it to
\begin{eqnarray}
\label{IabSph2}
\delta\hat{I}_{\alpha\beta}&=&
\int {d\Omega_\k\over 4\pi}\kh_\alpha\kh_\beta
~\ln\left[\hat{u}^2 \hat{k}_u^2 + \hat{v}^2 \hat{k}_v^2\right],\\
&=&\int_0^{2\pi} d\phi\int_{0}^\pi d\theta\sin\theta\kh_\alpha\kh_\beta
\times\\
&&\hspace{3cm}
\ln\left[\sin^2\theta(\hat{u}^2\cos^2\phi+\hat{v}^2\sin^2\phi)\right],
\nonumber\\
&=&A(\hat{u},\hat{v})\delta_{\alpha\beta} +
C(\hat{u},\hat{v})\hat{u}_\alpha\hat{u}_\beta +
C(\hat{v},\hat{u})\hat{v}_\alpha\hat{v}_\beta,
\end{eqnarray}
where in the last line we took advantage of the general tensor form
and $\hat{u}\leftrightarrow\hat{v}$ symmetry of
$\delta\hat{I}_{\alpha\beta}(\hat{u},\hat{v})$ to reduce its
computation to two functions $A(\hat{u},\hat{v})$ and
$C(\hat{u},\hat{v})$, that can be obtained by calculating any two
combinations of components of
$\delta\hat{I}_{\alpha\beta}(\hat{u},\hat{v})$. With this, we obtain
in the transverse gauge
\begin{eqnarray}
\label{IabSphFormula}
\delta\hat{I}_{\alpha\beta}&=&
{2 \over 3} \ln
\left[e^{-4/3}(\hat{u}+\hat{v})\right]\delta_{\alpha\beta}\\
&&+{2 \over 3\hat{u}(\hat{u}+\hat{v})}\hat{u}_\alpha \hat{u}_\beta
+ {2 \over 3 \hat{v} (\hat{u}+\hat{v})} \hat{v}_\alpha \hat{v}_\beta,
\nonumber
\end{eqnarray}
which when combined with \rfs{eq:IabLambda} gives the saddle-point
equation \rfs{sp_final} used in Sec.\ref{EgsPwave}.

By integrating this saddle-point equation,
\begin{eqnarray}
{\partial\ve_{GS}[\bB]\over\partial\Bb_\alpha}
=(\epsilon_\alpha - 2\mu)B_\alpha-\sum_\beta I_{\alpha\beta}[\bB] B_\beta=0,
\end{eqnarray}
over $\Bb_\alpha$ we can also obtain the ground-state energy density
$\ve_{GS}[\bB]$. Utilizing \rfs{IabSph} to integrate the last term we find
\begin{widetext}
\begin{eqnarray}
\hspace{-2.5cm}
\ve_{GS}[\bB]&=&(\omega_0-2\mu)(1+c_2)|\bB|^2
+\gamma_p c_2\frac{8\epsilon_F}{5n}\left[ \left(\bBb\cdot\bB\right)^2 +
\oh |\bB\cdot\bB|^2\right]\nonumber\\
&&+\theta(\mu)3\gamma_p\mu\sqrt{\mu\over\epsilon_F}\int {d\Omega_\k\over 4\pi}
(\bBb\cdot{\bf\kh})(\bB\cdot{\bf\kh})
\left(\ln\left[a_0(\bBb\cdot{\bf\kh})(\bB\cdot{\bf\kh})\right]-1\right),\\
&=&(\omega_0-2\mu)(1+c_2)|\bB|^2
+\gamma_p c_2\frac{8\epsilon_F}{5n}\left[ \left(\bBb\cdot\bB\right)^2 +
\oh |\bB\cdot\bB|^2\right]
+\theta(\mu)\gamma_p\mu\sqrt{\mu\over\epsilon_F}
\sum_{\alpha,\beta}\left(3\delta\hat{I}_{\alpha\beta}[\bB]
-\delta_{\alpha\beta}\right)\Bb_\alpha B_\beta,\nonumber\\
&&
\end{eqnarray}
\end{widetext}
which gives the result quoted in \rfs{EgsGaugeInvnt} of
Sec.\ref{EgsPwave} and used to study phase behavior of a $p$-wave
resonant Fermi gas.

\subsection{Finite temperature}

Above analysis can be easily extended to finite temperature, by analyzing
\begin{eqnarray}
\label{eq:IabTApp}
\hspace{-1cm}
I^{(T)}_{\alpha \beta} &=&g_p^2 \int {d^3 k \over (2 \pi)^3}
{ k_\alpha k_\beta \tanh\left({E_\k\over2T}\right)\over E_\k},\\
&=&g_p^2~ 2m|\mu| N(|\mu|)\left(\hat{I}^{(0)}_{\alpha\beta} +
\hat{I}^{(1)}_{\alpha\beta}\right),\nonumber
\end{eqnarray}
where
\begin{widetext}
\begin{eqnarray}
\label{eq:Iab01App}
\hspace{-1cm}
\hat{I}^{(0)}_{\alpha \beta} &=&\int {d\Omega_\k\over 4\pi}\kh_\alpha\kh_\beta
\int_0^{\El}\eh^{3/2}
{\tanh\left({|\mu|\over2T}|\eh-\mh|\right)\over|\eh-\mh|},\nonumber\\
&\approx&\theta(\mu)\frac{2}{3}\delta_{\alpha\beta}\left[\ln{\mu\over T}
+\frac{1}{3}\El^{3/2} + \El^{1/2}\right],\\
\hat{I}^{(1)}_{\alpha \beta} &=&\int {d\Omega_\k\over 4\pi}\kh_\alpha\kh_\beta
\int_0^{\El}{\eh^{5/2}\over2|\eh-\mh|^3}\left[{|\mu|\over2T}|\eh-1|
\sech^2\left({|\mu|\over2T}|\eh-1|\right)
-\tanh\left({|\mu|\over2T}|\eh-\mh|\right)\right]Q,\nonumber\\
&\approx&-\frac{8m g_p^2}{15|\mu|}\El^{1/2}
\left[\delta_{\alpha\beta}\bBb\cdot\bB
+ \Bb_\alpha B_\beta + \Bb_\beta B_\alpha\right],
\end{eqnarray}
\end{widetext}
and we have safely Taylor-expanded in $Q$ since Fermi-surface
divergences are cutoff by finite $T$. Combining these together we find
\begin{widetext}
\begin{eqnarray}
I^{(T)}_{\alpha\beta}&\approx& \left[c_1 + 2\mu c_2 +
\frac{2}{3} \Theta(\mu) g_p^2 2m\mu N(\mu)\ln{\mu\over
  T}\right]\delta_{\alpha\beta}
-\frac{8}{5}m g_p^2 c_2\left(\delta_{\alpha\beta}\bBb\cdot\bB
+ \Bb_\alpha B_\beta + \Bb_\beta B_\alpha\right),\nonumber\\
&=& I^\Lambda_{\alpha\beta} +
\frac{2}{3}\theta(\mu)g_p^2 2m\mu N(\mu)\ln{\mu\over T}\delta_{\alpha\beta},
\end{eqnarray}
\end{widetext}
where $I^\Lambda_{\alpha\beta}$ is given in \rfs{eq:IabLambda}, above.
As anticipated, the nonanalytic (Fermi surface, $a_1$) terms have been
replaced by $\delta_{\alpha\beta}\frac{2}{3}\Theta(\mu)g_p^2 2m\mu
N(\mu)\ln(\mu/T)=\delta_{\alpha\beta}\Theta(\mu)2\gamma_p^2\mu\sqrt{\mu/\epsilon_F}
\ln(\mu/T)$.  The resulting free-energy density is given by
\begin{widetext}
\begin{eqnarray}
\label{EgsGaugeInvntT}
\frac{f[\bB]}{1+c_2}= \sum_\alpha
\bigg(\tilde{\omega}_\alpha(T) - 2 \mu\bigg)|B_\alpha|^2
+ a_2\left[ \left(\bBb\cdot\bB\right)^2 +\oh |\bB\cdot\bB|^2\right],
\end{eqnarray}
\end{widetext}
with
\begin{equation}
\tilde{\omega}_\alpha(T)=\omega_\alpha-a_1\ln(\mu/T),
\end{equation}
determining $T_c^\alpha$ by $\tilde{\omega}_\alpha(T_c^\alpha)=2\mu$.

\bibliography{review}

\begin{thebibliography}{84}
\expandafter\ifx\csname natexlab\endcsname\relax\def\natexlab#1{#1}\fi
\expandafter\ifx\csname bibnamefont\endcsname\relax
  \def\bibnamefont#1{#1}\fi
\expandafter\ifx\csname bibfnamefont\endcsname\relax
  \def\bibfnamefont#1{#1}\fi
\expandafter\ifx\csname citenamefont\endcsname\relax
  \def\citenamefont#1{#1}\fi
\expandafter\ifx\csname url\endcsname\relax
  \def\url#1{\texttt{#1}}\fi
\expandafter\ifx\csname urlprefix\endcsname\relax\def\urlprefix{URL }\fi
\providecommand{\bibinfo}[2]{#2}
\providecommand{\eprint}[2][]{\url{#2}}

\bibitem[{\citenamefont{Schrieffer}(1989)}]{Schrieffer}
\bibinfo{author}{\bibfnamefont{R.}~\bibnamefont{Schrieffer}},
  \emph{\bibinfo{title}{Theory of Superconductivity}}
  (\bibinfo{publisher}{Perseus Books Group}, \bibinfo{address}{N.Y.},
  \bibinfo{year}{1989}).

\bibitem[{\citenamefont{Khalatnikov}(2000)}]{Khalatnikov}
\bibinfo{author}{\bibfnamefont{I.~M.} \bibnamefont{Khalatnikov}},
  \emph{\bibinfo{title}{An Introduction to the Theory of Superfluidity}}
  (\bibinfo{publisher}{Perseus Books Group}, \bibinfo{address}{N.Y.},
  \bibinfo{year}{2000}).

\bibitem[{\citenamefont{Eagles}(1969)}]{Eagles1969}
\bibinfo{author}{\bibfnamefont{D.~M.} \bibnamefont{Eagles}},
  \bibinfo{journal}{Phys. Rev.} \textbf{\bibinfo{volume}{186}},
  \bibinfo{pages}{456} (\bibinfo{year}{1969}).

\bibitem[{\citenamefont{Leggett}(1980)}]{Leggett1980}
\bibinfo{author}{\bibfnamefont{A.}~\bibnamefont{Leggett}}, in
  \emph{\bibinfo{booktitle}{Modern Trends in the Theory of Condensed Matter}}
  (\bibinfo{publisher}{Springer-Verlag}, \bibinfo{address}{Berlin},
  \bibinfo{year}{1980}), pp. \bibinfo{pages}{13--27}.

\bibitem[{\citenamefont{Nozi\`eres and Schmitt-Rink}(1985)}]{Nozieres1985}
\bibinfo{author}{\bibfnamefont{P.}~\bibnamefont{Nozi\`eres}} \bibnamefont{and}
  \bibinfo{author}{\bibfnamefont{S.}~\bibnamefont{Schmitt-Rink}},
  \bibinfo{journal}{J. Low Temp. Phys.} \textbf{\bibinfo{volume}{59}},
  \bibinfo{pages}{195} (\bibinfo{year}{1985}).

\bibitem[{\citenamefont{Levinsen and Gurarie}(2006)}]{Levinsen2006}
\bibinfo{author}{\bibfnamefont{J.}~\bibnamefont{Levinsen}} \bibnamefont{and}
  \bibinfo{author}{\bibfnamefont{V.}~\bibnamefont{Gurarie}},
  \bibinfo{journal}{Phys. Rev. A} \textbf{\bibinfo{volume}{73}},
  \bibinfo{pages}{053607} (\bibinfo{year}{2006}).

\bibitem[{com({\natexlab{a}})}]{commentBCSBEC}
\bibinfo{note}{While within mean-field theory it might appear that the
  distinctions between weakly and strongly-paired superfluids are qualitative,
  in a full treatment that includes fluctuations these can be shown to be
  merely quantitative differences. For example the separation between the
  transition temperature $T_c$ and the crossover temperature $T_*$ in principle
  exists in any system undergoing a continuous transition, but is usually quite
  small.}

\bibitem[{\citenamefont{{S\'a de Melo} et~al.}(1993)\citenamefont{{S\'a de
  Melo}, Randeria, and Engelbrecht}}]{deMelo1993}
\bibinfo{author}{\bibfnamefont{C.~A.~R.} \bibnamefont{{S\'a de Melo}}},
  \bibinfo{author}{\bibfnamefont{M.}~\bibnamefont{Randeria}}, \bibnamefont{and}
  \bibinfo{author}{\bibfnamefont{J.~R.} \bibnamefont{Engelbrecht}},
  \bibinfo{journal}{Phys. Rev. Lett.} \textbf{\bibinfo{volume}{71}},
  \bibinfo{pages}{3202} (\bibinfo{year}{1993}).

\bibitem[{\citenamefont{Chen et~al.}(2005)\citenamefont{Chen, Stajic, and
  Levin}}]{Levin2005}
\bibinfo{author}{\bibfnamefont{Q.}~\bibnamefont{Chen}},
  \bibinfo{author}{\bibfnamefont{J.}~\bibnamefont{Stajic}}, \bibnamefont{and}
  \bibinfo{author}{\bibfnamefont{K.}~\bibnamefont{Levin}},
  \bibinfo{journal}{Phys. Rep.} \textbf{\bibinfo{volume}{412}},
  \bibinfo{pages}{1} (\bibinfo{year}{2005}).

\bibitem[{\citenamefont{Volovik}(1992)}]{VolovikBook}
\bibinfo{author}{\bibfnamefont{G.~E.} \bibnamefont{Volovik}},
  \emph{\bibinfo{title}{Exotic Properties of Superfluid $^3$He}}
  (\bibinfo{publisher}{World Scientific}, \bibinfo{address}{Singapore},
  \bibinfo{year}{1992}).

\bibitem[{\citenamefont{Read and Green}(2000)}]{Read2000}
\bibinfo{author}{\bibfnamefont{N.}~\bibnamefont{Read}} \bibnamefont{and}
  \bibinfo{author}{\bibfnamefont{D.}~\bibnamefont{Green}},
  \bibinfo{journal}{Phys. Rev. B} \textbf{\bibinfo{volume}{61}},
  \bibinfo{pages}{10267} (\bibinfo{year}{2000}).

\bibitem[{\citenamefont{Volovik}(2003)}]{VolovikBook1}
\bibinfo{author}{\bibfnamefont{G.~E.} \bibnamefont{Volovik}},
  \emph{\bibinfo{title}{The Universe in a Helium Droplet}}
  (\bibinfo{publisher}{Oxford University Press}, \bibinfo{address}{Oxford},
  \bibinfo{year}{2003}).

\bibitem[{\citenamefont{DeMarco and Jin}(1999)}]{DeMarco1999}
\bibinfo{author}{\bibfnamefont{B.}~\bibnamefont{DeMarco}} \bibnamefont{and}
  \bibinfo{author}{\bibfnamefont{D.~S.} \bibnamefont{Jin}},
  \bibinfo{journal}{Science} \textbf{\bibinfo{volume}{285}},
  \bibinfo{pages}{1703} (\bibinfo{year}{1999}).

\bibitem[{\citenamefont{Strecker et~al.}(2003)\citenamefont{Strecker,
  Partridge, and Hulet}}]{Hulet2003}
\bibinfo{author}{\bibfnamefont{K.~E.} \bibnamefont{Strecker}},
  \bibinfo{author}{\bibfnamefont{G.~B.} \bibnamefont{Partridge}},
  \bibnamefont{and} \bibinfo{author}{\bibfnamefont{R.~G.} \bibnamefont{Hulet}},
  \bibinfo{journal}{Phys. Rev. Lett.} \textbf{\bibinfo{volume}{91}},
  \bibinfo{pages}{080406} (\bibinfo{year}{2003}).

\bibitem[{\citenamefont{Regal et~al.}(2004)\citenamefont{Regal, Greiner, and
  Jin}}]{Jin2004}
\bibinfo{author}{\bibfnamefont{A.}~\bibnamefont{Regal}},
  \bibinfo{author}{\bibfnamefont{M.}~\bibnamefont{Greiner}}, \bibnamefont{and}
  \bibinfo{author}{\bibfnamefont{D.~S.} \bibnamefont{Jin}},
  \bibinfo{journal}{Phys. Rev. Lett.} \textbf{\bibinfo{volume}{92}},
  \bibinfo{pages}{040403} (\bibinfo{year}{2004}).

\bibitem[{\citenamefont{Zwierlein et~al.}(2004)\citenamefont{Zwierlein, Stan,
  Schunck, Raupach, Kerman, and Ketterle}}]{Ketterle2004}
\bibinfo{author}{\bibfnamefont{M.~W.} \bibnamefont{Zwierlein}},
  \bibinfo{author}{\bibfnamefont{C.~A.} \bibnamefont{Stan}},
  \bibinfo{author}{\bibfnamefont{C.~H.} \bibnamefont{Schunck}},
  \bibinfo{author}{\bibfnamefont{S.~M.~F.} \bibnamefont{Raupach}},
  \bibinfo{author}{\bibfnamefont{A.~J.} \bibnamefont{Kerman}},
  \bibnamefont{and} \bibinfo{author}{\bibfnamefont{W.}~\bibnamefont{Ketterle}},
  \bibinfo{journal}{Phys. Rev. Lett.} \textbf{\bibinfo{volume}{92}},
  \bibinfo{pages}{120403} (\bibinfo{year}{2004}).

\bibitem[{\citenamefont{Feshbach}(1958)}]{Feshbach1959}
\bibinfo{author}{\bibfnamefont{H.}~\bibnamefont{Feshbach}},
  \bibinfo{journal}{Ann. Phys. (N.Y.)} \textbf{\bibinfo{volume}{5}},
  \bibinfo{pages}{357} (\bibinfo{year}{1958}).

\bibitem[{\citenamefont{Tiesinga et~al.}(1993)\citenamefont{Tiesinga, Verhaar,
  and Stoof}}]{StoofFeshbach}
\bibinfo{author}{\bibfnamefont{E.}~\bibnamefont{Tiesinga}},
  \bibinfo{author}{\bibfnamefont{B.~J.} \bibnamefont{Verhaar}},
  \bibnamefont{and} \bibinfo{author}{\bibfnamefont{H.~T.~C.}
  \bibnamefont{Stoof}}, \bibinfo{journal}{Phys. Rev. A.}
  \textbf{\bibinfo{volume}{47}}, \bibinfo{pages}{4114} (\bibinfo{year}{1993}).

\bibitem[{\citenamefont{Timmermans et~al.}(1999)\citenamefont{Timmermans,
  Tommasini, Hussein, and Kerman}}]{Timmermans1999}
\bibinfo{author}{\bibfnamefont{E.}~\bibnamefont{Timmermans}},
  \bibinfo{author}{\bibfnamefont{P.}~\bibnamefont{Tommasini}},
  \bibinfo{author}{\bibfnamefont{M.}~\bibnamefont{Hussein}}, \bibnamefont{and}
  \bibinfo{author}{\bibfnamefont{A.}~\bibnamefont{Kerman}},
  \bibinfo{journal}{Physics Reports} \textbf{\bibinfo{volume}{315}},
  \bibinfo{pages}{199} (\bibinfo{year}{1999}).

\bibitem[{sim()}]{simpleFR}
\bibinfo{note}{Here, for simplicity we use a highly oversimplied but
  qualitatively correct FR model in which a coupled multi-channel system is
  appproximated by two (nearly degenerate and therefore dominant) channels.}

\bibitem[{com({\natexlab{b}})}]{commentHF}
\bibinfo{note}{The spin-triplet and singlet channels are coupled by the
  hyperfine interaction corresponding to a singlet-triplet transition via
  electronic spin flip accompanied by a nuclear spin flip, such that the total
  spin remains unchanged.}

\bibitem[{\citenamefont{Inouye et~al.}(1998)\citenamefont{Inouye, Andrews,
  Stenger, Miesner, Stamper-Kurn, and Ketterle}}]{Ketterle1998}
\bibinfo{author}{\bibfnamefont{S.}~\bibnamefont{Inouye}},
  \bibinfo{author}{\bibfnamefont{M.~R.} \bibnamefont{Andrews}},
  \bibinfo{author}{\bibfnamefont{J.}~\bibnamefont{Stenger}},
  \bibinfo{author}{\bibfnamefont{H.-J.} \bibnamefont{Miesner}},
  \bibinfo{author}{\bibfnamefont{D.~M.} \bibnamefont{Stamper-Kurn}},
  \bibnamefont{and} \bibinfo{author}{\bibfnamefont{W.}~\bibnamefont{Ketterle}},
  \bibinfo{journal}{Nature} \textbf{\bibinfo{volume}{392}},
  \bibinfo{pages}{151} (\bibinfo{year}{1998}).

\bibitem[{\citenamefont{Moerdijk et~al.}(1995)\citenamefont{Moerdijk, Verhaar,
  and Axelsson}}]{Moerdijk1995}
\bibinfo{author}{\bibfnamefont{A.~J.} \bibnamefont{Moerdijk}},
  \bibinfo{author}{\bibfnamefont{B.~J.} \bibnamefont{Verhaar}},
  \bibnamefont{and} \bibinfo{author}{\bibfnamefont{A.}~\bibnamefont{Axelsson}},
  \bibinfo{journal}{Phys. Rev. A} \textbf{\bibinfo{volume}{51}},
  \bibinfo{pages}{4852} (\bibinfo{year}{1995}).

\bibitem[{com({\natexlab{c}})}]{commentExpWidth}
\bibinfo{note}{Unfortunately, experimentalists define the Feshbach resonance
  width $\Gamma_{exp}$ as the detuning window $\sim \mu_B B_w$ over which the
  resonant scattering length $a(B)$ exceeds the {\em background, nonresonant}
  scattering length $a_{bg}$ (which is natural for the experiments done in the
  dilute two-atom scattering limit). The latter being on the order of the
  interatomic potential, of the order 10s of Angstroms, therefore gives a $\sim
  1/(n^{1/3}a_{bg})\sim 100-1000$ larger $\gamma_{exp}\equiv\mu_B
  B/\epsilon_F$, than $\gamma_s=\sqrt{\Gamma_0/\epsilon_F}\sim(k_f
  a_{bg})(\mu_B B/\epsilon_F)\ll 1$ criterion relevant for the validity of a
  perturbative treatment of the condensed many-body system, see
  Sec.~\ref{s-wave2CMparameter}.}

\bibitem[{\citenamefont{Fano}(1961)}]{Fano1961}
\bibinfo{author}{\bibfnamefont{U.}~\bibnamefont{Fano}}, \bibinfo{journal}{Phys.
  Rev.} \textbf{\bibinfo{volume}{124}}, \bibinfo{pages}{1866}
  (\bibinfo{year}{1961}).

\bibitem[{\citenamefont{Donley et~al.}(2001)\citenamefont{Donley, Claussen,
  Cornish, Roberts, Cornell, and Wieman}}]{Wieman2001}
\bibinfo{author}{\bibfnamefont{E.}~\bibnamefont{Donley}},
  \bibinfo{author}{\bibfnamefont{N.}~\bibnamefont{Claussen}},
  \bibinfo{author}{\bibfnamefont{S.}~\bibnamefont{Cornish}},
  \bibinfo{author}{\bibfnamefont{J.}~\bibnamefont{Roberts}},
  \bibinfo{author}{\bibfnamefont{E.}~\bibnamefont{Cornell}}, \bibnamefont{and}
  \bibinfo{author}{\bibfnamefont{C.}~\bibnamefont{Wieman}},
  \bibinfo{journal}{Nature} \textbf{\bibinfo{volume}{412}},
  \bibinfo{pages}{295} (\bibinfo{year}{2001}).

\bibitem[{\citenamefont{Barankov and Levitov}(2004)}]{Levitov2004}
\bibinfo{author}{\bibfnamefont{R.~A.} \bibnamefont{Barankov}} \bibnamefont{and}
  \bibinfo{author}{\bibfnamefont{L.~S.} \bibnamefont{Levitov}},
  \bibinfo{journal}{Phys. Rev. Lett.} \textbf{\bibinfo{volume}{93}},
  \bibinfo{pages}{130403} (\bibinfo{year}{2004}).

\bibitem[{\citenamefont{Andreev et~al.}(2004)\citenamefont{Andreev, Gurarie,
  and Radzihovsky}}]{Andreev2004}
\bibinfo{author}{\bibfnamefont{A.~V.} \bibnamefont{Andreev}},
  \bibinfo{author}{\bibfnamefont{V.}~\bibnamefont{Gurarie}}, \bibnamefont{and}
  \bibinfo{author}{\bibfnamefont{L.}~\bibnamefont{Radzihovsky}},
  \bibinfo{journal}{Phys. Rev. Lett.} \textbf{\bibinfo{volume}{93}},
  \bibinfo{pages}{130402} (\bibinfo{year}{2004}).

\bibitem[{Res()}]{Resonance}
\bibinfo{note}{We emphasize the distinction between a {\sl resonance}, a long
  lived quasistationary state which eventually decays into the continuum (as
  used in particle physics where it describes an unstable particle) and the
  notion of {\sl resonant scattering} due to an intermediate state coming into
  resonance (coincident in energy) with a scattering state (a terminology
  popular in atomic physics). Namely, some resonant scatterings do {\em not}
  exhibit a resonance. For example, as can be seen in Fig.~\ref{Fig-polesEr},
  $s$-wave Feshbach resonance often occurs in the absence of any {\sl
  resonances}, quasistationary states (corresponding to a pole of a scattering
  amplitude with a negative imaginary part with a magnitude much smaller than
  its positive real part), as is the case in experiments on $s$-wave {\sl wide}
  Feshbach resonances, which take place in the presence of either bound states
  or virtual bound states, but not a quasistationary state. In contrast, narrow
  Feshbach resonances studied in this paper do exhibit a {\sl resonance}. It is
  somewhat unfortunate that these two distinct notions are referred to with
  similar names. In the absence of better terminology, we will use these terms
  but will try to be as clear as possible which usage we mean.}

\bibitem[{\citenamefont{Sheehy and
  Radzihovsky}(2006{\natexlab{a}})}]{SheehyDecouple}
\bibinfo{author}{\bibfnamefont{D.}~\bibnamefont{Sheehy}} \bibnamefont{and}
  \bibinfo{author}{\bibfnamefont{L.}~\bibnamefont{Radzihovsky}},
  \bibinfo{journal}{Phys. Rev. Lett.} \textbf{\bibinfo{volume}{96}},
  \bibinfo{pages}{060401} (\bibinfo{year}{2006}{\natexlab{a}}).

\bibitem[{Eps()}]{EpsLargeN}
\bibinfo{note}{Recently, a number of interesting studies have appeared. Guided
  by success in critical phenomena, these introduce a small parameter
  ($\epsilon=d-2$, $\epsilon=4-d$ or $1/N$, where $d$ is dimension of space and
  $N$ a number of fermion flavors) into a generalization of a single-channel
  model and can thereby treat a full crossover (including interesting unitary
  point) of even a broad resonance. See for
  example~\cite{Son2006a,Son2006,Sachdev2006,Veillette2006}.}

\bibitem[{\citenamefont{{Holland {\it et al.}}}(2001)}]{Holland2001}
\bibinfo{author}{\bibfnamefont{M.}~\bibnamefont{{Holland {\it et al.}}}},
  \bibinfo{journal}{Phys. Rev. Lett.} \textbf{\bibinfo{volume}{87}},
  \bibinfo{pages}{120406} (\bibinfo{year}{2001}).

\bibitem[{\citenamefont{Ohashi and Griffin}(2002)}]{Griffin2002}
\bibinfo{author}{\bibfnamefont{Y.}~\bibnamefont{Ohashi}} \bibnamefont{and}
  \bibinfo{author}{\bibfnamefont{A.}~\bibnamefont{Griffin}},
  \bibinfo{journal}{Phys. Rev. Lett.} \textbf{\bibinfo{volume}{89}},
  \bibinfo{pages}{130402} (\bibinfo{year}{2002}).

\bibitem[{\citenamefont{Bulgac et~al.}(2006)\citenamefont{Bulgac, Drut, and
  Magierski}}]{Bulgac2006}
\bibinfo{author}{\bibfnamefont{A.}~\bibnamefont{Bulgac}},
  \bibinfo{author}{\bibfnamefont{J.~E.} \bibnamefont{Drut}}, \bibnamefont{and}
  \bibinfo{author}{\bibfnamefont{P.}~\bibnamefont{Magierski}},
  \bibinfo{journal}{Phys. Rev. Lett.} \textbf{\bibinfo{volume}{96}},
  \bibinfo{pages}{090404} (\bibinfo{year}{2006}).

\bibitem[{\citenamefont{Petrov et~al.}(2005)\citenamefont{Petrov, Salomon, and
  Shlyapnikov}}]{Petrov2005}
\bibinfo{author}{\bibfnamefont{D.}~\bibnamefont{Petrov}},
  \bibinfo{author}{\bibfnamefont{C.}~\bibnamefont{Salomon}}, \bibnamefont{and}
  \bibinfo{author}{\bibfnamefont{G.}~\bibnamefont{Shlyapnikov}},
  \bibinfo{journal}{Phys. Rev. A} \textbf{\bibinfo{volume}{71}},
  \bibinfo{pages}{012708} (\bibinfo{year}{2005}).

\bibitem[{\citenamefont{Lifshitz and Pitaevskii}(1980)}]{LL9}
\bibinfo{author}{\bibfnamefont{E.~M.} \bibnamefont{Lifshitz}} \bibnamefont{and}
  \bibinfo{author}{\bibfnamefont{L.~P.} \bibnamefont{Pitaevskii}},
  \emph{\bibinfo{title}{Statistical Physics, Part 2}}
  (\bibinfo{publisher}{Butterworth-Heinemann}, \bibinfo{address}{Oxford, UK},
  \bibinfo{year}{1980}).

\bibitem[{\citenamefont{Ticknor et~al.}(2004)\citenamefont{Ticknor, Regal, Jin,
  and Bohn}}]{Ticknor2004}
\bibinfo{author}{\bibfnamefont{C.}~\bibnamefont{Ticknor}},
  \bibinfo{author}{\bibfnamefont{C.~A.} \bibnamefont{Regal}},
  \bibinfo{author}{\bibfnamefont{D.~S.} \bibnamefont{Jin}}, \bibnamefont{and}
  \bibinfo{author}{\bibfnamefont{J.~L.} \bibnamefont{Bohn}},
  \bibinfo{journal}{Phys. Rev. A} \textbf{\bibinfo{volume}{69}},
  \bibinfo{pages}{042712} (\bibinfo{year}{2004}).

\bibitem[{\citenamefont{Schunck et~al.}(2005)\citenamefont{Schunck, Zwierlein,
  Stan, Raupach, Ketterle, Simoni, Tiesinga, Williamsa, and
  Julienne}}]{Schunck2005}
\bibinfo{author}{\bibfnamefont{C.~H.} \bibnamefont{Schunck}},
  \bibinfo{author}{\bibfnamefont{M.~W.} \bibnamefont{Zwierlein}},
  \bibinfo{author}{\bibfnamefont{C.~A.} \bibnamefont{Stan}},
  \bibinfo{author}{\bibfnamefont{S.~M.~F.} \bibnamefont{Raupach}},
  \bibinfo{author}{\bibfnamefont{W.}~\bibnamefont{Ketterle}},
  \bibinfo{author}{\bibfnamefont{A.}~\bibnamefont{Simoni}},
  \bibinfo{author}{\bibfnamefont{E.}~\bibnamefont{Tiesinga}},
  \bibinfo{author}{\bibfnamefont{C.~J.} \bibnamefont{Williamsa}},
  \bibnamefont{and} \bibinfo{author}{\bibfnamefont{P.~S.}
  \bibnamefont{Julienne}}, \bibinfo{journal}{Phys. Rev. A}
  \textbf{\bibinfo{volume}{71}}, \bibinfo{pages}{045601}
  (\bibinfo{year}{2005}).

\bibitem[{\citenamefont{Ho and Diener}(2005)}]{Ho2005}
\bibinfo{author}{\bibfnamefont{T.-L.} \bibnamefont{Ho}} \bibnamefont{and}
  \bibinfo{author}{\bibfnamefont{R.}~\bibnamefont{Diener}},
  \bibinfo{journal}{Phys. Rev. Lett.} \textbf{\bibinfo{volume}{94}},
  \bibinfo{pages}{090402} (\bibinfo{year}{2005}).

\bibitem[{\citenamefont{Ohashi}(2005)}]{Ohashi2005}
\bibinfo{author}{\bibfnamefont{Y.}~\bibnamefont{Ohashi}},
  \bibinfo{journal}{Phys. Rev. Lett.} \textbf{\bibinfo{volume}{94}},
  \bibinfo{pages}{090403} (\bibinfo{year}{2005}).

\bibitem[{\citenamefont{Botelho and {S\'a de Melo}}(2005)}]{Botelho2005}
\bibinfo{author}{\bibfnamefont{S.~S.} \bibnamefont{Botelho}} \bibnamefont{and}
  \bibinfo{author}{\bibfnamefont{C.~A.~R.} \bibnamefont{{S\'a de Melo}}},
  \bibinfo{journal}{J. Low Temp Phys} \textbf{\bibinfo{volume}{140}},
  \bibinfo{pages}{409} (\bibinfo{year}{2005}).

\bibitem[{\citenamefont{Gurarie et~al.}(2005)\citenamefont{Gurarie,
  Radzihovsky, and Andreev}}]{Gurarie2005}
\bibinfo{author}{\bibfnamefont{V.}~\bibnamefont{Gurarie}},
  \bibinfo{author}{\bibfnamefont{L.}~\bibnamefont{Radzihovsky}},
  \bibnamefont{and} \bibinfo{author}{\bibfnamefont{A.~V.}
  \bibnamefont{Andreev}}, \bibinfo{journal}{Phys. Rev. Lett.}
  \textbf{\bibinfo{volume}{94}}, \bibinfo{pages}{230403}
  (\bibinfo{year}{2005}).

\bibitem[{\citenamefont{Cheng and Yip}(2005)}]{Yip2005}
\bibinfo{author}{\bibfnamefont{C.-H.} \bibnamefont{Cheng}} \bibnamefont{and}
  \bibinfo{author}{\bibfnamefont{S.-K.} \bibnamefont{Yip}},
  \bibinfo{journal}{Phys. Rev. Lett.} \textbf{\bibinfo{volume}{95}},
  \bibinfo{pages}{070404} (\bibinfo{year}{2005}).

\bibitem[{com({\natexlab{d}})}]{commentSvanish}
\bibinfo{note}{In the Schr\"odinger's equation two-particle formulation the
  vanishing of the $s$-wave scattering is due to destructive interference
  (cancellation) between scattering by $\theta$ and $\pi-\theta$. In the
  many-body language, as can be seen from spin and orbital channel
  decomposition of Sec.~\ref{Pscattering}, (see e.g., \rfs{eq:decomp}) this
  happens automatically because identical fermions can be considered to be in
  the flavor-triplet state $|\uparrow,\uparrow\rangle$ which therefore requires
  the orbital part to be antisymmetric, in particularly forbidding the $s$-wave
  channel interaction.}

\bibitem[{c2l()}]{c2largeComment}
\bibinfo{note}{As is clear from the analysis of
  Secs.\ref{Sec:PWaveTwoChannel},\ref{PWaveChapter}, in a perturbative study of
  a $p$-wave model, in addition to the dimensionless parameter $\gamma_p$,
  \rfs{gamma_p}, another dimensionless parameter $c_2$, \rfs{eq:c2} appears
  both in the scattering theory and in the analysis of the finite density gas.
  Since $c_2 \gg \gamma_p$, clearly for $p$-wave mean-field theory to be
  accurate $c_2$ must be small (which automatically ensures that $\gamma_p\ll
  1$). We are grateful to Y. Castin and collaborators for bringing this to our
  attention~\cite{Castin2006}. However, we observe that the dominant series of
  diagrams that dominate for a large $c_2$ are the same as those already
  appearing the two-body scattering problem of Sec.~\ref{TwoCM}. Consequently,
  we expect that once these diagrams are properly resummed, i.e., the large
  $c_2$ {\em two-body molecular} problem is solved, \cite{Levinsen2006} the
  many-body problem should be solvable as a saddle-point of the resulting
  nontrivial effective two-body action. This is analogous to the $s$-wave case,
  where away from the BCS regime, for large $\gamma_s$ the two-body molecular
  problem is extremely nontrivial but exactly solvable and once solved allows
  for a controlled treatment of the deep BEC regime where the gas parameter is
  small~\cite{Levinsen2006}.}

\bibitem[{\citenamefont{Klinkhamer and Volovik}(2004)}]{Volovik2004}
\bibinfo{author}{\bibfnamefont{F.~R.} \bibnamefont{Klinkhamer}}
  \bibnamefont{and} \bibinfo{author}{\bibfnamefont{G.~E.}
  \bibnamefont{Volovik}}, \bibinfo{journal}{Pisma Zh. Eksp. Teor. Fiz.}
  \textbf{\bibinfo{volume}{80}}, \bibinfo{pages}{389} (\bibinfo{year}{2004}).

\bibitem[{\citenamefont{Kitaev}(2003)}]{Kitaev2003}
\bibinfo{author}{\bibfnamefont{A.}~\bibnamefont{Kitaev}},
  \bibinfo{journal}{Ann. Phys.} \textbf{\bibinfo{volume}{303}},
  \bibinfo{pages}{2} (\bibinfo{year}{2003}).

\bibitem[{\citenamefont{Holland and Kokkelmans}(2002)}]{Holland2002}
\bibinfo{author}{\bibfnamefont{M.}~\bibnamefont{Holland}} \bibnamefont{and}
  \bibinfo{author}{\bibfnamefont{S.~J. J. M.~F.} \bibnamefont{Kokkelmans}},
  \bibinfo{journal}{Phys. Rev. Lett.} \textbf{\bibinfo{volume}{89}},
  \bibinfo{pages}{180401} (\bibinfo{year}{2002}).

\bibitem[{\citenamefont{Landau and Lifshitz}(1981)}]{LL}
\bibinfo{author}{\bibfnamefont{L.~D.} \bibnamefont{Landau}} \bibnamefont{and}
  \bibinfo{author}{\bibfnamefont{E.~M.} \bibnamefont{Lifshitz}},
  \emph{\bibinfo{title}{Quantum Mechanics}}
  (\bibinfo{publisher}{Butterworth-Heinemann}, \bibinfo{address}{Oxford, UK},
  \bibinfo{year}{1981}).

\bibitem[{\citenamefont{Levitov and Barankov}(2004)}]{LevitovUnpublished}
\bibinfo{author}{\bibfnamefont{L.}~\bibnamefont{Levitov}} \bibnamefont{and}
  \bibinfo{author}{\bibfnamefont{R.}~\bibnamefont{Barankov}},
  \bibinfo{journal}{unpublished}  (\bibinfo{year}{2004}).

\bibitem[{\citenamefont{Andersen}(2004)}]{Andersen2004}
\bibinfo{author}{\bibfnamefont{J.~O.} \bibnamefont{Andersen}},
  \bibinfo{journal}{Rev. Mod. Phys.} \textbf{\bibinfo{volume}{76}},
  \bibinfo{pages}{599} (\bibinfo{year}{2004}).

\bibitem[{\citenamefont{Anderson and Morel}(1961)}]{Anderson1961}
\bibinfo{author}{\bibfnamefont{P.~W.} \bibnamefont{Anderson}} \bibnamefont{and}
  \bibinfo{author}{\bibfnamefont{P.}~\bibnamefont{Morel}},
  \bibinfo{journal}{Phys. Rev.} \textbf{\bibinfo{volume}{123}},
  \bibinfo{pages}{1911} (\bibinfo{year}{1961}).

\bibitem[{com({\natexlab{e}})}]{commentYip}
\bibinfo{note}{The prediction of the $p_x$- to $p_x+i p_y$-superfluid phase
  transition was first made in our original manuscript cond-mat/0410620v1.
  However, in the original version of this paper, for the intermediate regime
  of dipolar splitting $\delta$ {\em only}, we made an error that reversed the
  two phases, a mistake that was subsequently corrected by C.-H. Cheng and
  S.-K. Yip, cond-mat/0504278, Ref.~\cite{Yip2005}.}

\bibitem[{\citenamefont{Volovik}(2006)}]{VolovikReview}
\bibinfo{author}{\bibfnamefont{G.}~\bibnamefont{Volovik}}
  (\bibinfo{year}{2006}), \eprint{cond-mat/0601372}.

\bibitem[{\citenamefont{Gurarie and Radzihovsky}()}]{Gurarie2006}
\bibinfo{author}{\bibfnamefont{V.}~\bibnamefont{Gurarie}} \bibnamefont{and}
  \bibinfo{author}{\bibfnamefont{L.}~\bibnamefont{Radzihovsky}},
  \eprint{cond-mat/0610094}.

\bibitem[{\citenamefont{Tewari et~al.}(2006{\natexlab{a}})\citenamefont{Tewari,
  Sarma, Nayak, Zhang, and Zoller}}]{Nayak2006a}
\bibinfo{author}{\bibfnamefont{S.}~\bibnamefont{Tewari}},
  \bibinfo{author}{\bibfnamefont{S.~D.} \bibnamefont{Sarma}},
  \bibinfo{author}{\bibfnamefont{C.}~\bibnamefont{Nayak}},
  \bibinfo{author}{\bibfnamefont{C.}~\bibnamefont{Zhang}}, \bibnamefont{and}
  \bibinfo{author}{\bibfnamefont{P.}~\bibnamefont{Zoller}}
  (\bibinfo{year}{2006}{\natexlab{a}}), \eprint{cond-mat/0606101}.

\bibitem[{com({\natexlab{f}})}]{commentVanish_delta}
\bibinfo{note}{It is easy to see that the $s$-wave scattering vanishes
  identically for strict repulsive delta-function interactions. Indeed,
  scattering off an infinite wall potential of radius $r_0$ leads to $a=r_0$.
  Taking the limit $r_0 \rightarrow 0$, to mimic delta-function, gives $a=0$,
  i.e. no scattering.}

\bibitem[{\citenamefont{Landau}(1956)}]{LandauFL}
\bibinfo{author}{\bibfnamefont{L.}~\bibnamefont{Landau}},
  \bibinfo{journal}{Sov. Phys. ZhETP} \textbf{\bibinfo{volume}{30}},
  \bibinfo{pages}{1058} (\bibinfo{year}{1956}).

\bibitem[{\citenamefont{Brodsky et~al.}(2005)\citenamefont{Brodsky, Klaptsov,
  Kagan, Combescot, and Leyronas}}]{Brodsky2005}
\bibinfo{author}{\bibfnamefont{I.~V.} \bibnamefont{Brodsky}},
  \bibinfo{author}{\bibfnamefont{A.~V.} \bibnamefont{Klaptsov}},
  \bibinfo{author}{\bibfnamefont{M.~Y.} \bibnamefont{Kagan}},
  \bibinfo{author}{\bibfnamefont{R.}~\bibnamefont{Combescot}},
  \bibnamefont{and} \bibinfo{author}{\bibfnamefont{X.}~\bibnamefont{Leyronas}},
  \bibinfo{journal}{JETP Lett.} \textbf{\bibinfo{volume}{82}},
  \bibinfo{pages}{273} (\bibinfo{year}{2005}).

\bibitem[{\citenamefont{Castin}()}]{Castin2006}
\bibinfo{author}{\bibfnamefont{Y.}~\bibnamefont{Castin}},
  \bibinfo{note}{private communication}.

\bibitem[{\citenamefont{Hugenholtz and Pines}(1959)}]{Hugenholtz1959}
\bibinfo{author}{\bibfnamefont{N.~M.} \bibnamefont{Hugenholtz}}
  \bibnamefont{and} \bibinfo{author}{\bibfnamefont{D.}~\bibnamefont{Pines}},
  \bibinfo{journal}{Phys. Rev.} \textbf{\bibinfo{volume}{116}},
  \bibinfo{pages}{489} (\bibinfo{year}{1959}).

\bibitem[{\citenamefont{Nambu}(1960)}]{Nambu1960}
\bibinfo{author}{\bibfnamefont{Y.}~\bibnamefont{Nambu}},
  \bibinfo{journal}{Phys. Rev.} \textbf{\bibinfo{volume}{117}},
  \bibinfo{pages}{648} (\bibinfo{year}{1960}).

\bibitem[{\citenamefont{Bogoliubov et~al.}(1958)\citenamefont{Bogoliubov,
  Tolmachev, and Shirkov}}]{Bogoliubov}
\bibinfo{author}{\bibfnamefont{N.~N.} \bibnamefont{Bogoliubov}},
  \bibinfo{author}{\bibfnamefont{V.~V.} \bibnamefont{Tolmachev}},
  \bibnamefont{and} \bibinfo{author}{\bibfnamefont{D.~V.}
  \bibnamefont{Shirkov}}, \emph{\bibinfo{title}{A New Method in the Theory of
  Superconductivity}} (\bibinfo{publisher}{Academy of Sciences, USSR},
  \bibinfo{address}{Moscow}, \bibinfo{year}{1958}).

\bibitem[{pwa()}]{pwaveNames}
\bibinfo{note}{Throughout the paper, we use a nomenclature ``$p_x$-wave'' to
  refer to a $p$-wave superfluid with a vanishing projection of angular
  momentum along an axis (not necessarily just $x$), that is more generally
  characterized by $\bu,\bv$ with one (but not both) of them vanishing, or
  equivalently $\bu||\bv$. Also we generically refer to a superfluid with a
  definite projection of angular momentum $\pm 1$ along an axis as a ``$p_x+i
  p_y$-wave'' superfluid, that is also characterized by nonparallel $\bu$ and
  $\bv$. Finally, for anisotropic resonance, a state that breaks time-reversal
  symmetry is characterized by $u\neq v>0$ and therefore is an anisotropic $p_x
  + i \alpha p_y$ state that is a linear combination of a pure $p_x + i p_y$
  and a $p_x$ states. Nevertheless, for simplicity we will refer to it as a
  $p_x+i p_y$-superfluid.}

\bibitem[{\citenamefont{Sheehy and
  Radzihovsky}(2006{\natexlab{b}})}]{Sheehy2006a}
\bibinfo{author}{\bibfnamefont{D.}~\bibnamefont{Sheehy}} \bibnamefont{and}
  \bibinfo{author}{\bibfnamefont{L.}~\bibnamefont{Radzihovsky}}
  (\bibinfo{year}{2006}{\natexlab{b}}), \eprint{cond-mat/0607803}.

\bibitem[{\citenamefont{Zinn-Jistin}(2002)}]{ZinnJustin}
\bibinfo{author}{\bibfnamefont{J.}~\bibnamefont{Zinn-Jistin}},
  \emph{\bibinfo{title}{Quantum Field Theory and Critical Phenomena}}
  (\bibinfo{publisher}{Oxford University Press}, \bibinfo{address}{Oxford, UK},
  \bibinfo{year}{2002}).

\bibitem[{\citenamefont{Calabrese et~al.}(2004)\citenamefont{Calabrese,
  Parruccini, Pelissetto, and Vicari}}]{Vicari2004}
\bibinfo{author}{\bibfnamefont{P.}~\bibnamefont{Calabrese}},
  \bibinfo{author}{\bibfnamefont{P.}~\bibnamefont{Parruccini}},
  \bibinfo{author}{\bibfnamefont{A.}~\bibnamefont{Pelissetto}},
  \bibnamefont{and} \bibinfo{author}{\bibfnamefont{E.}~\bibnamefont{Vicari}},
  \bibinfo{journal}{Phys. Rev. B} \textbf{\bibinfo{volume}{70}},
  \bibinfo{pages}{174439} (\bibinfo{year}{2004}).

\bibitem[{\citenamefont{Chaikin and Lubensky}(2000)}]{ChaikinLubensky}
\bibinfo{author}{\bibfnamefont{P.~M.} \bibnamefont{Chaikin}} \bibnamefont{and}
  \bibinfo{author}{\bibfnamefont{T.~C.} \bibnamefont{Lubensky}},
  \emph{\bibinfo{title}{Principles of Condensed Matter Physics}}
  (\bibinfo{publisher}{Cambridge University Press},
  \bibinfo{address}{Cambridge, UK}, \bibinfo{year}{2000}).

\bibitem[{\citenamefont{Moore and Read}(1991)}]{MR}
\bibinfo{author}{\bibfnamefont{G.}~\bibnamefont{Moore}} \bibnamefont{and}
  \bibinfo{author}{\bibfnamefont{N.}~\bibnamefont{Read}},
  \bibinfo{journal}{Nucl. Phys. B} \textbf{\bibinfo{volume}{360}},
  \bibinfo{pages}{362} (\bibinfo{year}{1991}).

\bibitem[{\citenamefont{Wilkin and Gunn}(2000)}]{Gunn2000}
\bibinfo{author}{\bibfnamefont{N.~K.} \bibnamefont{Wilkin}} \bibnamefont{and}
  \bibinfo{author}{\bibfnamefont{J.~M.~F.} \bibnamefont{Gunn}},
  \bibinfo{journal}{Phys. Rev. Lett.} \textbf{\bibinfo{volume}{84}},
  \bibinfo{pages}{6} (\bibinfo{year}{2000}).

\bibitem[{\citenamefont{Cooper et~al.}(2001)\citenamefont{Cooper, Wilkin, and
  Gunn}}]{Gunn2001}
\bibinfo{author}{\bibfnamefont{N.~R.} \bibnamefont{Cooper}},
  \bibinfo{author}{\bibfnamefont{N.~K.} \bibnamefont{Wilkin}},
  \bibnamefont{and} \bibinfo{author}{\bibfnamefont{J.~M.~F.}
  \bibnamefont{Gunn}}, \bibinfo{journal}{Phys. Rev. Lett.}
  \textbf{\bibinfo{volume}{87}}, \bibinfo{pages}{120405}
  (\bibinfo{year}{2001}).

\bibitem[{\citenamefont{Schweikhard et~al.}(2004)\citenamefont{Schweikhard,
  Coddington, Engels, Mogendorff, and Cornell}}]{Cornell2004}
\bibinfo{author}{\bibfnamefont{V.}~\bibnamefont{Schweikhard}},
  \bibinfo{author}{\bibfnamefont{I.}~\bibnamefont{Coddington}},
  \bibinfo{author}{\bibfnamefont{P.}~\bibnamefont{Engels}},
  \bibinfo{author}{\bibfnamefont{V.~P.} \bibnamefont{Mogendorff}},
  \bibnamefont{and} \bibinfo{author}{\bibfnamefont{E.~A.}
  \bibnamefont{Cornell}}, \bibinfo{journal}{Phys. Rev. Lett.}
  \textbf{\bibinfo{volume}{92}}, \bibinfo{pages}{040404}
  (\bibinfo{year}{2004}).

\bibitem[{\citenamefont{Altland and Zirnbauer}(1997)}]{Altland1997}
\bibinfo{author}{\bibfnamefont{A.}~\bibnamefont{Altland}} \bibnamefont{and}
  \bibinfo{author}{\bibfnamefont{M.}~\bibnamefont{Zirnbauer}},
  \bibinfo{journal}{Phys. Rev. B} \textbf{\bibinfo{volume}{55}},
  \bibinfo{pages}{1142} (\bibinfo{year}{1997}).

\bibitem[{\citenamefont{Zirnbauer}()}]{ZirnbauerPrivate}
\bibinfo{author}{\bibfnamefont{M.}~\bibnamefont{Zirnbauer}},
  \bibinfo{note}{private communication}.

\bibitem[{\citenamefont{Read}()}]{NickReadUnpublished}
\bibinfo{author}{\bibfnamefont{N.}~\bibnamefont{Read}},
  \bibinfo{note}{unpublished}.

\bibitem[{\citenamefont{Kopnin and Salomaa}(1991)}]{Kopnin1991}
\bibinfo{author}{\bibfnamefont{N.~B.} \bibnamefont{Kopnin}} \bibnamefont{and}
  \bibinfo{author}{\bibfnamefont{M.~M.} \bibnamefont{Salomaa}},
  \bibinfo{journal}{Phys. Rev. B} \textbf{\bibinfo{volume}{44}},
  \bibinfo{pages}{9667} (\bibinfo{year}{1991}).

\bibitem[{\citenamefont{Tewari et~al.}(2006{\natexlab{b}})\citenamefont{Tewari,
  Sarma, and Lee}}]{Tewari2006}
\bibinfo{author}{\bibfnamefont{S.}~\bibnamefont{Tewari}},
  \bibinfo{author}{\bibfnamefont{S.~D.} \bibnamefont{Sarma}}, \bibnamefont{and}
  \bibinfo{author}{\bibfnamefont{D.-H.} \bibnamefont{Lee}}
  (\bibinfo{year}{2006}{\natexlab{b}}), \eprint{cond-mat/0609556}.

\bibitem[{\citenamefont{Ivanov}(2001)}]{Ivanov2001}
\bibinfo{author}{\bibfnamefont{D.~A.} \bibnamefont{Ivanov}},
  \bibinfo{journal}{Phys. Rev. Lett.} \textbf{\bibinfo{volume}{86}},
  \bibinfo{pages}{268} (\bibinfo{year}{2001}).

\bibitem[{\citenamefont{Mahan}(2000)}]{Mahan}
\bibinfo{author}{\bibfnamefont{G.~D.} \bibnamefont{Mahan}},
  \emph{\bibinfo{title}{Many Particle Physics}}
  (\bibinfo{publisher}{Springer-Verlag}, \bibinfo{address}{Berlin, Germany},
  \bibinfo{year}{2000}).

\bibitem[{\citenamefont{Gurarie}(2006)}]{Gurarie2006a}
\bibinfo{author}{\bibfnamefont{V.}~\bibnamefont{Gurarie}},
  \bibinfo{journal}{Phys. Rev. A} \textbf{\bibinfo{volume}{73}},
  \bibinfo{pages}{033612} (\bibinfo{year}{2006}).

\bibitem[{\citenamefont{Nishida and Son}(2006)}]{Son2006a}
\bibinfo{author}{\bibfnamefont{Y.}~\bibnamefont{Nishida}} \bibnamefont{and}
  \bibinfo{author}{\bibfnamefont{D.~T.} \bibnamefont{Son}},
  \bibinfo{journal}{Phys. Rev. Lett.} \textbf{\bibinfo{volume}{97}},
  \bibinfo{pages}{050403} (\bibinfo{year}{2006}).

\bibitem[{\citenamefont{Arnold et~al.}()\citenamefont{Arnold, Drut, and
  Son}}]{Son2006}
\bibinfo{author}{\bibfnamefont{P.}~\bibnamefont{Arnold}},
  \bibinfo{author}{\bibfnamefont{J.~E.} \bibnamefont{Drut}}, \bibnamefont{and}
  \bibinfo{author}{\bibfnamefont{D.~T.} \bibnamefont{Son}},
  \eprint{cond-mat/0608477}.

\bibitem[{\citenamefont{Nikoli\'c and Sachdev}()}]{Sachdev2006}
\bibinfo{author}{\bibfnamefont{P.}~\bibnamefont{Nikoli\'c}} \bibnamefont{and}
  \bibinfo{author}{\bibfnamefont{S.}~\bibnamefont{Sachdev}},
  \eprint{cond-mat/0609106}.

\bibitem[{\citenamefont{Veillette et~al.}()\citenamefont{Veillette, Sheehy, and
  Radzihovsky}}]{Veillette2006}
\bibinfo{author}{\bibfnamefont{M.}~\bibnamefont{Veillette}},
  \bibinfo{author}{\bibfnamefont{D.}~\bibnamefont{Sheehy}}, \bibnamefont{and}
  \bibinfo{author}{\bibfnamefont{L.}~\bibnamefont{Radzihovsky}},
  \eprint{cond-mat/0610798}.

\end{thebibliography}

\end{document}